\pdfoutput=1
\newcommand*{\ATLASLATEXPATH}{latex/}
\documentclass[cernpreprint,texlive=2011,txfonts,UKenglish]{\ATLASLATEXPATH atlasdoc}

\usepackage[subfigure=true]{\ATLASLATEXPATH atlaspackage}
\usepackage{\ATLASLATEXPATH atlasbiblatex}

\usepackage{\ATLASLATEXPATH atlasphysics}

\graphicspath{{logos/}{figures/}}

\usepackage{paper-defs}

\usepackage{pdflscape}
\usepackage{multirow}

\AtlasTitle{Identification of high transverse momentum top quarks in $pp$ collisions at $\sqrt{s} = 8 \TeV$ with the ATLAS detector}

\author{The ATLAS Collaboration}

\PreprintIdNumber{CERN-EP-2016-010}

\AtlasJournal{JHEP}

\AtlasAbstract{%
This paper presents studies of the performance of several jet-substructure techniques,
which are used to identify hadronically decaying top quarks with high transverse
momentum contained in large-radius jets.
The efficiency of identifying top quarks is measured using a sample of top-quark pairs
and the rate of wrongly identifying jets from other quarks or gluons as top quarks
is measured using multijet events
collected with the ATLAS experiment in
$20.3\invfb$ of $8\TeV$ proton--proton collisions at the Large
Hadron Collider. Predictions from Monte Carlo simulations are found
to provide an accurate description of the performance. The techniques are
compared in terms of signal efficiency and background rejection using
simulations, covering a larger range in jet transverse momenta than
accessible in the dataset. Additionally, a novel technique is developed that
is optimized to reconstruct top quarks in events with many jets.
}

\hypersetup{pdftitle={ATLAS document},pdfauthor={The ATLAS Collaboration}}

\begin{document}

\maketitle

\tableofcontents

\clearpage

\section{Introduction}
Conventional top-quark identification methods reconstruct
the products of a hadronic top-quark decay ($t \to bW \to b q'\bar{q}$) as jets with a small radius parameter $R$ (typically
$R=0.4$ or 0.5).\footnote{
  The ATLAS experiment uses a right-handed coordinate system with its origin at the nominal interaction point (IP) in the centre
  of the detector and the $z$-axis along the beam line. The $x$-axis points from the IP to the centre of the LHC ring, and the
  $y$-axis points upwards. Cylindrical coordinates $(r, \phi)$ are used in the transverse plane, $\phi$ being the azimuthal angle
  around the beam line. Observables labelled ``transverse'' are projected into the $x$--$y$ plane. The pseudorapidity is defined
  in terms of the polar angle $\theta$ as $\eta=-\ln\tan \theta/2$. The transverse momentum is defined as $\pt=p\sin\theta=p/\cosh\eta$,
  and the transverse energy $\et$ has an analogous definition.
  The distance in $\eta$--$\phi$ space is referred to as $\Delta R = \sqrt{(\Delta\eta)^2+(\Delta\phi)^2}$.
  The rapidity of a particle is defined as $y = \frac{1}{2} \ln\frac{E + p_z}{E - p_z}$, in which $E$ and $p_z$ are the
  energy and momentum $z$-component of the particle.
  The jet radius parameter $R$ sets the range in $y$--$\phi$ space over which clustering to form jets occurs.
}
There are usually several of these \smallR jets in a high-energy, hard proton--proton ($pp$) collision event
at the Large Hadron Collider (LHC). Hadronic top-quark decays are reconstructed by taking those jets
which, when combined, best fit the kinematic properties of the top-quark decay,
such as the top-quark mass and the $W$-boson mass. These kinematic constraints
may also be fulfilled for a collection of jets which do not all originate
from the same top-quark decay chain.

In analyses of LHC $pp$ collisions,
conventional top-quark identification methods are inefficient at high top-quark energies
because the top-quark decay products are collimated and the probability of resolving separate \smallR jets is reduced.
Top quarks with
high transverse momentum ($\pt \gtrsim 200 \GeV$)
may instead be reconstructed as a jet with
large radius parameter, $R\ge 0.8$ (\largeR
jet)~\cite{Aad:2012raa,Aad:2014xra,Aad:2015fna,Aad:2015pfx,Aad:2015hna,Aad:2015lgx,Chatrchyan:2013wfa,Khachatryan:2015wza,Khachatryan:2015sma,Khachatryan:2015oba,Khachatryan:2015edz,Khachatryan:2015mta,Khachatryan:2016oia}.
An analysis of the internal jet structure is then performed
to identify and reconstruct hadronically decaying top quarks (top tagging).

Since a single jet that contains all of the decay products of a massive particle has different properties from a jet
of the same transverse momentum originating from a light quark or gluon, it is possible to use the substructure of \largeR jets to distinguish
top quarks with high \pt from jets from other sources, for example from multijet production.
These differences in the jet substructure can
be better resolved after contributions from soft gluon radiation or
from additional \pp interactions in the same or adjacent bunch crossings (pile-up) are removed from the jets.
Such methods are referred to as \textit{jet grooming} and consist of either an adaptive modification of the jet algorithm or a selective
removal of soft radiation during the process of iterative recombination in jet reconstruction~\cite{Butterworth:2008iy,Ellis:2009su,Krohn:2009th}.

The jet-substructure approach aims to reduce combinatorial background
from assigning \smallR jets to top-quark candidates
in order to achieve a more precise reconstruction of the top-quark four-momentum
and a higher background rejection. In searches for
top--anti-top quark (\ttbar) resonances, the improved kinematic reconstruction
leads to a better mass resolution for large resonance masses ($\ge 1\TeV$) compared to
the conventional approach, resulting in an increased sensitivity to physics beyond
the Standard Model (SM)~\cite{Aad:2013nca}.

ATLAS has published performance studies of jet-substructure methods for top tagging
at a $pp$ centre-of-mass energy of $\sqrt{s}=7\TeV$~\cite{Aad:2013gja}.
In the paper presented here, the performance of several approaches to top
tagging at $\sqrt{s}=8\TeV$ is documented.
Top tagging based on the combination of jet-substructure variables,
\sd~\cite{Soper:2011cr,Soper:2012pb},
and the \htt~\cite{Plehn:2009rk,Plehn:2010st}
is studied, as described in \secref{techniques}.
A new method, \httofour, is introduced.
Optimised for top tagging in events with many jets, it uses a preselection of \smallR jets as
input to the \htt algorithm.

Monte-Carlo (MC) simulation is used to compare
the efficiencies and misidentification rates of all approaches over a large
kinematic range.
The performance of the different methods is studied in data using two different
event samples: a signal sample
enriched with top quarks and a background sample dominated by multijet production.
The signal sample is used to measure top-tagging efficiencies from data, which are
compared to the predictions obtained from MC simulations.
Quantifying the degree to which MC simulations
correctly model the top-tagging efficiency observed in data
is crucial for any physics analysis in which top-tagging methods
are used because MC simulations are commonly used to model signal and
background processes.
The signal sample is also used to determine the energy
scale of subjets in situ from the reconstructed top-quark mass distribution.
Top-tagging misidentification rates are measured in the background sample
and are also compared to the prediction of MC simulations.

\section{The ATLAS detector}
\label{sec:exp_setup}

The ATLAS detector consists of an inner tracking detector system (ID),
which is surrounded by electromagnetic (EM) and hadronic calorimeters and
a muon spectrometer (MS).
The ID consists of silicon pixel and strip detectors
and a transition-radiation tracker covering $|\eta|<2.5$, and
it is immersed in a $2\;\mathrm{T}$ axial magnetic field.
The EM calorimeters use lead/liquid argon (LAr) technology to provide calorimetry
for $|\eta|<3.2$, with
copper/LAr used in the forward region $3.1 < |\eta| < 4.9$.
In the region $|\eta|<1.7$, hadron calorimetry is provided by steel/scintillator
calorimeters.
In the forward region, copper/LAr and tungsten/LAr calorimeters are used for
$1.5 < |\eta| < 3.2$ and $3.1 < |\eta| < 4.9$, respectively.
The MS surrounds the calorimeter system and consists of multiple layers of trigger and
tracking chambers within a toroidal magnetic field generated by air-core superconducting magnets, which
allows for the measurement of muon momenta for $|\eta| < 2.7$.
ATLAS uses a three-level trigger system~\cite{Aad:2012xs}
with a hardware-based first-level trigger, which is
followed by two software-based trigger levels with an increasingly fine-grained
selection of events at lower rates.
A detailed description of the ATLAS detector is given in
Ref.~\cite{PERF-2007-01}.

\section{Monte-Carlo simulation}
\label{sec:mc}

MC simulations are used to model different
SM contributions to the signal and background samples.
They are also used
to study and compare the performance of top-tagging algorithms
over a larger kinematic range than accessible in the data samples.

Top-quark pair production is simulated with \PowhegBox
r2330.3~\cite{Frixione:2007nw,Nason:2004rx,Frixione:2007vw,Alioli:2010xd}
interfaced with \Pythia v6.426~\cite{Sjostrand:2006za} with the set of tuned parameters (tune)
Perugia 2011C~\cite{Skands:2010ak} and the CT10~\cite{Lai:2010vv}
set of parton distribution functions (PDFs).
The \hdamp parameter, which effectively regulates the high-\pT gluon radiation in
\Powheg, is left at the default value of $\hdamp = \infty$.
This MC sample is referred to as the
\PowhegPythia \ttbar sample.
Alternative $\ttbar$ samples are used to evaluate systematic uncertainties.
A sample generated with
\Mcatnlo v4.01~\cite{Frixione:2002ik,Frixione:2003ei} interfaced to \Herwig
v6.520~\cite{Corcella:2000bw} and \Jimmy v4.31~\cite{Butterworth:1996zw}
with the AUET2 tune~\cite{ATL-PHYS-PUB-2011-008},
again simulated using the CT10 PDF set, is used to estimate the uncertainty related
to the choice of generator.
To evaluate the impact of variations in the parton shower and hadronization models,
a sample is generated with \PowhegBox interfaced to \Herwig and \Jimmy.
The effects of variations in the QCD (quantum chromodynamics) initial- and final-state radiation
(ISR and FSR) modelling are estimated with samples generated with
\Acermc v3.8~\cite{Kersevan:2004yg} interfaced to \Pythia v.6.426
with the AUET2B tune and the CTEQ6L1 PDF set~\cite{Pumplin:2002vw},
where the parton-shower parameters are varied in the range allowed
by data~\cite{ATLAS:2012al}.
For the study of systematic uncertainties on kinematic distributions resulting
from PDF uncertainties, a sample is generated using \PowhegBox interfaced
with \Pythia v.6.427 and using the HERAPDF set~\cite{Aaron:2009aa}. For
all $\ttbar$ samples, a top-quark mass of $172.5\gev$ is used.

The \ttbar cross section for $pp$ collisions at a centre-of-mass energy
of $\sqrt{s} = 8 \tev$ is $\sigma_{\ttbar}= 253^{+13}_{-15}$~pb for a
top-quark mass of $172.5 \gev$. It has been calculated at next-to-next-to-leading order (NNLO) in QCD including resummation of next-to-next-to-leading
logarithmic (NNLL) soft gluon terms with top++2.0~\cite{Cacciari:2011hy,
Beneke:2011mq,Baernreuther:2012ws,Czakon:2012zr,Czakon:2012pz,Czakon:2013goa,
Czakon:2011xx}. The PDF and $\alpha_\mathrm{s}$ uncertainties were calculated using the
PDF4LHC prescription~\cite{Botje:2011sn}
with the MSTW2008 68\% CL NNLO~\cite{Martin:2009iq,Martin:2009bu}, CT10
NNLO~\cite{Lai:2010vv,Gao:2013xoa} and NNPDF2.3 5f FFN~\cite{Ball:2012cx} PDF
sets, and their effect is added in quadrature to the effect of factorization- and
renormalization-scale uncertainties. The NNLO+NNLL
value is about 3\% larger than the exact NNLO prediction, as implemented in
Hathor 1.5~\cite{Aliev:2010zk}.

In measurements of the differential $\ttbar$ production cross section as a function
of the top-quark \pt, a discrepancy between data and MC
predictions was observed in $7\tev$ data~\cite{Aad:2014zka}.
Based on this measurement, a method of
sequential reweighting of the top-quark-$\pt$ and $\ttbar$-system-$\pt$ distributions
was developed~\cite{Aad:2015gra}, which gives better agreement between the MC predictions
and $8\tev$ data.
In this paper, this reweighting technique is applied to the \PowhegPythia \ttbar sample,
for which the technique was developed.
The predicted total \ttbar cross section at NNLO+NNLL is not changed by the
reweighting procedure.

Single-top-quark production in the $s$- and $Wt$-channel is modelled with \PowhegBox and the CT10 PDF
set interfaced to \Pythia v6.426 using Perugia 2011C.
Single-top-quark production in the $t$-channel is generated with \PowhegBox in the four-flavour scheme
(in which
$b$-quarks are generated in the hard scatter and the PDF does not contain $b$-quarks) using
the four-flavour CT10 PDF set interfaced to \Pythia v6.427.
The overlap between $Wt$ production and $\ttbar$ production is removed
with the diagram-removal scheme~\cite{Frixione:2008yi} and the different
single-top-production processes are normalized to the approximate NNLO cross-section
predictions~\cite{Kidonakis:2010tc,Kidonakis:2011wy,Kidonakis:2010ux}.

Events with a $W$ or a $Z$ boson produced in association with jets (\Wjets or $Z$+jets) are generated with
\Alpgen~\cite{Mangano:2002ea} interfaced to \Pythia v6.426 using the CTEQ6L1 PDF set
and Perugia 2011C.
Up to five additional partons are included in the calculation of the matrix element,
as well as additional $c$-quarks, $c\bar{c}$-quark pairs, and $b\bar{b}$-quark pairs, taking into
account the masses of these heavy quarks.
The \Wjets contribution is normalized using the charge asymmetry in $W$-boson
production in data~\cite{ATLAS:2012an,Aad:2012hg} by selecting $\mu$+jets events
and comparing to the prediction from MC simulations.
The $Z$+jets contribution is normalized to the calculation of the inclusive
cross section at NNLO in QCD obtained with FEWZ~\cite{Gavin:2012sy}.

For the comparison of the different top-tagging techniques using MC simulation only,
multijet samples are generated with \Pythia v8.160 with the CT10 PDF set and AU2.
As a source of high-transverse-momentum
top quarks, samples of events with a hypothetical massive $Z^\prime$ resonance decaying to top-quark pairs, $Z^\prime\rightarrow\ttbar$,
are generated
with resonance masses ranging from $400\gev$ to $3000\gev$ and a resonance width of
1.2\% of the resonance mass~\cite{Harris:1999ya} using \Pythia v8.175 with the MSTW2008 68\%
CL LO PDF set~\cite{Martin:2009iq,Martin:2009bu} and AU2.

For a study of top-quark reconstruction in a final state with many jets, the
process\footnote{The process $pp\rightarrow H^-t(\bar{b})\rightarrow \bar{t}bt(\bar{b})$ is also simulated. For simplicity only the
positively charged Higgs boson is indicated explicitly in this paper, but it should be understood to denote both signs of the electric charge.}
$pp\rightarrow H^+\bar{t}(b)\rightarrow t\bar{b}\bar{t}(b)$ is generated in a type-II 2HDM model~\cite{Branco:2011iw}
with a mass of $1400\gev$ of the charged Higgs boson using \PowhegBox interfaced to \Pythia v8.165
with AU2 and the CT10 PDF set.
The width of the charged Higgs boson is set to zero and the five-flavour scheme is used.
The additional $b$-quark (in parentheses above) can be present or not, depending on whether
the underlying process is $gg\rightarrow H^+\bar{t}b$ or $g\bar{b}\rightarrow H^+\bar{t}$.

All MC samples are passed through a full simulation of the ATLAS
detector~\cite{Aad:2010ah} based on GEANT4~\cite{Agostinelli:2002hh},
except for the \ttbar samples used to estimate systematic uncertainties due to the choice of MC
generator, parton shower, and amount of ISR/FSR, which are passed
through a faster detector simulation with reduced complexity in the description of the
calorimeters~\cite{ATLAS:1300517}.
All MC samples are reconstructed using the same algorithms as used for data
and have minimum-bias events simulated with \Pythia v8.1~\cite{Sjostrand:2007gs} overlaid
to match the pile-up conditions of the collision data sample.

\section{Object reconstruction and event selection}
\label{sec:objandeventsel}

\subsection{Object reconstruction}
\label{sec:obj}

Electron candidates are reconstructed~\cite{Aad:2014fxa,ATLAS-CONF-2014-032} from clusters in the EM calorimeter
and are required to have a track in the ID, associated with
the main primary vertex~\cite{ATLAS-CONF-2010-069}, which
is defined as the one with the largest $\sum p^2_\textrm{T,track}$.
They must have $\et > 25\gev$ and $|\eta_{\rm cluster}| < 2.47$ excluding the
barrel/end-cap-calorimeter transition region $1.37 < |\eta_{\rm cluster}| < 1.52$,
where $\eta_{\rm cluster}$ is the pseudorapidity of the cluster in the EM calorimeter.
The shape of the cluster in the calorimeter must be consistent with the typical energy deposition
of an electron and the electron candidate must satisfy the
{\em mini-isolation}~\cite{Rehermann:2010vq,Aad:2013nca} requirement
to reduce background contributions from non-prompt electrons and hadronic showers:
the scalar sum of track transverse momenta within a cone of size
$\DeltaR=10\GeV/\ET^{\rm el}$ around the electron track
must be less than $5\%$ of the electron transverse energy
$\ET^{\rm el}$ (only tracks with $\pt > 1\gev$ are considered in the sum, excluding the track
matched to the electron cluster).

Muons are reconstructed~\cite{Aad:2014rra} using both the ID and the MS and must be
associated with the main primary vertex of the event.
Muons are required to have $\pT>25\GeV$ and $|\eta|<2.5$ and are required to be isolated with
requirements similar to those used for electron candidates:
the scalar sum of the track transverse momenta within a cone of size
$\DeltaR=10\GeV/\pt^\mu$ around the muon track must be less than $5\%$ of $\pt^\mu$,
where $\pt^\mu$ is the transverse momentum of the muon.

Jets are built~\cite{Cacciari:2011ma} from topological clusters of calorimeter cells, which
are calibrated to the hadronic energy scale~\cite{Aad:2011he}
using a local cell-weighting scheme~\cite{Issever:2004qh}.
The clusters are treated as massless and are combined
by adding their four-momenta, leading to massive jets.
The reconstructed jet energy is calibrated using energy- and $\eta$-dependent
corrections obtained from MC simulations.
These corrections are obtained by comparing reconstructed jets with
geometrically matched jets built from stable particles (particle level).
The corrections are validated using in situ measurements of \smallR jets~\cite{Aad:2014bia}.

Jets reconstructed with the \akt~\cite{Cacciari:2008gp} algorithm using a radius parameter
$R=0.4$ must satisfy $\pt > 25\gev$ and $|\eta|<2.5$.
The jet vertex fraction (JVF) uses the tracks matched to a jet and is defined as the
ratio of the scalar sum of the transverse momenta of
tracks from the main primary vertex
to that of all matched tracks. A jet without any matched track is assigned a JVF value of $-1$.
For \akt $R=0.4$ jets with $\pt < 50\gev$ and $|\eta|<2.4$, the JVF
must be larger than 0.5~\cite{Aad:2015ina} to suppress jets from pile-up.

\LargeR jets are reconstructed with the \akt algorithm using $R=1.0$
and with the \ca algorithm~\cite{Dokshitzer:1997in} (\CamKt) using
$R=1.5$.
\Akt $R=1.0$ jets are groomed using a trimming procedure~\cite{Krohn:2009th}: the constituents of the \akt $R=1.0$ jet
are reclustered using the \kt algorithm~\cite{Catani:1993hr} with $R=0.3$.
Subjets with a \pt of less than 5\% of the \largeR jet \pt are removed~\cite{Aad:2013gja}.
The properties of the trimmed jet are recalculated from the constituents of the remaining subjets.
The trimmed jet mass, \pt, and pseudorapidity are corrected to be, on average, equal to the
particle-level jet mass, \pt, and pseudorapidity using MC simulations~\cite{ATLAS:2012am,Aad:2013gja}.
An illustration of trimming is given in Figure 4 of Ref.~\cite{Aad:2013gja}.

The \CamKt $R=1.5$ jets are required to satisfy $\pt > 200\gev$. These jets are
used as input to the \htt, which employs an internal pile-up suppression, and are
therefore left ungroomed.
For
trimmed \akt $R=1.0$ jets, the minimum
\pt is raised to $350\GeV$ to reduce the fraction of jets not
containing all top-quark decay products due to the smaller jet radius parameter.
All \largeR jets must satisfy $|\eta|<2.0$.

The missing transverse momentum is calculated from the vector sum of
the transverse energy of clusters in the calorimeters,
and it is corrected for identified electrons, muons and \akt $R=0.4$ jets, for which specific
object-identification criteria are applied~\cite{ATLAS-CONF-2012-101}.
The magnitude of the missing transverse momentum is denoted by \met.

\subsection{Event selection}
\label{sec:matched}
The data used in this paper were taken in 2012
at a centre-of-mass-energy $\sqrt{s} = 8\tev$
and correspond
to an integrated luminosity of $20.3\;\ifb$ \cite{Aad:2013ucp}.
Data are used only if all subsystems of the
detector as well as the trigger system were fully functional.
Baseline quality criteria are imposed to reject
contamination from detector noise, non-collision beam backgrounds,
and other spurious effects.
Events are required to have at least one reconstructed primary vertex with at least five associated ID tracks, each with a $\pt$ larger than $400\MeV$.
This vertex must be consistent with the LHC beam spot~\cite{ATLAS-CONF-2010-069}.
In addition, all \akt $R=0.4$ jets in the event which have $\pt>20\GeV$
are required to satisfy the ``looser'' quality criteria discussed in detail
in Ref.~\cite{Aad:2014bia}, otherwise the event is rejected.

Two different event samples are used to study the performance of top-tagging
algorithms in data: a signal sample enriched in hadronically decaying
top quarks and a background sample consisting mainly of multijet events.

\subsubsection{Signal sample}
\label{sec:signalsample}

For the signal sample, a selection of \ttbar events in the lepton+jets channel is used,
in which one of the \W bosons from $\ttbar \to W^+bW^-\bar{b}$
decays hadronically and the other
\W boson decays leptonically. The selection is
performed in the muon channel and the electron channel.

The selection criteria for the muon and electron channels differ only in the requirements
imposed on the reconstructed leptons.
For the muon channel, the events are required to pass at least one of two muon triggers,
where one is optimized to select isolated muons with a transverse
momentum of at least $24\GeV$ and the other selects muons with at
least $36\GeV$ without the isolation requirement.
Exactly one muon with $\pt>25\gev$ is required as defined in
\secref{obj}.
Muons are rejected if they are close to an \akt $R=0.4$ jet that has $\pt>25\GeV$.
The rejection occurs if $\Delta R(\mu, {\rm jet})<(0.04+10\GeV/\pt^\mu)$. Events in the muon channel are rejected if they
contain an additional electron candidate.

For the electron channel, events are required to pass at least one of two triggers. The first
is designed for isolated electrons with $\pt > 24\GeV$ and the
second trigger requires electrons with $\pt > 60\GeV$ without the isolation requirement.
Exactly one electron is required with $\ET>25\GeV$ as defined in \secref{obj}.
An electron--jet overlap removal is applied based on the observation that
the electron \pt contributes a significant fraction of the \pt of close-by \akt $R=0.4$ jets.
Therefore, the electron momentum is subtracted from the jet momentum before kinematic requirements
are applied to the jet, so that jets close to an electron often fall below
the jet \pt threshold. If the electron-subtracted jet still fulfils the
kinematic requirements for \akt $R=0.4$ jets and the electron
is still close, the electron is considered not isolated.
In this case, the electron is
removed from the event and the original non-subtracted jet is
kept. Events in the electron channel are rejected if they also
contain a muon candidate.

To select events with a leptonically decaying \W boson, the following requirements
are imposed. The events are required to have missing transverse momentum $\Etmiss>20\GeV$.
Additionally, the scalar sum of \Etmiss and the transverse mass of the
leptonic \W-boson candidate must satisfy $\met+\MTW>60\GeV$, where
$\MTW=\sqrt{2\pt^\ell\met(1-\cos\Delta\phi)}$ is calculated from the transverse momentum of
the lepton, $\pt^\ell$, and
\met in the event. The variable $\Delta\phi$ is the azimuthal angle between the lepton
momentum and the \Etmiss direction.

To reduce contamination from \Wjets events, each event must contain at least
two $b$-tagged \akt $R=0.4$ jets with $\pt>25\GeV$ and $|\eta|<2.5$.
A neural-network-based $b$-tagging algorithm~\cite{Aad:2015ydr}
is employed,
which uses information on the impact
parameters of the tracks associated with the jet, the secondary vertex, and
the decay topology as its input.
The operating point chosen for this analysis corresponds to
a $b$-tagging identification efficiency of 70\% in simulated $\ttbar$ events.
In \ttbar events with high-momentum top quarks, the direction of the
$b$-quark from the leptonic decay of a top quark is often close to the lepton direction.
Hence, at least one $b$-tagged jet is required to be within \mbox{$\DeltaR = 1.5$} of the lepton
direction.
A second $b$-tag away from the lepton is required that fulfils
\mbox{$\Delta R($lepton, $b$-tag$)>1.5$.} This $b$-tagged jet
is expected to originate from the $b$-quark
from the hadronic top-quark decay, and is expected to be well separated from the
decay products of the leptonically decaying top quark.

Each event is required to contain at least one \largeR jet that fulfils
the requirement $\Delta R(\textrm{lepton, \largeR jet})>1.5$. This criterion
increases the probability that the \largeR jet originates from a hadronically
decaying top quark.
The \largeR jet has to fulfil $|\eta|<2$ and exceed a \pt threshold.
The jet algorithm, the radius parameter, and the \pt threshold
depend on the top tagger under study. An overview is given in
\tabref{taggerFatjets}. The top taggers are introduced in \secref{techniques}
where also the choice of particular \largeR jet types is motivated.
If several \largeR jets in an event satisfy the
mentioned criteria, only the jet with the highest \pt is considered.
This choice does not bias the measurements presented in this paper,
because the top-tagging efficiencies and misidentification rates
are measured as a function of the \largeR jet kinematics.

\begin{table}[b]
\begin{center}
\begin{tabular}{|c|c|c|c|c|c|}
      \hline
Tagger & Jet algorithm & Grooming & Radius parameter & \pt range & $|\eta|$ range\\
      \hline
Tagger I--V & \multirow{3}{*}{\akt} & trimming & \multirow{3}{*}{$R=1.0$} & \multirow{3}{*}{$>350\GeV$} & \multirow{3}{*}{$<2$}\\
\WPT & & ($R_\textrm{sub}=0.3$, & & & \\
\sd & & $f_\textrm{cut}=0.05)$ & & & \\
\hline
\htt & \CamKt & none & $R=1.5$ & $>200\GeV$ & $<2$ \\
\hline
\end{tabular}
\caption{Definitions of \largeR jets and their \pt thresholds used as input to
the different top taggers.}
\label{tab:taggerFatjets}
\end{center}
\end{table}

In simulated events containing top quarks, \largeR jets are classified as {\em matched}
or {\em not matched} to a hadronically decaying top quark.
The classification is based on the distance $\Delta R$ between
the axis of the \largeR jet and the flight direction of a generated hadronically decaying
top quark.
The top-quark flight direction at the top-quark decay vertex is chosen,
so as to take into account radiation from the top quark changing its direction.
Matched jets are those with $\Delta R$ smaller than a predefined value $ R_\textrm{match}$, while not-matched jets are those with
$\Delta R > R_\textrm{match}$.
The radius $R_\textrm{match}$ is $0.75$ for the \akt $R=1.0$
jets
and $1.0$ for the \CamKt $R=1.5$ jets. Changing $R_\textrm{match}$ to $1.0$ for the \akt $R=1.0$ jets
has a negligible impact on the size of the not-matched \ttbar contribution (less than 1\%).
Alternative matching schemes were tested but did not show improved matching properties, such as
a higher matching efficiency.

Distributions for the signal selection with
at least one trimmed \akt $R=1.0$ jet with $\pt>350\GeV$
are shown in \figref{ctrl_akt_pretag_ept_fj}.
The top-quark purity in this sample is 97\%,
with a small background contribution from \Wjets production (3\%).
Single-top production accounts for 4\% of the event yield and
the \ttbar prediction accounts for 93\% (62\% from matched and
31\% from not-matched events).
Not-matched \ttbar events are an intrinsic feature of the signal selection. With different selection
criteria the fraction of not-matched \ttbar events varies, as does the total number of selected
events. The chosen signal selection in the lepton+jets channel was found to be a good compromise between a
reduced fraction of not-matched \ttbar events and a sizeable number of selected events.

The mass and the transverse momentum of the highest-\pt trimmed \akt $R=1.0$ jet
are shown in \figsref{ctrl_akt_pretag_fjm}{ctrl_akt_pretag_fjpt}, respectively.
The systematic uncertainties shown in these plots are described
in detail in \secref{systematics}.
The mass distribution shows three peaks: one at the top-quark mass, a second at
the $W$-boson mass and a third around $35\GeV$. According to simulation, which
describes the measured distribution within uncertainties,
the top-quark purity in the region near the top-quark mass is very high, with the largest
contribution being matched \ttbar.
The peak at the position of the $W$-boson mass originates from hadronically decaying top
quarks where the $b$-jet from the decay is not contained in the \largeR jet.
Even smaller masses are obtained if one of the decay products of the hadronically
decaying \W boson is not contained in the \largeR jet or if only one top-quark-decay product
is captured in the \largeR jet.
In these cases,
a small mass is obtained due to the kinematic requirements imposed during trimming.
The fraction of not-matched \ttbar increases for decreasing \largeR jet mass indicating a
decreasing fraction of jets with a close-by hadronically decaying top quark.
Only a small fraction of the peak at small mass is due to matched \ttbar.
The \largeR jet \pt exhibits a falling spectrum, and the application of the sequential \pt reweighting
to the simulation (cf.~\secref{mc}) yields a good description of the data.

The dominant systematic uncertainties in \figref{ctrl_akt_pretag_ept_fj}
result from uncertainties in the \largeR jet energy scale (JES), the PDF, and the \ttbar generator.
The contributions from these sources
are approximately equal in size, except for
\largeR jets with $\pt>500\GeV$
where the choice of \ttbar generator dominates.
These uncertainties affect mostly the
normalization of the distributions.
For the PDF and \ttbar generator uncertainties, this normalization uncertainty
comes about as follows: while the total \ttbar cross section is fixed when the
different MC event samples are compared, the \pt dependence of the
cross section varies from sample to sample, leading to a change in normalization
for the phase space considered here ($\pt>350\GeV$).

\begin{figure}[!h]
\begin{centering}
\subfigure[]{
\label{fig:ctrl_akt_pretag_fjm}
\includegraphics[width=0.48\textwidth]{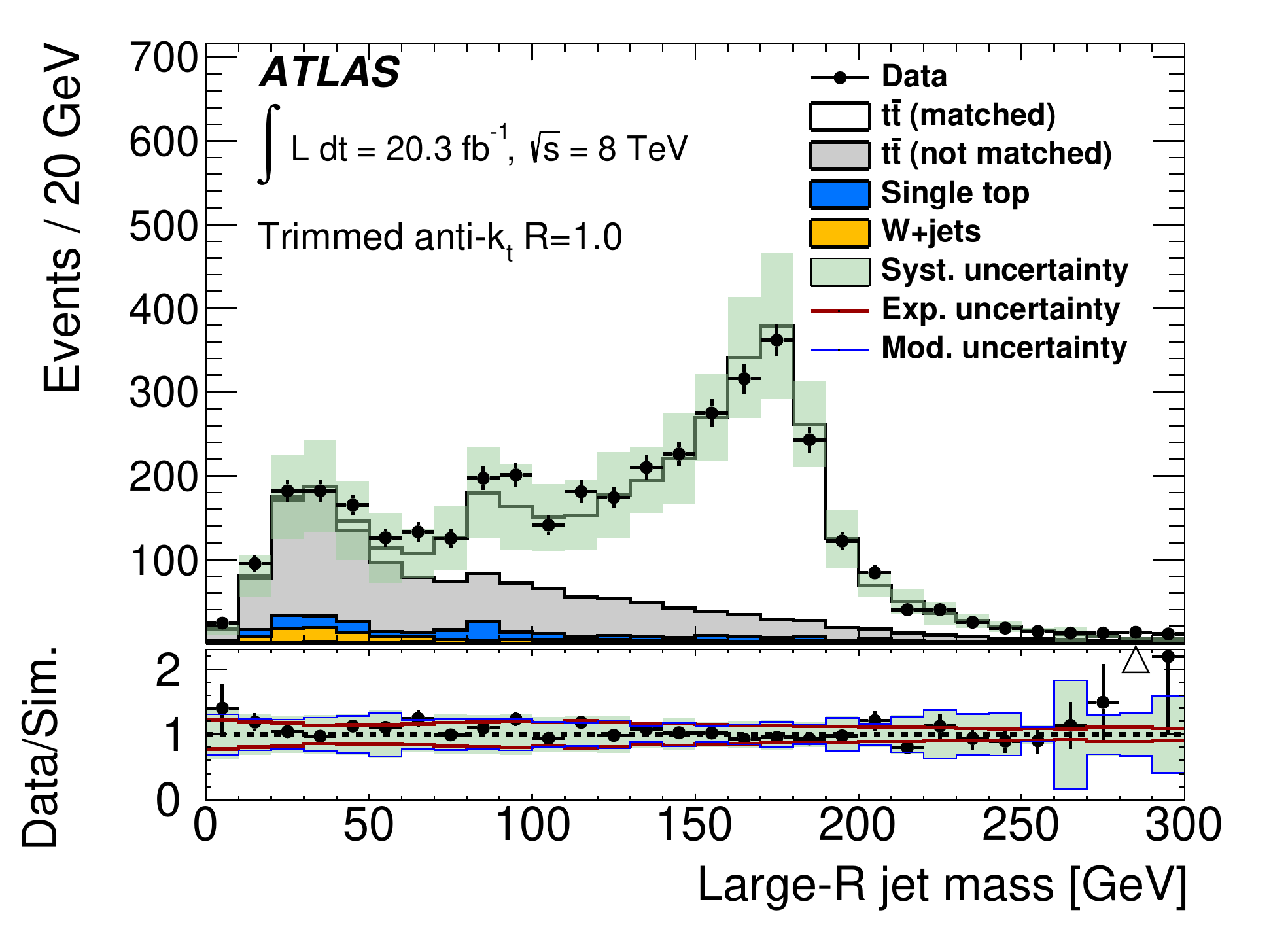}
}
\subfigure[]{
\label{fig:ctrl_akt_pretag_fjpt}
\includegraphics[width=0.48\textwidth]{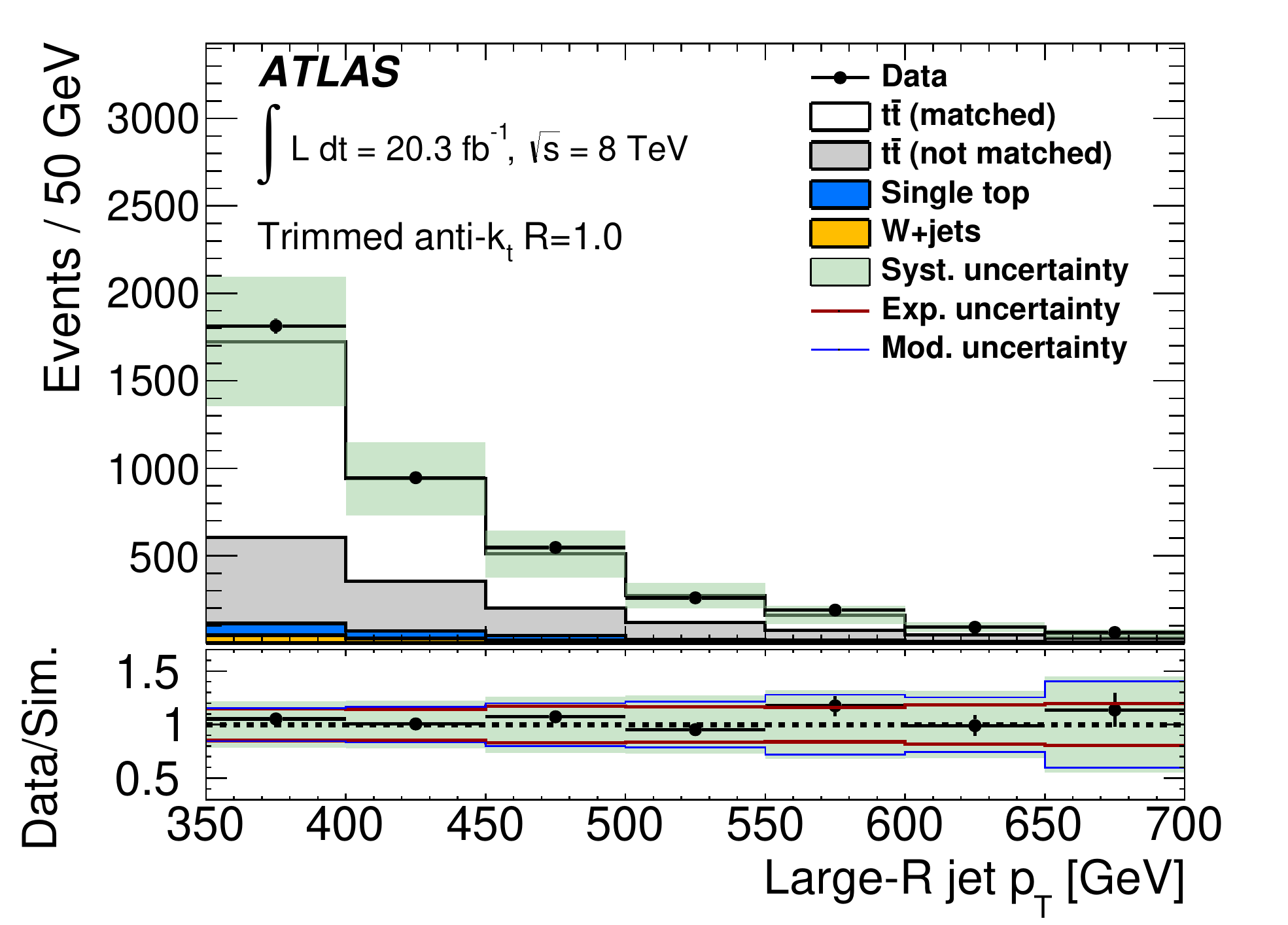}
}
\caption{Detector-level distributions of variables reconstructed in events
passing the signal-sample selection (\ttbar) with
at least one trimmed \akt $R=1.0$ jet with $\pt>350\GeV$.
Shown in (a) is the mass and in (b) the transverse momentum of the highest-\pt
\akt $R=1.0$ jet.
The vertical error bar indicates the statistical uncertainty
of the measurement. Also shown are distributions for simulated SM contributions
with systematic uncertainties (described in \secref{systematics}) indicated as a band.
The \ttbar prediction is split into a {\em matched} part for which the \largeR
jet axis is within $\Delta R = 0.75$ of the flight direction of a hadronically decaying
top quark and a {\em not matched} part for which this criterion does not hold.
The ratio of measurement to
prediction is shown at the bottom of each subfigure and the error bar and band
give the statistical and systematic uncertainties of the ratio, respectively.
The impacts of experimental and \ttbar modelling uncertainties are
shown separately for the ratio.
}
\label{fig:ctrl_akt_pretag_ept_fj}
\end{centering}
\end{figure}

Distributions for events fulfilling the signal selection with
at least one \CamKt $R=1.5$ jet with $\pt>200\GeV$, to be used in the \htt studies,
are shown in \figref{ctrl_HTT_pretag_mtw_fj}.
According to the simulation, the top quark purity in this sample is 97\%.
The only non-negligible background process is \Wjets production (3\%).
The \ttbar prediction is split into a matched part (59\%)
and a not-matched part (29\%).
Single-top production contributes 9\% to the total event yield.
The mass of the highest-\pt \CamKt $R=1.5$ jet with $\pt>200\GeV$ is shown in
\figref{ctrl_HTT_pretag_fjm} and it exhibits a broad peak around $190\GeV$.
The \largeR-jet mass distributions from not-matched \ttbar, single-top production, and \Wjets
production have their maxima at
smaller values than the distribution from matched \ttbar.
No distinct $W$-boson peak is visible, because the \CamKt $R=1.5$ jets are ungroomed.
The \pt spectrum of the highest-\pt \CamKt $R=1.5$ jet is smoothly falling
and well described by simulation after the sequential \pt
reweighting is applied (\figref{ctrl_HTT_pretag_fjpt}).

The \CamKt $R=1.5$ jet distributions are described by the
simulation within the uncertainties. The systematic uncertainties are slightly smaller than
those in the distributions shown in \figref{ctrl_akt_pretag_ept_fj} for \akt $R=1.0$ jets with $\pt>350\GeV$ because
the \ttbar modelling uncertainties increase with \largeR jet \pt.
The uncertainties in the \largeR JES, the
$b$-tagging efficiency, the prediction of the \ttbar cross section, and \ttbar modelling uncertainties
from the choice of generator, parton shower, and PDF set
all contribute to the systematic uncertainty in the \largeR-jet mass distribution.
The uncertainty from the choice of generator increases
in the high-mass tail,
which is particularly sensitive to additional radiation close to the
hadronically decaying top quark.
The modelling uncertainties for the \largeR-jet \pt distribution increase with \pt due to increasing
uncertainties from the \largeR JES, the $b$-tagging efficiency, and the \ttbar modelling uncertainties.
The increase of the \ttbar modelling uncertainty with \largeR-jet \pt is
an observation consistent with \figref{ctrl_akt_pretag_fjpt}.

Distributions of other kinematic variables are also well described by the simulation
and are shown in \appref{incldistrib}.

\begin{figure}[!h]
\begin{centering}
\subfigure[]{
\label{fig:ctrl_HTT_pretag_fjm}
\includegraphics[width=0.48\textwidth]{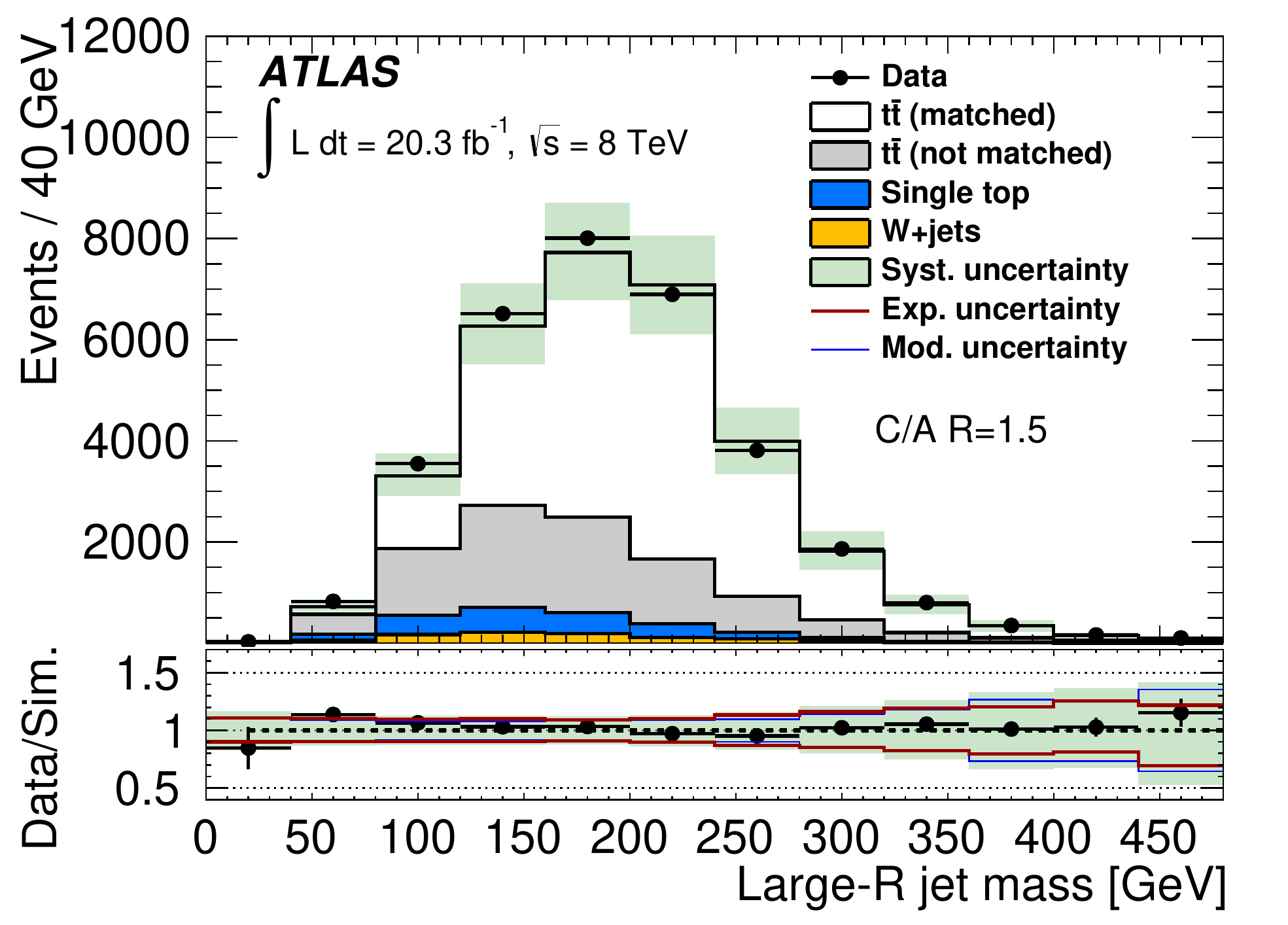}
}
\subfigure[]{
\label{fig:ctrl_HTT_pretag_fjpt}
\includegraphics[width=0.48\textwidth]{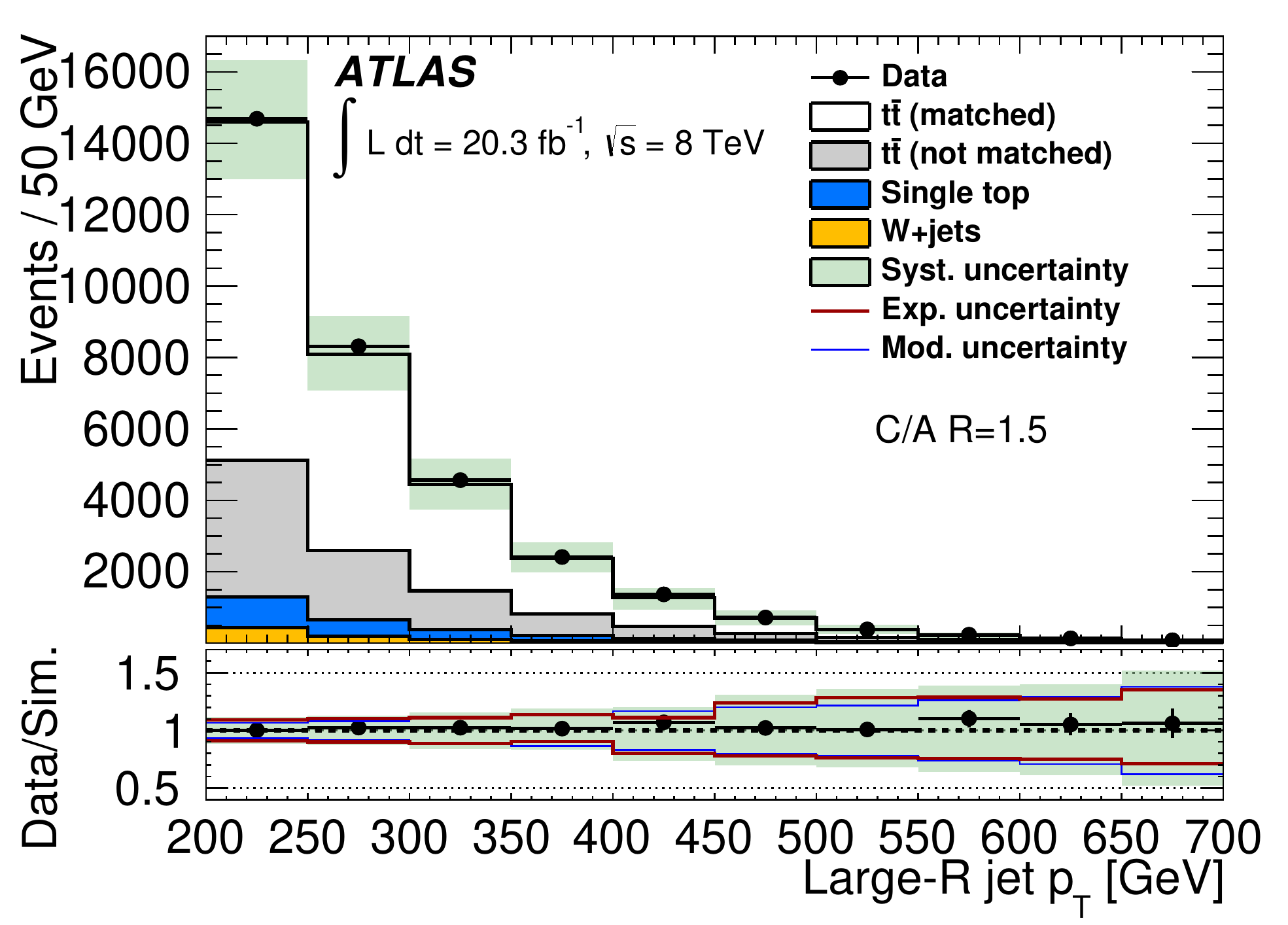}
}
\caption{Detector-level distributions of (a) the mass and (b) the transverse momentum of the highest-\pt
\CamKt $R=1.5$ jet in events
passing the signal-sample selection (\ttbar) with
at least one \CamKt $R=1.5$ jet with $\pt>200\GeV$.
The vertical error bar indicates the statistical uncertainty
of the measurement. Also shown are distributions for simulated SM contributions
with systematic uncertainties (described in \secref{systematics}) indicated as a band.
The \ttbar prediction is split into a {\em matched} part for which the \largeR
jet axis is within $\Delta R = 1.0$ of the flight direction of a hadronically decaying
top quark and a {\em not matched} part for which this criterion does not hold.
The ratio of measurement to
prediction is shown at the bottom of each subfigure and the error bar and band
give the statistical and systematic uncertainties of the ratio, respectively.
The impacts of experimental and \ttbar modelling uncertainties are
shown separately for the ratio.
}
\label{fig:ctrl_HTT_pretag_mtw_fj}
\end{centering}
\end{figure}

\subsubsection{Background sample}
\label{sec:backgroundsample}

Due to the high threshold of the unprescaled jet triggers, such triggers do not provide an unbiased background sample of \largeR jets from multijet production.
Therefore, the misidentification rate is measured in a multijet sample collected with single-electron triggers,
where the event is triggered by an object which in the detailed offline analysis
fails the electron-identification requirements.

For the electron candidate used at the trigger level,
the requirements on the pseudorapidity of the cluster of calorimeter cells are the same
as for reconstructed electrons (cf.~\secref{obj}).
Events with an offline reconstructed electron satisfying loose identification requirements~\cite{ATLAS-CONF-2014-032} (these loose identification requirements do not include isolation criteria) are rejected to reduce contributions from electroweak processes.
Only \largeR jets well separated from the electron-trigger candidate are studied.
This selection provides a sample that is largely dominated by
multijet production, for which the electron-trigger candidate is a jet misidentified as an electron.
Events are required to be selected by the trigger
for electrons with $\pt > 60\GeV$ and not by the trigger for isolated electrons with
a threshold of $24\GeV$ (described in \secref{signalsample}). Not using the isolated electron trigger reduces top-quark
contamination in the selected jet sample.
The fraction of $\ttbar$ events before requiring a tagged top candidate is negligible.
After requiring a tagged top candidate, the $\ttbar$ events are subtracted
for the top taggers for which they present a non-negligible part of the sample, as
detailed in \secref{misid}.

At least one \largeR jet is required with a jet axis separated from the
electron-trigger object by $\Delta R > 1.5$.
The algorithm, radius parameter, and \pt threshold of the jet depend on the
particular top-tagging algorithm under study (see \tabref{taggerFatjets}). If several
\largeR jets satisfying these criteria are found, only the
jet with the highest \pt is considered for the study of the misidentification rate.
This choice does not bias the measurements, because the misidentification
rate is measured as a function of the \largeR-jet \pt.

\section{Top-tagging techniques}
\label{sec:techniques}

Top tagging classifies a given \largeR jet as a top jet if its substructure satisfies
certain criteria. This paper examines several top-tagging methods, which differ
in their substructure analysis and which are described in the following subsections.

Due to the different substructure criteria applied, the methods have different
efficiencies for tagging signal jets and different misidentification rates for
background jets. High efficiency is obtained for loose criteria and implies
a high misidentification rate. The performance of the taggers in terms of efficiencies
and misidentification rates is provided in \secref{compare}.

\subsection{Substructure-variable taggers}

The choice of trimmed \akt $R=1.0$ jets (as defined in \secref{obj})
for substructure-based analyses has been previously studied in detail~\cite{Aad:2013gja},
including comparisons of different grooming techniques and parameters. The following
jet-substructure variables are used for top tagging in this analysis:

\begin{itemize}
\item \textit{trimmed mass} - The mass, $m$, of the trimmed \akt $R=1.0$ jets
is less susceptible to energy depositions from
\pileup and the underlying event
than the mass of the untrimmed jet.
On average, \largeR
jets containing top-quark decay products have a larger mass than background jets.

\item \textit{$k_{t}$ splitting scales} - The \kt splitting scales~\cite{Butterworth:2002tt}
are a measure of the scale of the last recombination steps in the \kt
algorithm, which clusters high-momentum and large-angle proto-jets last.
Hence, the \kt splitting scales are sensitive to whether the last recombination
steps correspond to the merging of the decay products of massive particles.
They are determined by reclustering
the constituents of the trimmed \largeR jet with the \kt algorithm
and are defined as
\begin{equation}
  \Dij = \text{min}(p_{\mathrm{T}i}, p_{\mathrm{T}j})\times \Delta R_{ij} \quad ,
  \label{eq:grooming_ktsplitting}
\end{equation}
\noindent
in which $\Delta R_{ij}$ is the distance between two subjets $i$ and $j$
in $\eta$--$\phi$ space, and $p_{\mathrm{T}i}$ and $p_{\mathrm{T}j}$ are the
corresponding subjet transverse momenta. Subjets merged in the last \kt
clustering step provide the \DOneTwo observable, and
\DTwoThr is the splitting scale of the second-to-last merging.
The expected value of the first splitting scale \DOneTwo for hadronic top-quark
decays captured fully in a \largeR jet is approximately $m_t/2$, where $m_t$
is the top quark mass.
The second splitting scale \DTwoThr targets the hadronic decay of the $W$ boson
with an expected value of approximately $m_{W}/2$. The use of the splitting
scale for \W-boson tagging in $8\TeV$ ATLAS data is explored in Ref.~\cite{bosontagging}.
Background jets initiated by hard gluons or light quarks tend to have smaller
values of the splitting scales and exhibit a steeply falling spectrum.

\item \textit{\Nsj} - The \Nsj variables \tauN~\cite{Thaler:2010tr, Thaler:2011gf}
quantify how well jets can be described as containing $N$ or fewer subjets.
The $N$ subjets found by an exclusive \kt clustering of the constituents of the trimmed \largeR jet
define axes within the jet. The quantity \tauN is given by the \pt-weighted sum
of the distances of the constituents from the subjet axes:
\begin{equation}
  \tauN = \frac{1}{d_{0}} \sum_k p_{\mathrm{T}k} \times \Delta R^\textrm{min}_{k} \quad~~~ \textrm{with} \quad
  ~~~d_{0}\equiv\sum_{k} p_{\mathrm{T}k}\times R \quad ,
  \label{eq:grooming_nsubj}
\end{equation}
in which $p_{\mathrm{T}k}$ is the transverse momentum of constituent $k$,
$\Delta R^\textrm{min}_k$ is the distance between constituent $k$ and the axis
of the closest subjet, and $R$ is the radius parameter of the \largeR jet.
The ratio $\tau_3/\tau_2$ (denoted \tauThrTwo) provides discrimination between
\largeR jets formed from hadronically decaying top quarks with high
transverse momentum (top jets) which
have a 3-prong subjet structure (small values of \tauThrTwo) and non-top jets
with two or fewer subjets (large values of \tauThrTwo).
Similarly, the ratio $\tau_2/\tau_1 \equiv \tauTwoOne$ is used to separate
\largeR jets with a 2-prong structure (hadronic decays of \Z or \W bosons)
from jets with only one hard subjet, such as those produced from light quarks or gluons.
The variable \tauTwoOne is studied in the context of \W-boson tagging with
the ATLAS and CMS detectors in Ref.~\cite{bosontagging} and Ref.~\cite{Khachatryan:2014vla,},
respectively. A method that distinguishes
hadronically decaying high-\pt \Z bosons from \W bosons is studied in Ref.~\cite{WZdiscrimination}.
\end{itemize}

Distributions of the \kt splitting scales and \Nsj variables
for \largeR jets in a top-quark-enriched event sample (cf. \secref{signalsample})
are shown in
\figref{ctrl_akt_pretag_substr}.
The \DOneTwo distribution shows a broad shoulder at
values above $40\GeV$ and the matched \ttbar contribution exhibits a peak near
$m_t/2$ as expected. For the not-matched \ttbar contribution and the \Wjets process,
\DOneTwo takes on smaller values and the requirement of a minimum value of \DOneTwo
can be used to increase the ratio of top-quark signal to background ($S/B$).
For the second splitting scale \DTwoThr, signal and background are less well
separated than for \DOneTwo, but \DTwoThr also provides signal--background discrimination.
The distribution of \tauThrTwo shows the expected behaviour, with the matched
\ttbar contribution having small values, because the hadronic top-quark
decay is better described by a three-subjet structure than by two subjets.
For not-matched \ttbar and \Wjets production, the distribution peaks at $\approx\!0.75$.
Requiring a maximum value of \tauThrTwo increases the signal-to-background
ratio.
For \tauTwoOne, the separation of signal and background is less pronounced,
but values above $0.8$ are obtained primarily for background.
Thus, \tauTwoOne also provides signal--background discrimination.

The distributions are well described by the simulation of SM processes within systematic uncertainties, which
are described in \secref{systematics}.
For all distributions shown, the \largeR JES, \ttbar generator, and parton-shower uncertainties give sizeable contributions,
as do the uncertainties of the modelling of the respective substructure variables shown.
The uncertainties for \DOneTwo and \DTwoThr are dominated by the \ttbar generator and ISR/FSR
uncertainties, respectively, for low values of the substructure variable.
Low values of these variables are mainly present for not-matched \ttbar, for which the
modelling is particularly sensitive to the amount of high-\pt radiation in addition to \ttbar,
because these \largeR jets do not primarily originate from hadronically decaying top quarks.
The modelling of additional radiation in \ttbar events is also an important uncertainty for
the number of events at low values of \tauThrTwo and \tauTwoOne, for which
the \ttbar ISR/FSR uncertainties dominate the total uncertainty.
The modelling of the substructure variables themselves dominates for
high values of \DOneTwo, \DTwoThr, \tauThrTwo, and \tauTwoOne.

\begin{figure}[p]
\begin{centering}
\subfigure[]{
\includegraphics[width=0.48\textwidth]{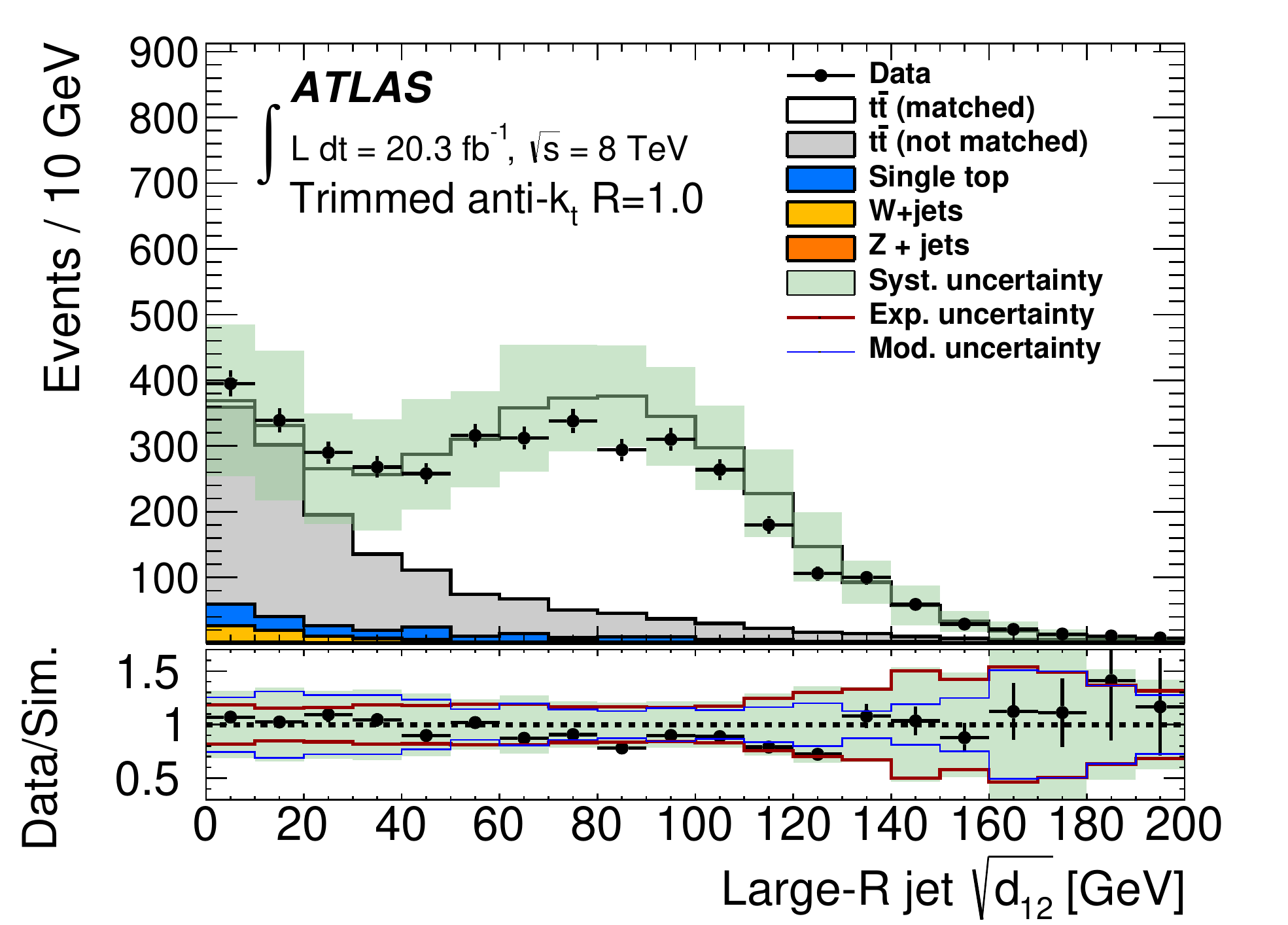}
}
\subfigure[]{
\includegraphics[width=0.48\textwidth]{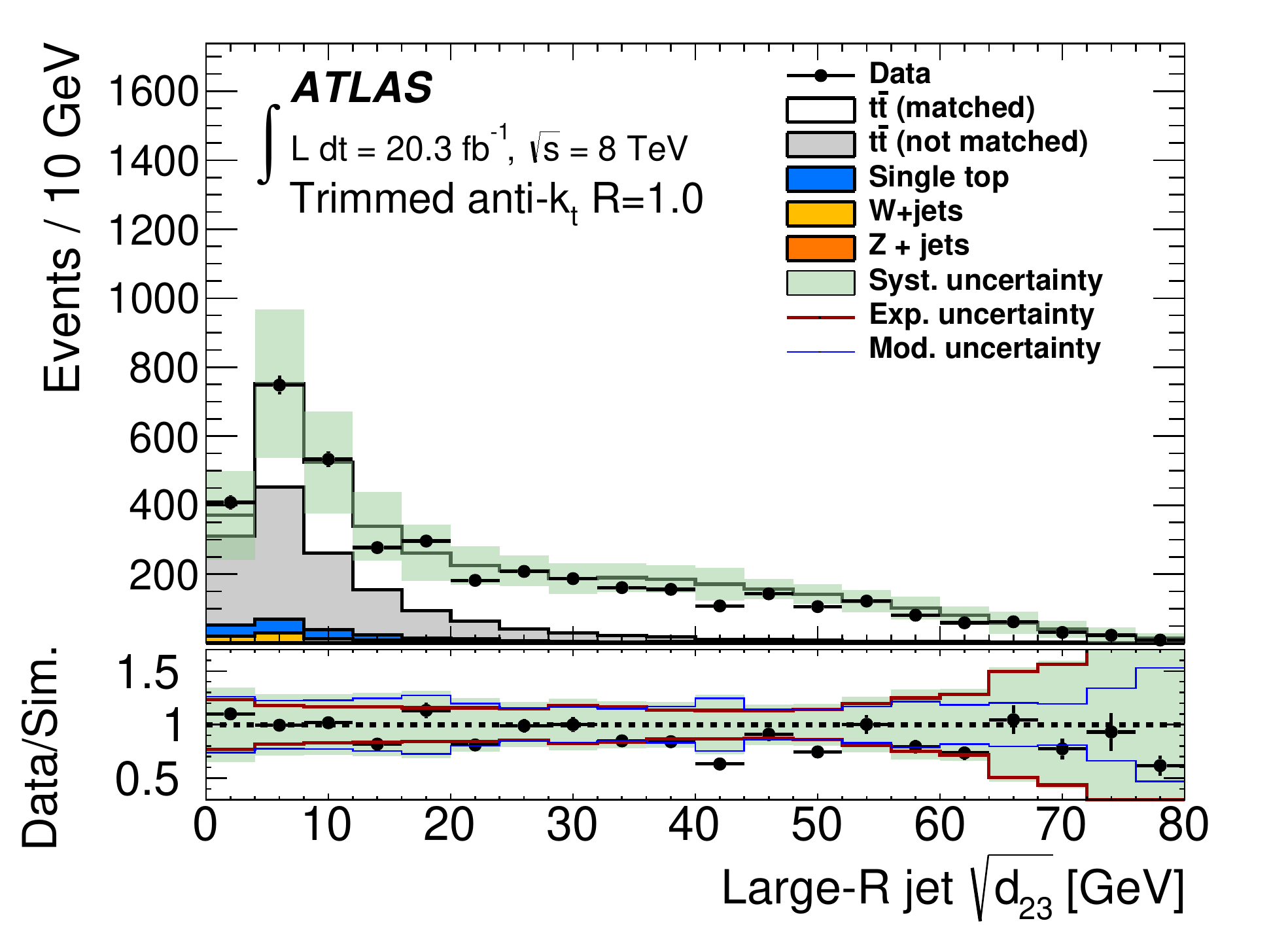}
} \\
\subfigure[]{
\includegraphics[width=0.48\textwidth]{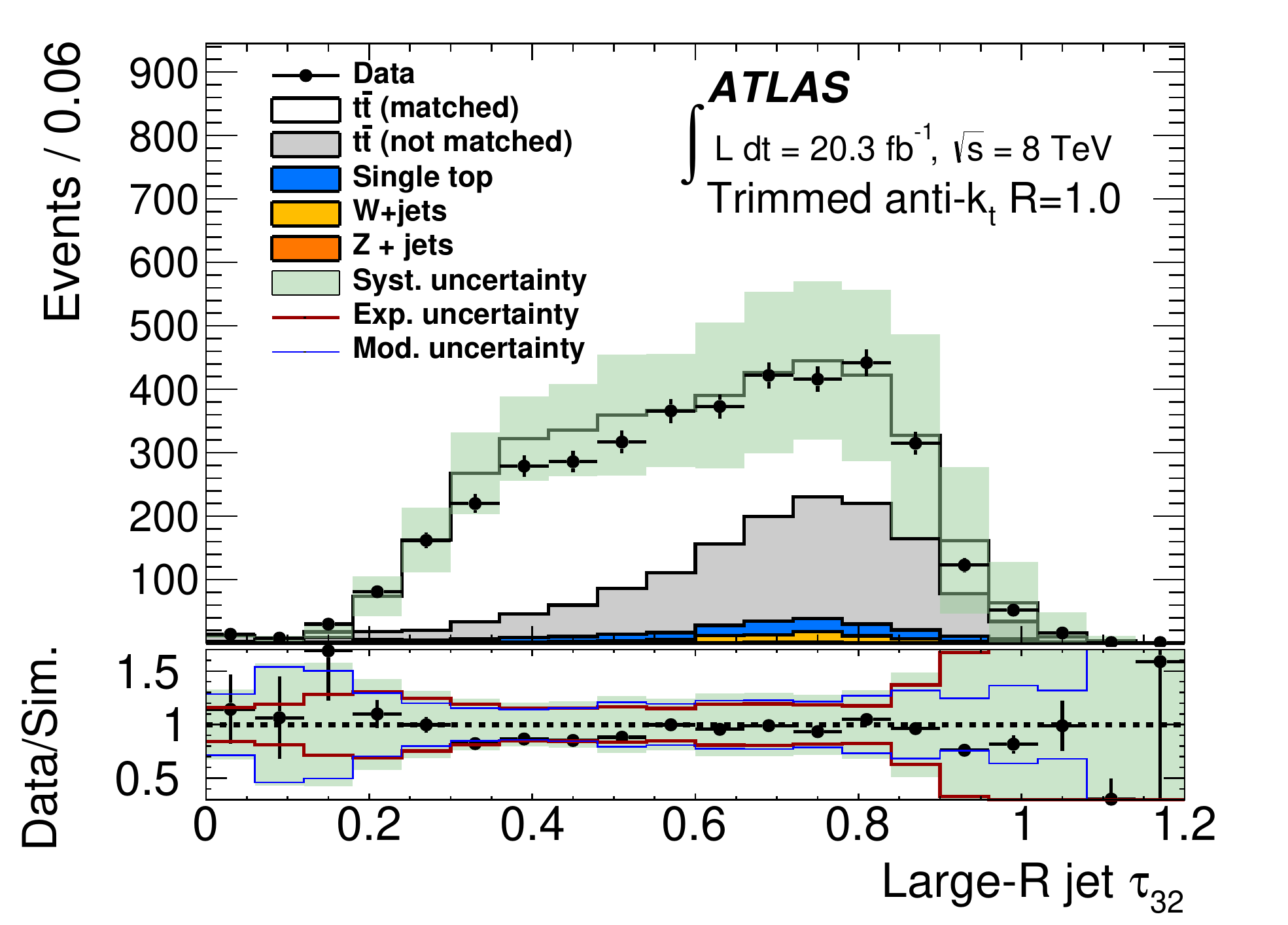}
}
\subfigure[]{
\includegraphics[width=0.48\textwidth]{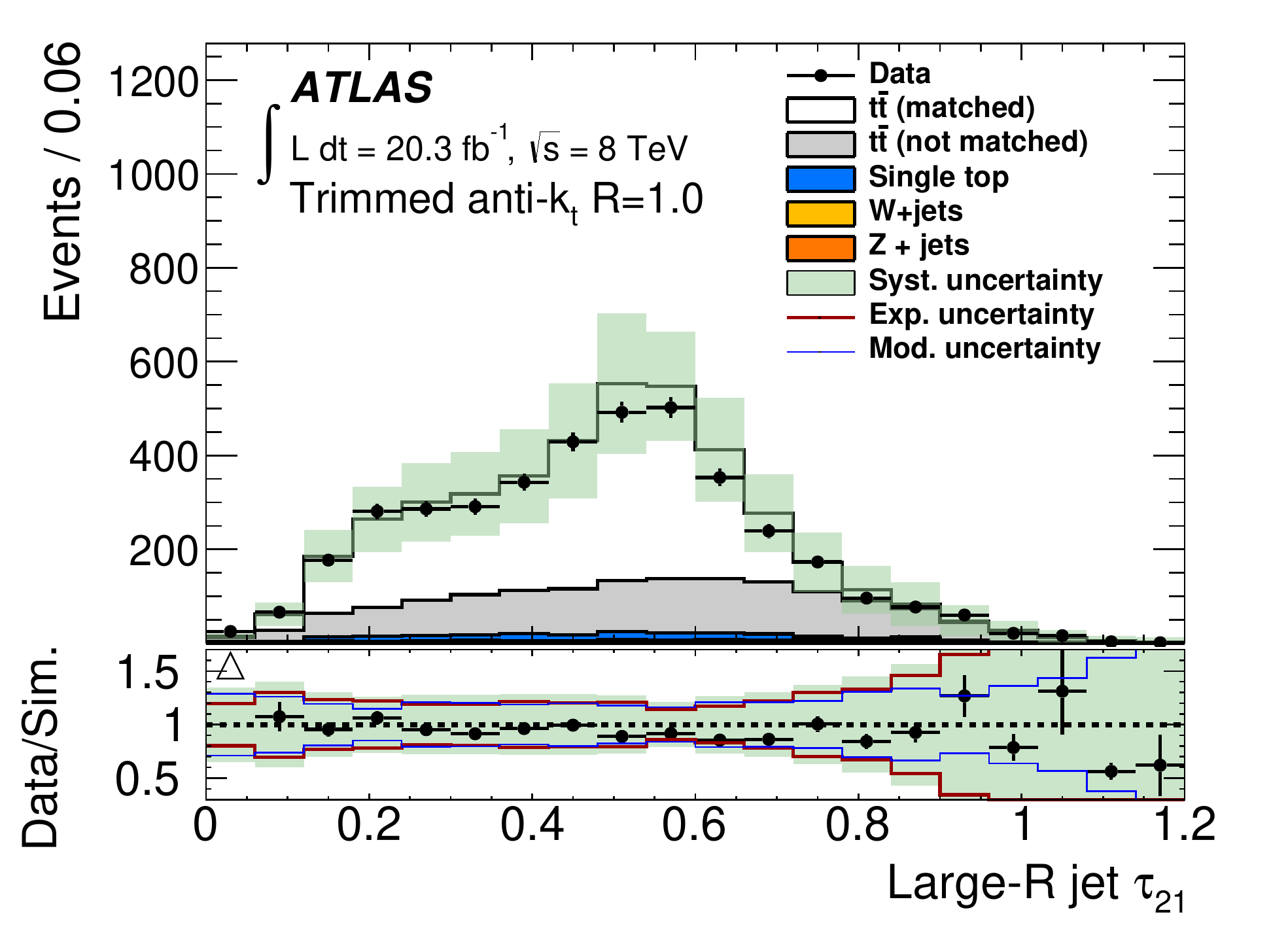}
}
\caption{Detector-level distribution of substructure variables of the highest-\pt trimmed
\akt $R=1.0$ jet with $\pt>350\GeV$ in events passing the
signal selection.
The splitting scales (a) \DOneTwo and (b) \DTwoThr and
the \Nsj ratios (c) \tauThrTwo and (d) \tauTwoOne are shown.
The vertical error bar indicates the statistical uncertainty
of the measurement. Also shown are distributions for simulated SM contributions
with systematic uncertainties (described in \secref{systematics}) indicated as a band.
The \ttbar prediction is split into a {\em matched} part for which the \largeR
jet axis is within $\Delta R = 0.75$ of the flight direction of a hadronically decaying
top quark and a {\em not matched} part for which this criterion does not hold.
The ratio of measurement to
prediction is shown at the bottom of each subfigure and the error bar and band
give the statistical and systematic uncertainties of the ratio, respectively.
The impacts of experimental and \ttbar modelling uncertainties are
shown separately for the ratio.
}
\label{fig:ctrl_akt_pretag_substr}
\end{centering}
\end{figure}

Different top taggers, based on these substructure variables,
are defined (\tabref{substrtaggers}).
A \largeR jet is tagged as a top jet by the corresponding tagger if the top-tagging criteria are
fulfilled.
Substructure tagger III was optimized for a search for $t\bar{t}$
resonances in the single-lepton channel~\cite{Aad:2013nca}. Compared to
other taggers, it has a rather high efficiency and misidentification rate because the
analysis required only little background rejection, as the background was already
much reduced by a lepton requirement.
Removing the mass requirement or the requirement on \DOneTwo further increases the efficiency
(taggers I and II).
The \WPT was optimized for a search for $tb$ resonances ($W^\prime$) in the
fully-hadronic decay mode~\cite{Aad:2014xra}, where a high background suppression
is required. The efficiency of this tagger is therefore lower than
that of taggers I to III. Taggers IV and V are introduced to study the effect
of a requirement on \DTwoThr in addition to the requirements of tagger III.

\begin{table}[p]
  \begin{center}
    \begin{tabular}{|c|l|}
      \hline
      Tagger & Top-tagging criterion \\
      \hline
 Substructure tagger I & \DOneTwo $> 40\GeV$ \\
 Substructure tagger II & $m > 100\GeV$ \\
 Substructure tagger III & $m > 100\GeV$ and \DOneTwo $> 40\GeV$ \\
 Substructure tagger IV & $m> 100\GeV$ and \DOneTwo $> 40\GeV$ and \DTwoThr $> 10\GeV$ \\
 Substructure tagger V & $m> 100\GeV$ and \DOneTwo $> 40\GeV$ and \DTwoThr $> 20\GeV$ \\
 \WPT & \DOneTwo $> 40\GeV$ and $0.4 <$ \tauTwoOne $< 0.9$ and \tauThrTwo $< 0.65$ \\
      \hline
    \end{tabular}
    \caption{Top taggers based on substructure variables of trimmed \akt $R=1.0$ jets.}
  \label{tab:substrtaggers}
  \end{center}
\end{table}

Distributions of the \pt and mass of trimmed \akt $R=1.0$ jets after applying
the six different taggers based on substructure variables are shown in
\figsref{ctrl_akt_posttag_substrPt}{ctrl_akt_posttag_substrM}, respectively, for
events passing the full signal selection of \secref{signalsample}.
While the \pt spectra look similar after tagging by the different taggers,
the mass spectra differ significantly due to the different substructure-variable
requirements imposed by the taggers. Taggers II to V require the
mass to be greater than $100\GeV$, and this cut-off is visible in the distributions.
The mass distribution after the $\DOneTwo>40\GeV$ requirement of Tagger I (\figref{ctrl_akt_posttag_substrM_I})
differs from that of the pre-tag distribution (\figref{ctrl_akt_pretag_fjm}),
because \DOneTwo is strongly correlated with the trimmed mass.
The impact of the $\DOneTwo>40\GeV$ requirement plus the \Nsj requirements of the \WPT
on the mass spectrum is visible by comparing \figref{ctrl_akt_posttag_substrM_WPT}
with the pre-tag distribution (\figref{ctrl_akt_pretag_fjm}).
The prominent peak around the top-quark mass shows that the sample after tagging is pure
in jets which contain all three decay products of the hadronic top-quark decay.

All distributions are described by the MC simulation
within uncertainties, indicating that the kinematics and the substructure of
tagged \largeR jets are well modelled by simulation.
The uncertainty in the \largeR jet \pt requiring a top tag is dominated by
the \largeR JES and the  parton-shower and \ttbar generator uncertainties.
Hence, the same uncertainties dominate in the different regions of the \pt spectrum
as before requiring a top tag (\secref{signalsample}).
The uncertainty on the \largeR-jet mass distributions is dominated by the
jet-mass scale uncertainty for all substructure taggers. The \largeR JES as well
as \ttbar modelling uncertainties also contribute, but have a smaller impact.
For all substructure taggers, the uncertainties in the substructure variables
used in the respective taggers have a non-negligible impact, in particular
for low \largeR jet masses, i.e.\ in the regime which is sensitive to the
modelling of not-matched \ttbar and extra radiation.

\begin{figure}[p]
\begin{centering}
\subfigure[Tagger I]{
\includegraphics[width=0.48\textwidth]{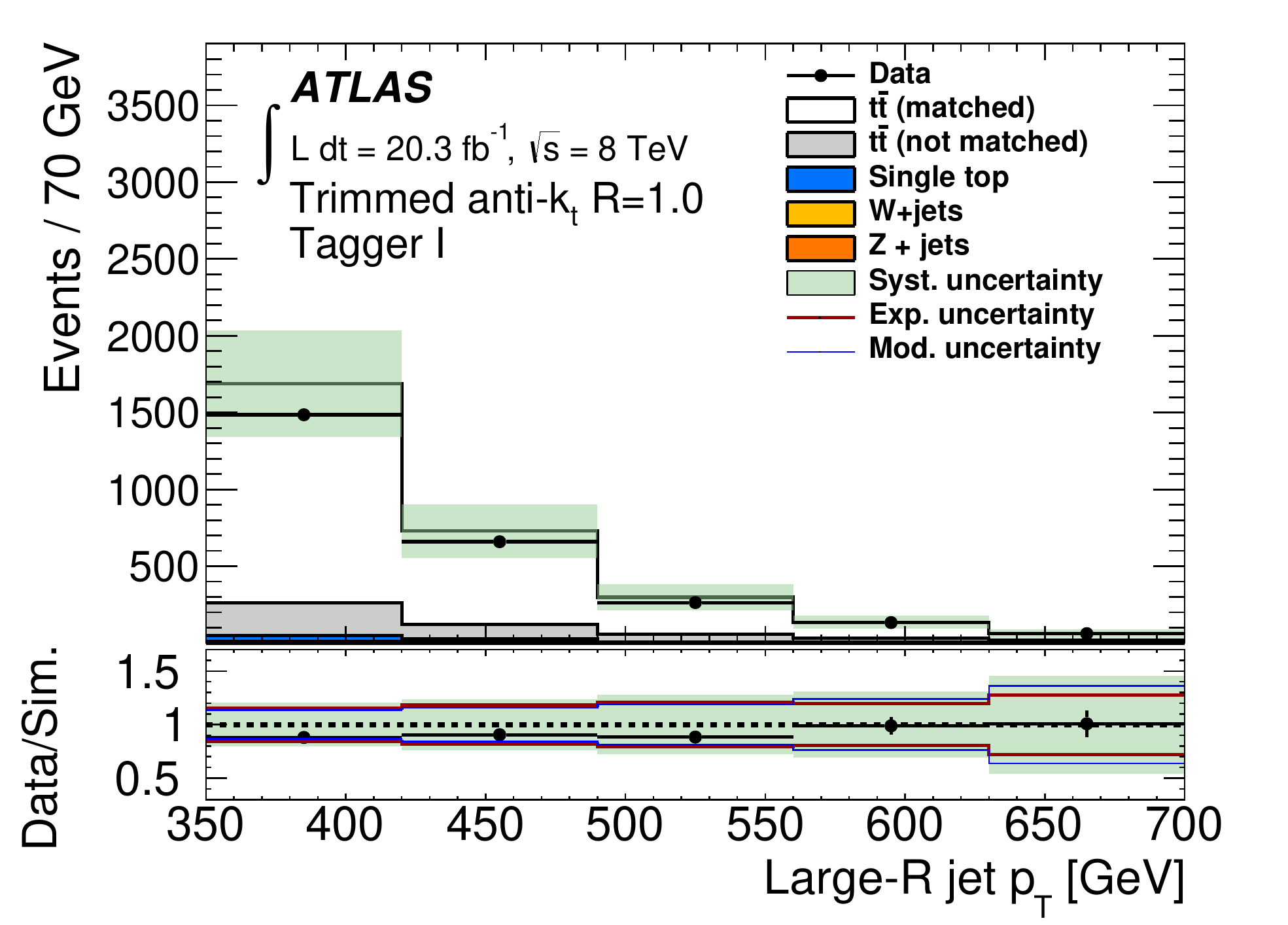}
}
\subfigure[Tagger II]{
\includegraphics[width=0.48\textwidth]{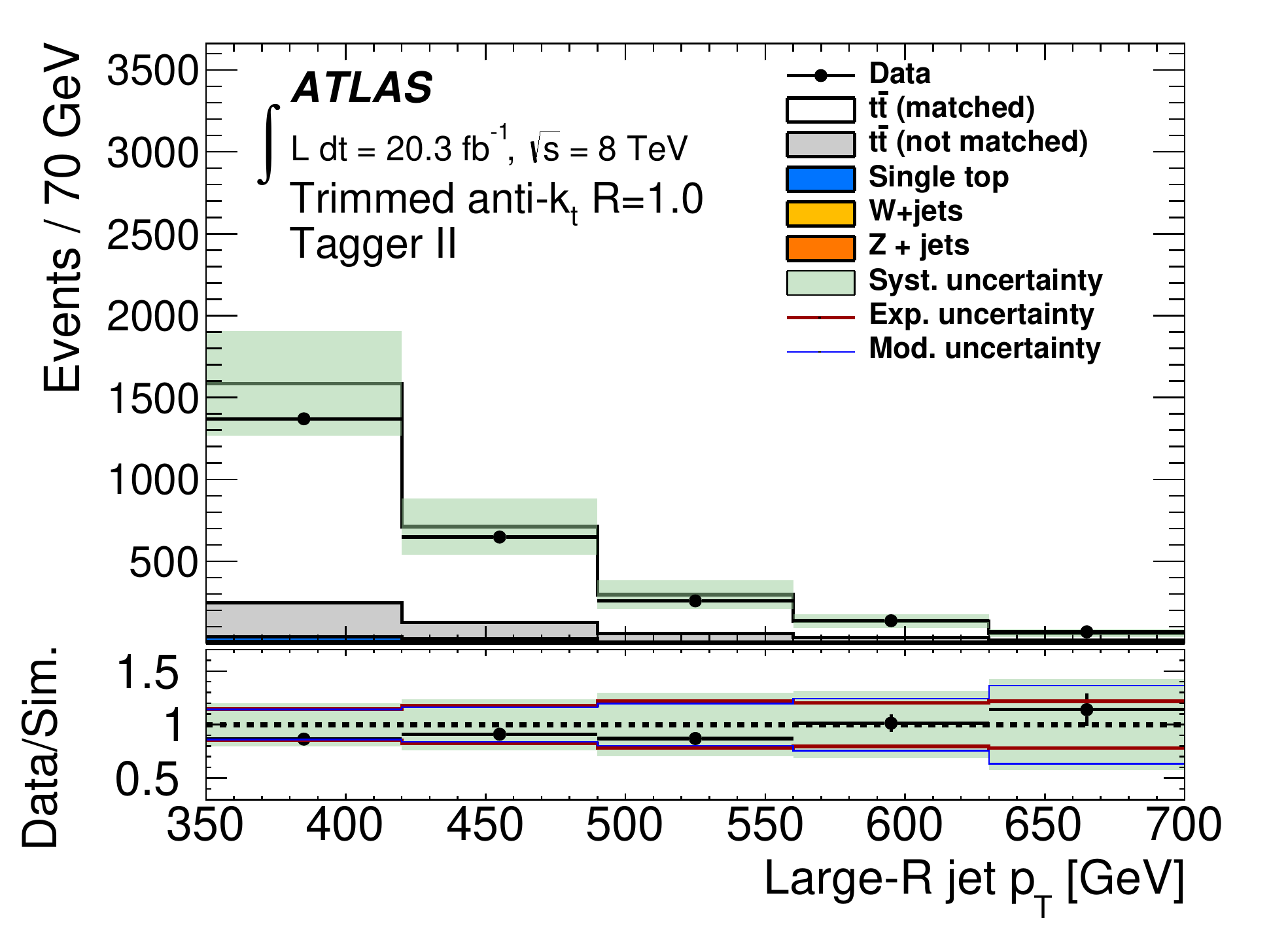}
} \\
\subfigure[Tagger III]{
\includegraphics[width=0.48\textwidth]{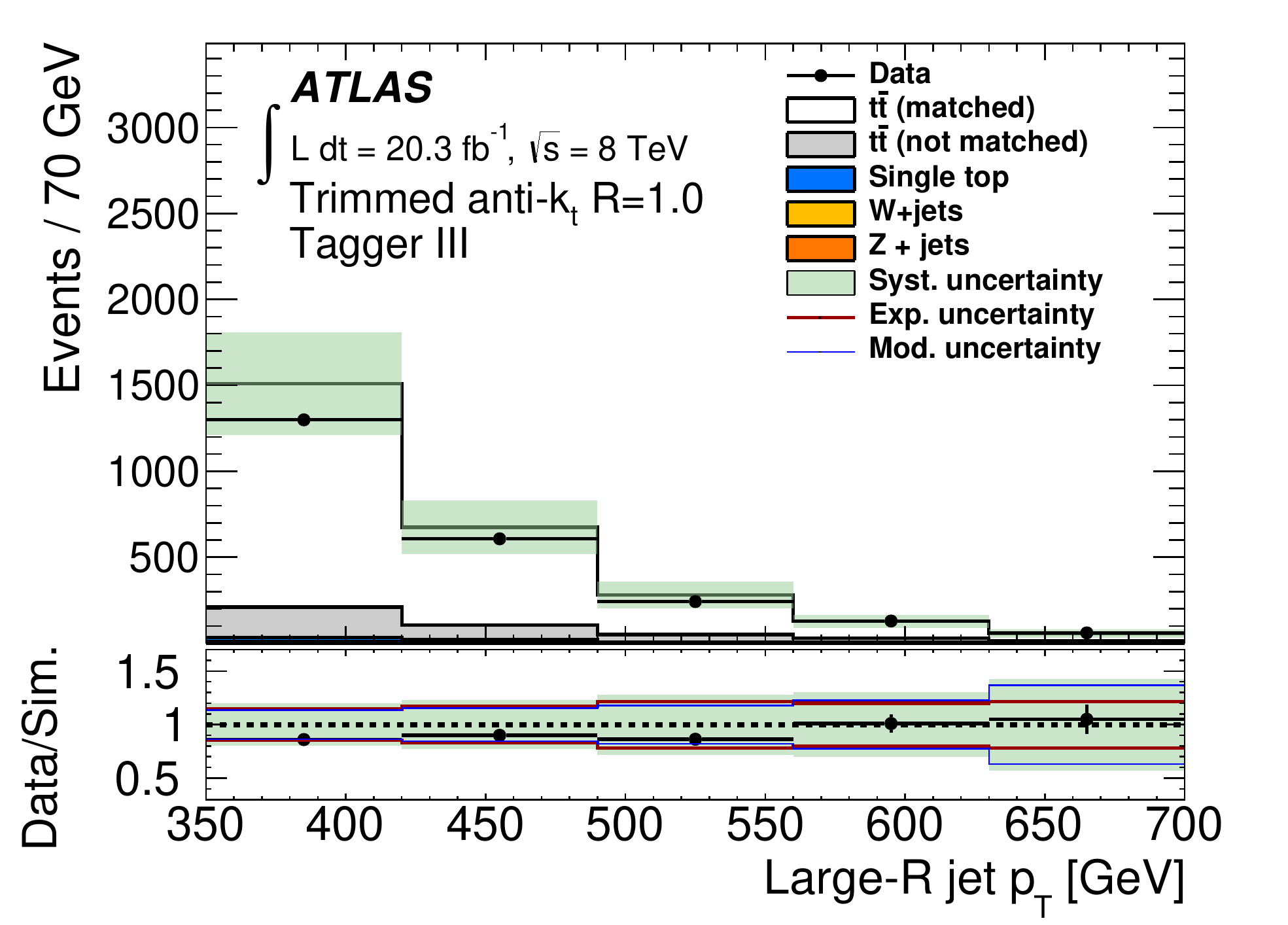}
}
\subfigure[Tagger IV]{
\includegraphics[width=0.48\textwidth]{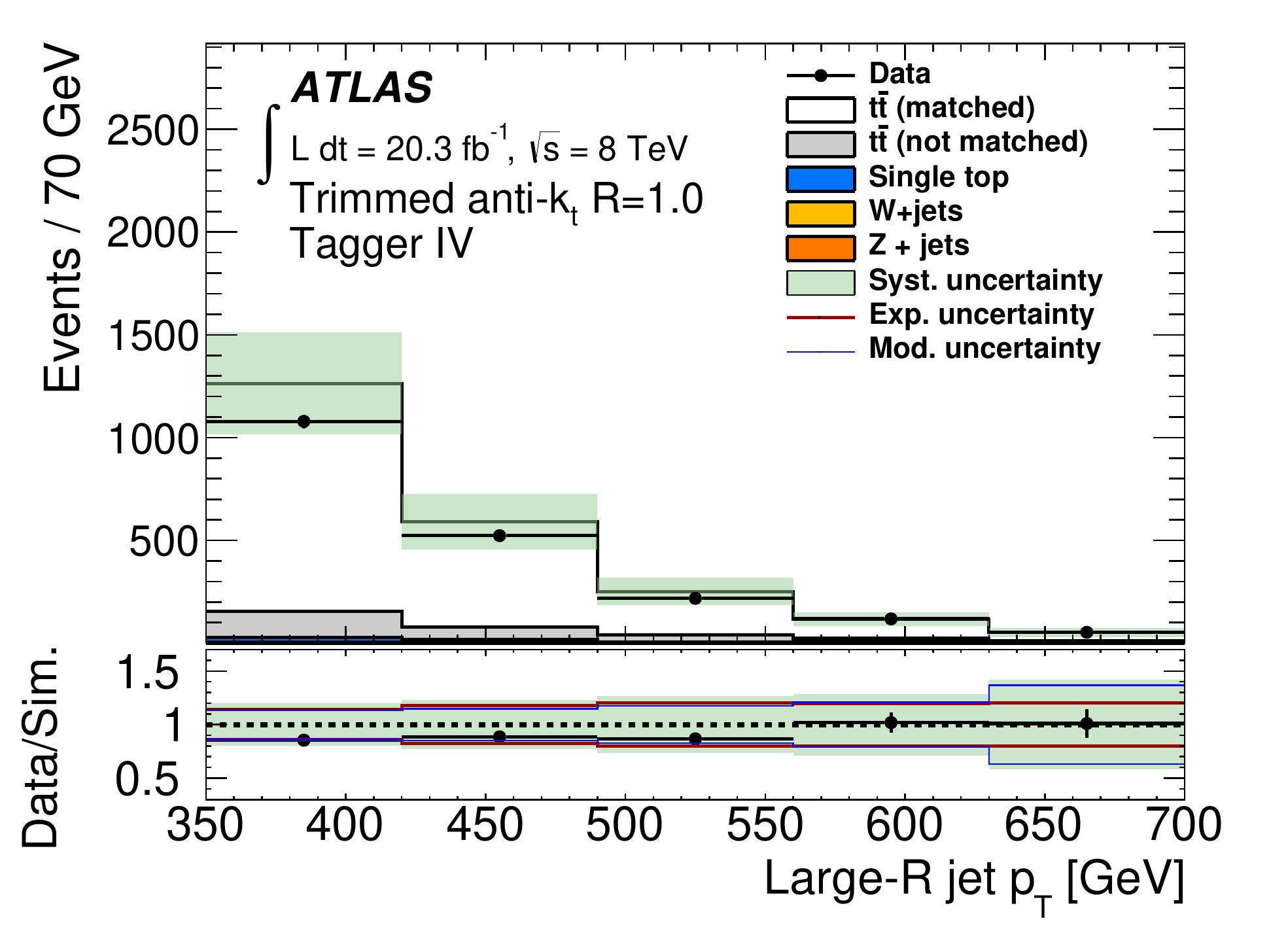}
} \\
\subfigure[Tagger V]{
\includegraphics[width=0.48\textwidth]{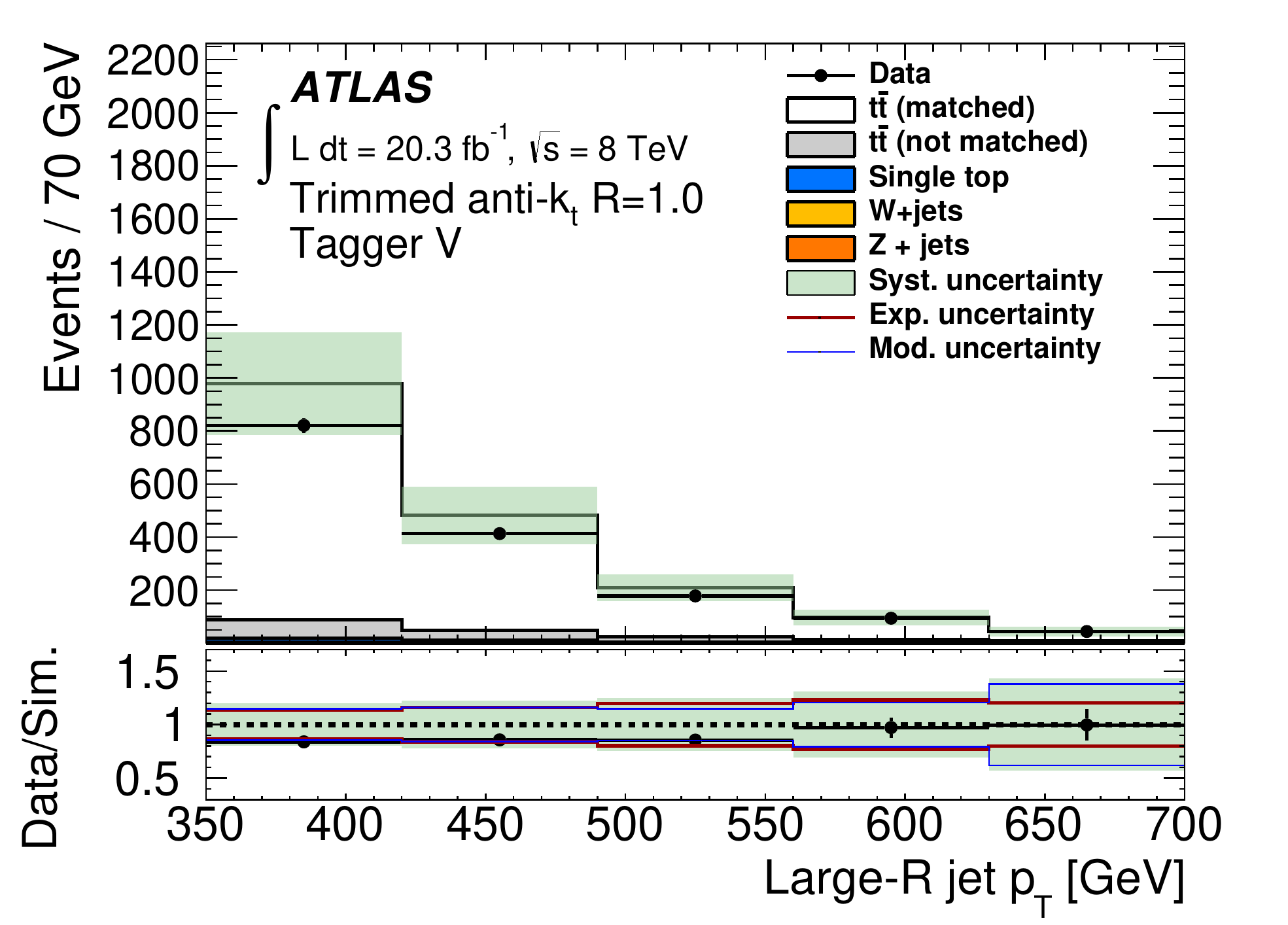}
}
\subfigure[\WPT]{
\includegraphics[width=0.48\textwidth]{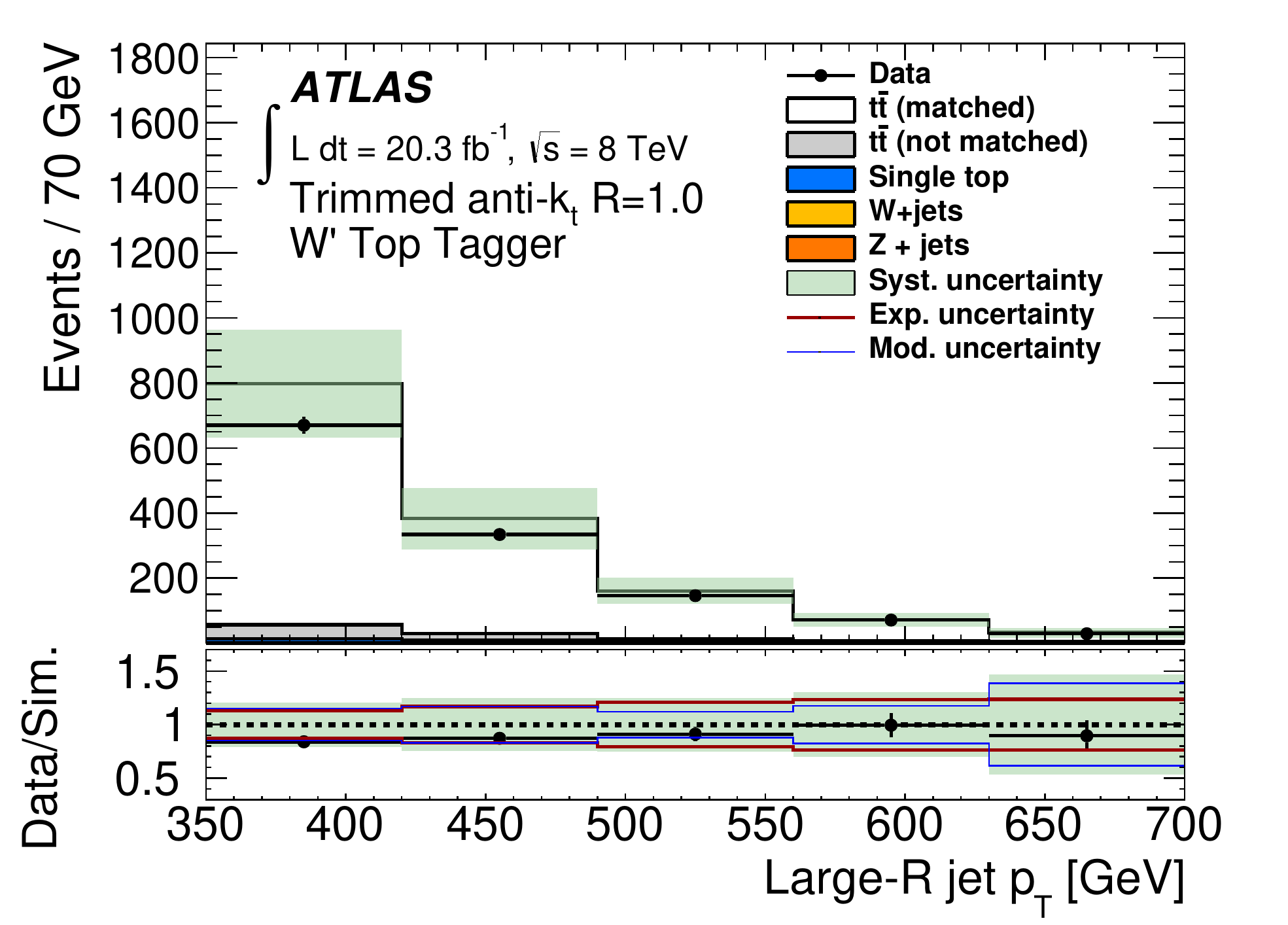}
}
\caption{Detector-level distributions of the \pt of the highest-\pt trimmed
\akt $R=1.0$ jet after tagging with different top taggers based on
substructure variables in events passing the
signal selection.
The vertical error bar indicates the statistical uncertainty
of the measurement. Also shown are distributions for simulated SM contributions
with systematic uncertainties (described in \secref{systematics}) indicated as a band.
The \ttbar prediction is split into a {\em matched} part for which the \largeR
jet axis is within $\Delta R = 0.75$ of the flight direction of a hadronically decaying
top quark and a {\em not matched} part for which this criterion does not hold.
The ratio of measurement to
prediction is shown at the bottom of each subfigure and the error bar and band
give the statistical and systematic uncertainties of the ratio, respectively.
The impacts of experimental and \ttbar modelling uncertainties are
shown separately for the ratio.
}
\label{fig:ctrl_akt_posttag_substrPt}
\end{centering}
\end{figure}

\begin{figure}[p]
\begin{centering}
\subfigure[Tagger I]{
\label{fig:ctrl_akt_posttag_substrM_I}
\includegraphics[width=0.48\textwidth]{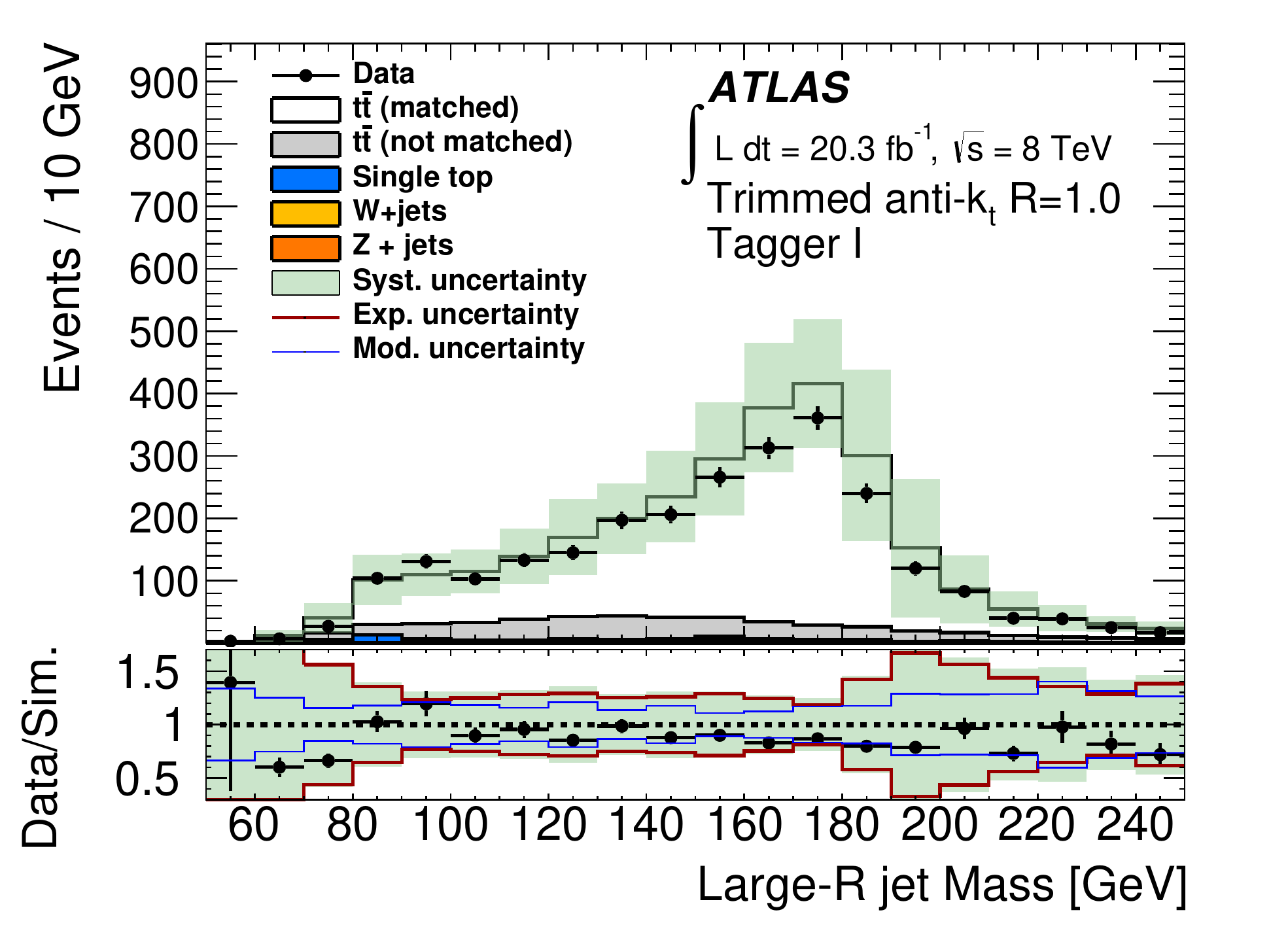}
}
\subfigure[Tagger II]{
\includegraphics[width=0.48\textwidth]{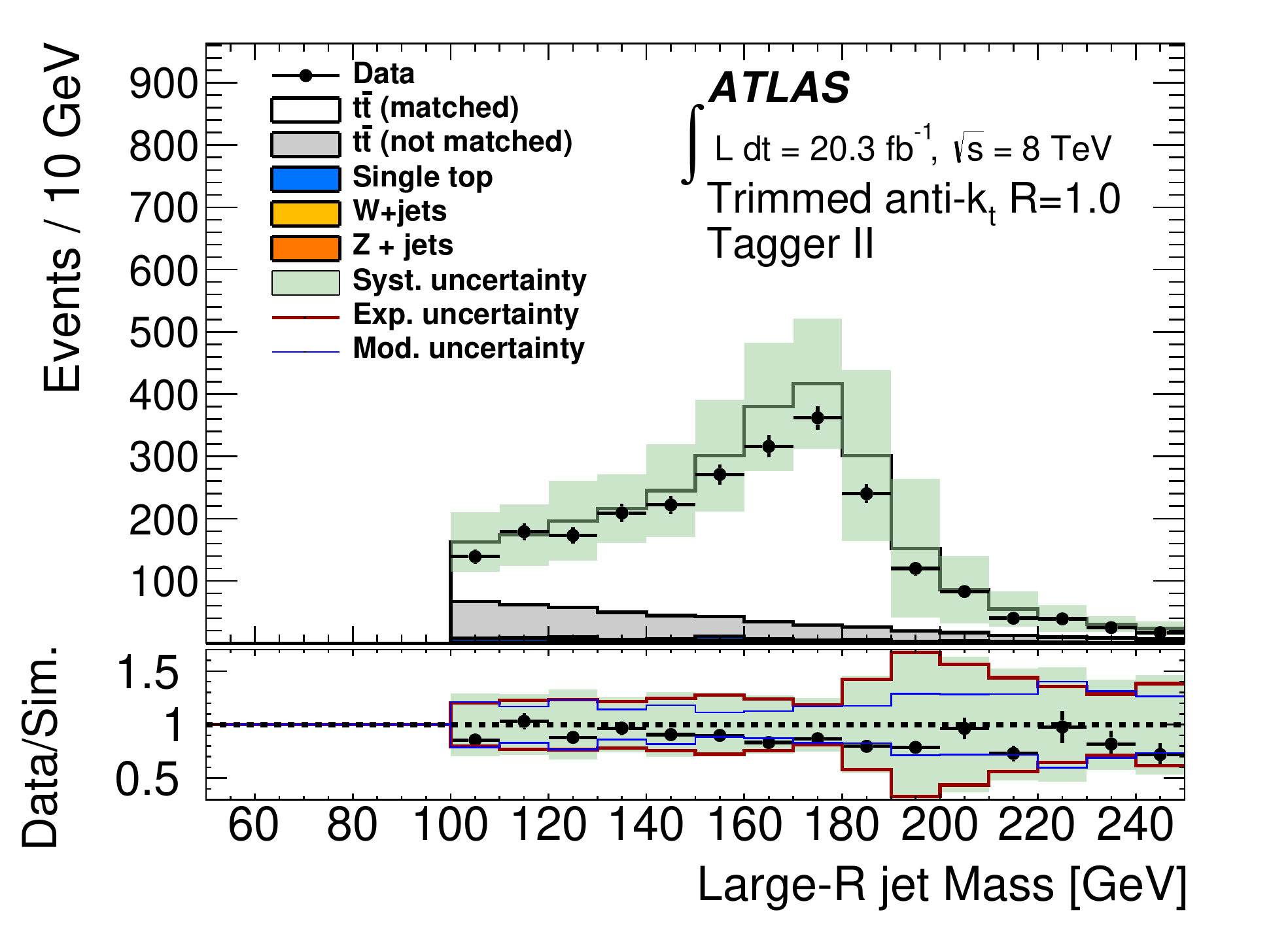}
} \\
\subfigure[Tagger III]{
\includegraphics[width=0.48\textwidth]{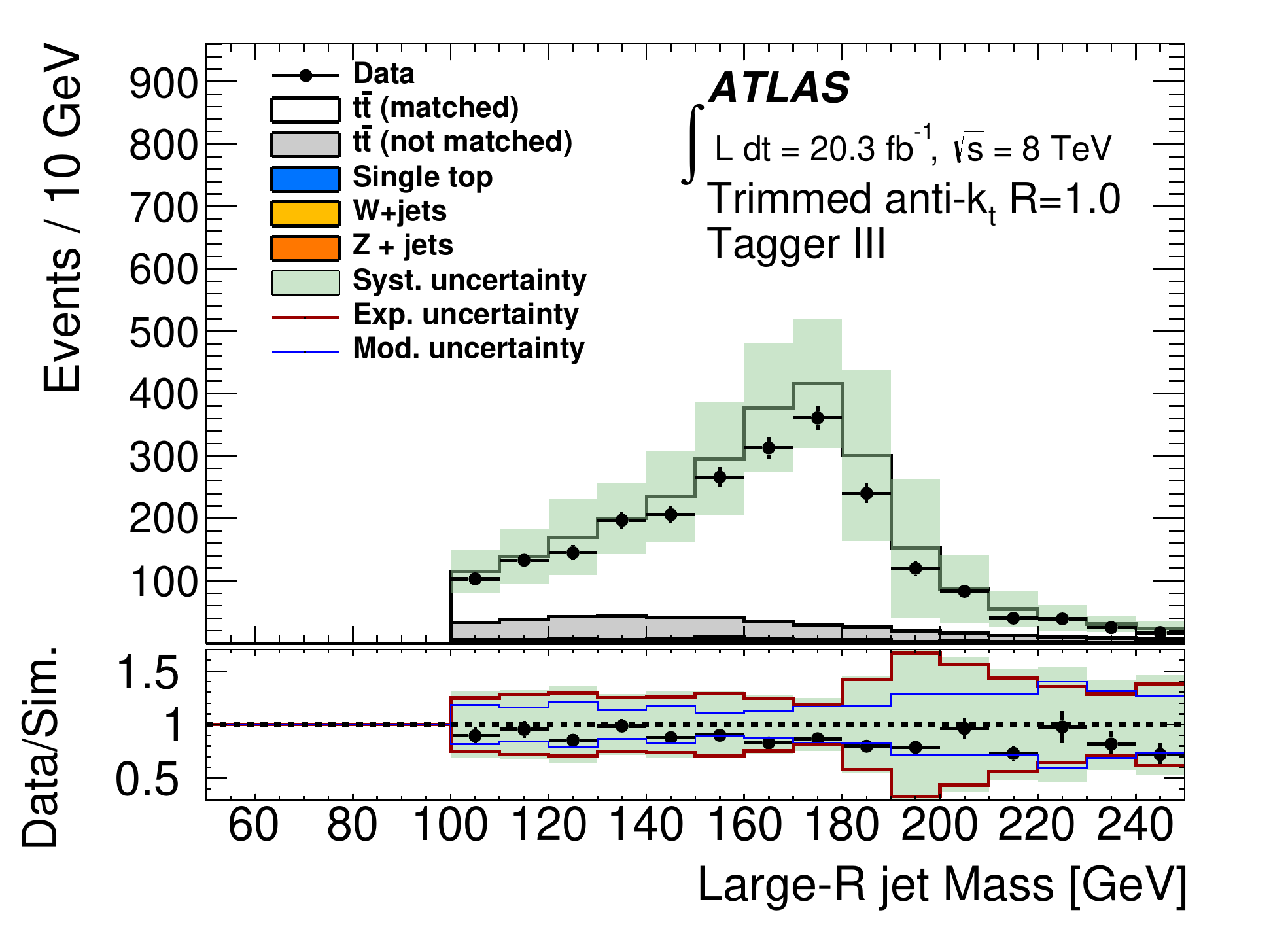}
}
\subfigure[Tagger IV]{
\includegraphics[width=0.48\textwidth]{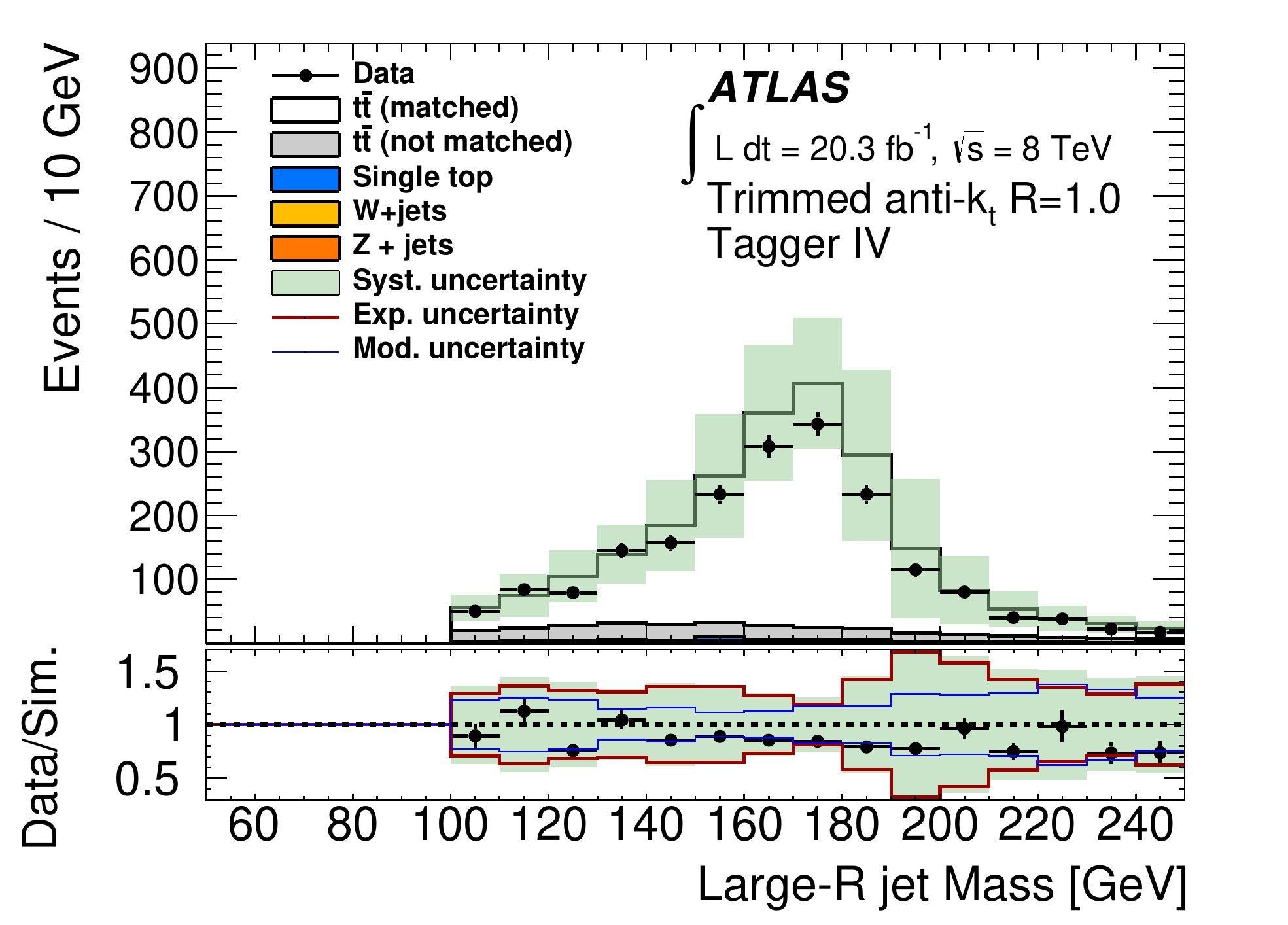}
} \\
\subfigure[Tagger V]{
\includegraphics[width=0.48\textwidth]{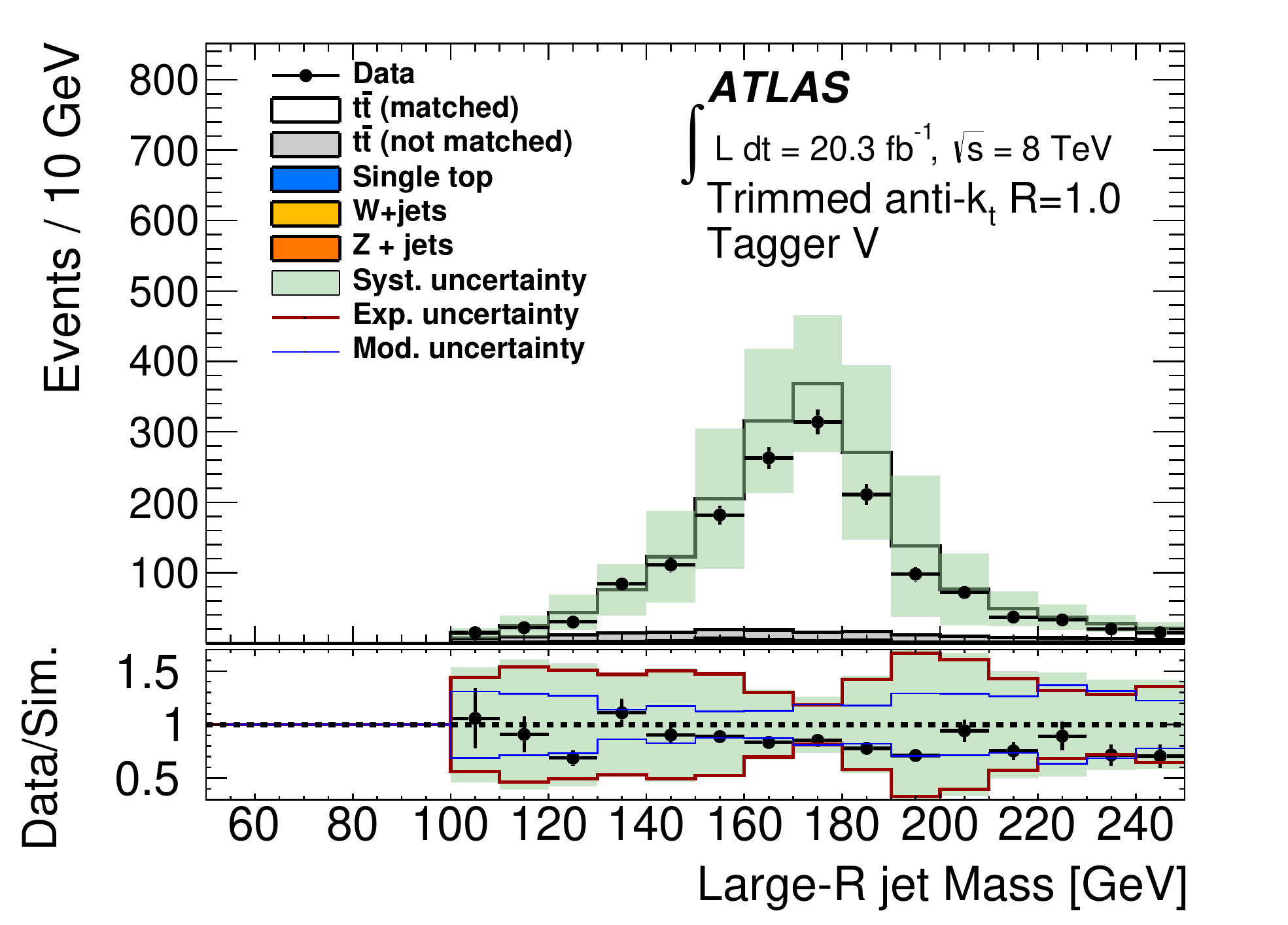}
}
\subfigure[\WPT]{
\label{fig:ctrl_akt_posttag_substrM_WPT}
\includegraphics[width=0.48\textwidth]{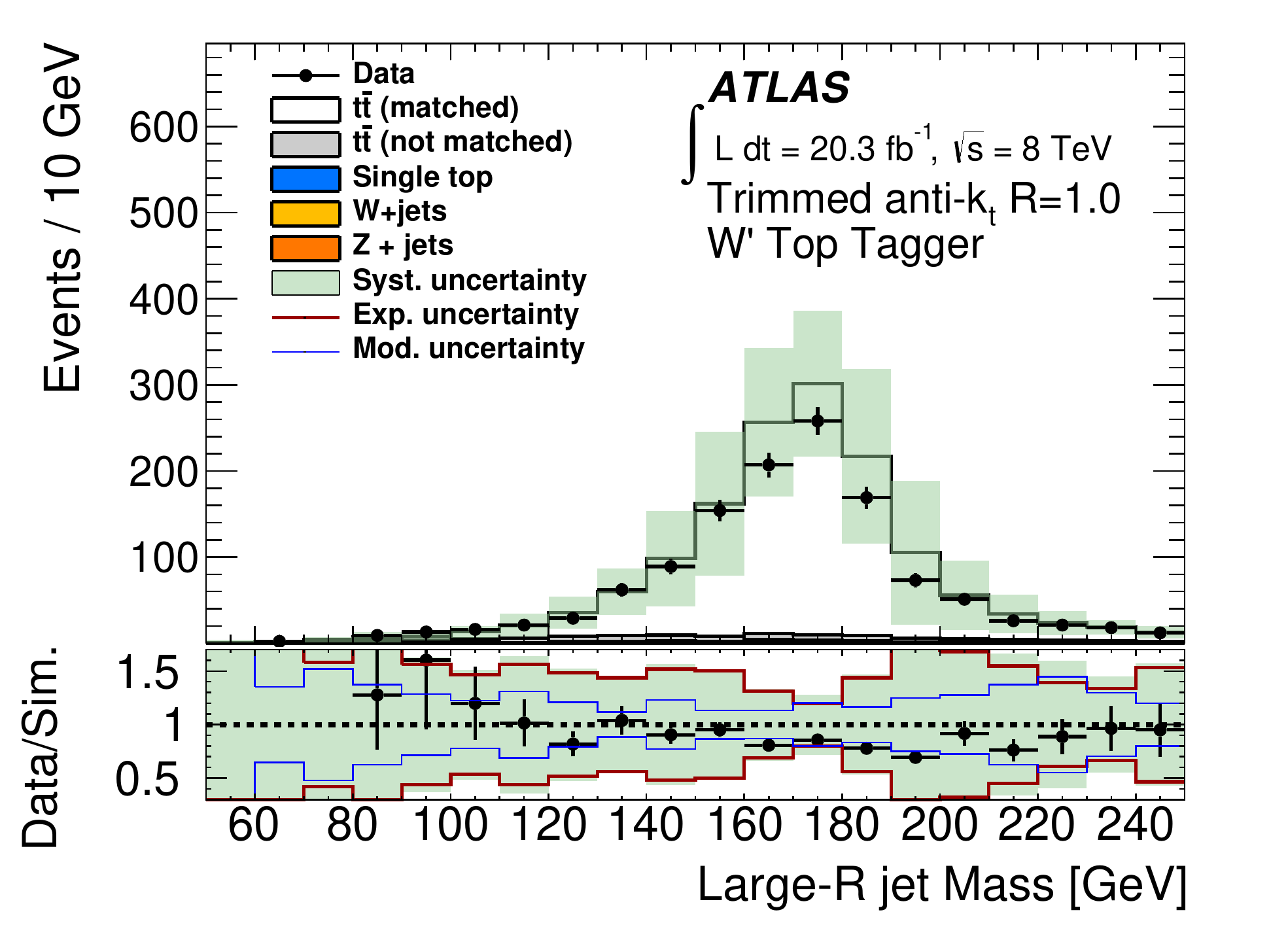}
}
\caption{Distribution of the mass of the highest-\pt trimmed
\akt $R=1.0$ jet after tagging with different top taggers based on
substructure variables in events passing the
signal selection.
The vertical error bar indicates the statistical uncertainty
of the measurement.
Also shown are distributions for simulated SM contributions
with systematic uncertainties (described in \secref{systematics}) indicated as a band.
The \ttbar prediction is split into a {\em matched} part for which the \largeR
jet axis is within $\Delta R = 0.75$ of the flight direction of a hadronically decaying
top quark and a {\em not matched} part for which this criterion does not hold.
The ratio of measurement to
prediction is shown at the bottom of each subfigure and the error bar and band
give the statistical and systematic uncertainties of the ratio, respectively.
The impacts of experimental and \ttbar modelling uncertainties are
shown separately for the ratio.
}
\label{fig:ctrl_akt_posttag_substrM}
\end{centering}
\end{figure}

\subsection{\sd}
\label{sec:sd}
In \textit{\sd} (SD)~\cite{Soper:2011cr,Soper:2012pb}, likelihoods are
separately calculated for the scenario that a given
\largeR jet originates from a hadronic top-quark decay and for the scenario that it originates from a background process.
The likelihoods are calculated from theoretical hypotheses, which for the
application in this paper correspond to the SM. The signal process
is the hadronic decay of a top quark and for the background process, the splitting
of hard gluons into $q\bar{q}$ is considered.
For signal and background, the effect of the parton
shower is included in the calculation of the likelihood. Subjets of the \largeR
jet are used as proxies for partons in the underlying model
and a weight is calculated for each
possible shower that leads to the observed subjet configuration.
This weight is proportional to the probability that the assumed initial
particle generates the final configuration, taking into account the SM amplitude
for the underlying hard process and the Sudakov form factors for the parton shower.
A discriminating variable $\chi$ is calculated as the ratio of the sum
of the signal-hypothesis weights to the sum of the background-hypothesis weights.
For a set $\{p_i^\kappa\}$ of $N$ observed subjet four-momenta $p_i^\kappa$, in which $i \in [1, N]$, the
value of $\chi$ is given by
\begin{equation}
\label{eq:SDchi}
\chi(\{p^\kappa_i\}) = \frac{\sum_{\rm perm.}\textrm{P}(\{p^\kappa_i\} | \textrm{signal})}{\sum_{\rm perm.}\textrm{P}(\{p^\kappa_i\} | \textrm{background})} \quad ,
\end{equation}
with $\textrm{P}(\{p^\kappa_i\} | \textrm{signal})$ being the weight for the
hypothesis that a signal process leads to the observed configuration $\{p^\kappa_i\}$
and the sum in the numerator is over all showers,
in which signal processes lead to this configuration. Similarly, the denominator sums the
weights for the background processes.
If $\chi$ is larger than a certain cut value, the \largeR jet is tagged as a
top jet. By adjusting the threshold value for $\chi$, the tagging efficiency can be
changed continuously.

The inputs to SD are the four-momenta of the subjets in the \largeR jet.
SD has an internal mechanism to suppress pile-up, which is based on the fact that the weights
of the likelihood ratio contain the probability that a subset of the subjets did
not originate from the hard interaction but are the result of pile-up. Details
can be found in Refs.~\cite{Soper:2011cr,Soper:2012pb}.
In this paper, trimmed \akt $R=1.0$ jets are used as input to SD, but the subjets of the
untrimmed jet are fed to the SD algorithm, and the kinematic properties (\pt, $\eta$)
of the trimmed jet are only used to preselect the signal sample.
This procedure avoids interference of the trimming with the
SD-internal pile-up suppression.

To obtain the best SD performance, the smallest structures in the flow of particles
should be resolved by the subjets used as input to SD.
Therefore, \CamKt $R=0.2$ subjets are used, as they
are the jets with the smallest radius parameter for which ATLAS calibrations
and calibration uncertainties have been derived~\cite{Aad:2013gja,Aad:2011he}.
Only the nine hardest subjets of the \largeR jet are used in the present study
to reduce the processing time per event, which grows with the number of subjets
considered in the calculation.
The signal weight is zero for \largeR jets with fewer than three subjets
because a finite signal weight requires the existence of at least three subjets
which are identified with the three partons from the top-quark decay.
To speed up the computation of the signal weights, the signal weight is set to
zero if no combination of at least three subjets can be found
that has an invariant mass within a certain range around the top-quark mass.
The rationale for this mass requirement is that subjet combinations outside of
this mass range would receive only a very small (but finite) weight due to the
Breit--Wigner distribution assumed for the signal hypothesis.
Similarly, a subset of the subjets which have a combined invariant mass
close to the top-quark mass must give an invariant mass within a given range around the $W$-boson
mass. Due to detector effects, the values of
these ranges around the top-quark mass and the \W-boson mass
must be tuned to optimize the performance and cannot be extracted
directly from the model. The values used in this study are a range of $40\GeV$
around a top-quark mass of $172\GeV$ and a range of $20\GeV$ around a \W-boson
mass of $80.4\GeV$.
For the background hypothesis,
no constraint on the subjet multiplicity is present and also no mass-range
requirements are imposed.

Distributions of the multiplicity and \pt of \CamKt $R=0.2$ subjets found in
the untrimmed \akt $R=1.0$ jets from the signal selection are shown
in \figref{ctrl_akt_pretag_subjets}.
These subjets are used as input to SD and must satisfy the kinematic
constraints $\pt>20\GeV$ and $|\eta|<2.1$.
The subjet multiplicity of the \largeR jet is shown in \figref{sd_subjetn}.
Most of the \largeR jets have two or three subjets and only a small fraction
have more than four subjets.
Of the \largeR jets, $41\%$ have fewer than three subjets and are hence
assigned a SD signal weight of zero.
The simulation describes the data within
statistical and systematic uncertainties indicating that the
input to the SD algorithm, the subjet multiplicity and kinematics, are well described.
For two and three subjets, the uncertainty is dominated by uncertainties
in the \largeR JES and the PDF. For one subjet and for four or more
subjets, as well, the uncertainty is dominated by the subjet energy-resolution
uncertainty.
The source of most events with only one subjet is not-matched \ttbar, for which
the modelling of additional low-\pt radiation exceeding the minimum
subjet \pt depends on the precision of the subjet energy scale and resolution.
The same effect is present for four or more subjets, because hadronically
decaying top quarks are expected to give rise to a distinct three-subjet
structure and additional subjets may be due to additional low-\pt radiation
close to the top quark.

\begin{figure}[!h]
\begin{centering}
\subfigure[]{
\label{fig:sd_subjetn}
\includegraphics[width=0.48\textwidth]{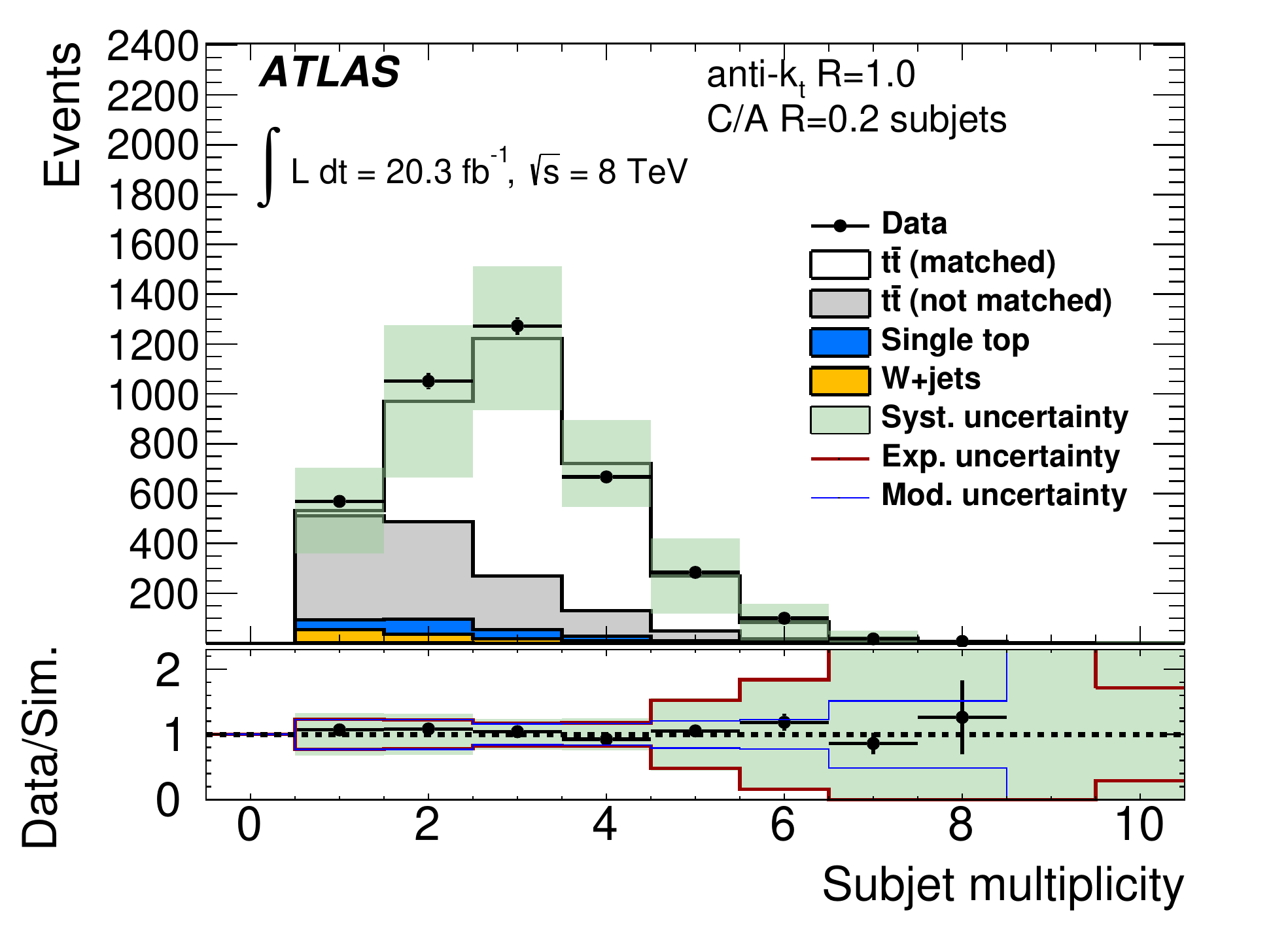}
}
\subfigure[]{
\label{fig:sd_subjetpt1}
\includegraphics[width=0.48\textwidth]{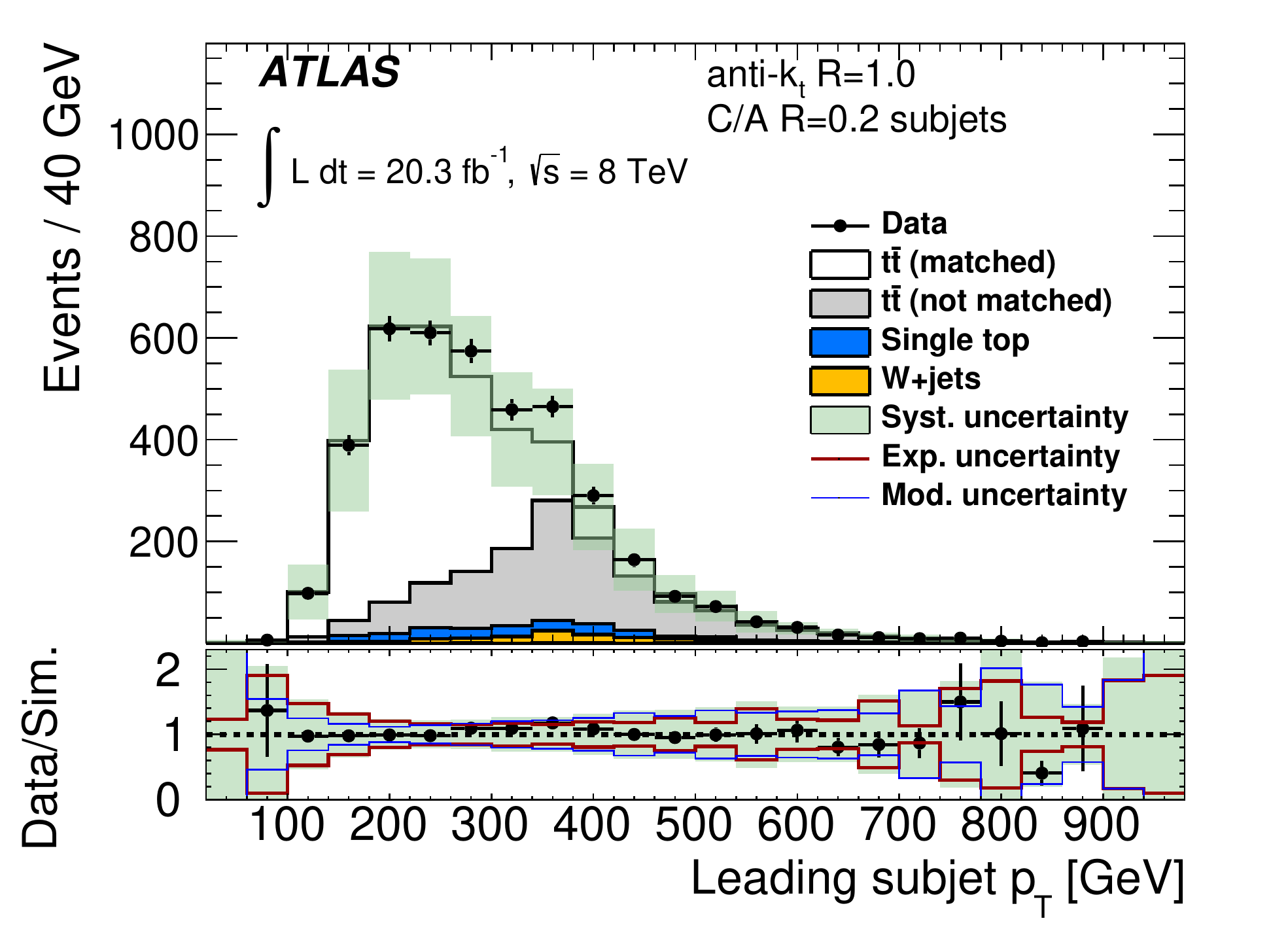}
} \\
\subfigure[]{
\label{fig:sd_subjetpt2}
\includegraphics[width=0.48\textwidth]{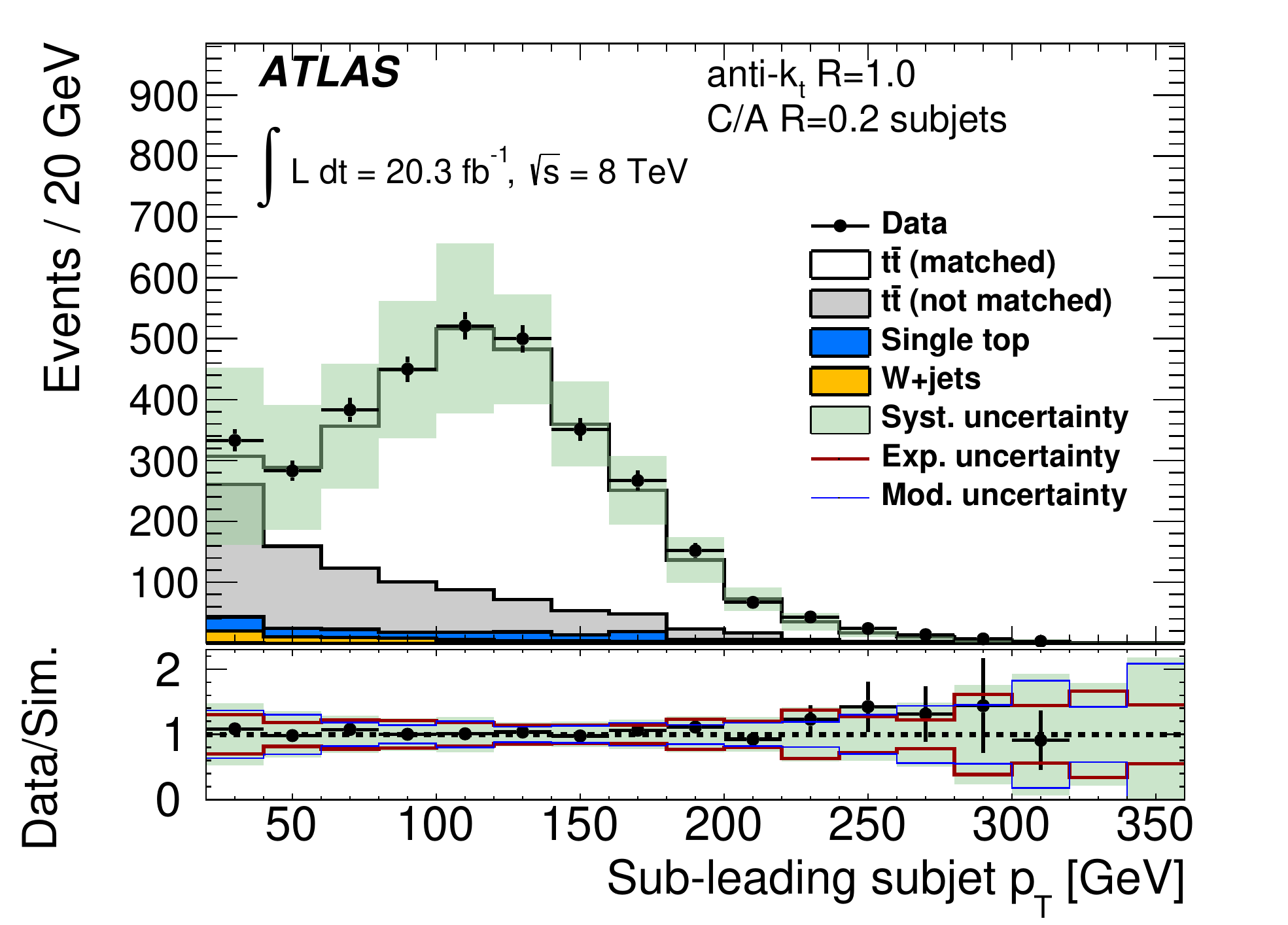}
}
\subfigure[]{
\label{fig:sd_subjetpt3}
\includegraphics[width=0.48\textwidth]{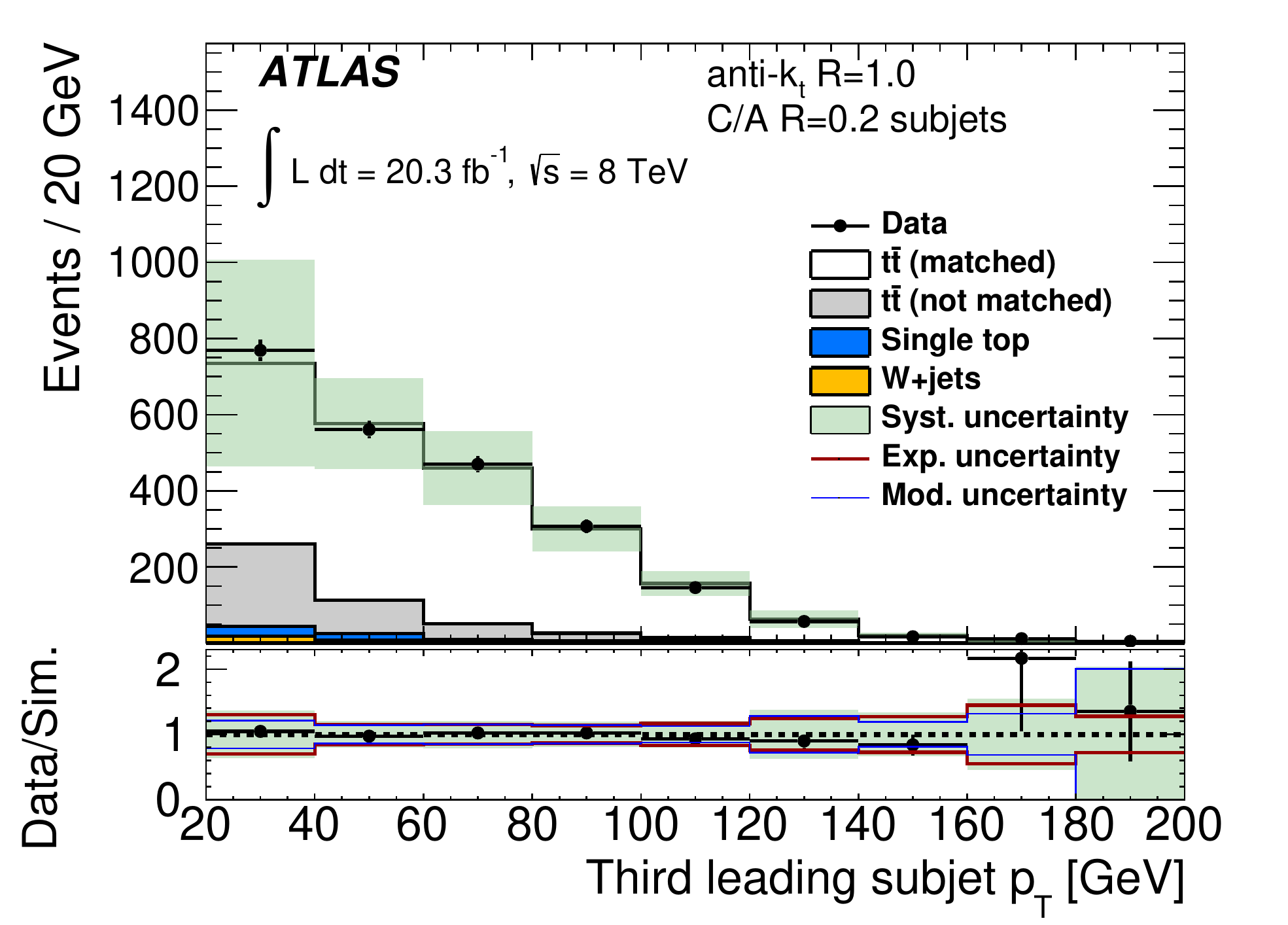}
}
\caption{Detector-level distributions of \CamKt $R=0.2$ subjets found in the untrimmed \akt
$R=1.0$ jet corresponding to the highest-\pt trimmed \akt $R=1.0$ jet with $\pt>350\GeV$
in the signal selection:
(a) the subjet multiplicity, and (b)
the \pt of the highest-\pt subjet, (c) the second-highest-\pt subjet,
and (d) the third-highest-\pt subjet.
The vertical error bar indicates the statistical uncertainty
of the measurement. Also shown are distributions for simulated SM contributions
with systematic uncertainties (described in \secref{systematics}) indicated as a band.
The \ttbar prediction is split into a {\em matched} part for which the \largeR
jet axis is within $\Delta R = 0.75$ of the flight direction of a hadronically decaying
top quark and a {\em not matched} part for which this criterion does not hold.
The ratio of measurement to
prediction is shown at the bottom of each subfigure and the error bar and band
give the statistical and systematic uncertainties of the ratio, respectively.
The impacts of experimental and \ttbar modelling uncertainties are
shown separately for the ratio.
}
\label{fig:ctrl_akt_pretag_subjets}
\end{centering}
\end{figure}

The \pt distributions of the three hardest subjets are shown in \figrange{sd_subjetpt1}{sd_subjetpt3}.
The \pt of the highest-\pt subjet is larger than $\approx\!100\GeV$ and has a broad peak from $200$ to $400\GeV$.
The shoulder at $370\GeV$ is caused by \largeR jets from not-matched \ttbar and \Wjets
background, as many of these jets have only one subjet, as shown in \figref{sd_subjetn},
and in that case the single subjet carries most of the momentum of the
\largeR jet, i.e.\ most of the momentum is concentrated in the core of the jet.
Therefore, the shoulder at $370\GeV$ is due to the requirement $\pt>350\GeV$ for the \largeR jet.
The systematic uncertainty in
the region mainly
populated by jets with one dominant subjet
($\pt > 350\GeV$) or by jets with many subjets ($100<\pt<150\GeV$)
in \figref{sd_subjetn}
has sizeable
contributions from the modelling of the subjet properties, here the subjet
energy scale.
While the \largeR JES also contributes for $100<\pt<150\GeV$, it is
dominant for jets mainly showing the expected distinct two-subjet or
three-subjet structure ($150<\pt<350\GeV$).
For $\pt>500\GeV$, the largest uncertainty results from the difference
between the \ttbar generators, as this is the main source of uncertainties
for the modelling of \ttbar events in the upper range of the \pt spectrum studied.

For the second-highest subjet \pt, the background distribution peaks near the $20\GeV$
threshold. These are subjets in \largeR jets with only two subjets
where the highest-\pt subjet carries most of the \largeR jet momentum. These
asymmetric configurations, where the highest-\pt subjet carries a much larger
\pt than the second-highest-\pt subjet, are seen mainly for the not-matched \ttbar and
\Wjets processes.
The acceptance limit at $20\GeV$ cuts into the \pt distributions of all but the
highest-\pt subjet, as also seen for the distribution of the third-highest-\pt subjet.
The uncertainties in the distributions of the second-highest-\pt and
third-highest-\pt subjet are again dominated by the uncertainty of the
subjet modelling, i.e.\ the subjet energy-resolution and energy-scale modelling, for low values of
\pt (mostly populated by not-matched \ttbar events) and for high values of \pt.
For intermediate values ($60$--$150\GeV$ for the second-highest-\pt subjet
and $40$--$100\GeV$ for the third-highest-\pt subjet), where jets with a distinct
top-like subjet structure dominate the distributions, the \largeR JES uncertainty dominates.
If $40<\pt<60\GeV$ for the second-highest subjet, the \largeR JES uncertainty contributes
significantly, but does not dominate due to significant contributions from the PDF and generator
uncertainties.

The following invariant masses of combinations of the \CamKt $R=0.2$ subjets are shown in
\figref{ctrl_akt_pretag_subjetmasses} for events fulfilling the
signal selection: the mass of the two highest-\pt subjets, $m_{12}$,
the mass of the second-highest-\pt and third-highest-\pt subjet, $m_{23}$, and the mass of the
three hardest subjets, $m_{123}$.
These distributions illustrate some of the masses built from subjet combinations which are
used by SD to reject subjet combinations that lead to masses outside the top-quark
and \W-boson mass ranges.
Also these distributions are described by the simulation within statistical and
systematic uncertainties and give further confidence in the description of the inputs
to the SD algorithm.
The uncertainty for large values of $m_{12}$, $m_{23}$ and $m_{123}$,
i.e.\ for values larger than $140\GeV$, $120\GeV$ and $165\GeV$, respectively,
is dominated by the subjet energy-scale uncertainty, consistent with this
uncertainty also being dominant for large values of the subjet transverse momenta
(\figref{ctrl_akt_pretag_subjets}).
The parts of the distributions which are populated with jets showing primarily
a distinct top-like substructure again show large contributions from
the \largeR JES uncertainty ($60<m_{12}<140\GeV$, $80<m_{23}<120\GeV$,
$135<m_{123}<165\GeV$), where the ISR/FSR and the subjet JES uncertainties also contribute for
$m_{23}$.
For lower values, the three different invariant masses are all sensitive to radiation
effects in a region populated by not-matched \ttbar events, i.e.\ jets which do not originate from
a hadronically decaying top quark.
ISR/FSR uncertainties contribute to $20<m_{12}<30\GeV$,
the subjet energy resolution contributes significantly to $m_{23}<60\GeV$
and $m_{123}<135\GeV$, and also the PDF uncertainty has an increasing effect with
increasing $m_{23}$ for $10<m_{23}<60\GeV$ with the uncertainty from the subjet energy
resolution decreasing with increasing $m_{23}$.
For $20<m_{12}<30\GeV$, the \largeR JES uncertainty dominates the total uncertainty together with the ISR/FSR uncertainty.
For $m_{23}<10\GeV$, the uncertainty is dominated by the uncertainty on the
subjet energy resolution and the differences between the \ttbar generators.
For $30<m_{12}<60\GeV$, the choice of \ttbar generator and the \largeR JES dominate the total uncertainty.

\begin{figure}[!h]
\begin{centering}
\subfigure[]{
\includegraphics[width=0.48\textwidth]{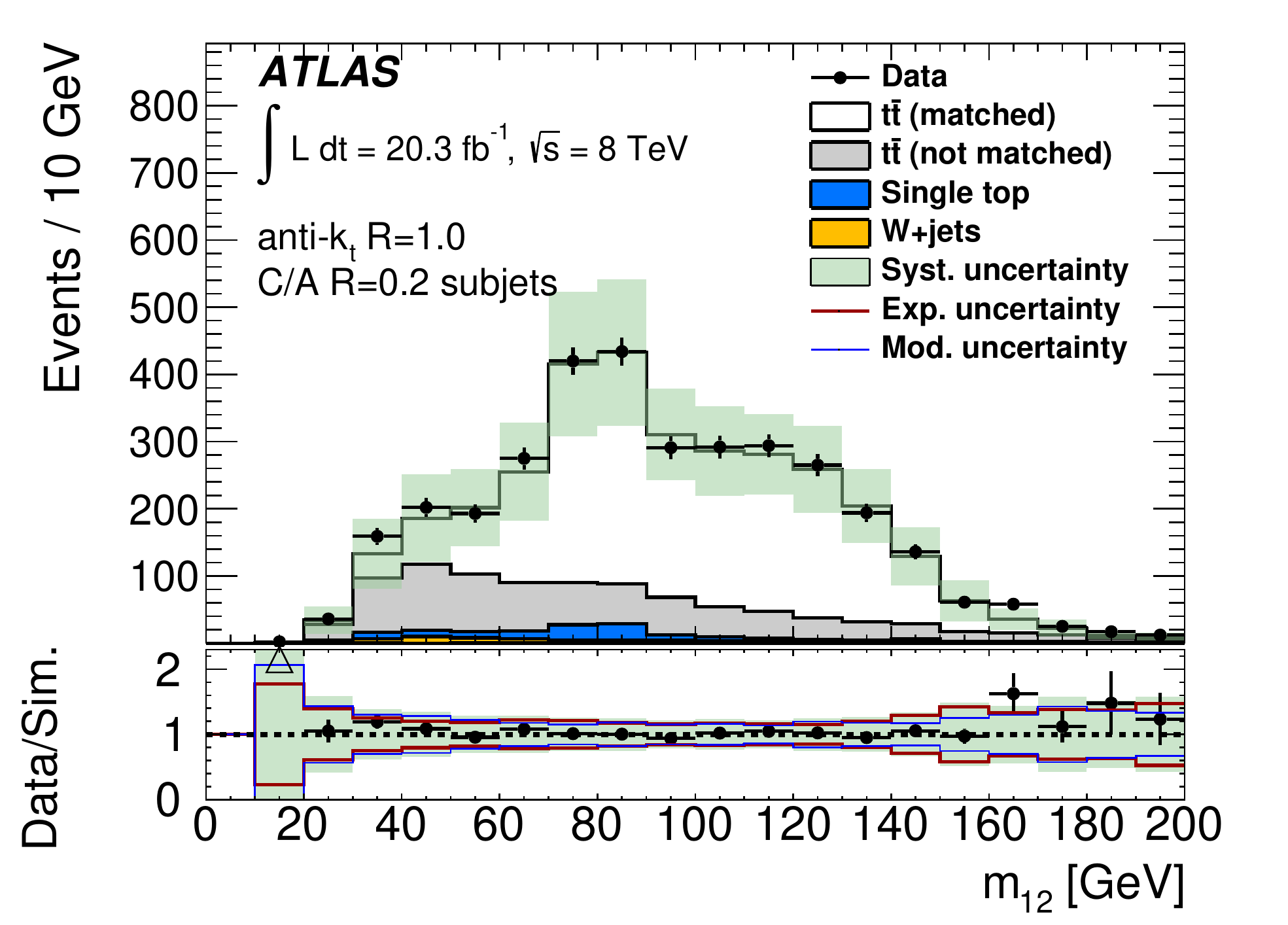}
}
\subfigure[]{
\includegraphics[width=0.48\textwidth]{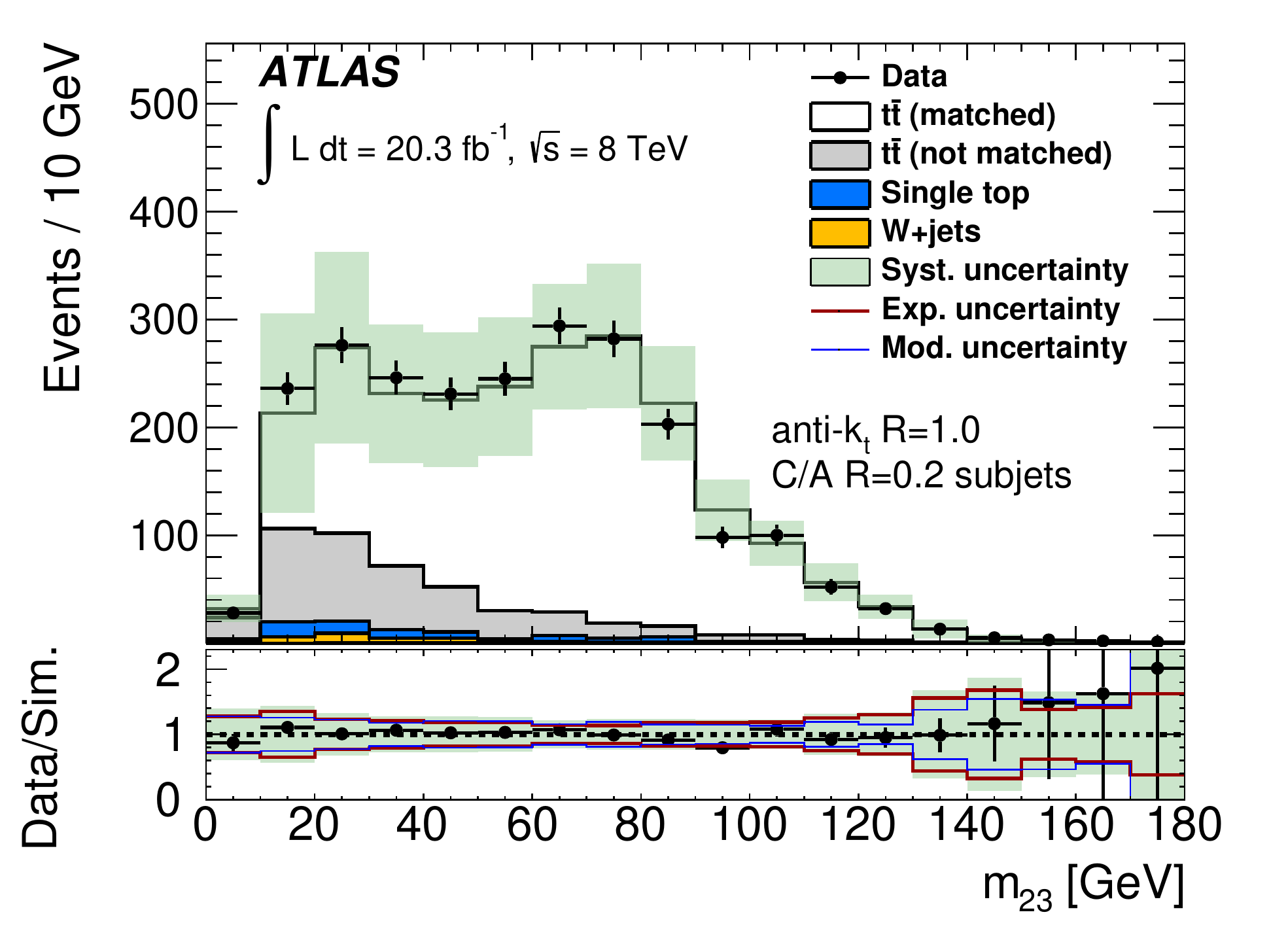}
} \\
\subfigure[]{
\includegraphics[width=0.48\textwidth]{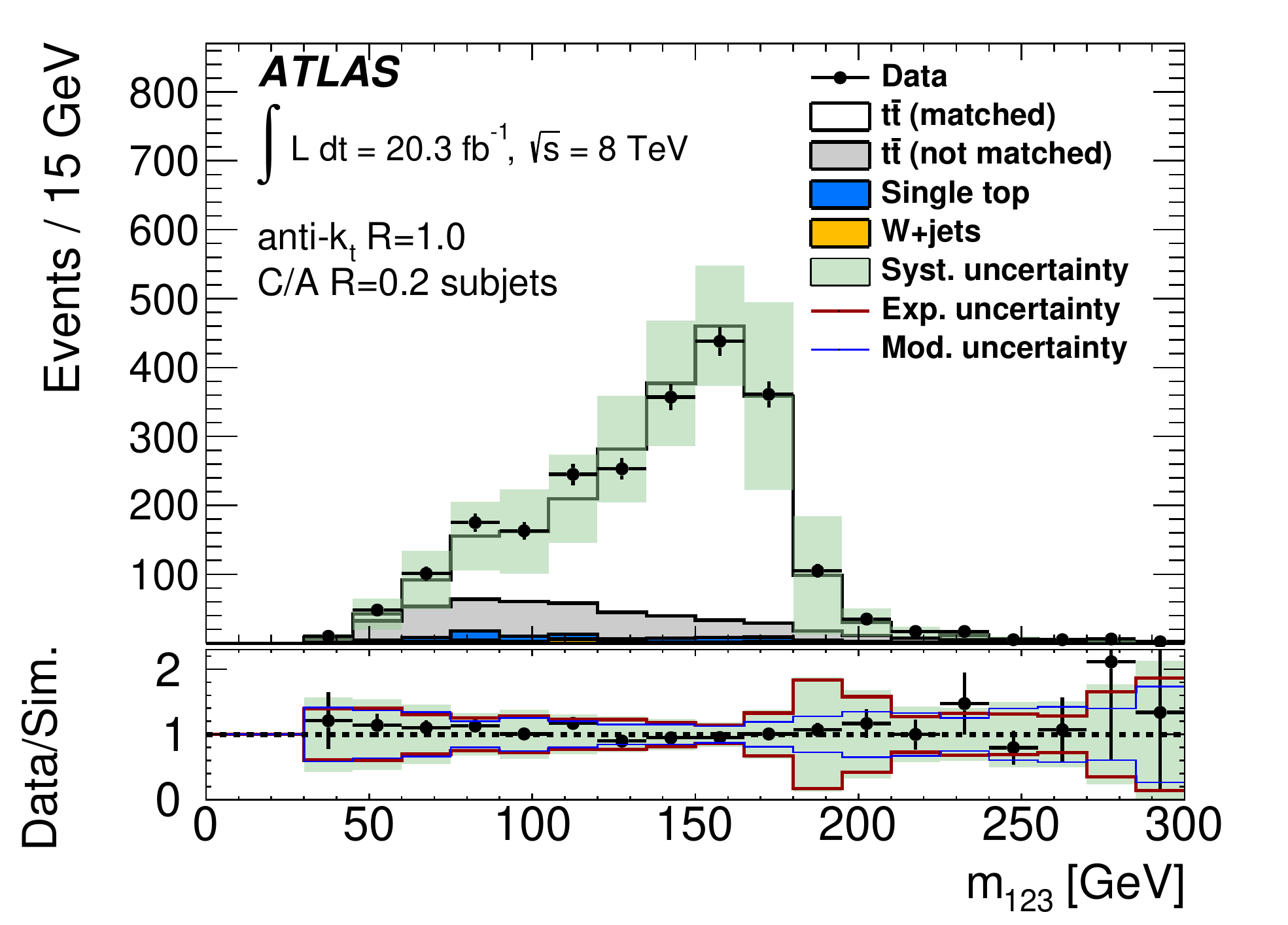}
}
\caption{
Distributions of invariant masses of combinations of \CamKt $R=0.2$ subjets
found in the untrimmed \akt $R=1.0$ jet
corresponding to the highest-\pt \akt $R=1.0$ trimmed jet with $\pt>350\GeV$
in the signal selection:
(a) the invariant mass of the highest-\pt subjet and the second-highest-\pt subjet, (b)
the mass of the second- and third-highest-\pt subjets, (c) the mass of the three
highest-\pt subjets.
The vertical error bar indicates the statistical uncertainty
of the measurement. Also shown are distributions for simulated SM contributions
with systematic uncertainties (described in \secref{systematics}) indicated as a band.
The \ttbar prediction is split into a {\em matched} part for which the \largeR
jet axis is within $\Delta R = 0.75$ of the flight direction of a hadronically decaying
top quark and a {\em not matched} part for which this criterion does not hold.
The ratio of measurement to
prediction is shown at the bottom of each subfigure and the error bar and band
give the statistical and systematic uncertainties of the ratio, respectively.
The impacts of experimental and \ttbar modelling uncertainties are
shown separately for the ratio.
}
\label{fig:ctrl_akt_pretag_subjetmasses}
\end{centering}
\end{figure}

The distributions of the SD weights and the ratio of the weights, i.e.\ the final
discriminant $\chi$ (\equref{SDchi}), are shown in
\figref{ctrl_akt_pretag_sdweights} for events fulfilling the signal-selection criteria.
For $\approx\!60\%$ of the \largeR jets, the signal weight is zero because
there are fewer than three subjets or the top-quark or \W-boson mass-window
requirements are not met. These cases are not shown
in \figref{ctrl_akt_pretag_sdweights}.
The natural logarithm of the sum
$\sum_{\rm perm.}\textrm{P}(\{p^\kappa_i\}| \textrm{signal})$ of all weights
obtained with the assumption that the subjet configuration in the \largeR jet
is the result of a hadronic top-quark decay is shown in \figref{sd_signal}.
The logarithm of the sum of all weights for the background hypothesis is shown in
\figref{sd_background}. For the signal hypothesis the distribution peaks
between $-23$ and $-21$, while for the background hypothesis the peak is at lower
values, between $-26$ and $-25$.
The logarithm of the ratio of the sums of the weights $\chi$, is shown in \figref{sd_chi}.
The $\ln\chi$ distribution is also shown in \figref{sd_chipt} for
\largeR jet $\pt>550\GeV$, which defines a different kinematic regime
for which the probability to contain all top-quark decay products in the
\largeR jet is higher than for the lower threshold of $350\gev$.
All distributions of SD output variables are described by simulation within
the statistical and systematic uncertainties.
The subjet energy-resolution uncertainty dominates for low values of the
logarithm of the SD signal weight (region $<-26$), the logarithm of
the SD background weight (region $<-30$) and $\ln\chi$ (region
$<1$ in \figref{sd_chi}). Hence, this uncertainty dominates, consistent
with the observations in previous figures, in the phase space not primarily
populated by jets from hadronically decaying top quarks.
The \largeR JES contributes significantly for the central parts of the
signal-weight distribution, i.e.\ from $-26$ to $-23$ in \figref{sd_signal}, and $\ln\chi$,
i.e.\ from $1$ to $5$ in \figref{sd_chi}.
In the region, $1 < \ln\chi < 5$, there are equally large contributions to the total
uncertainty from the subjet energy resolution, ISR/FSR, and the parton-shower modelling uncertainties.
For larger values of the signal weight, from $-23$ to $-21$ in \figref{sd_signal}, there are sizeable contributions from the
subjet energy-resolution uncertainty. The uncertainty
from the \largeR JES dominates in the highest bins of the distribution ($>-20$).
ISR/FSR uncertainties and the uncertainty in the subjet energy scale dominate for
$\ln\chi > 5$ in \figref{sd_chi}.
The uncertainties in the bulk of the background-weight distribution
(\figref{sd_background}) are dominated by the subjet energy-scale and energy-resolution
uncertainties (from $-30$ to $-28$), the PDF and parton-shower uncertainties
(from $-28$ to $-25$) and for larger values ($>-25$) by the uncertainties from the \largeR JES and the
subjet energy scale.

\begin{figure}[!h]
\begin{centering}
\subfigure[]{
\label{fig:sd_signal}
\includegraphics[width=0.48\textwidth]{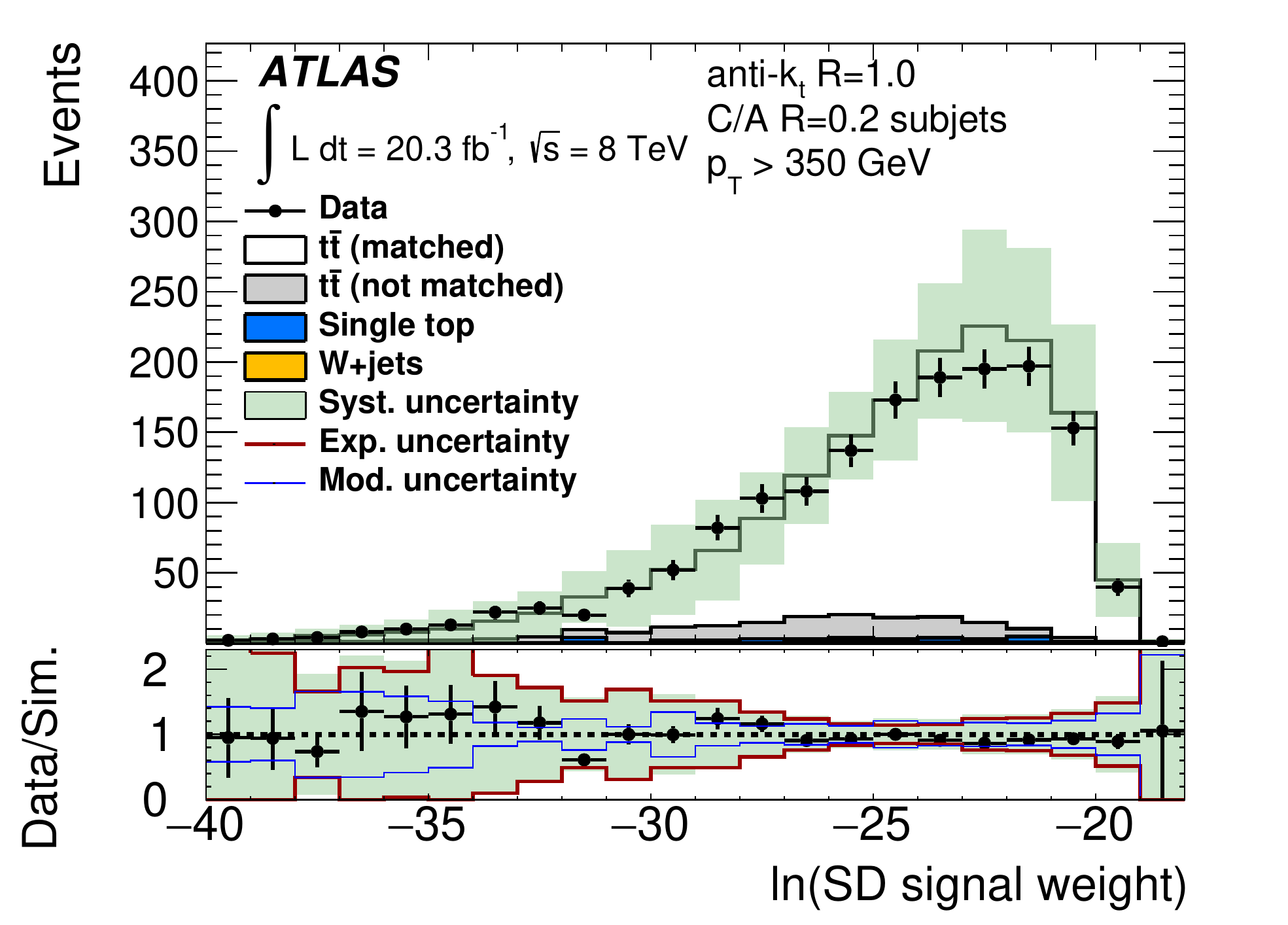}
}
\subfigure[]{
\label{fig:sd_background}
\includegraphics[width=0.48\textwidth]{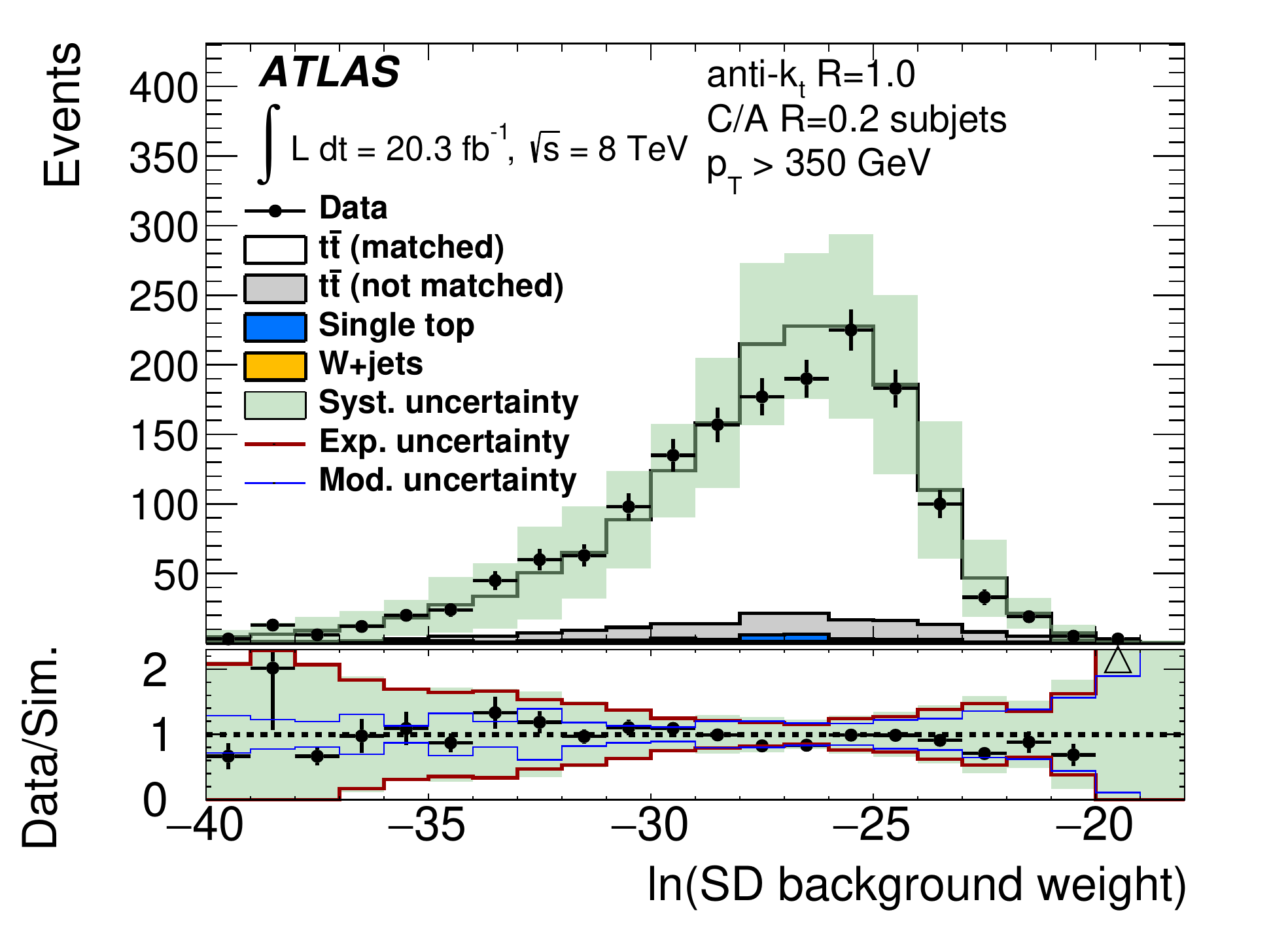}
} \\
\subfigure[]{
\label{fig:sd_chi}
\includegraphics[width=0.48\textwidth]{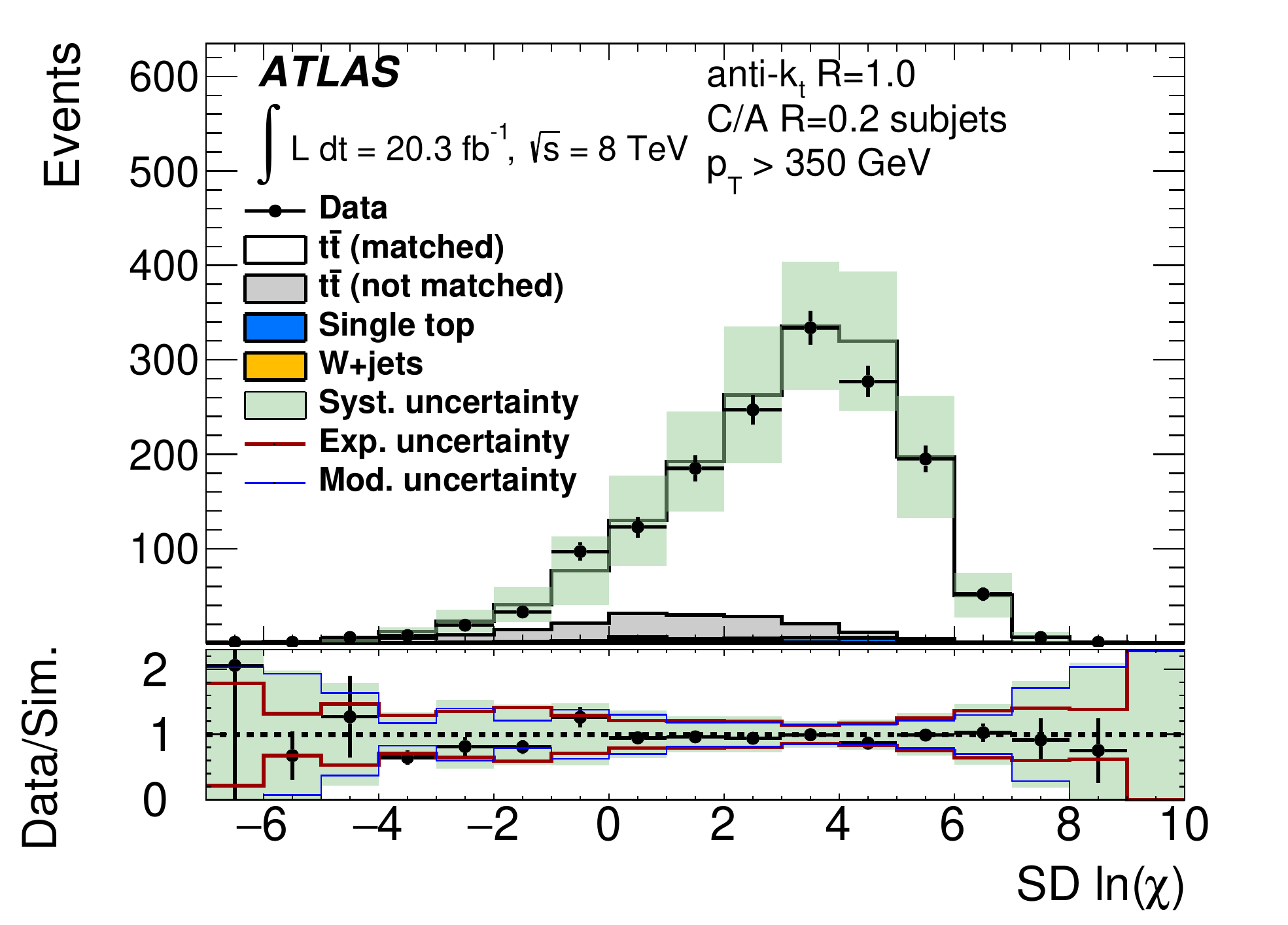}
}
\subfigure[]{
\label{fig:sd_chipt}
\includegraphics[width=0.48\textwidth]{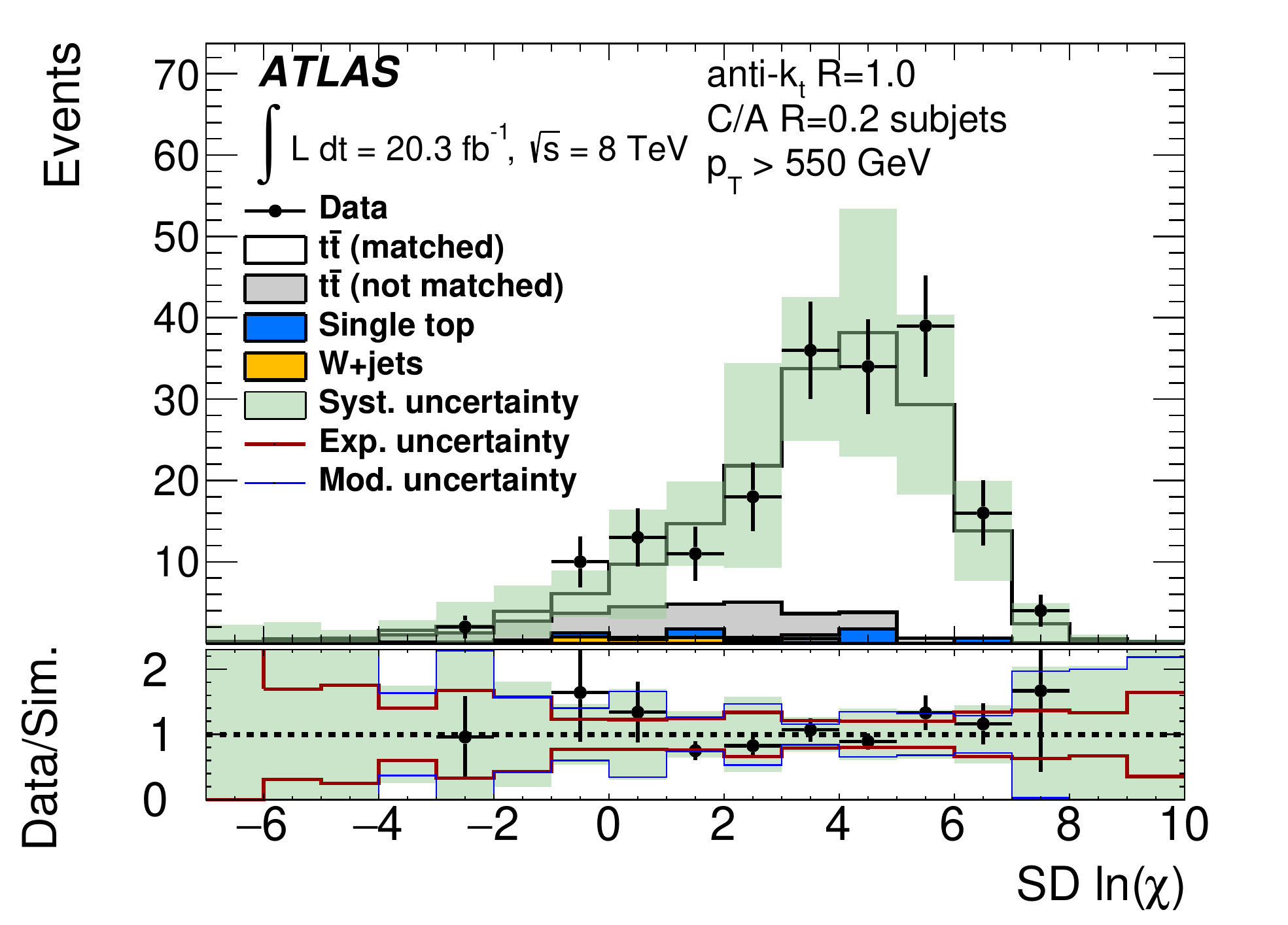}
}
\caption{
Distributions of \sd weights and the likelihood ratio $\chi$ in the
signal selection:
untrimmed \akt $R=1.0$ jets corresponding to the highest-\pt trimmed \akt $R=1.0$
jet with $\pt > 350\GeV$.
Cases in which the signal weight is zero because
there are fewer than three subjets or the top-quark- or \W-boson mass-window
requirements are not met (cf.~\secref{sd}) are not shown.
(a) Natural logarithm of the sum of all
weights obtained under the assumption that the subjet configuration in the
\largeR jet is the result of hadronic top-quark decay.
(b) Natural logarithm of the sum of all weights obtained for the background hypothesis.
(c) Distribution of the natural logarithm of the \sd likelihood ratio $\chi$.
(d) The same distribution as in (c) but for the requirement that the
trimmed \largeR jet \pt be larger than $550\GeV$.
The vertical error bar indicates the statistical uncertainty
of the measurement. Also shown are distributions for simulated SM contributions
with systematic uncertainties (described in \secref{systematics}) indicated as a band.
The \ttbar prediction is split into a {\em matched} part for which the \largeR
jet axis is within $\Delta R = 0.75$ of the flight direction of a hadronically decaying
top quark and a {\em not matched} part for which this criterion does not hold.
The ratio of measurement to
prediction is shown at the bottom of each subfigure and the error bar and band
give the statistical and systematic uncertainties of the ratio, respectively.
The impacts of experimental and \ttbar modelling uncertainties are
shown separately for the ratio.
}
\label{fig:ctrl_akt_pretag_sdweights}
\end{centering}
\end{figure}

\begin{figure}[!h]
\begin{centering}
\subfigure[]{
\label{fig:ctrl_sd_posttag_fjpt}
\includegraphics[width=0.48\textwidth]{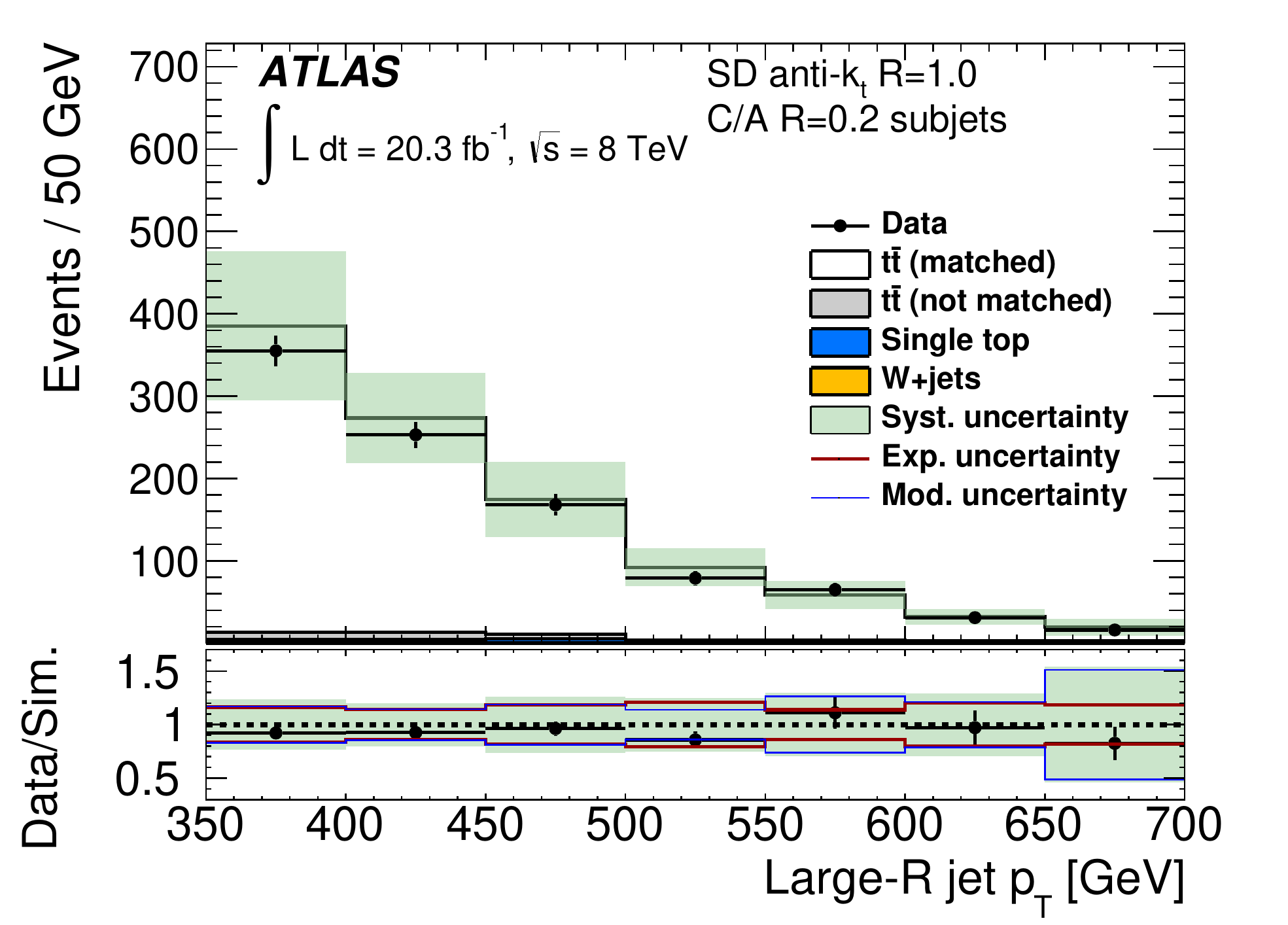}
} \\
\subfigure[]{
\label{fig:ctrl_sd_posttag_fjm}
\includegraphics[width=0.48\textwidth]{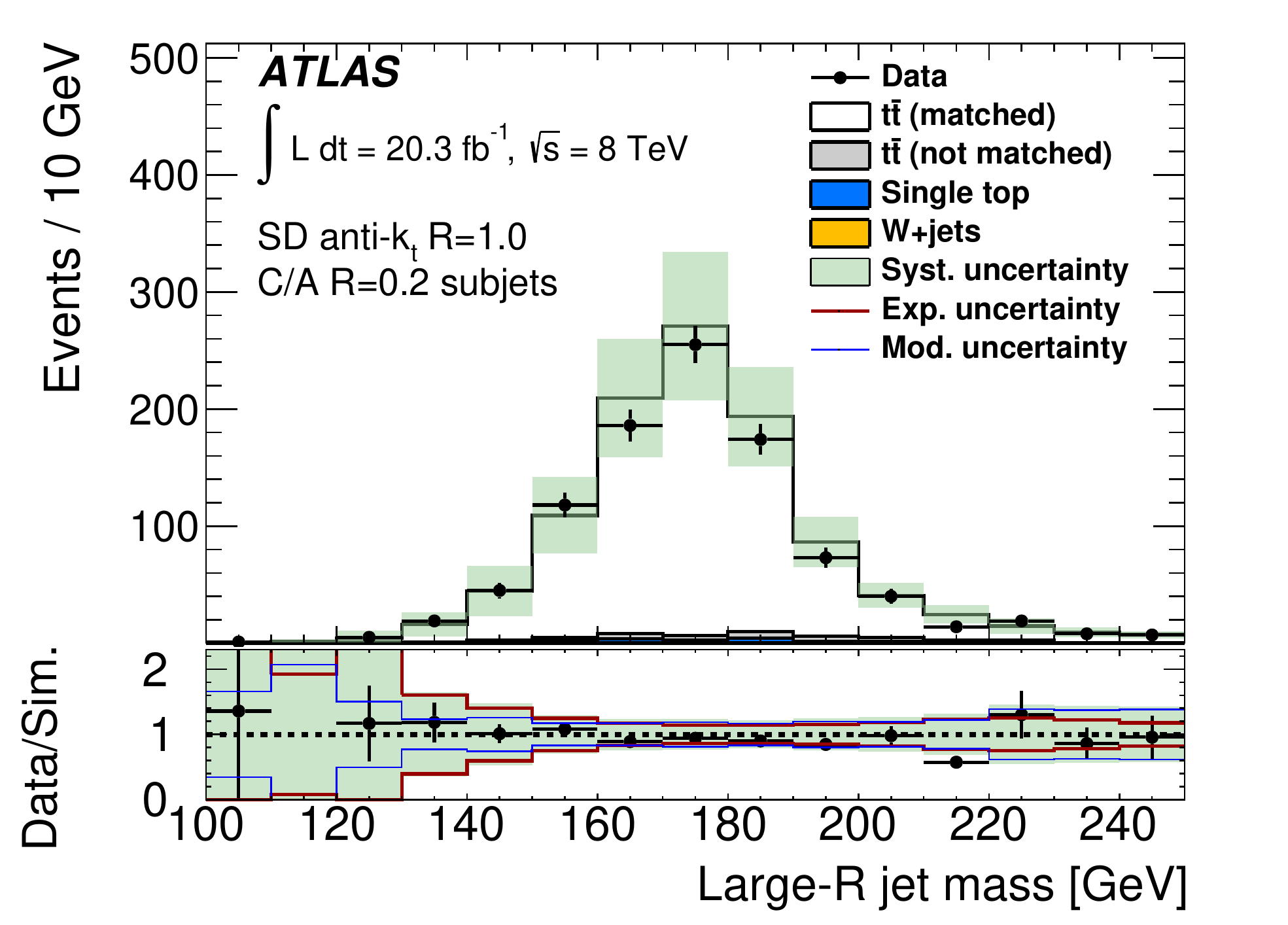}
}
\subfigure[]{
\label{fig:ctrl_sd_posttag_fjweightedm}
\includegraphics[width=0.48\textwidth]{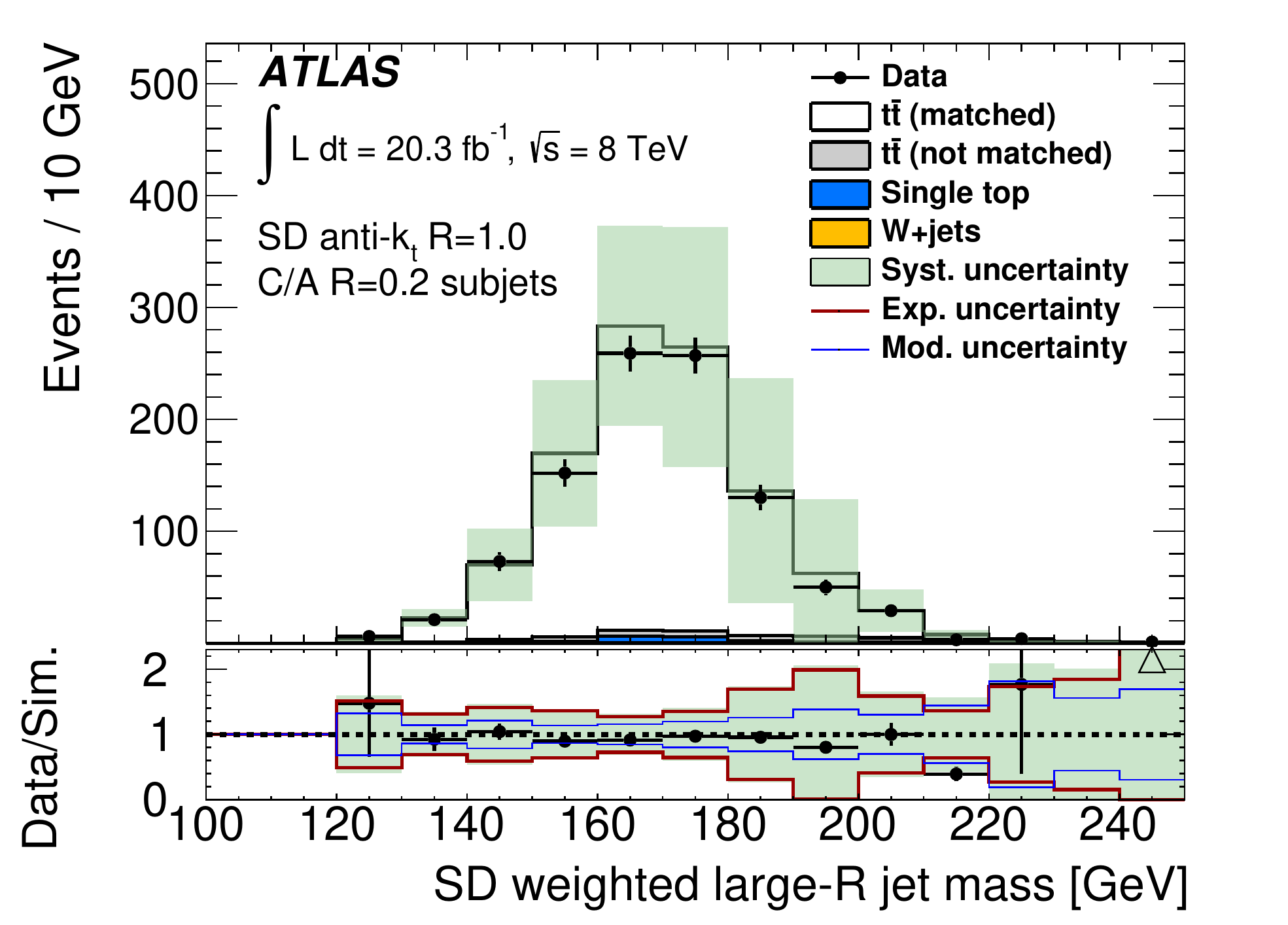}
}
\caption{
Distributions for \largeR jets which are top tagged by \sd using the requirement
$\ln(\chi)>2.5$ in events passing the signal selection.
(a) The transverse momentum and (b) the mass of trimmed \akt $R=1.0$ jets for which the
corresponding untrimmed \akt $R=1.0$ jet is tagged.
(c) The mass of the top-quark candidate, where the four-momentum is
calculated by taking the weighted average of each signal-hypothesis four-momentum.
The vertical error bar indicates the statistical uncertainty
of the measurement. Also shown are distributions for simulated SM contributions
with systematic uncertainties (described in \secref{systematics}) indicated as a band.
The \ttbar prediction is split into a {\em matched} part for which the \largeR
jet axis is within $\Delta R = 0.75$ of the flight direction of a hadronically decaying
top quark and a {\em not matched} part for which this criterion does not hold.
The ratio of measurement to
prediction is shown at the bottom of each subfigure and the error bar and band
give the statistical and systematic uncertainties of the ratio, respectively.
The impacts of experimental and \ttbar modelling uncertainties are
shown separately for the ratio.
}
\label{fig:ctrl_sd_posttag_fj}
\end{centering}
\end{figure}

Distributions of the \pt and the mass of \akt $R=1.0$ jets tagged as top jets by
SD using the requirement $\ln(\chi)>2.5$ are shown in \figref{ctrl_sd_posttag_fj}
for events passing the signal selection.
The \pt (\figref{ctrl_sd_posttag_fjpt}) and the mass (\figref{ctrl_sd_posttag_fjm}) are shown for the trimmed version of
the \akt $R=1.0$ jet. The \pt spectrum is smoothly falling
and the mass spectrum is peaked at $m_t$. Both distributions are described
by the simulation within the uncertainties.
The uncertainty of the simulation for $\pt<400\GeV$ is dominated by the uncertainties
in the subjet energy scale and on the PDF. From $400$ to $500\GeV$, important
contributions come from the PDF, ISR/FSR, the \largeR JES,
and the parton shower.
Between $500$ and $550\GeV$, the \largeR JES gives the largest contribution.
For $\pt>550\GeV$, the dominant uncertainties are the ones on the PDF and the \largeR JES.
For masses below $160\GeV$, the uncertainty is dominated by
the uncertainties in the subjet energy scale and resolution. For masses
greater than $210\GeV$, the differences between the generators and the PDF
uncertainty dominate,
consistent with previous figures, where the \largeR jet mass distribution
receives significant contributions from the generator uncertainty for
high mass values.
In the mass region $160$--$210\GeV$, multiple sources contribute
significantly to the uncertainty.

A top-quark mass distribution can be constructed differently, making use of the SD weights.
The signal weights are related to the likelihood of a set of subjets
to originate from a top-quark decay. For each set of subjets, a combined four-momentum
is built by adding the four-momenta of all subjets in the set.
A top-quark four-momentum is then reconstructed as a weighted average of the
four-momenta of all possible subjet combinations:
\begin{equation}
\displaystyle
p_\textrm{SD}^{\kappa} = \frac{\sum_{\textrm{all possible sets of subjets S}} \textrm{P}(\{p^{\kappa}(i), i \in S\} | \textrm{signal large-}R\textrm{ jet}) \times \sum_{i} p^{\kappa}(i)}{\sum_{\textrm{all possible sets of subjets S}} \textrm{P}(\{p^{\kappa}(i), i \in S\} | \textrm{signal large-}R\textrm{ jet})},
\label{eq:SDweightedMass}
\end{equation}
where $p^{\kappa}(i)$ is the four-momentum
of the $i$-th subjet.
The mass $\displaystyle \sqrt{p^2_\textrm{SD}}$ is shown in \figref{ctrl_sd_posttag_fjweightedm}.
For the background, this mass takes on values closer to the top-quark mass than
in \figref{ctrl_sd_posttag_fjm} because of the use of the signal weights in \eqref{SDweightedMass}.
Although not directly used in the SD tagging decision, this mass offers a glimpse
into the inner workings of SD. The distribution is similar to the
distribution of the trimmed jet mass. While the width in the central peak region
from $140$ to $200\GeV$ is similar, outliers in the weighted mass are significantly reduced. The distribution is well described by the simulation
within statistical and systematic uncertainties. The systematic
uncertainties are dominated by the uncertainties in the subjet energy scale
and resolution.

\subsection{\htt}

\CamKt $R=1.5$ jets are analysed with
the \textit{\htt} algorithm~\cite{Plehn:2009rk,Plehn:2010st}, which identifies the hard
jet substructure and tests it for compatibility with the 3-prong pattern of
hadronic top-quark decays.
This tagger was developed to find top quarks with $\pt>200\GeV$ and to
achieve a high rejection of background, which is largest for low-\pt \largeR jets.
The \htt studied in this paper is the original algorithm which does not employ multivariate techniques. An extended version, HEPTopTagger2, has been developed in Ref.~\cite{Kasieczka:2015jma}.
The algorithm makes use of the fact that in \CamKt jets, large-angle proto-jets are clustered last.
The \htt has internal parameters that can be changed to optimize the
performance, and the settings used in this paper are given in \tabref{HTTsettings}
and are introduced in the following brief summary of the algorithm.

In the first step, the \largeR jet is iteratively
broken down into hard substructure objects using a mass-drop criterion~\cite{Butterworth:2008iy}.
The procedure stops when all substructure objects have a mass below the value $\mcut$.
In the second phase, all combinations of three substructure objects are tested
for kinematic compatibility with a hadronic top-quark decay.
Energy contributions from underlying event and \pileup are removed using a filtering
procedure:
small distance parameter $\CamKt$ jets are built from the constituents of the substructure objects
using a radius parameter that depends on the distance between these objects but
has at most the value $\Rfilt^{\rm max}$.
The constituents of the $\Nfilt$ highest-\pt jets found in this way (filter jets) are then clustered
into three top-quark subjets using the exclusive \CamKt algorithm.
In the final step, kinematic requirements are applied to differentiate hadronic top-quark decays
from background. One of the criteria is that one pair of subjets must have an invariant mass
in the range $80.4\GeV \times (1\pm f_W)$ around the $W$-boson mass, with $f_W$ being a parameter of the algorithm.
If all criteria are met, the top-quark candidate is built by adding the four-momenta of the $\Nfilt$ highest-\pt filter jets.
The \largeR jet is considered to be tagged if the top-quark-candidate mass is between $140$ and $210\GeV$ and the
top-quark-candidate \pt is larger than $200\GeV$.
An illustration of the \htt algorithm is given in Figure 6 of Ref.~\cite{Aad:2013gja}.

\begin{table}[b]
  \begin{center}
    \begin{tabular}{|c|c|}
      \hline
      Parameter & Value \\
      \hline
      \mcut & $50\GeV$ \\
      \hline
      $\Rfilt^{\rm max}$ & 0.25 \\
      \hline
      \Nfilt & 5 \\
      \hline
      $f_W$ & 15\% \\
      \hline
    \end{tabular}
    \caption{The \htt parameter settings used in this study.}
  \label{tab:HTTsettings}
  \end{center}
\end{table}

Distributions of the \htt substructure variables after requiring a top tag
are shown in \figref{ctrl_HTT_posttag_substr_cand}, together with
the \pt and mass distributions of the top-quark candidate
for events passing the signal selection.
The purity of processes with top quarks (\ttbar and single-top production) in this
sample is more than 99\%.
The variable $m_{12}$ ($m_{23}$) is the invariant mass of the highest-\pt (second-highest-\pt)
and the second-highest-\pt (third-highest-\pt) subjet found in the final, i.e.\ exclusive,
subjet clustering step. The variable $m_{13}$ is defined analogously, and the variable
$m_{123}$ is the mass of the three exclusive subjets.
The ratio $m_{23}/m_{123}$ is used internally in the \htt algorithm and is
displayed in \figref{ctrl_HTT_posttag_m23m123}. It shows a peak at $m_W/m_t$, which
indicates that in most of the cases, the highest-\pt subjet corresponds to the $b$-quark.
The inverse tangent of the ratio $m_{13}/m_{12}$ is also used internally in the \htt algorithm and its
distribution is shown in \figref{ctrl_HTT_posttag_atan1312}.
The \htt top-quark-candidate \pt (\figref{ctrl_HTT_posttag_pt}) is peaked at
$\approx\!250\GeV$ and falls smoothly at higher \pt. At around $200\GeV$, the tagging efficiency
increases strongly with \pt (cf.~\secref{eff}) and therefore there are fewer
entries in the lowest \pt interval from $200$ to $250\GeV$ than would be expected from a falling \pt distribution.
The \htt top-quark-candidate mass (\figref{ctrl_HTT_posttag_m}) is peaked near the top-quark
mass with tails to lower and higher values.
To be considered as \htt-tagged, the top-quark candidate must have a mass between
$140$ and $210\GeV$.

The distributions of $m_{23}/m_{123}$ and $\arctan(m_{13}/m_{12})$, as well as the top-quark-candidate \pt
and mass are well described by the simulation within statistical and systematic uncertainties.
For the two ratios of subjet invariant masses, important sources of systematic uncertainty are
the subjet JES, the $b$-tagging efficiency and the \ttbar modelling uncertainties from the choice
of the PDF set and the ISR/FSR settings. The choice of PDF set dominates the uncertainty for
$m_{23}/m_{123}$ for very low and very high values of the ratio.
These uncertainties also contribute to the modelling of the top-quark-candidate \pt and $\eta$.
The uncertainty in the top-quark-candidate \pt increases with \pt due to increasing uncertainties from the
subjet JES, the $b$-tagging efficiency and the choice of PDF set, as well as from additional
\ttbar modelling uncertainties due to the choice of generator and parton shower.

\begin{figure}[!h]
\begin{centering}
\subfigure[]{
\label{fig:ctrl_HTT_posttag_m23m123}
\includegraphics[width=0.48\textwidth]{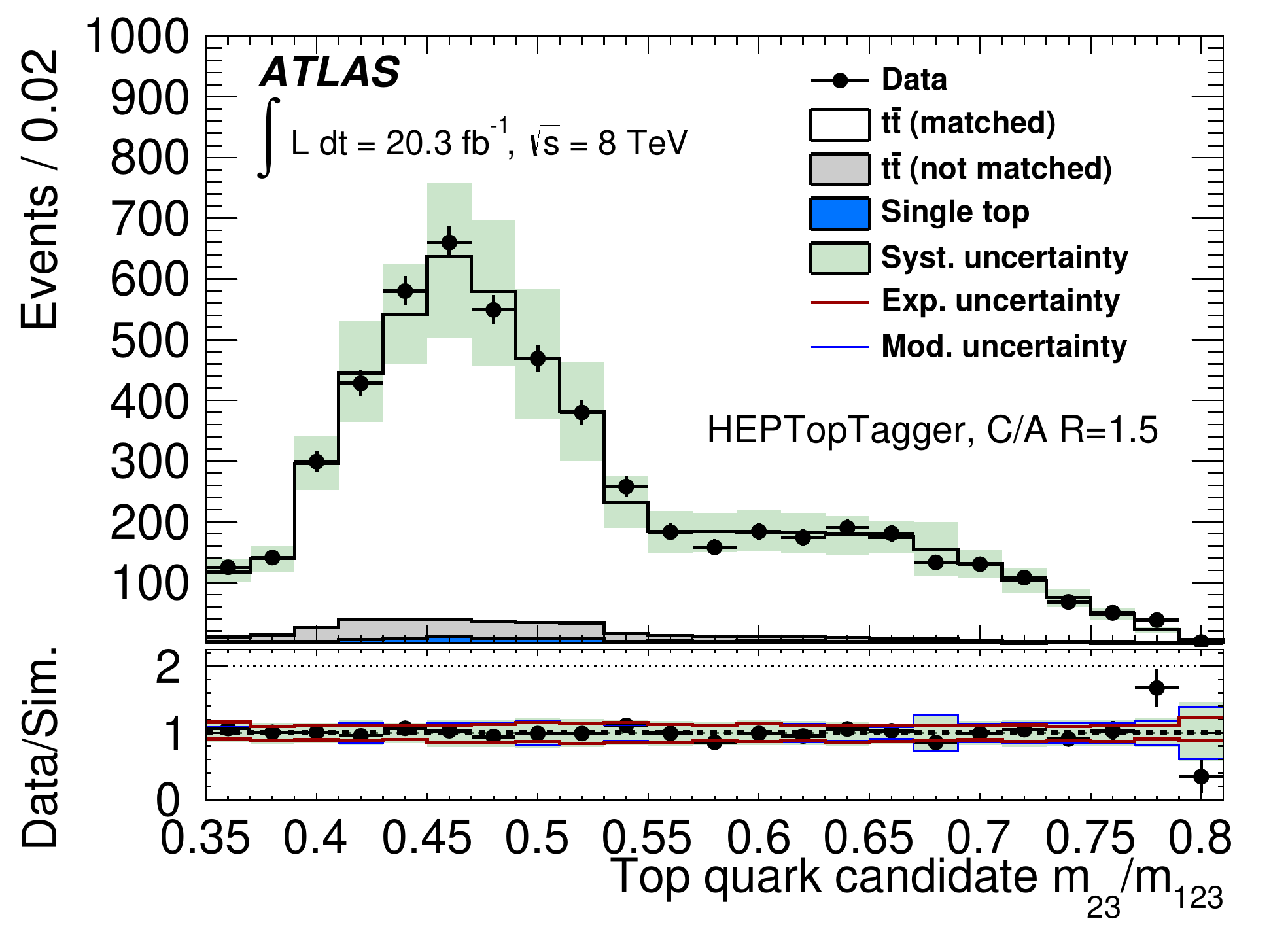}
}
\subfigure[]{
\label{fig:ctrl_HTT_posttag_atan1312}
\includegraphics[width=0.48\textwidth]{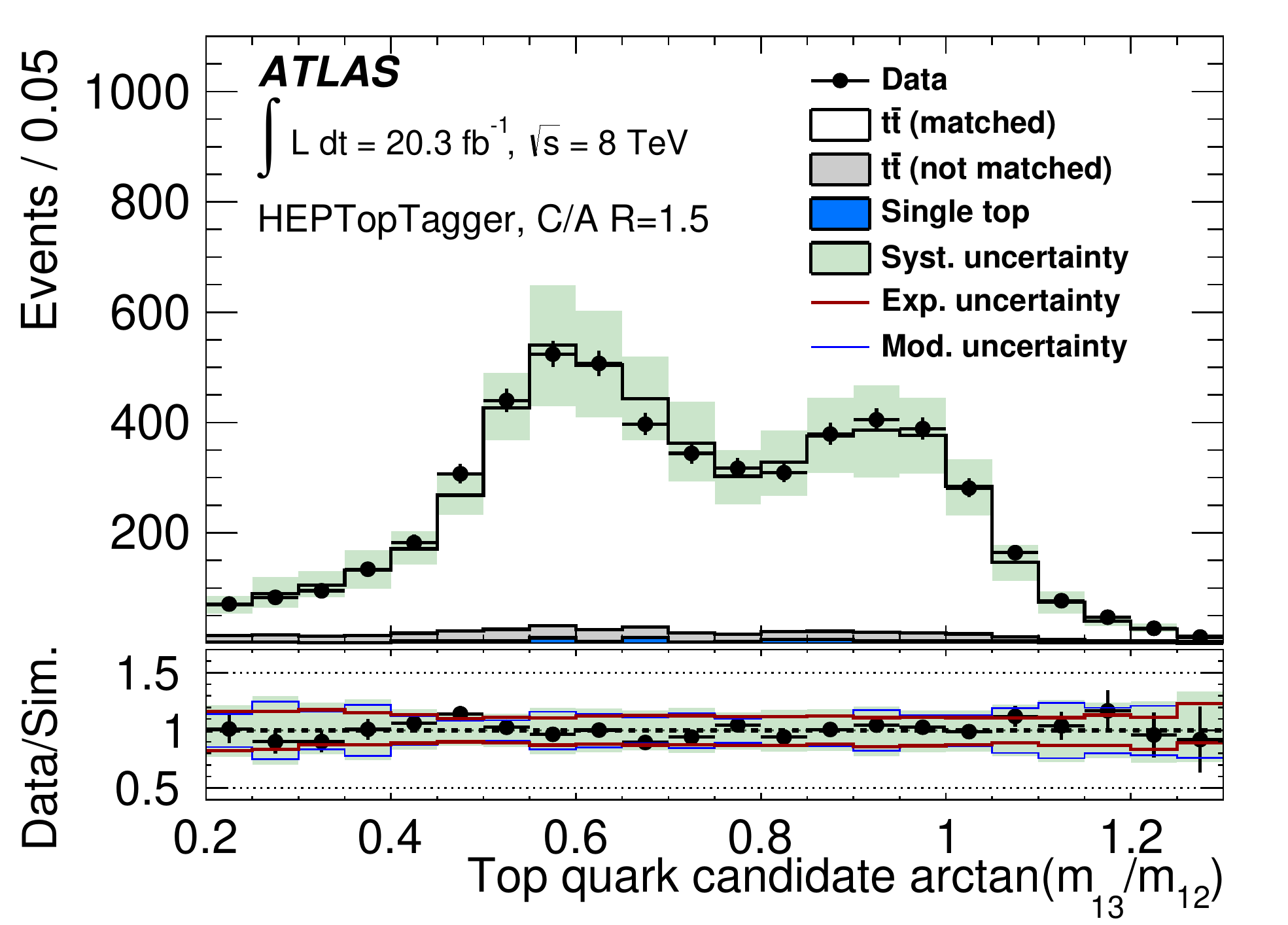}
} \\
\subfigure[]{
\label{fig:ctrl_HTT_posttag_pt}
\includegraphics[width=0.48\textwidth]{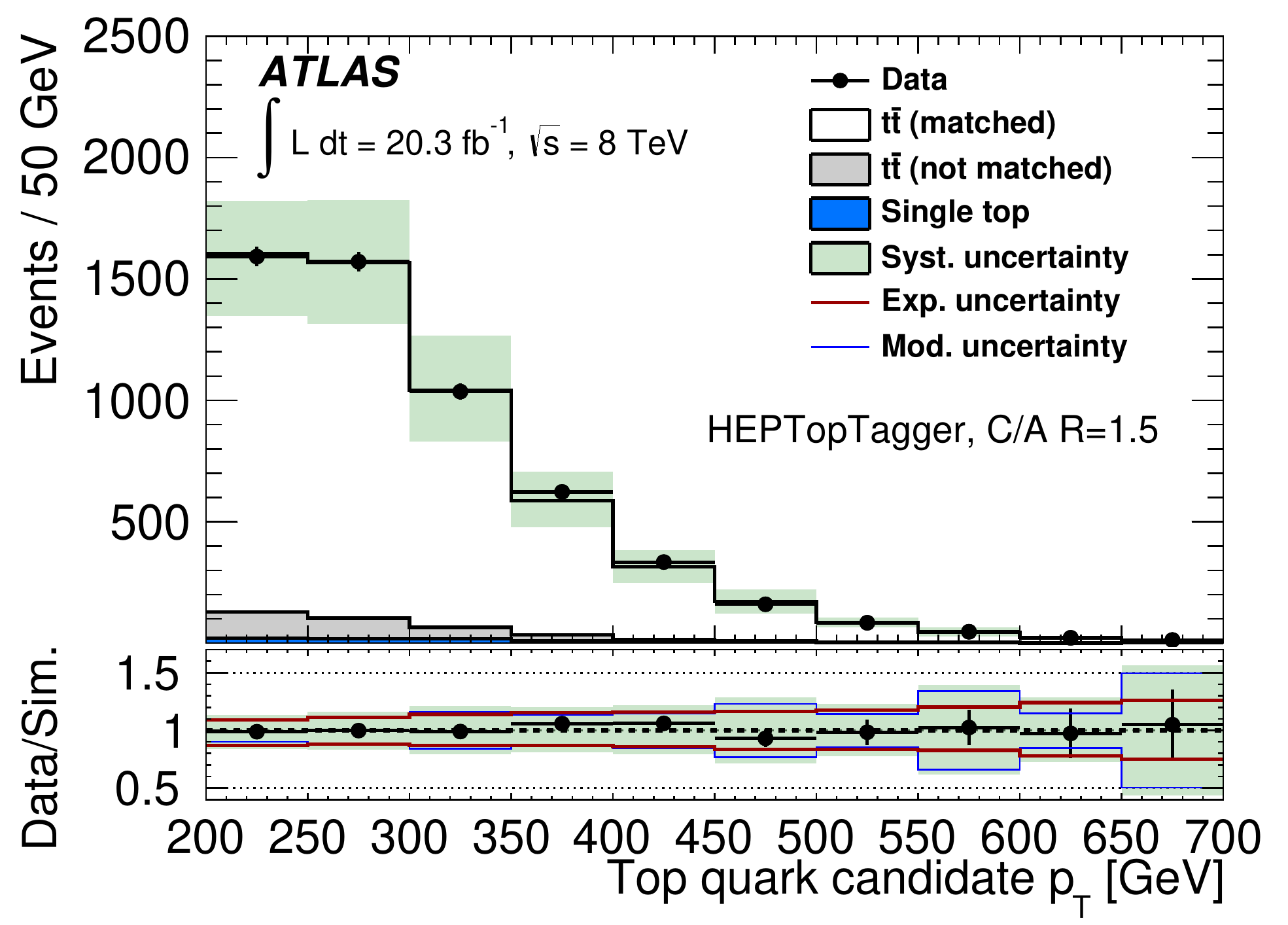}
}
\subfigure[]{
\label{fig:ctrl_HTT_posttag_m}
\includegraphics[width=0.48\textwidth]{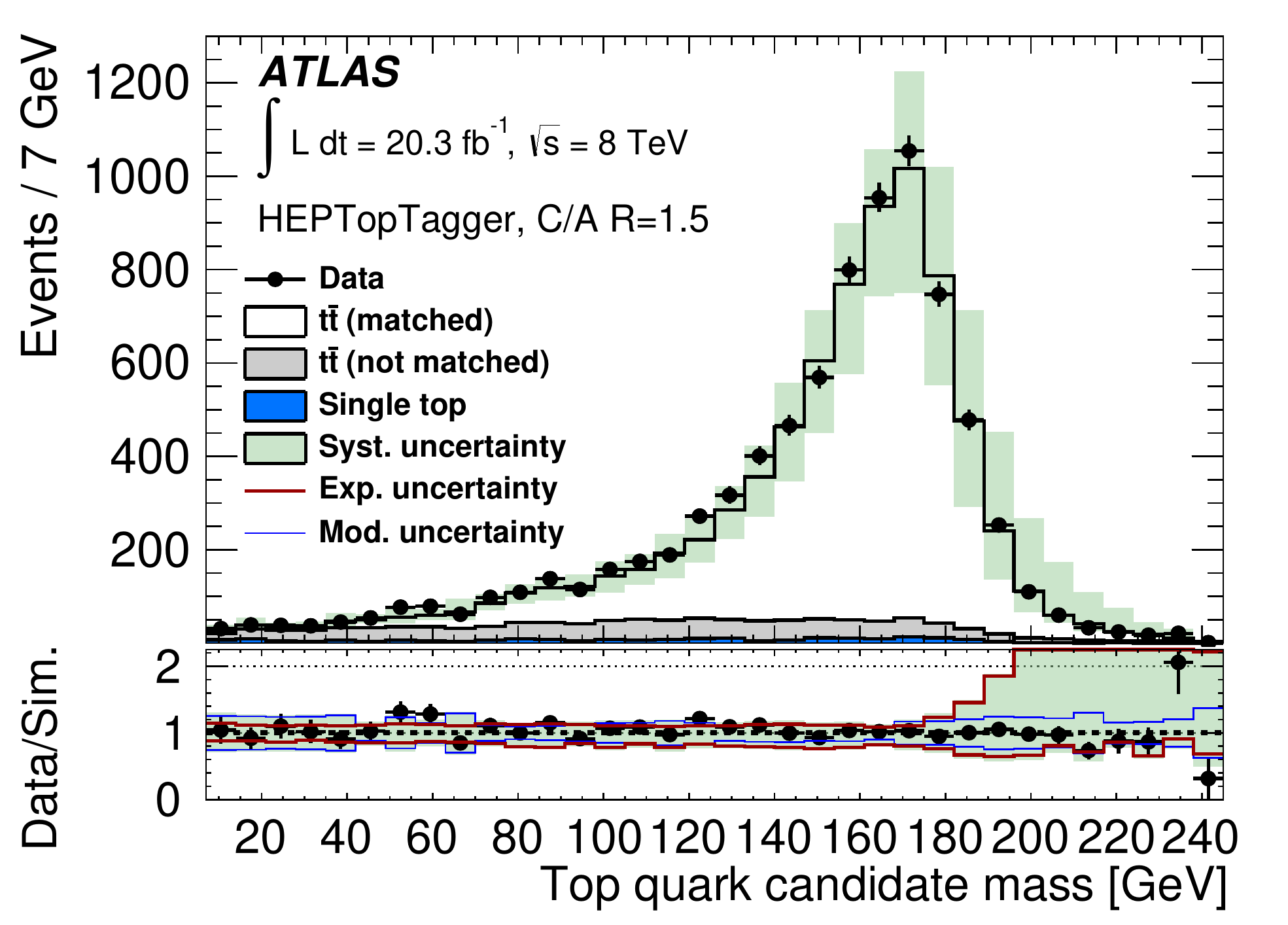}
}
\caption{Distributions of \htt substructure variables ((a) and (b)) for
\htt-tagged highest-\pt \CamKt $R=1.5$ jets in events passing the
signal selection:
Shown in (c) and (d) are the \pt and mass of the top-quark candidate,
respectively.
The vertical error bar indicates the statistical uncertainty
of the measurement. Also shown are distributions for simulated SM contributions
with systematic uncertainties (described in \secref{systematics}) indicated as a band.
The \ttbar prediction is split into a {\em matched} part for which the \largeR
jet axis is within $\Delta R = 1.0$ of the flight direction of a hadronically decaying
top quark and a {\em not matched} part for which this criterion does not hold.
The ratio of measurement to
prediction is shown at the bottom of each subfigure and the error bar and band
give the statistical and systematic uncertainties of the ratio, respectively.
The impacts of experimental and \ttbar modelling uncertainties are
shown separately for the ratio.
}
\label{fig:ctrl_HTT_posttag_substr_cand}
\end{centering}
\end{figure}

A variant of the \htt has been developed
that uses a collection of \smallR jets as input,
instead of \largeR jets. This variant is referred to as \textit{\httofour}, because it is
based on \smallR jets with $R=0.4$.
This approach can be useful when aiming for a full event reconstruction
in final states with many jets in events in which the top quarks have only a moderately
high transverse momentum ($\pt > 180\gev$).
The advantages of the method are explained using the performance in MC
simulation in \secref{httofour}.

The \httofour technique proceeds as follows. All sets of up to three \akt $R = 0.4$ jets (\smallR jets in the following) are
considered, and an early top-quark candidate
(not to be confused with the \htt candidate) is built by adding the
four-momenta of these jets.
Only sets with $m_{\rm candidate} > m_{\rm min}$ and
$p_{\rm T, candidate} > p_{\rm T, min}$ are kept and
all \smallR jets in the set must satisfy
$\Delta R_{i,{\rm candidate}} < \Delta R_{\rm max}$.
The values of these parameters are given in \tabref{HTT04settings}.
The constituents of the selected \smallR jets are then passed to the
\htt algorithm to be tested with being compatible with a hadronically decaying top quark.
The same parameters as given in~\tabref{HTTsettings} are used.
If a top-quark candidate is found with the \htt algorithm based on the \smallR jets'
constituents, it is called a \httofour top-quark candidate.
If more than one \httofour top-quark candidate is found in an event, they are all kept if they
do not share a common input jet.
In the case that top-quark candidates share \smallR input jets, the largest possible set of
top-quark candidates which do not share input jets is chosen.
If multiple such sets exist, the set for which the average top-quark-candidate mass is closest
to the top-quark mass is selected.

\begin{table}[b]
  \begin{center}
    \begin{tabular}{|c|c|}
      \hline
      Parameter & Value \\
      \hline
      $m_{\rm min}$ & $100\GeV$ \\
      \hline
      $p_{\rm T, min}$ & $140\GeV$ \\
      \hline
      $\Delta R_{\rm max}$ & 1.1 \\
      \hline
    \end{tabular}
    \caption{The parameters used in the \httofour technique to build an early top-quark candidate from
    up to three \akt $R = 0.4$ jets.}
  \label{tab:HTT04settings}
  \end{center}
\end{table}

Post-tag distributions from the \httofour approach for events passing the signal
selection (but omitting all requirements related to a \largeR jet) are shown in
\figref{ctrl_HTTofour_posttag_substr_cand} and show features similar to the ones
described for the \htt.
Events are classified as matched or not-matched based on the angular distance between
hadronically decaying top quarks and the top-quark candidate, and not the \largeR
jet as in the other tagging techniques, because for the \httofour no \largeR
jet exists.
The distributions are well described by the simulation
within statistical and systematic uncertainties.
The systematic uncertainty of the predicted event yield after tagging is approximately 16\%,
with the largest contributions from the subjet energy scale (8.1\%),
the uncertainty in initial-state and final-state radiation (8.9\%),
the \ttbar cross-section normalization (6.2\%),
the PDF uncertainty (5.2\%),
and the uncertainty in the $b$-tagging efficiency (5.1\%).
The uncertainties related to the \akt $R=0.4$ jets used
as input to the \httofour method have a negligible impact ($<\!1\%$),
as the \akt $R=0.4$ jet energies are only used to select the early top-quark candidate
in the \httofour procedure and the \htt algorithm is run on the constituents
of these \akt $R=0.4$ jets.

\begin{figure}[!h]
\begin{centering}
\subfigure[]{
\includegraphics[width=0.48\textwidth]{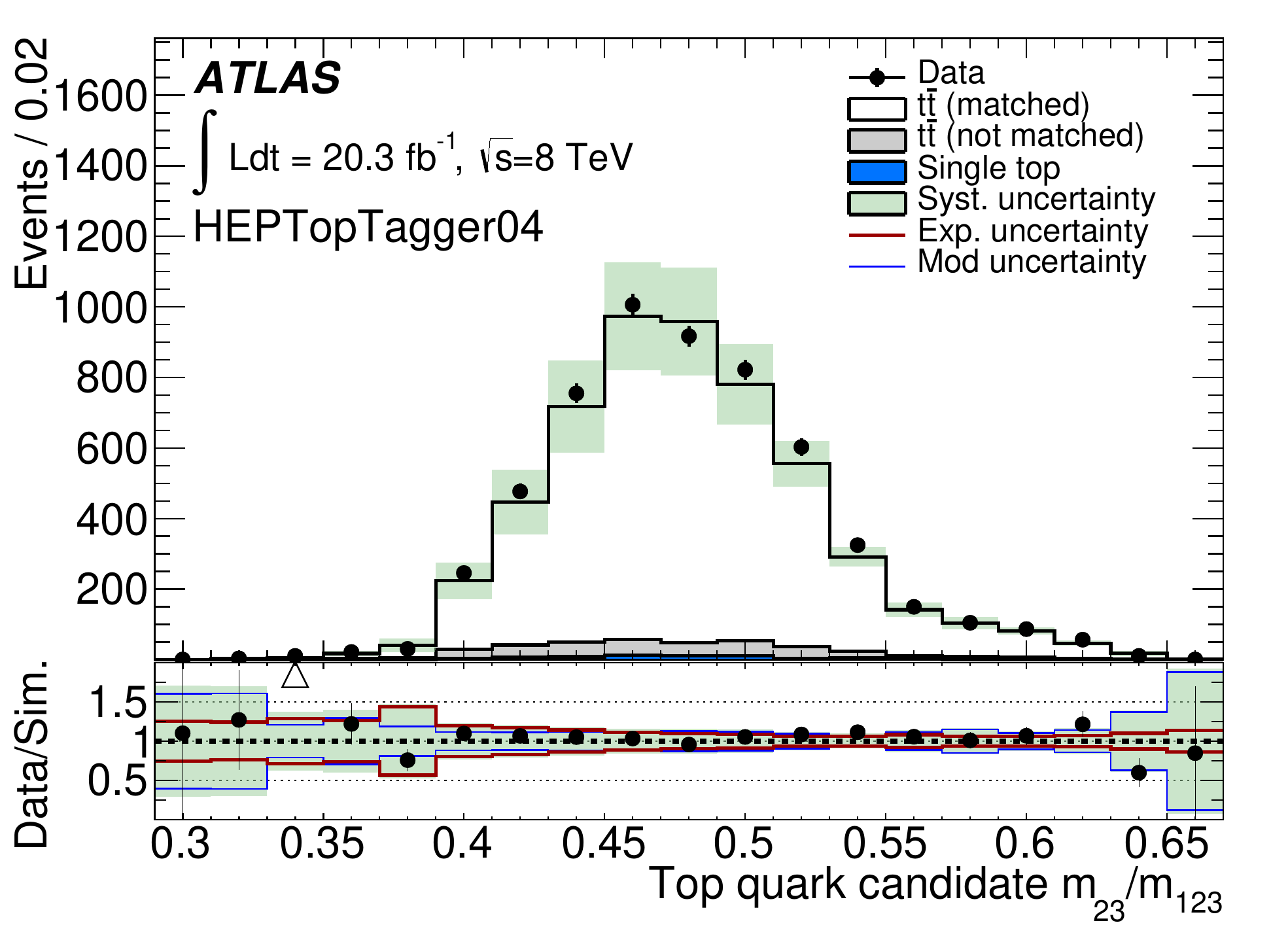}
}
\subfigure[]{
\includegraphics[width=0.48\textwidth]{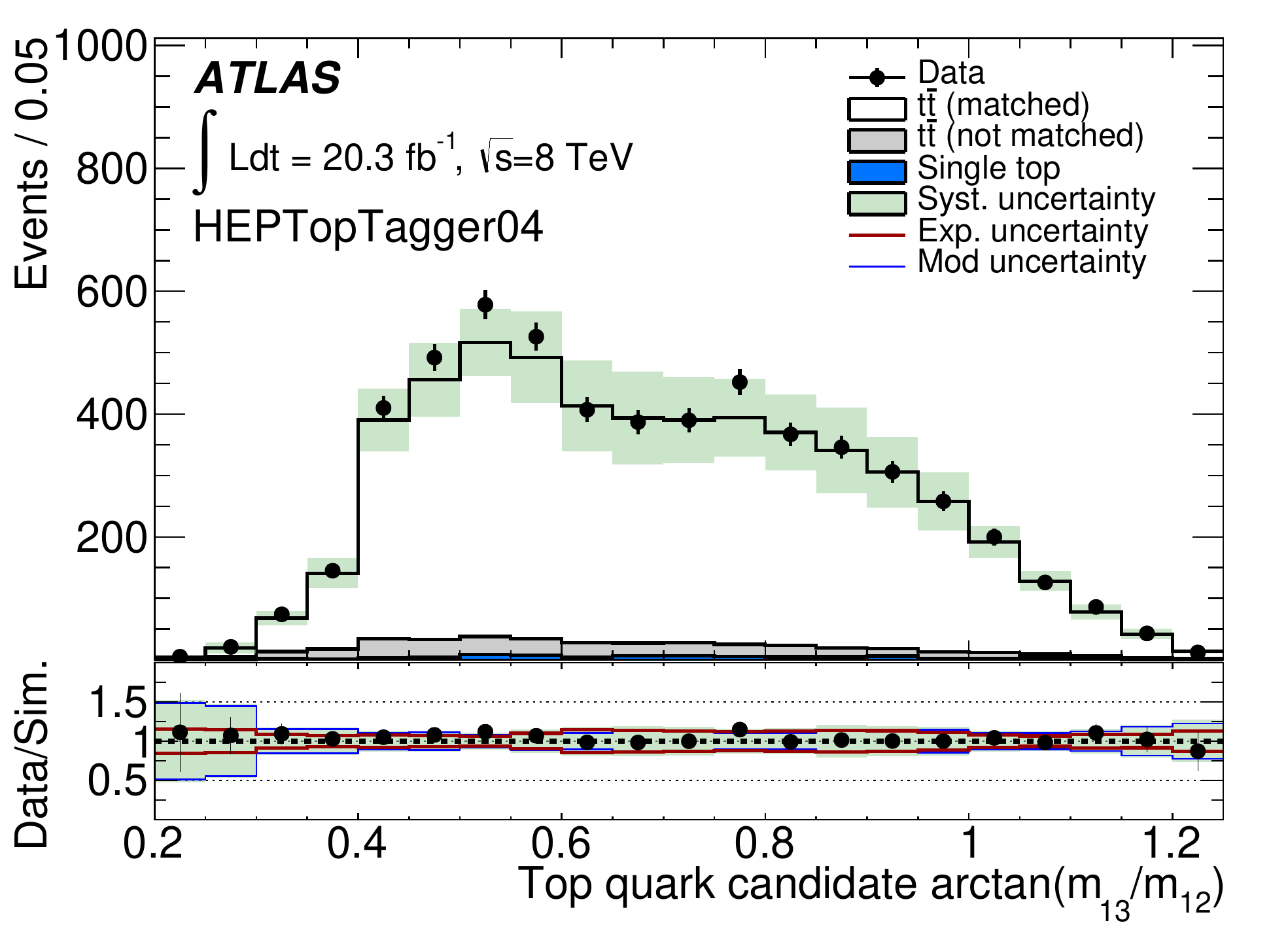}
} \\
\subfigure[]{
\includegraphics[width=0.48\textwidth]{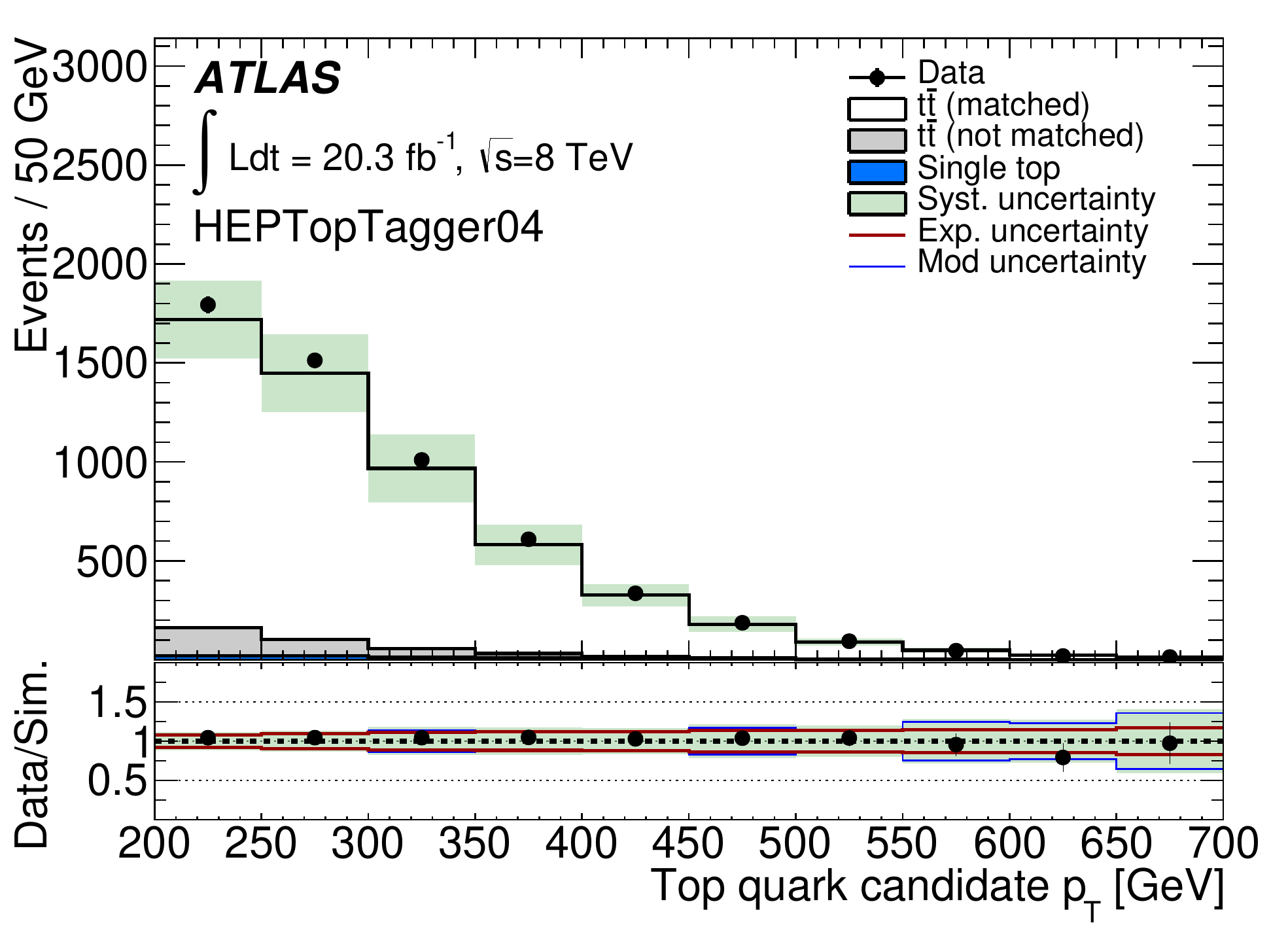}
}
\subfigure[]{
\includegraphics[width=0.48\textwidth]{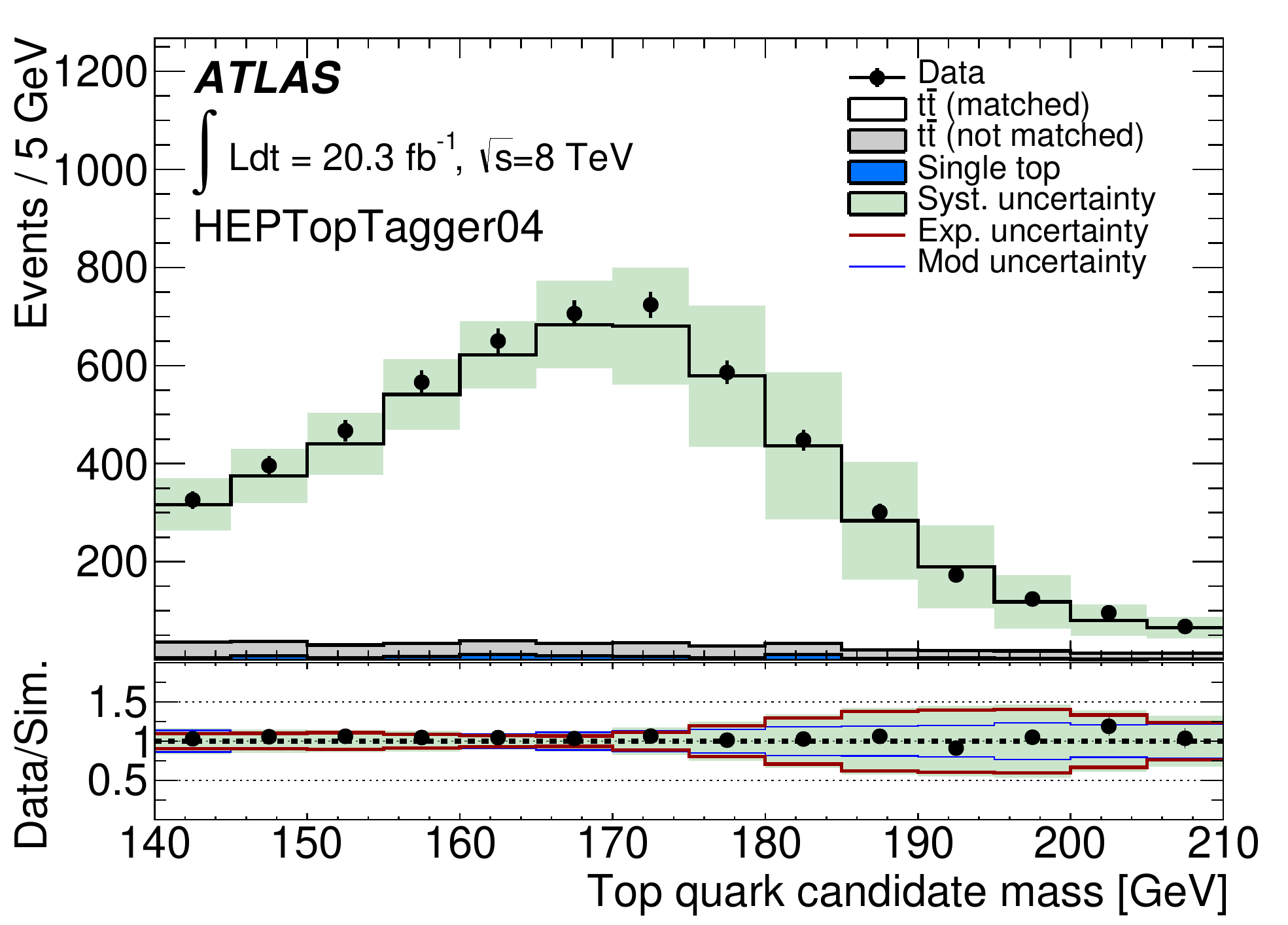}
}
\caption{Distributions from the \httofour approach for top tags in events passing the
signal selection. (a) and (b) show the \htt substructure variables;
(c) and (d) show the \pt and mass of the top-quark candidate,
respectively.
The vertical error bar indicates the statistical uncertainty
of the measurement. Also shown are distributions for simulated SM contributions
with systematic uncertainties (described in \secref{systematics}) indicated as a band.
The \ttbar prediction is split into a {\em matched} part for which the top-quark candidate
axis is within $\Delta R = 1.0$ of the flight direction of a hadronically decaying
top quark and a {\em not matched} part for which this criterion does not hold.
The ratio of measurement to
prediction is shown at the bottom of each subfigure and the error bar
corresponds to the statistical uncertainty from the measurement and
the bands give the statistical and systematic uncertainties of the prediction.
The impacts of experimental and \ttbar modelling uncertainties are
shown separately for the ratio.
}
\label{fig:ctrl_HTTofour_posttag_substr_cand}
\end{centering}
\end{figure}

\section{Systematic uncertainties}
\label{sec:systematics}

The measurements presented in this paper are performed at the detector level,
i.e.\ differential in reconstructed kinematic quantities and not corrected for
detector effects such as limited efficiency and resolution. The measured
distributions are compared with SM predictions obtained from MC-generated events
which have been passed through a simulation of the detector and are reconstructed
in the same way as the data. Systematic uncertainties of the predictions
can be grouped into different categories:
uncertainties related to the simulation of the detector
response and the luminosity measurement, and uncertainties related
to the modelling of the physics processes (production cross sections, parton shower,
hadronization, etc.).

Systematic uncertainties in the results presented in this paper are obtained
by varying parameters of the simulation (one parameter at a time) and
repeating the analysis with this varied simulation to determine its impact.
The change from the nominal prediction is taken
as the $1\sigma$ uncertainty related to the uncertainty in the varied
parameter. The systematic uncertainties are considered uncorrelated unless
otherwise specified.

\subsection{Experimental uncertainties}

The uncertainty in the integrated luminosity is $2.8\%$.
It is derived from a calibration of the luminosity scale derived
from beam-separation scans, following the methodology
detailed in Ref.~\cite{Aad:2013ucp}.

The $b$-tagging efficiency is measured using fits to the observed
$b$-tag multiplicity in \ttbar events~\cite{Aad:2015ydr,ATLAS-CONF-2014-004} and from jets
containing muons~\cite{Aad:2015ydr}. The rate at which jets from
charm and light quarks are classified as $b$-jets (mistag rate) is determined
from the distributions of the signed impact parameter and the signed decay
length in multijet events~\cite{Aad:2015ydr,ATLAS-CONF-2014-046}.
Uncertainties in the $b$-tagging efficiency and mistag rate in simulation
are obtained by comparing the predictions with the measurements.
The uncertainty in the mistag rate has a negligible impact on the
results presented here.

The uncertainties in the lepton trigger,
reconstruction and identification efficiencies are determined from
$Z \to ee$~\cite{Aad:2014fxa,ATLAS-CONF-2014-032} and $Z \to \mu \mu$~\cite{Aad:2014rra} events.
Also considered, but found to have negligible impact in the present analysis, are
uncertainties in the scale and resolution of the lepton energy
and in the \met reconstruction.

Systematic uncertainties related to jet reconstruction are considered as follows.
The uncertainty in the energy scale of \akt $R=0.4$ jets is determined
using a combination of in situ techniques exploiting the transverse-momentum
balance between a jet and a reference object such as a photon or a
\Z boson~\cite{Aad:2014bia}. The uncertainty in the energy resolution of
\akt $R=0.4$ jets is found to have negligible impact for the results
presented here.

The \largeR jets and subjets used in this analysis are reconstructed from calorimeter information.
Systematic uncertainties related to the modelling of the calorimeter response
in simulation are estimated by comparing these jets to tracks
which are matched to the jets~\cite{Aad:2013gja}.
Uncertainties in the following quantities are estimated in this way:
the energy scale of the \largeR jets; the \kt splitting scales, the
\Nsj ratios, and the mass of trimmed \akt $R=1.0$ jets;
the subjet energy scale for SD.
For $\pt<900\GeV$ of trimmed \akt $R=1.0$ jets, the uncertainty is not derived
from the track-jet method, but using $\gamma$+jet events and an additional uncertainty based on
the difference between the calorimeter's response to QCD jets and jets from $\ttbar$ decays.
The uncertainties in the \kt splitting scales, the \Nsj ratios and the trimmed mass
are $4$--$7\%$ for \pt between $350$ and $700\GeV$, depending on the jet \pt, \eta
and the ratio $m/\pt$. For values of $m/\pt<0.1$, the uncertainties are larger
and reach values of up to 10\%.
The subjet energy-scale uncertainty for the \htt is determined in situ from
the reconstructed top-quark mass peak as described in \secref{insitu}.
The correlations between the uncertainties in the substructure variables used by
taggers I--V and the \WPT have not been determined;
the largest observed variations are used
based on testing different combinations of zero and
full (anti-)correlation of the systematic uncertainties of the different substructure variables.

The energy-resolution uncertainties for \CamKt $R=1.5$ jets and for subjets
used by SD and the \htt are determined using the \pt balance in dijet
events~\cite{Aad:2013gja}.
To determine the impact of the energy-resolution uncertainty for trimmed \akt jets with $R=1.0$,
the energy resolution in simulation is scaled by 1.2.
The impact of the mass-resolution uncertainty for trimmed \akt $R=1.0$ jets is
estimated analogously.

\subsection{In situ determination of the subjet energy scale for the \htt}
\label{sec:insitu}
The top-quark candidates identified with the \htt in the $\mu$+jets channel
of the signal selection are used to determine the subjet energy scale
for the \htt.
For this study, the signal selection with only the $b$-tag close to the
lepton is used and the second $b$-tag requirement with $\Delta R > 1.5$ from the
lepton direction is omitted.
With this change, the $\mu$+jets channel alone provides sufficient events to perform this study.
The four-momentum of the top-quark candidate is obtained in the \htt by
combining the calibrated subjet four-momenta. A change in the subjet \pt is
therefore reflected in a change of the top-quark-candidate momentum. The top-quark
peak in the distribution of the top-quark-candidate mass can be used to
constrain the energy-scale uncertainty of the subjets as suggested
in Ref.~\cite{Schaetzel:2013vka}.
The method consists of varying the energy scale of the calibrated subjets in
simulation and comparing the resulting top-quark mass distribution to the one from data.
A higher (lower) subjet energy scale shifts the predicted distribution to larger
(smaller) masses. This shift is constrained by the necessity to describe the
measured mass peak within uncertainties.

The subjet energy-scale uncertainty is
determined by calculating a $\chi^2$ value for different variations of the
energy scale.
The $\chi^2$ is calculated in the mass window from $133$ to $210\GeV$, in 11
bins of width $7\GeV$. The statistical uncertainties of
the measured and predicted number of top-quark candidates in each bin are taken
into account, as well as all systematic
uncertainties other than that of the subjet energy scale itself.
The systematic uncertainties due to the imperfect modelling of the physics
processes (\secref{modellingsyst}) are considered, including
a systematic uncertainty in the top-quark mass of $\pm1\GeV$.

Variations of the subjet energy scale are considered by
raising or lowering all subjet transverse momenta in a correlated way:
\begin{equation}
\pt \to \pt \times \left( 1 \pm f \right) \, ,
\end{equation}
in which $f$ is a function which specifies the relative variation. Three
different scenarios for the dependence of $f$ on the subjet \pt are considered
(the parameters $k_i$ are constants):
\begin{itemize}
   \item  $f = k_1 \sqrt{\pt}$ (larger variation for high-energy subjets),
   \item  $f = k_2 /\pt$ (larger variation for low-energy subjets),
   \item  $f = k_3$ (no \pt dependence, variation by a constant factor).
\end{itemize}

Separate $\chi^2$ values are determined for all three functional forms and
for different values of the parameters $k_i$.
The \htt top-quark-candidate mass distribution is shown in \figref{insituJES_chi_dist}.
The simulation is shown for the nominal energy scale and, as an example, for the
case of the variation with
$f = k_2/\pt$ with $k_2 = 1\GeV$.
For subjets with $\pt=100\GeV$, this corresponds
to a relative change of the transverse momentum of $\pm 1\%$.
The description of the measured distribution is improved by the $+1\%$ variation.
The level of agreement between the measured and predicted distributions is quantified
in terms of the $\chi^2$ value shown in \figref{insituJES_chi_chi}
for different values of $k_2$. The variation is expressed as the relative
\pt change for subjets with $\pt = 100\GeV$ (JES shift). A parabola is
fitted to the $\chi^2$ values as a function of the JES shift. The best agreement
is obtained for a JES shift of $+1\%$, which leads to the smallest $\chi^2$, $\chi^2_\textrm{min}$.
This result can be used to correct the subjet \pt scale in the simulation. This is left to future studies.
Here, an uncertainty in the \pt scale is determined as follows.
From the two JES-shift values that correspond to $\chi^2 = \chi^2_\textrm{min}+1$, the
larger absolute value is used as the $1\sigma$ systematic uncertainty of the \pt scale.
In \figref{insituJES_chi_chi} this uncertainty is 2.2\%.

\begin{figure}[!h]
\centering
\subfigure[\label{fig:insituJES_chi_dist}]{
   \includegraphics[width=0.48\textwidth]{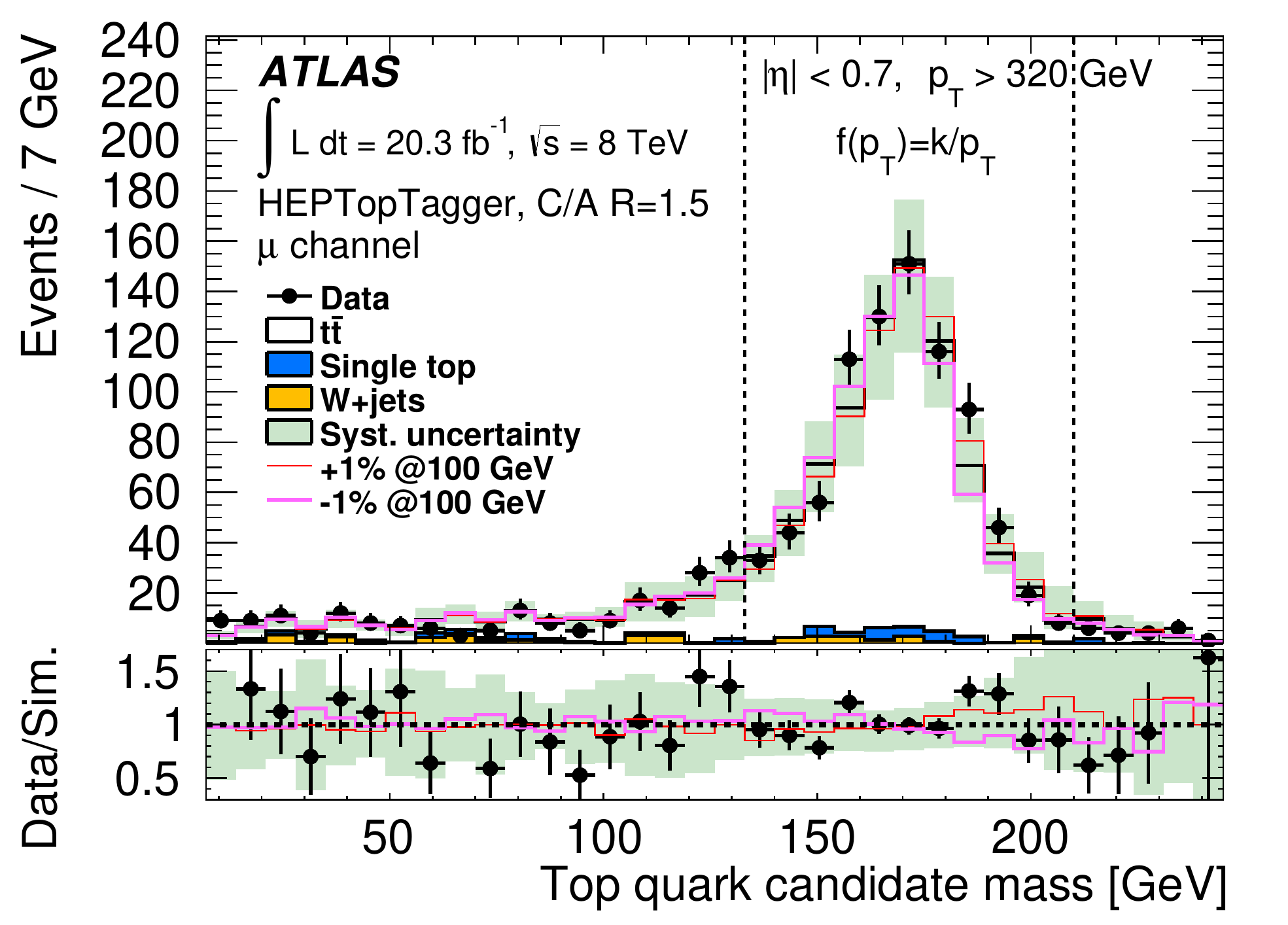}
}
\subfigure[\label{fig:insituJES_chi_chi}]{
   \includegraphics[width=0.48\textwidth]{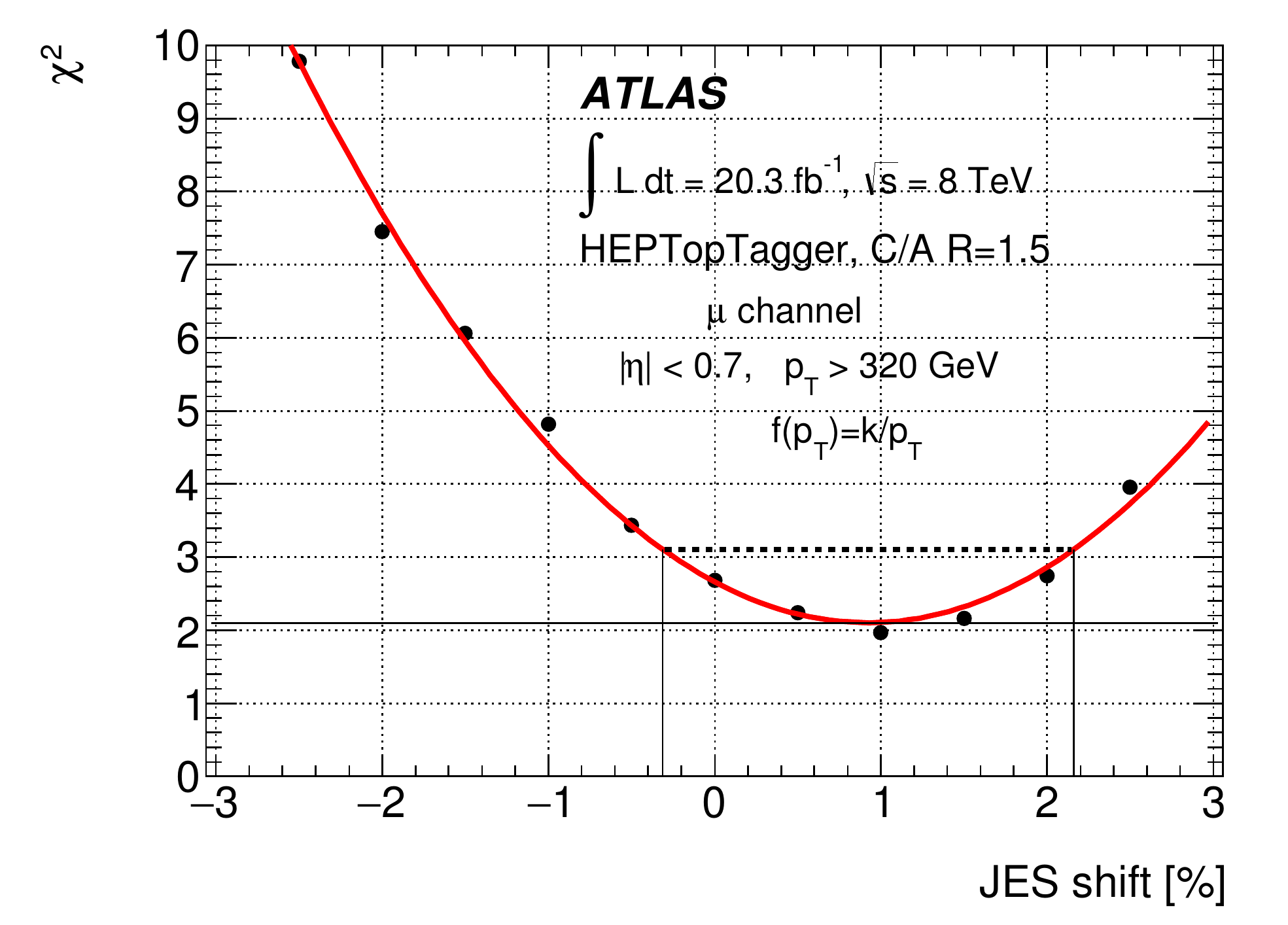}
}
\caption{(a) The \htt top-quark candidate mass distribution reconstructed in the
$\mu$+jets signal selection from \CamKt $R=1.5$ jets with
$\pt>320\GeV$ and $|\eta|<0.7$.
Only one $b$-tag within $\Delta R < 1.5$ of the lepton is required and the
second $b$-tag requirement is omitted.
Also shown are predictions for \ttbar, single top, and $W$+jets production with
the nominal subjet energy scale and with the subjet \pt multiplied by
$1+f$ (label `+1\%@100 $\GeV$') and $1-f$ (`-1\%@100 $\GeV$') with
$f = 1\GeV/\pt$, corresponding to shifts of $\pm 1\%$ for subjets with $\pt = 100\GeV$.
(b) The $\chi^2$ calculated from the measured top-quark candidate mass distribution
in the mass window from $133$ to $210\GeV$ as a function of different variations
of the simulated subjet energy scale of the form $f = k_2 /\pt$.
The variation is expressed as the relative \pt change for subjets with
$\pt = 100\GeV$ (JES shift). The nominal energy scale coincides with no
JES shift. The `1\%@100 $\GeV$' variation in
(a) corresponds to a JES shift of $+1\%$ and leads to the smallest $\chi^2$, $\chi^2_\textrm{min}$.
The distribution is fitted with a parabola and the positive and negative JES-shift
values at $\chi^2_\textrm{min}+1$ are indicated.
}
\label{fig:insituJES_chi}
\end{figure}

The subjet energy-scale uncertainty is determined in two bins of \largeR-jet
\pt ($<320\GeV, >320\GeV$) and two bins of \largeR jet
pseudorapidity ($|\eta|<0.7, 0.7<|\eta|<2.0$).
The results are shown in \figref{HTT_JESsjresultfunctions}.
The largest relative uncertainty is 10\% at a subjet \pt of $20\GeV$, dropping with
$1/\pt$ to 2.5\% at $90\GeV$ and then rising proportionally to $\sqrt{\pt}$,
reaching $3.5$--$4.0\%$ at $200\GeV$. The uncertainty depends weakly on the
\largeR jet \pt and \eta.

\begin{figure}[!h]
\centering
\subfigure[$|\eta|<0.7$]{
   \includegraphics[width=0.48\textwidth]{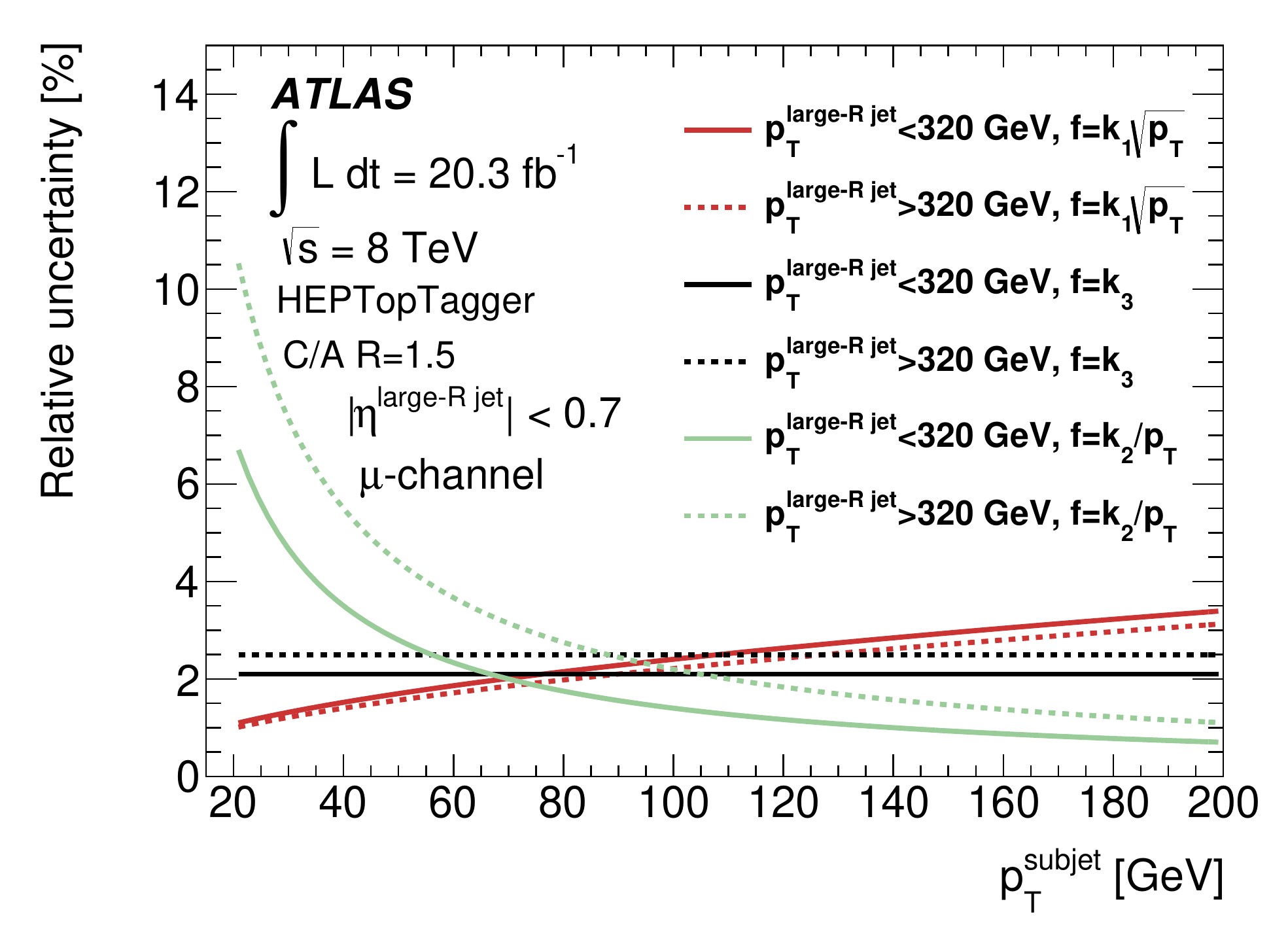}
}
\subfigure[$0.7<|\eta|<2.0$]{
   \includegraphics[width=0.48\textwidth]{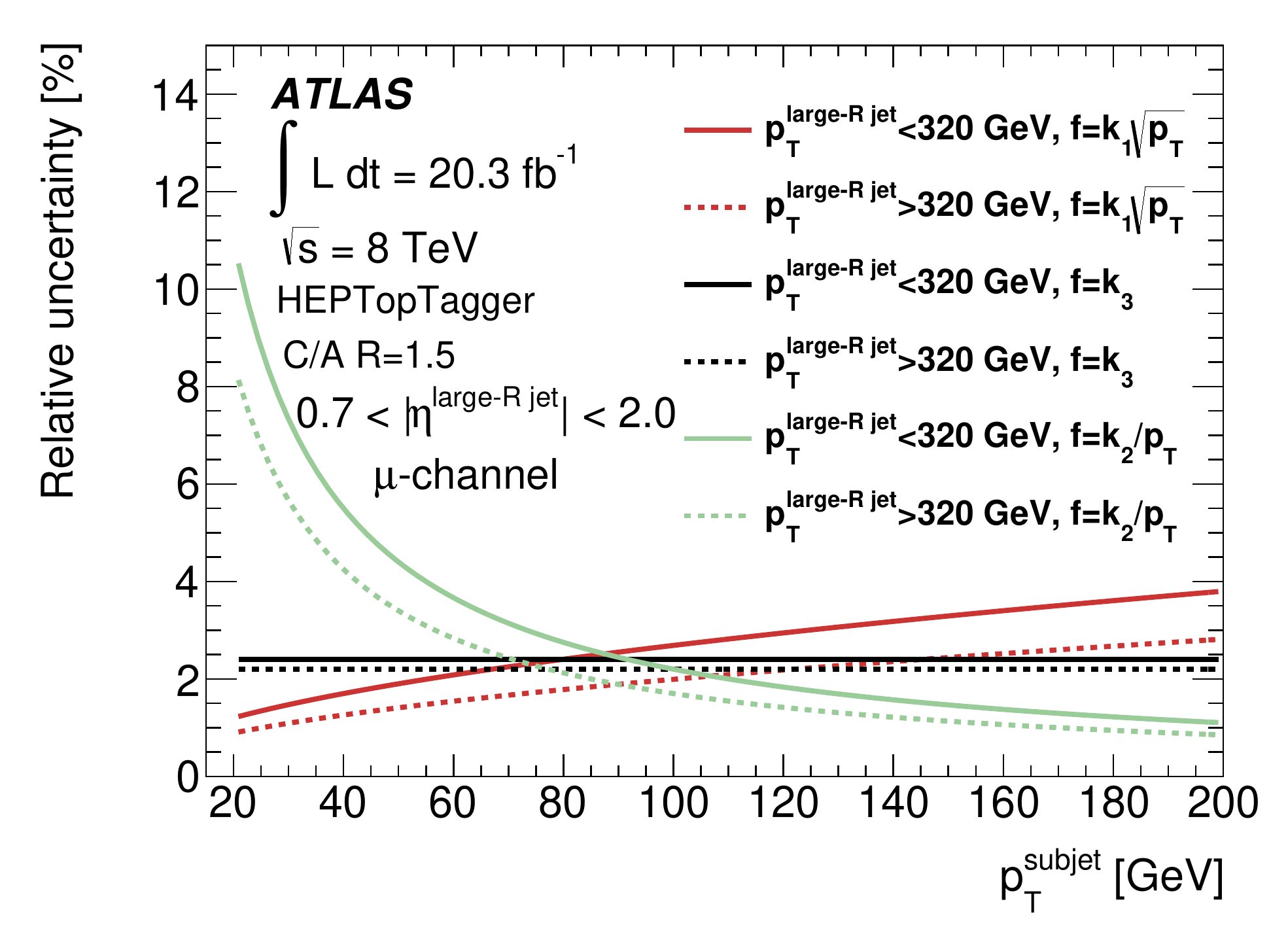}
}
\caption{Relative subjet energy-scale uncertainty as a function of the \htt subjet \pt
    for three functional forms of the relative \pt variation.
    The uncertainty is shown for two pseudorapidity intervals
    of the \CamKt $R=1.5$ jets in which the subjets are found: (a) $|\eta|<0.7$ and (b)
    $0.7<|\eta|<2.0$. The uncertainty is shown in two bins of
    the \CamKt $R=1.5$ jet \pt.}
\label{fig:HTT_JESsjresultfunctions}
\end{figure}

In the \htt analysis, the impact on each studied quantity (the number of
tagged \largeR jets, the tagging efficiency, and the mistag rate) is
determined for all three functional forms. The largest of the three
changes in the quantity is then used as the uncertainty related to the
imperfectly known subjet energy scale.

\subsection{Uncertainties in the modelling of physics processes}
\label{sec:modellingsyst}
Uncertainties related to the \ttbar simulation are taken into account
as follows. If the uncertainties are estimated from samples not generated
with the nominal \ttbar generator \PowPythia, then the sequential \pt reweighting
mentioned in \secref{mc} is not applied, because the reweighting used only
applies to \PowPythia: the nominal
\PowPythia prediction without reweighting is compared to the prediction
from the alternative simulation without reweighting.

The \ttbar cross-section uncertainty of $^{+13}_{-15}$~pb quoted in
\secref{mc} is used and an additional
normalization uncertainty of $^{+7.6}_{-7.3}$~pb from a
variation of the top-quark mass by $ \pm 1.0 \GeV$ is added in quadrature, leading
to a total relative normalization uncertainty of $^{+5.9\%}_{-6.6\%}$. For the
evaluation of the other \ttbar modelling uncertainties mentioned below, the total
\ttbar cross section of the generated event samples is set to the value given in \secref{mc},
so that no double-counting of normalization uncertainties occurs.

To account for uncertainties in the parton shower, the prediction from \PowHerwig
is compared to the prediction from \PowPythia.
Uncertainties in the choice of \ttbar generator are estimated by comparing
the prediction from \McatnloHerwig with the prediction from \PowHerwig.
The uncertainty in the amount of ISR and FSR is estimated
using two \AcermcPythia \ttbar samples with increased and decreased radiation.

PDF uncertainties affect the normalization of the total \ttbar cross
section and this is taken into account as described in \secref{mc}. They additionally affect
the \ttbar cross section in the phase space examined by this analysis and
the distributions of kinematic variables. These effects are determined by comparing
the prediction based on CT10 to the prediction based on HERAPDF1.5.
The cross-section difference obtained when comparing these two PDF sets was found to
match the difference due to the CT10 PDF uncertainty~\cite{Aad:2014zka}
for this region of phase space.

The factorization and renormalization scales are varied by factors two and one half
and the impact on the total \ttbar cross section is included in the
cross-section uncertainty. The impact in the phase space examined by this analysis
and on the distributions of kinematic variables is evaluated by comparing
dedicated \ttbar samples in which the two scales are varied independently.
The variation of the renormalization scale has a significant impact, while
the analysis is not sensitive to variations of the factorization scale beyond
the change of the total \ttbar cross section.

The impact of variations on the top-quark-candidate mass peak
of varying the top-quark mass in the generator by $\pm 1.0\GeV$
is taken into account
for the in situ determination of the subjet energy scale in \secref{insitu}.
For the efficiency and misidentification-rate measurements this uncertainty is negligible
compared to other sources of systematic uncertainty.

The uncertainties on the normalization of the single top,
\Wjets, and \Zjets background contributions were found to have a
negligible impact.

\section{Study of top-tagging performance using Monte-Carlo simulation}
\label{sec:MCcomparison}
\subsection{Comparison of top-tagging performance}
\label{sec:compare}

The performance of the different top-tagging approaches is compared using MC
simulations to relate the different \largeR jets used by the taggers and to
extend the comparison in \largeR jet \pt beyond the kinematic reach of the $8\tev$ data samples.

The performance is studied in terms of the efficiency for tagging signal
\largeR jets and the background rejection, defined as the reciprocal of the
tagging rate for background \largeR jets.
Signal jets are obtained from $\Zprime \to \ttbar$ events and background jets
are obtained from multijet events.
Multijets typically pose the largest background in \ttbar analyses in the
fully hadronic channel. The \Wjets background, where the \W boson decays
hadronically, is less important because of the smaller cross section. Also,
in the kinematic region considered in the comparison presented here, it was
shown for the \htt that the mistag rate is similar for multijet background
and background from $W\to q'\bar{q}$~\cite{Aad:2013gja}. In the lepton+jets
channel, \Wjets tends to be the most important background if the \W boson
decays leptonically, and then the background from the additional jets is
very similar to the multijets case. The conclusions drawn in this section
can therefore be extended to the context of this \Wjets background.

Stable-particle jets are built in all MC events using the \akt algorithm
and a radius parameter $R=1.0$. These jets are trimmed with the same parameters
as described in \secref{obj} for the detector-level jets.
These particle-level jets are used to relate the different jet types used at reconstruction
level.
The different types of \largeR jets used by the tagging algorithms
are listed in \tabref{taggerFatjets}.
Each reconstructed \largeR jet must be geometrically matched to a particle-level jet
within $\Delta R = 0.75$ for the trimmed \akt $R=1.0$ jets, and within $\Delta R = 1.0$ for the
\CamKt $R=1.5$ jets.
The fraction of reconstructed \largeR jets with no matching particle-level jet
is negligible.
In addition, particle-level jets in the signal sample must be geometrically
matched to a hadronically decaying top quark within $\Delta R = 0.75$.
The top-quark flight direction at the top-quark decay vertex is chosen, consistent
with the matching procedure discussed in \secref{signalsample}.
The particle-level jet \pt spectrum of the signal sample is reweighted to the \pt spectrum
of the background sample to remove the dependence on a specific signal model.
However, since the results in this section are given for different ranges of \pt, the
conclusions are believed to hold, approximately independently of the choice of
specific underlying \pt spectrum.

The comparison is performed in bins of the \pt of the
particle-level jet, \pttrue, in the range $350<\pttrue<1500\GeV$ in which
all taggers are studied.
For the performance comparison, the statistical uncertainties of the
simulated efficiencies and rejections are taken into account,
while no systematic uncertainties are considered.

\begin{figure}[p]
\begin{centering}
\subfigure[]{
\includegraphics[width=0.8\textwidth]{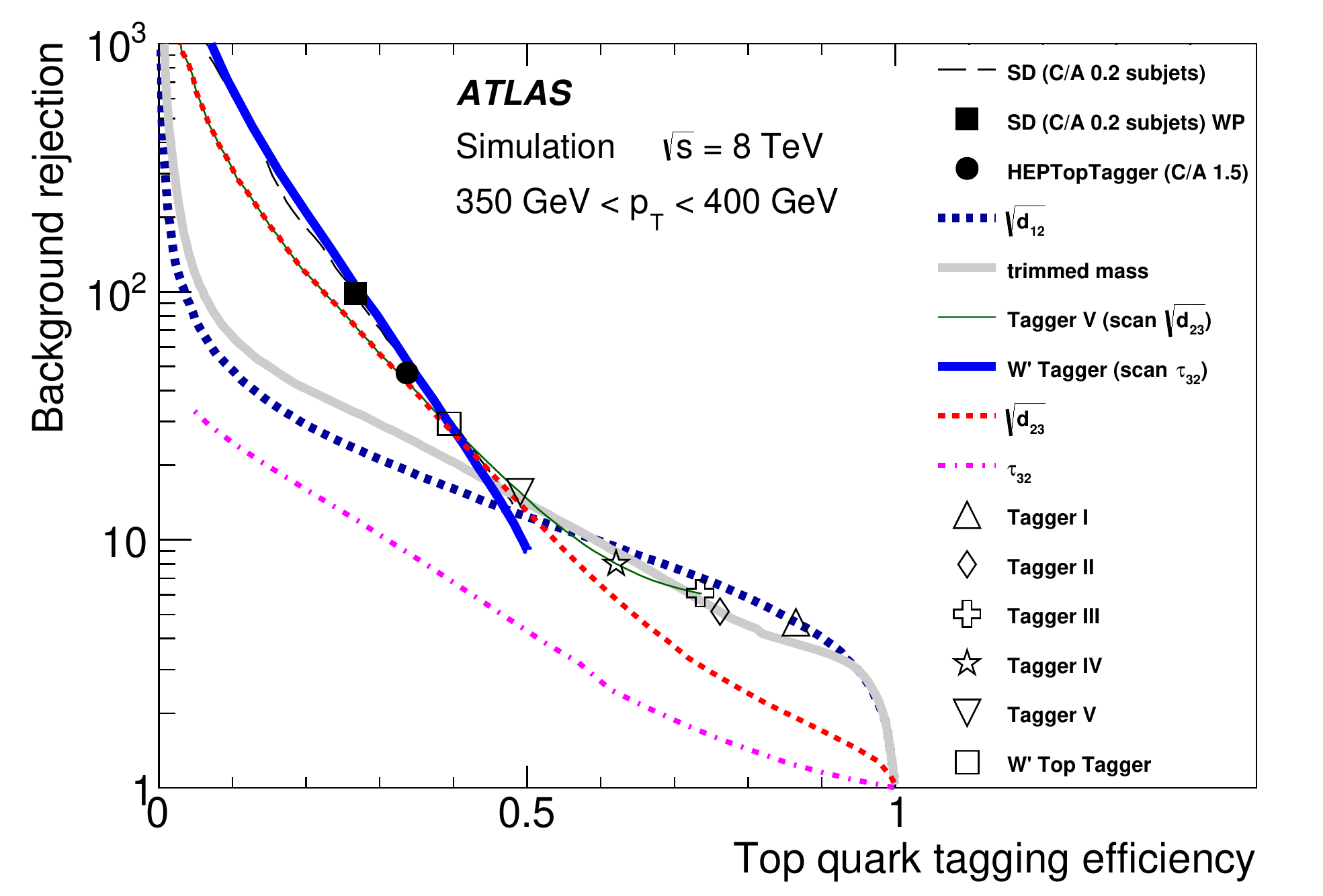}
}
\subfigure[]{
\includegraphics[width=0.8\textwidth]{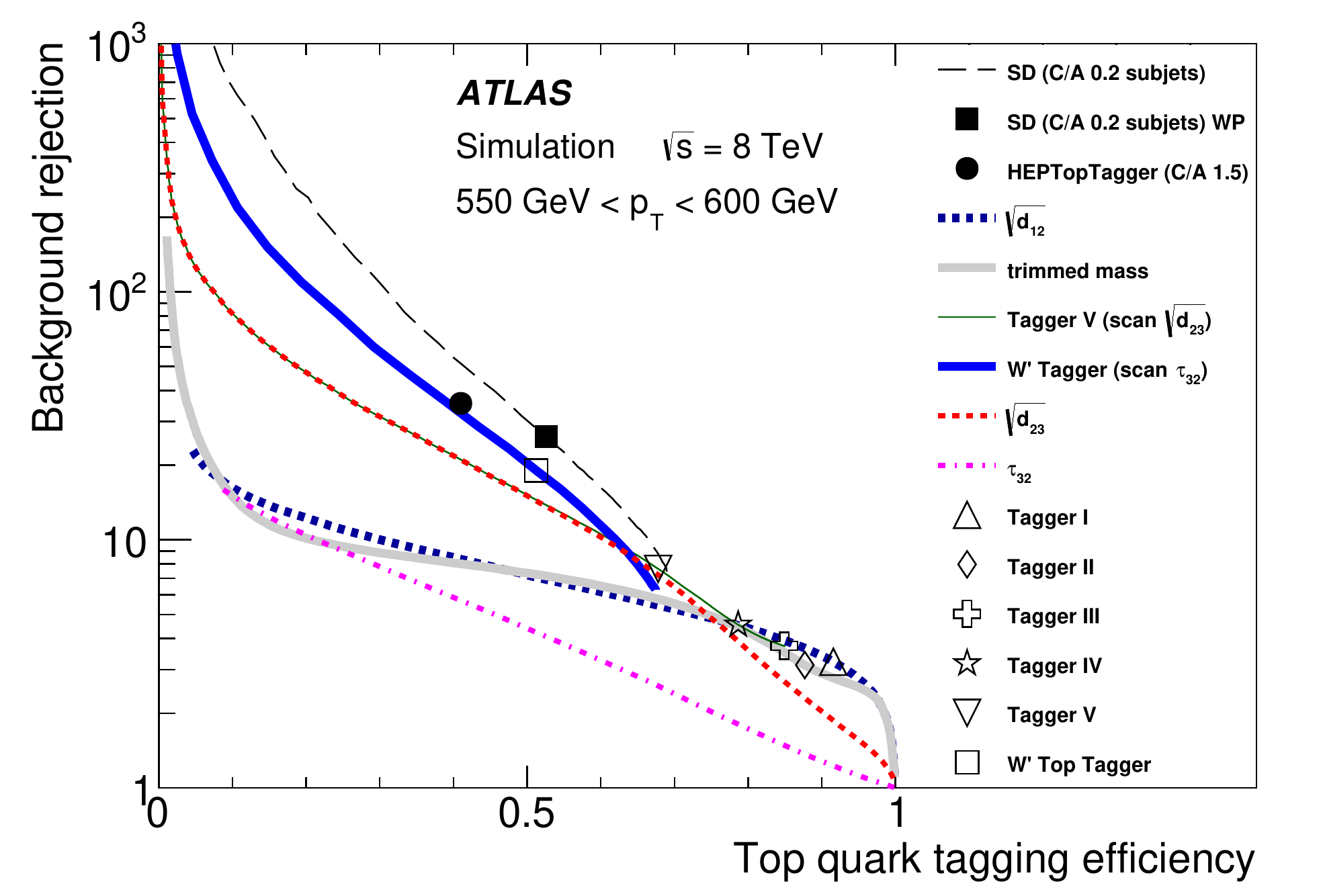}
}
\caption{
The background rejection as a function of the tagging efficiency of
\largeR jets, as obtained from MC simulations
for $350\gev<\pt<400\gev$ and $550\gev<\pt<600\gev$ for
trimmed \akt $R=1.0$ particle-level jets to which the \largeR jets are geometrically matched.
The \htt uses \CamKt $R=1.5$ jets; the other taggers use trimmed \akt $R=1.0$
jets.
For SD, the cut value of the discriminant $\ln\chi$ is scanned over.
Substructure-variable-based taggers are also shown including single scans over the trimmed mass,
\DOneTwo, \DTwoThr, \tauThrTwo and scans over cuts on \DTwoThr and \tauThrTwo for substructure
tagger V and the \WPT, respectively.
The curves are not shown if the background efficiency is higher than the
signal efficiency, which for some substructure-variable scans occurs for very
low signal efficiencies, i.e.\ for scans in the tails of the distributions.
The statistical uncertainty from the simulation is smaller than the symbols for the
different working points and it is no larger than the width of the lines shown.
}
\label{fig:compare1}
\end{centering}
\end{figure}

\begin{figure}[p]
\begin{centering}
\subfigure[]{
\includegraphics[width=0.8\textwidth]{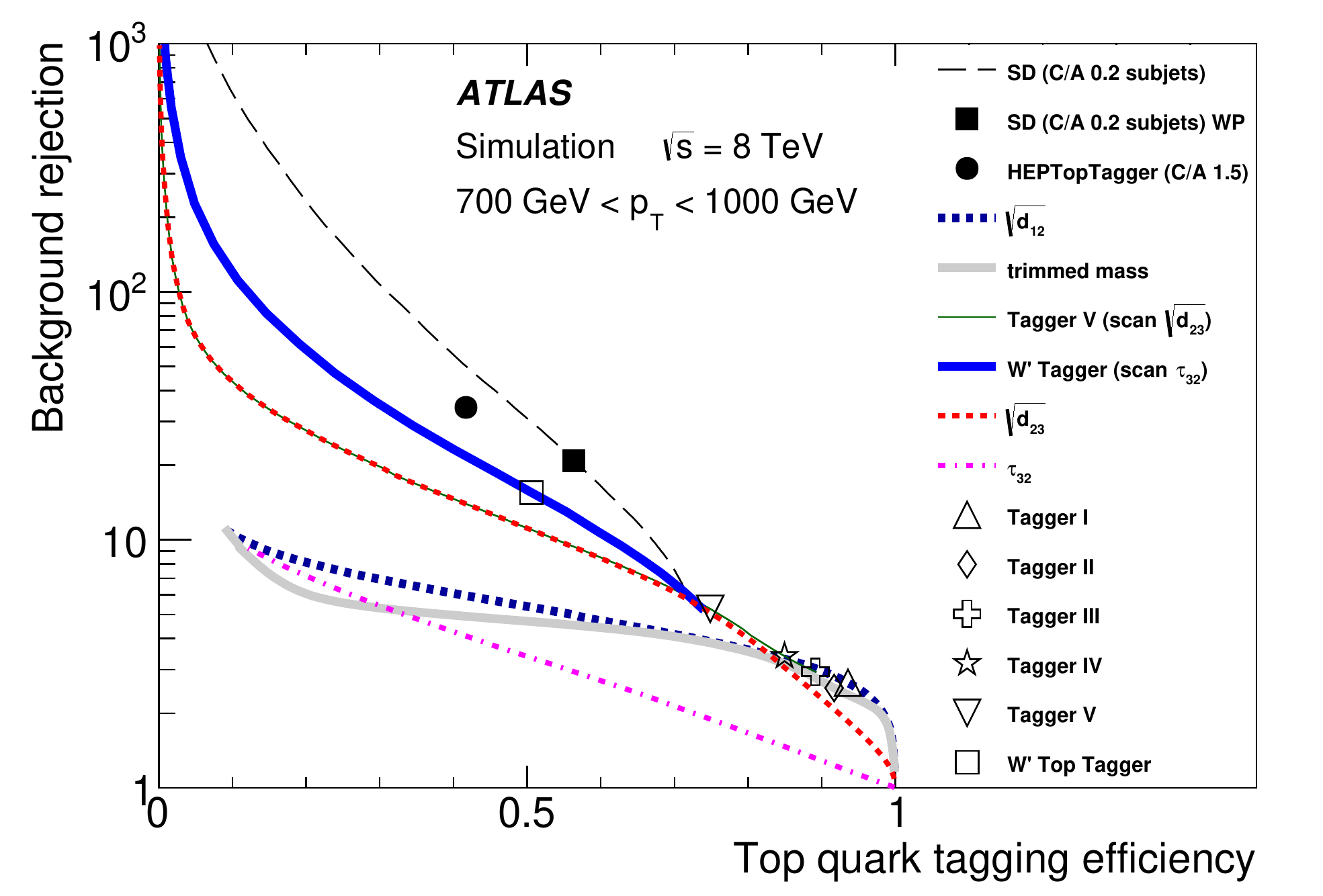}
}
\subfigure[]{
\includegraphics[width=0.8\textwidth]{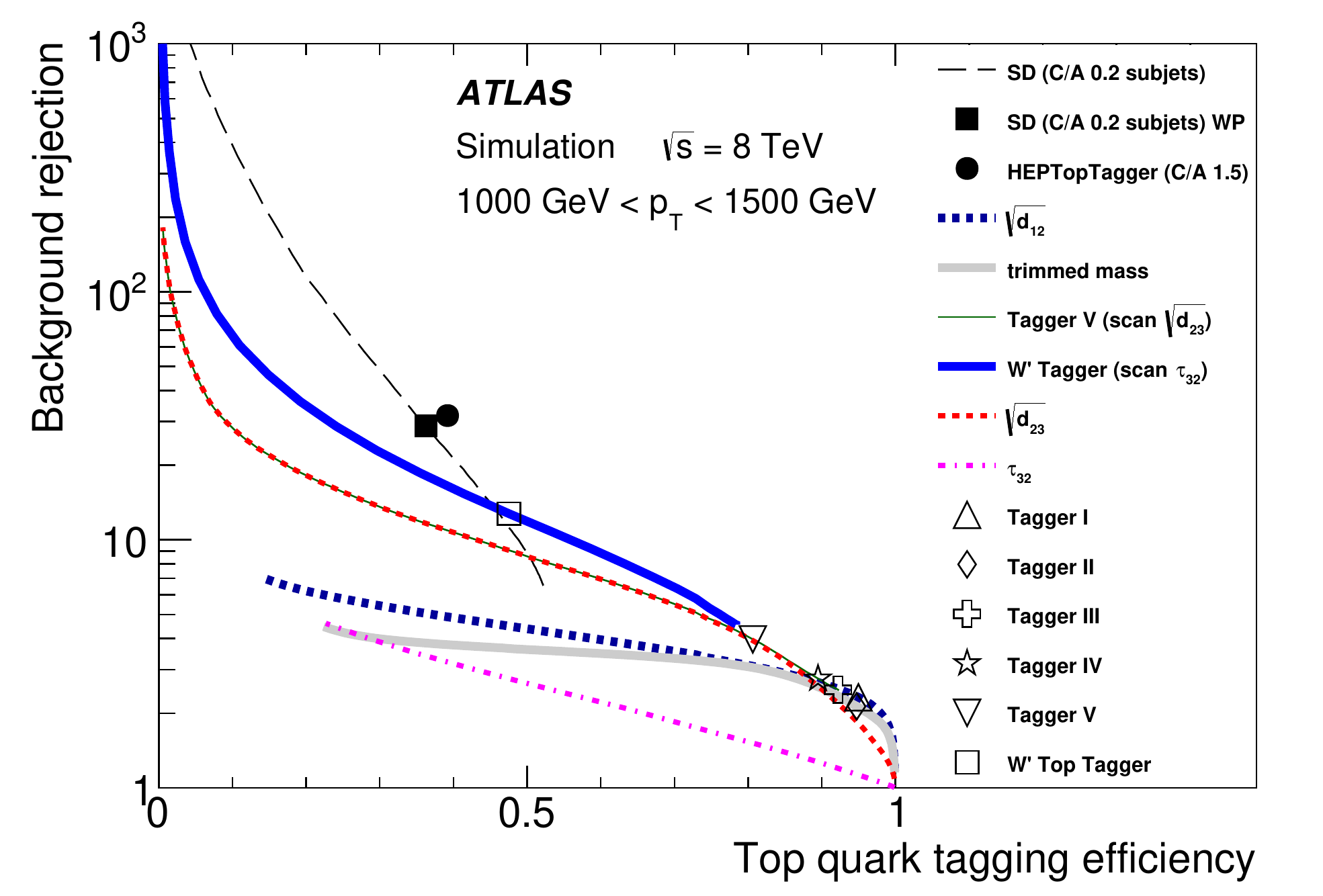}
}
\caption{
The background rejection as a function of the tagging efficiency of
\largeR jets, as obtained from MC simulations
for $700\gev<\pt<1000\gev$ and $1000\gev<\pt<1500\gev$ for
trimmed \akt $R=1.0$ particle-level jets to which the \largeR jets are geometrically matched.
The \htt uses \CamKt $R=1.5$ jets; the other taggers use trimmed \akt $R=1.0$
jets.
For SD, the cut value of the discriminant $\ln\chi$ is scanned over.
Substructure-variable-based taggers are also shown including single scans over the trimmed mass,
\DOneTwo, \DTwoThr, \tauThrTwo and scans over cuts on \DTwoThr and \tauThrTwo for substructure
tagger V and the \WPT, respectively.
The curves are not shown if the background efficiency is higher than the
signal efficiency, which for some substructure-variable scans occurs for very
low signal efficiencies, i.e.\ for scans in the tails of the distributions.
The statistical uncertainty from the simulation is smaller than the symbols for the
different working points and it is no larger than the width of the lines shown.
}
\label{fig:compare2}
\end{centering}
\end{figure}

The background rejection is shown as a function of the tagging efficiency in
\figsref{compare1}{compare2} in four bins of \pttrue:
$350$--$400\gev$, $550$--$600\gev$, $700$--$1000\gev$, and $1000$--$1500\gev$.
Curves in the efficiency--rejection plane are obtained by varying the values
of cuts in the tagger definitions.
For the taggers based on substructure variables,
scans over the cut values of the trimmed mass, \DOneTwo, \DTwoThr, and \tauThrTwo are shown,
and in addition scans over the cut values of \DTwoThr in substructure tagger V and of \tauThrTwo
in the \WPT, for which the cuts on the other variables are kept at their nominal values.
The cuts on the trimmed mass and splitting scales are single-sided lower bounds, and the
cut on \tauThrTwo is a single-sided upper bound.

When using only a single substructure-variable cut, the best performing variables
in all studied \pttrue intervals
are the splitting scale \DOneTwo at high efficiency and \DTwoThr at lower efficiency.
At an efficiency of 80\%, a cut on \DOneTwo achieves a background rejection of $\approx\!3$--$6$ over
the full range in \pttrue.
At an efficiency of 40\%, a cut on \DTwoThr achieves a rejection of $\approx\!25$ for lower
values of \pttrue, decreasing to a rejection of $15$ for $700<\pttrue<1000\gev$
and $11$ for $1000<\pttrue<1500\gev$, respectively.
The efficiency at which the rejection of a cut on \DTwoThr is higher than the rejection for the
trimmed-mass cut depends on \pttrue: it is $\approx\!45\%$ for $350<\pttrue<400\GeV$
and increases to 90\% for $1000<\pttrue<1500\GeV$.
A cut on the trimmed mass performs similarly to the \DOneTwo cut. A cut on \tauThrTwo
performs significantly worse.
For high efficiencies and the ranges of lower \pttrue (e.g. $\approx\!60$--$90\%$ for
$350<\pttrue<400\gev$), the cut on the trimmed mass shows only a small increase in the
rejection with decreasing signal efficiency. For lower efficiencies, the rejection
increases more strongly with decreasing signal efficiency. This is due to the two distinct
$W$-boson and top-quark mass peaks in signal, as exemplified in \figref{ctrl_akt_pretag_fjm}.
Adding the cuts on the mass and \DOneTwo to the cut on \DTwoThr (Tagger V (scan \DTwoThr))
does not significantly improve the performance over a cut on \DTwoThr alone,
since for high enough cuts on \DTwoThr, the other cuts are automatically satisfied
because of the relation $m>\DOneTwo>\DTwoThr$.

A combination of \Nsj and splitting-scale information, as used in the \WPT,
gives the best performance of all studied substructure-variable-based approaches
for efficiencies below a certain threshold efficiency. This threshold efficiency
is $\approx\!40\%$ for $350<\pttrue<400\GeV$ and it increases to $\approx\!80\%$
for $1000<\pttrue<1500\GeV$. By varying the \tauThrTwo requirement in the \WPT,
rejections close to the ones of SD and the \htt can be achieved at the
same efficiency.

For SD, the cut value of the discriminant $\ln\chi$ is varied. The maximum efficiency
is $\approx\!50\%$ in the lowest \pt bin studied ($350<\pttrue<400\gev$).
For higher \pt, the efficiency rises up to 70\%.
The maximum efficiency is determined by the requirement of having at least three subjets
which combine to an invariant mass near the top-quark mass and a subset of these
subjets to give a mass near the \W-boson mass.
The increase of the maximum efficiency from approximately 50\%
at $350$--$400\gev$ to approximately 70\% at $550$--$1000\gev$ is a result of the larger average
containment of the top-quark decay products in the \largeR jet at higher \pt.
At the highest \pt values ($1000$--$1500\gev$),
the use of $R=0.2$ subjets limits the efficiency as
the top-quark decay products cannot be fully resolved for an increasing fraction
of \largeR jets, resulting in a maximum efficiency of $\approx\!50\%$.

For $350<\pttrue<400\gev$, the \htt has an efficiency of 34\% at a
rejection of 47. For $\pttrue>550\GeV$, the efficiency is $\approx\!40\%$
and the rejection is $\approx\!35$, approximately independent of \pttrue.
The \htt performance was also investigated for $200<\pttrue<350\gev$ (not shown):
efficiency and rejection are 18\% and 300, respectively, for $200<\pttrue<250\gev$,
22\% and 130 for $250<\pttrue<300\gev$, and 28\% and 65 for $300<\pttrue<350\gev$.

For $350<\pttrue<450\GeV$, the performance of SD, the \htt, and the \WPT are comparable.
For $450<\pttrue<1000\GeV$, SD offers the best rejection in simulation, up to its maximum efficiency.
Top tagging efficiencies above 70\% can be achieved with cuts on substructure variables,
where, depending on \pttrue, optimal or close-to-optimal performance can be achieved
with a requirement on \DOneTwo alone.
For $1000<\pttrue<1500\gev$, of all the top-tagging methods studied, the
\htt offers the best rejection ($\approx\!30$) at an efficiency of $\approx\!40\%$,
making it a viable option for high-\pt searches despite not having been optimized
for this \pt regime.
The only tagger studied for $200<\pttrue<350\gev$ is the \htt.

\subsection{\httofour performance}
\label{sec:httofour}

The efficiencies for hadronically decaying top quarks to be reconstructed as top-quark candidates with
the \httofour and \htt methods are shown in \figref{HTT04_MCc} as a function
of the true \pt of the top quark in simulated \ttbar events. The events are
selected according to the criteria described in \secref{signalsample}, except
that all requirements related to \largeR jets are not applied in the case of
\httofour.
For these efficiencies, a top quark is considered tagged if a top-quark
candidate is reconstructed with a momentum direction within $\Delta R=1.0$ of
the top-quark momentum direction. The definition of the efficiency is therefore
different from the \largeR-jet-based one used in \secref{compare},
where also a different event selection and different matching criteria are applied.
The efficiency of the \httofour method increases with the \pt of the top quark and reaches values of
$\approx\!50\%$ for $\pt > 500\gev$.
The efficiency of the \httofour method is lower than the efficiency of the \htt, but follows the trend
of the \htt efficiency closely.
The \htt efficiency reaches higher values than in \secref{compare} primarily
because the event selection here requires two $b$-tagged jets.

This efficiency, however, does not take into account the specific needs of event reconstruction
in final states with top quarks and many additional jets, for which the \httofour was designed.
An example of such a topology in an extension of the SM is the associated
production of a top quark and a charged Higgs boson, $H^+$,
decaying to $t\bar{b}$, i.e.\ $pp\rightarrow H^+ \bar{t}(b)\rightarrow t\bar{b}\bar{t}(b)$.
After the decay of the top quarks, the final state contains three or four $b$-quarks.
Up to two $b$-jets not associated with a top-quark decay can in principle be reconstructed,
and they should not be part of the reconstructed top-quark candidates.

\begin{figure}[!h]
\centering
\includegraphics[width=0.48\textwidth]{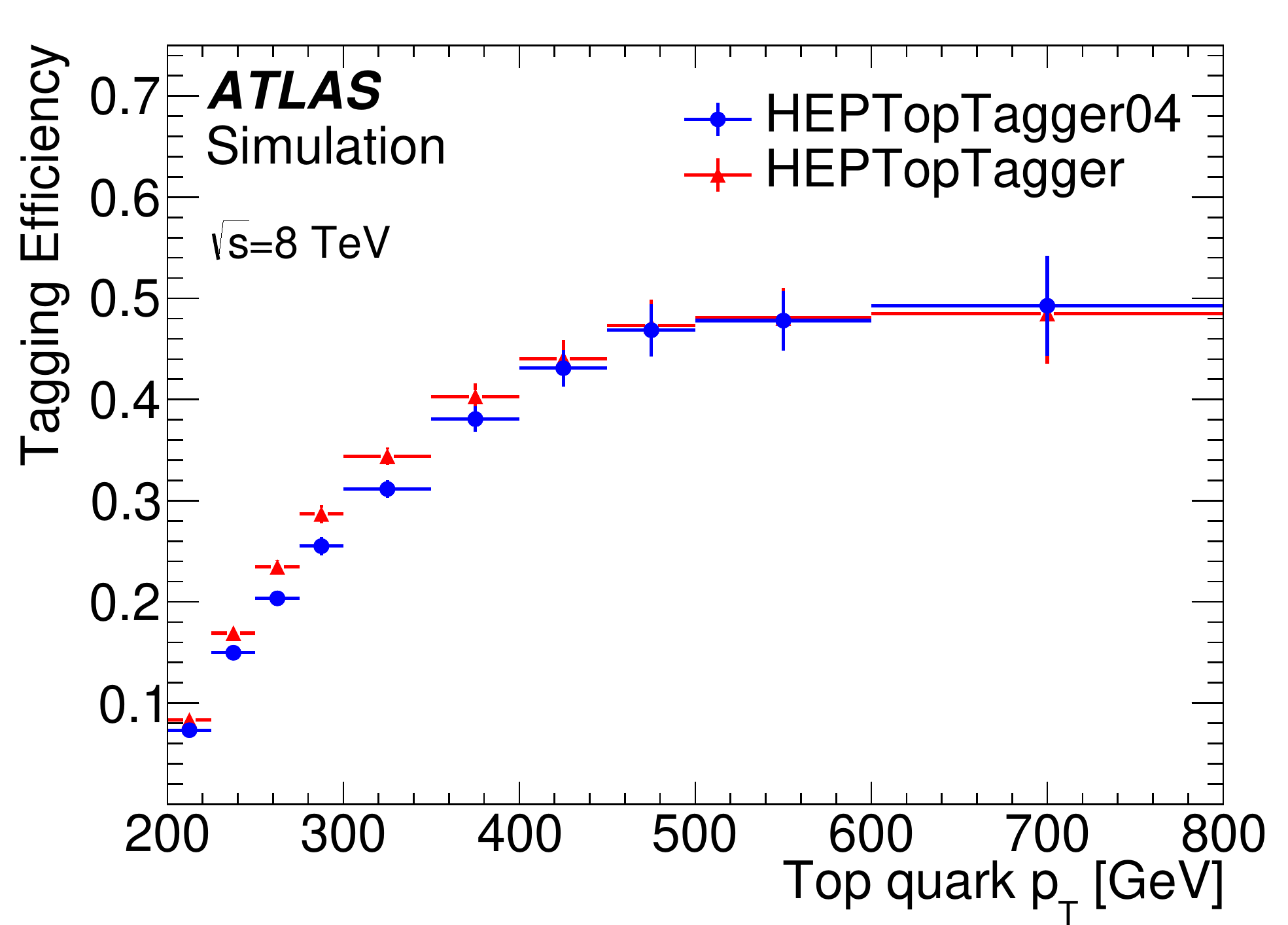}
\caption{
  Efficiency to reconstruct and identify a hadronically decaying top quark with the \httofour (blue circles) and the \htt
  (red triangles) as a function of the \pt of the top quark for events passing the signal selection
  described in \secref{signalsample}. A top quark is considered tagged if
  a top-quark candidate is reconstructed with a momentum direction within $\Delta R=1.0$ of
  the top-quark momentum direction.
}
\label{fig:HTT04_MCc}
\end{figure}

In ATLAS, $b$-jets are usually reconstructed using the \akt algorithm with $R=0.4$.
For large $H^+$ masses, for which the top quarks from its decay may have large \pt,
ensuring no overlap between the top-quark candidates and the unassociated $b$-jets may not be trivial.
In this case, hadronically decaying top quarks may be reconstructed with \largeR jet substructure analysis.
The reconstruction of \akt $R=0.4$ and \largeR jets, however, proceeds independently,
so that the same clusters may be present in \akt $R=0.4$ and \largeR jets.
If the \akt $R=0.4$ jet and the \largeR jet overlap, the $b$-tagged \akt $R=0.4$ jet might also originate from the hadronic top quark
decay, which prevents an unambiguous reconstruction of the final state. Moreover, clusters
included in both objects may lead to a double-counting of deposited energy, which is an issue
if for example an invariant mass is formed from the tagged top and a close-by $b$-jet targeting
the $H^+ \rightarrow t\bar{b}$ decay.

In the case of the \htt, subjets of the \largeR jet are explicitly
reconstructed, and it would be an option
to only consider \akt $R=0.4$ jets not matched to one of the three subjets
which form the top-quark candidate as being not associated with a hadronically decaying top.
This approach, however, is not straightforward because of the different jet algorithms and
jet radii used for \htt subjets and $b$-tagging. A simple approach is to require
an angular separation $\DeltaR$ between the top-quark candidate
and the \akt $R=0.4$ jets in the event, denoted \httDR in the following.
The \httofour is therefore compared to \httDR, using the latter as a benchmark.

In \figref{HTT04_MCa}, the energy shared by \akt $R=0.4$ jets and
\CamKt $R=1.5$ jets is shown for simulated \ttbar events.
The shared energy is calculated from the clusters of calorimeter cells included as constituents
in the \smallR and \largeR jets.
The \CamKt jets are required to fulfil $|\eta|<2.1$ and $\pt>180\gev$, and the \akt jets
must fulfil $|\eta|<2.5$ and $\pt>25\gev$.
All combinations of \largeR \CamKt jets and \smallR \akt jets in each event are shown.
The shared energy is normalized to the total energy of the \smallR jet
and this shared energy fraction is shown as a function of the angular separation $\DeltaR$
of the \smallR and \largeR jets. The region of small angular separation is populated
by combinations where a large fraction of the energy of the \smallR jet is included
in the \largeR jet, i.e.\ where the two jets originate from the same object.
However, for larger values of $\Delta R$, a significant fraction of the energy of the \smallR jet can still be shared with the \largeR jet.

\begin{figure}[!h]
\centering
\subfigure[]{
\label{fig:HTT04_MCa}
   \includegraphics[width=0.48\textwidth]{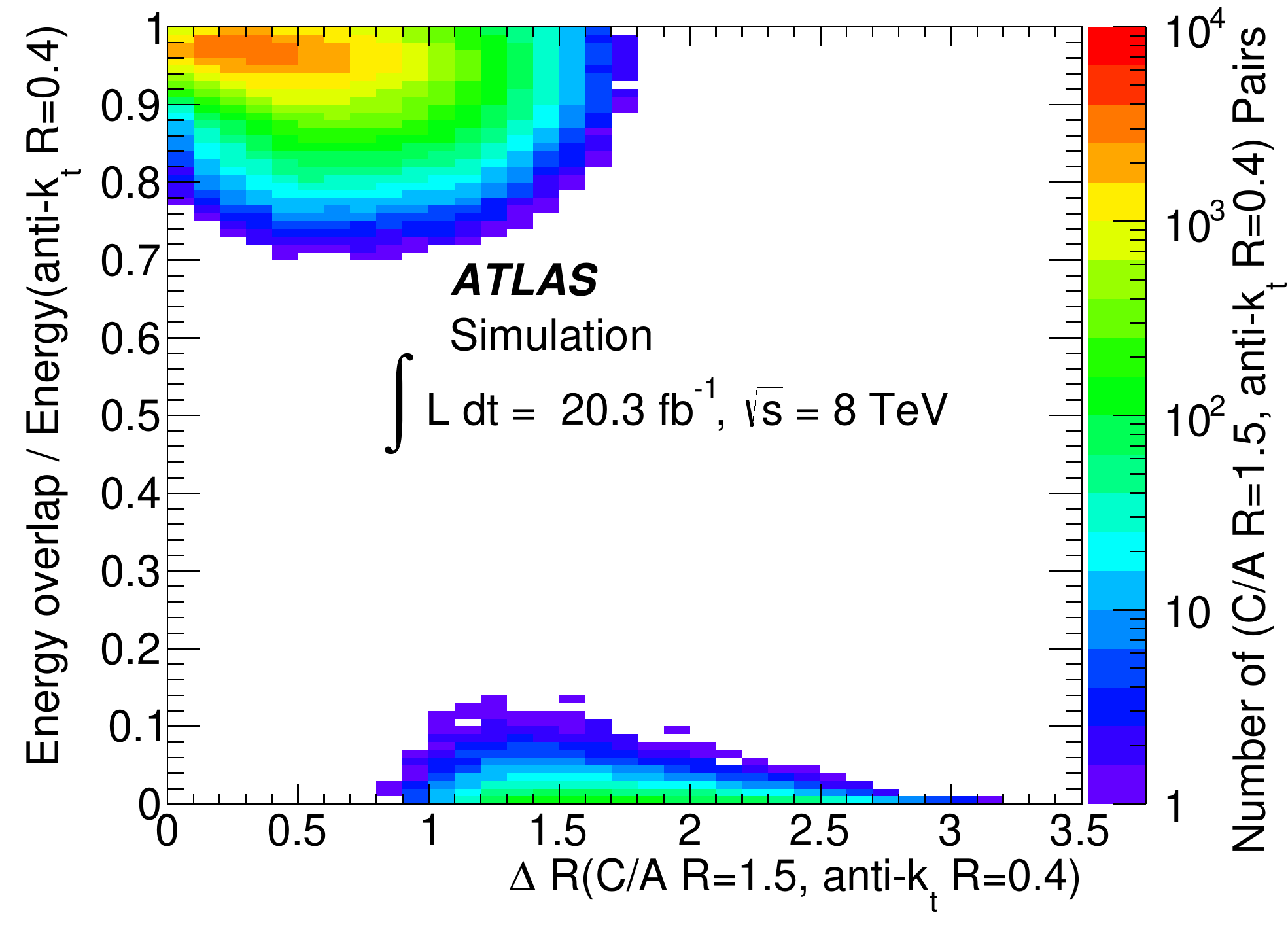}
}
\subfigure[]{
\label{fig:HTT04_MCb}
   \includegraphics[width=0.48\textwidth]{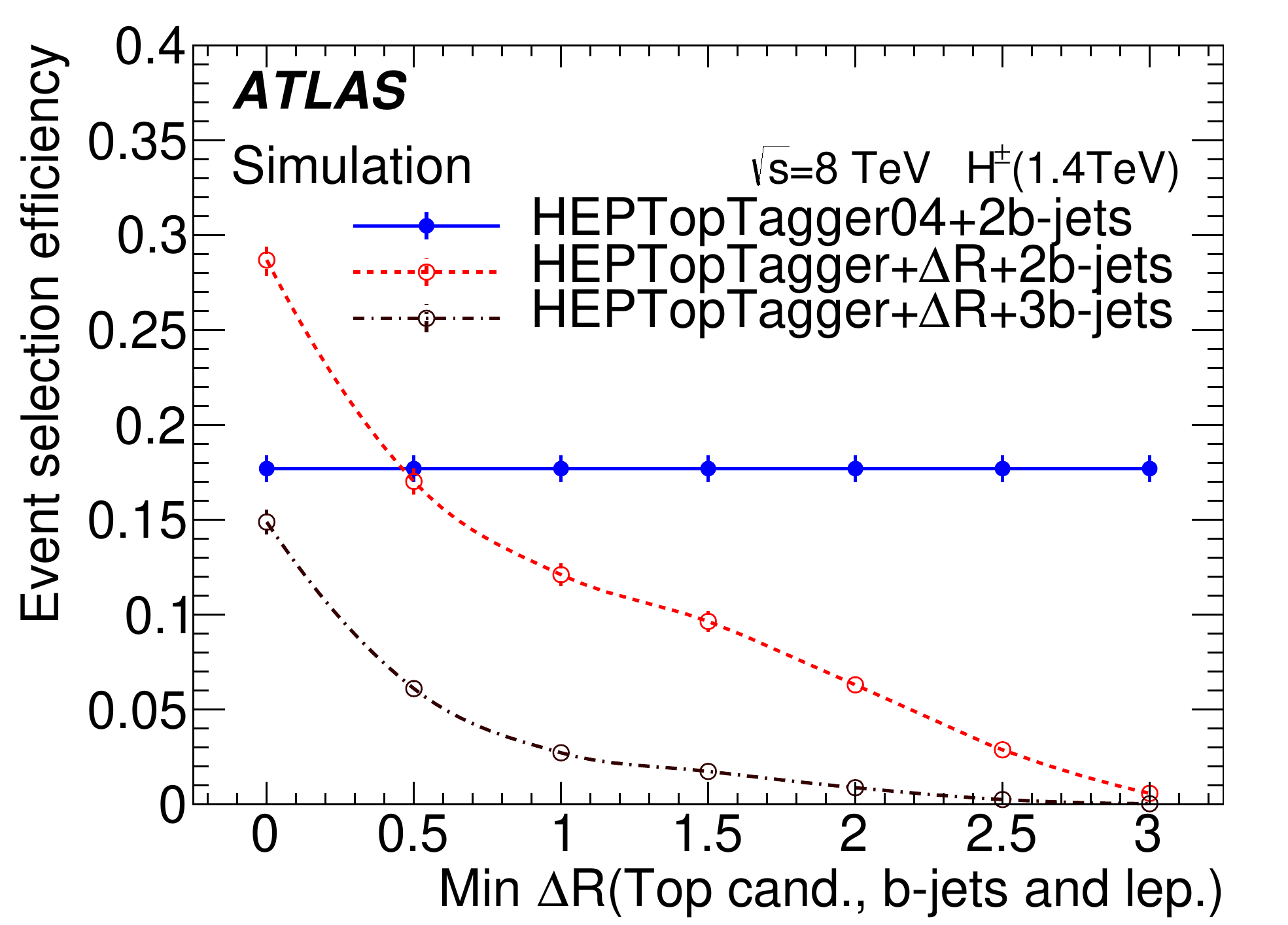}
}
\caption{(a) Energy fraction of clusters included in \akt jets with $R=0.4$ also included in
  C/A jets with $R=1.5$
  in \ttbar MC simulation as a function of the angular separation
  of the two jets. The C/A jets have to fulfil $|\eta|<2.1$ and $\pt>180\gev$,
  and all combinations of \largeR and \smallR jets in each event are shown.
  (b) Efficiency for the $H^+$ selection for the \httofour method for a
  $1400~\gev$ $H^+$ signal (blue, full circles)
  and for \htt for which an angular separation
  $\Delta R$ is required between the top-quark candidate
  and the closest anti-$k_t$ $R=0.4$ jet (or lepton) in the event (red open circles),
  \httDR.
  The efficiency of an alternative $H^+$ selection with three $b$-tagged \akt $R=0.4$ jets is
  shown in addition for \httDR.
  For \httDR, the efficiency is shown as a function of $\Delta R$,
  while the \httofour algorithm is independent of $\Delta R$.
}
\label{fig:HTT04_MC}
\end{figure}

The \httofour approach solves the issue of overlap between \largeR and \smallR jets by passing only the constituents of a set
of \smallR jets to the \htt algorithm and by removing these \smallR jets
from the list of jets considered for the remaining event reconstruction, i.e.\ the identification of extra $b$-jets.

The charged-Higgs-boson process mentioned above is used to illustrate the
advantage of the \httofour approach. A basic event selection for events with an $H^+$ boson
is introduced in order to study the performance of the \httofour in this topology
using simulated events only.
It consists of the signal selection for \ttbar events
as detailed in \secref{signalsample} requiring
at least one top-quark candidate reconstructed with the \httofour method
and two $b$-tagged \akt $R=0.4$ jets not considered as part of the \httofour candidate ($H^+$ selection).
The $b$-tagged \akt $R=0.4$ jets are allowed to be identical to the $b$-tagged jets required in the
signal selection, if these jets are not part of the \httofour candidate.

The \httofour method is compared with \httDR in the $H^+$ selection.
Only those $b$-tagged \akt $R=0.4$ jets that are more than $\Delta R$ away from
the top-quark candidate are considered in the $H^+$ selection for \httDR.
Moreover, the top-quark candidate is required to be separated from the reconstructed
lepton by at least $\Delta R$.
\figref{HTT04_MCb} shows the efficiency of the $H^+$ selection for a $1400\gev$ $H^+$ signal MC sample
for \httDR as a function of $\Delta R$, and for the \httofour method, which is independent
of $\Delta R$.
The \httofour leads to a higher efficiency than the simple \httDR benchmark for values of $\Delta R > 0.5$. In order to
avoid energy sharing, larger values of $\Delta R$ would be appropriate (cf. \figref{HTT04_MCa}).
For small values of $\Delta R$, \httDR shows a higher efficiency than the \httofour
method, because at least one $b$-tagged jet largely overlaps with the top-quark candidate and can be
identified with the $b$-quark from the top-quark decay and not with one of the additional $b$-quarks from the
$pp\rightarrow H^+ \bar{t}(b)\rightarrow t\bar{b}\bar{t}(b)$ process.
An additional $b$-tagged \akt $R=0.4$ jet can be required in the event selection
for \httDR to address this issue,
which leads to a lower efficiency for \httDR than for the \httofour method for all values of $\Delta R$.

In order to determine the optimal method for a particular application, mistag-rate comparisons
of the two approaches are important to evaluate using the exact selection of that analysis due
to the critical dependence on the dominant background composition and kinematic region.

\section{Measurement of the top-tagging efficiency and mistag rate}
\label{sec:eff_misid}
In this section, the signal and background samples introduced in
\secsref{signalsample}{backgroundsample} are used to study the
top tagging efficiency and the mistag rate for the different
top taggers introduced in \secref{techniques}.

\subsection{Top-tagging efficiency}
\label{sec:eff}
The \largeR jets in the signal selection are identified with a high-\pt hadronically
decaying top quark in lepton+jets \ttbar events and are therefore
used to measure the top-tagging efficiency in data as a function of the
kinematic properties of the \largeR jet (\pt, $\eta$).
The tagging efficiency is given by the fraction of tagged \largeR jets after
background has been statistically subtracted using simulation. In each
\largeR jet \pt and $\eta$ bin $i$, the efficiency is defined as
\begin{equation}
\fdatai = \left( \frac{N_{\textrm{data}}^\textrm{tag} -
N_{\textrm{\ttbar not~matched}}^\textrm{tag}-N_{\textrm{non-\ttbar}}^\textrm{tag}}
{N_{\textrm{data}}-N_{\textrm{\ttbar not~matched}}-N_{\textrm{non-\ttbar}}} \right)_i \ ,
 \label{eq:HTTtaggedfjfraction}
\end{equation}
in which
\begin{itemize}
\item $N_{\textrm{data}}^\textrm{(tag)}$ is the number of measured (tagged) \largeR jets;
\item $N_{\textrm{\ttbar not~matched}}^\textrm{(tag)}$ is the number of
(tagged) not-matched \largeR jets, i.e.\ jets not matched to a hadronically
decaying top quark (cf. \secref{matched}), according to the \PowhegPythia simulation;
\item $N_\textrm{non-\ttbar}^\textrm{(tag)}$ is the number of (tagged) \largeR jets
predicted by simulation to arise from other background contributions, such as \Wjets, \Zjets and single-top production.
\end{itemize}
Systematic uncertainties affecting the numerator and the denominator do not
fully cancel in the ratio, because in particular the amount of not-matched
\ttbar production is much reduced after requiring a top-tagged jet, but before
the top-tagging requirement the number of not-matched \ttbar events is non-negligible.

The measurement is shown for \pt bins in which the relative statistical uncertainty
of the efficiency is less than 30\% and the relative systematic uncertainty
is less than 65\%.
Two regions in \largeR jet pseudorapidity are chosen,
$|\eta|<0.7$ and $0.7<|\eta|<2.0$,
in which approximately equal numbers of events are expected.

The measured efficiency is compared to the efficiency in simulated \ttbar events, which is defined as
\begin{equation}
\fMCi = \left( \frac{N_\textrm{MC}^\textrm{tag}}{N_\textrm{MC}} \right)_i \ ,
\label{eq:f_MC}
\end{equation}
in which $N_\textrm{MC}^\textrm{(tag)}$ is the number of (tagged) \largeR jets
in matched \ttbar events which pass the signal selection.

\subsubsection{Efficiency of the substructure-variable taggers}

The measured and predicted top-tagging efficiencies for the top taggers I--V and the \WPT\
are studied as a function of the \pt of the trimmed
\akt $R=1.0$ jet in the two pseudorapidity regions.
In \figsref{eff_substructure_lowEta}{eff_substructure_lowEtaTwo},
the efficiencies in the lower $|\eta|$ region
are shown.
The efficiencies of the different top taggers are similar in the two $\eta$ regions,
as seen in \figref{eff_substructure_highEta},
in which the efficiencies of tagger III and the \WPT in the higher $|\eta|$ region are
shown.

\begin{figure}[p]
\begin{centering}
\subfigure[]{
\includegraphics[width=0.48\textwidth]{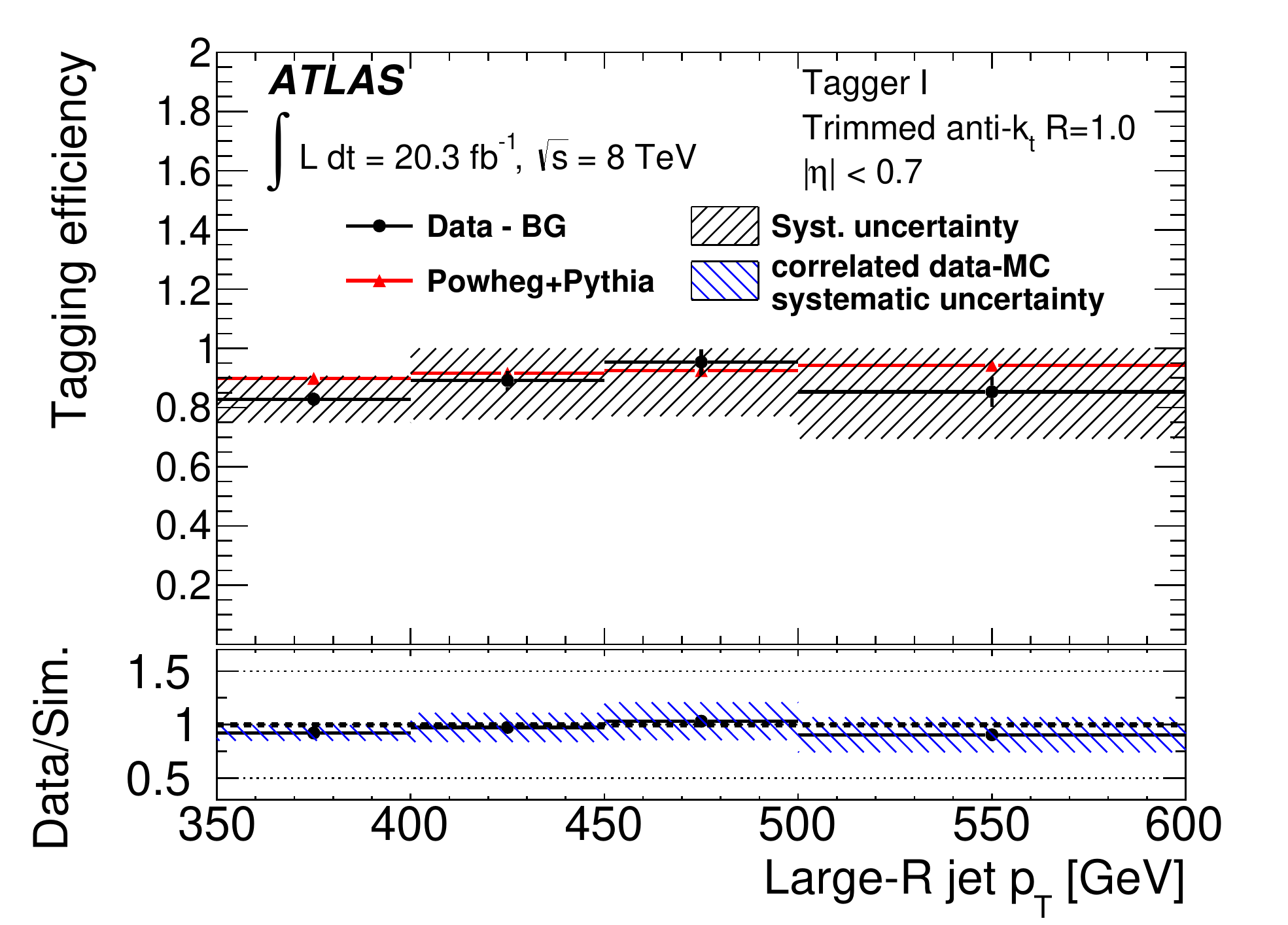}
}
\subfigure[]{
\includegraphics[width=0.48\textwidth]{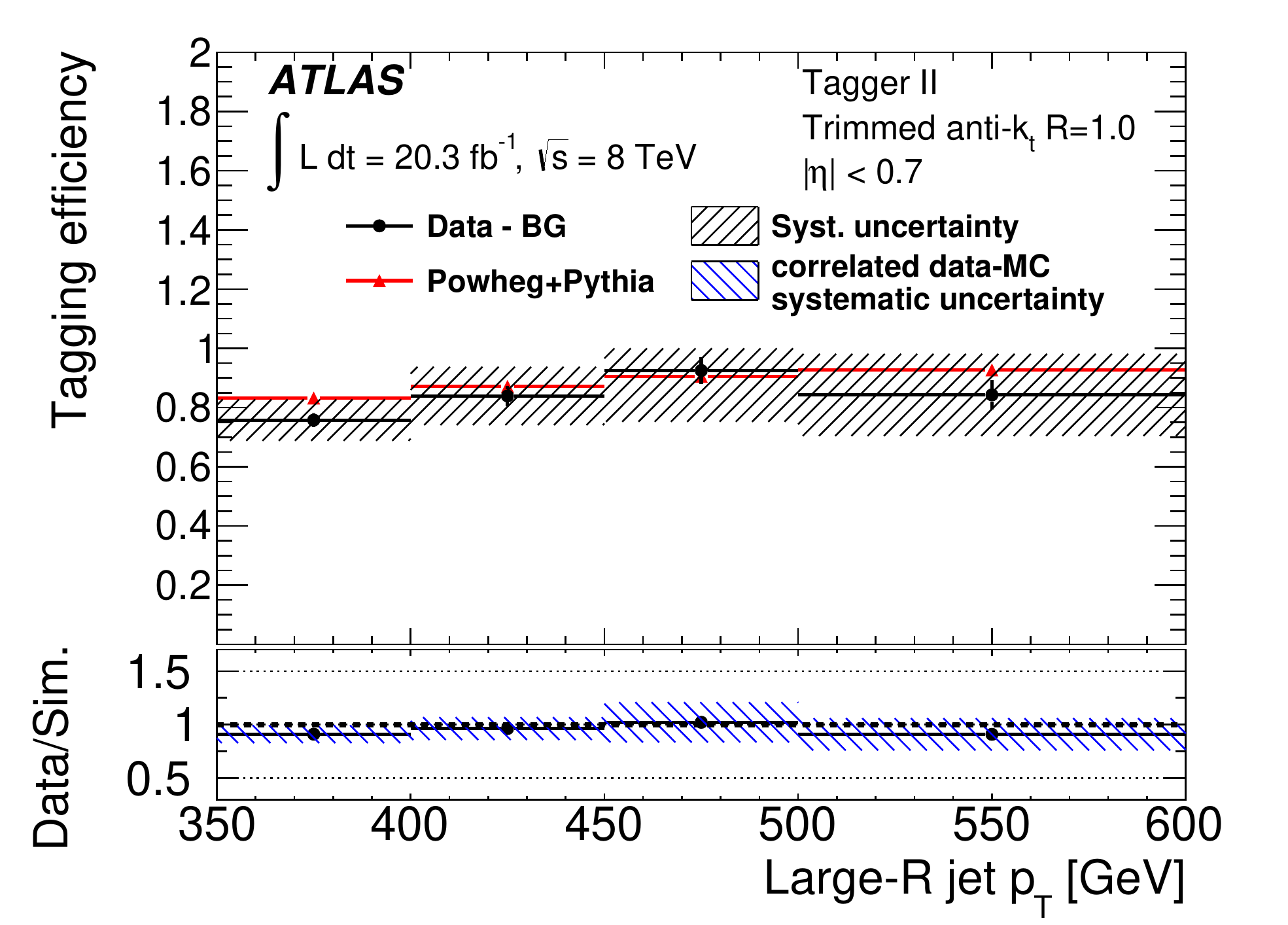}
} \\
\subfigure[]{
\includegraphics[width=0.48\textwidth]{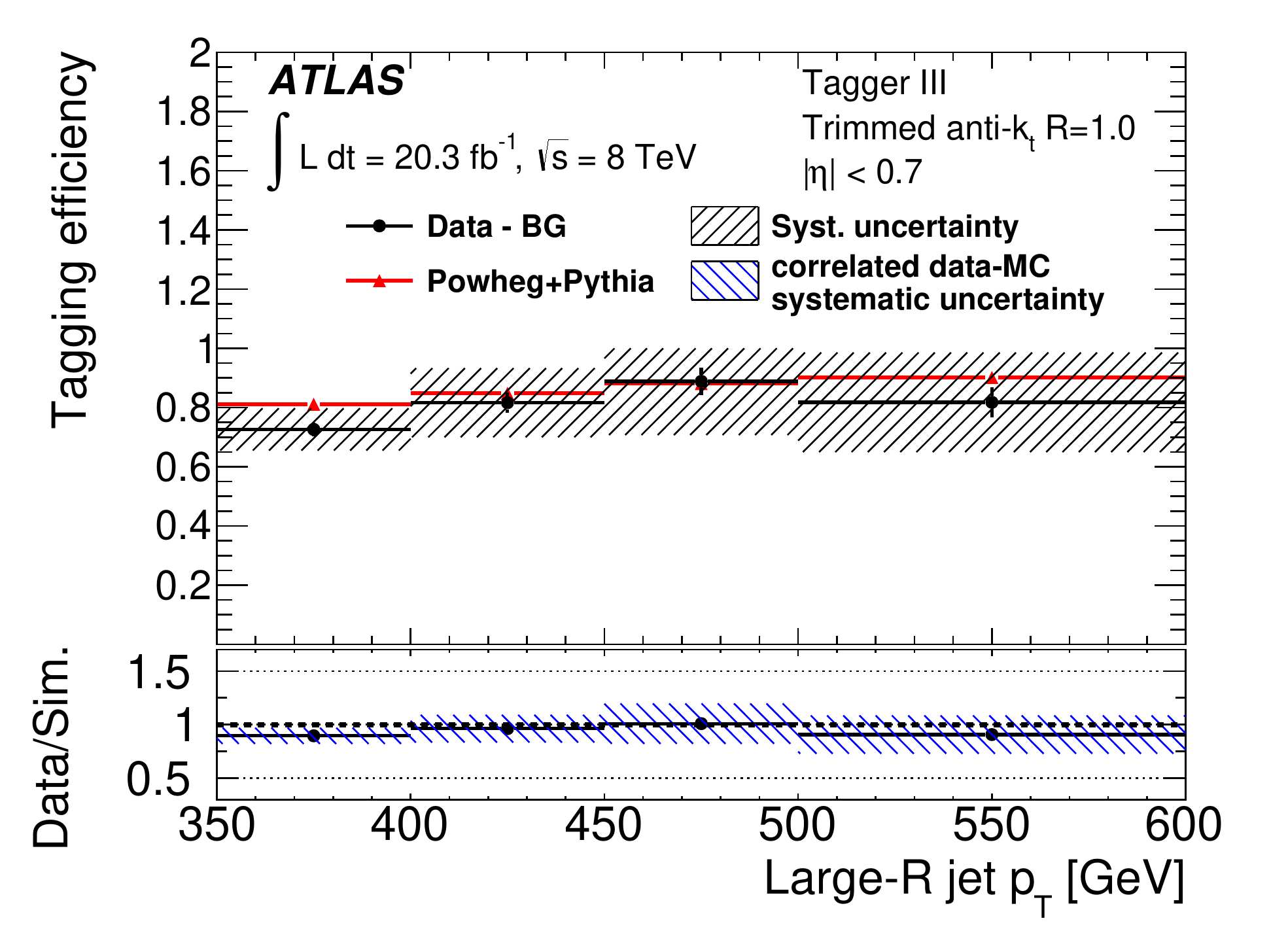}
}
\subfigure[]{
\includegraphics[width=0.48\textwidth]{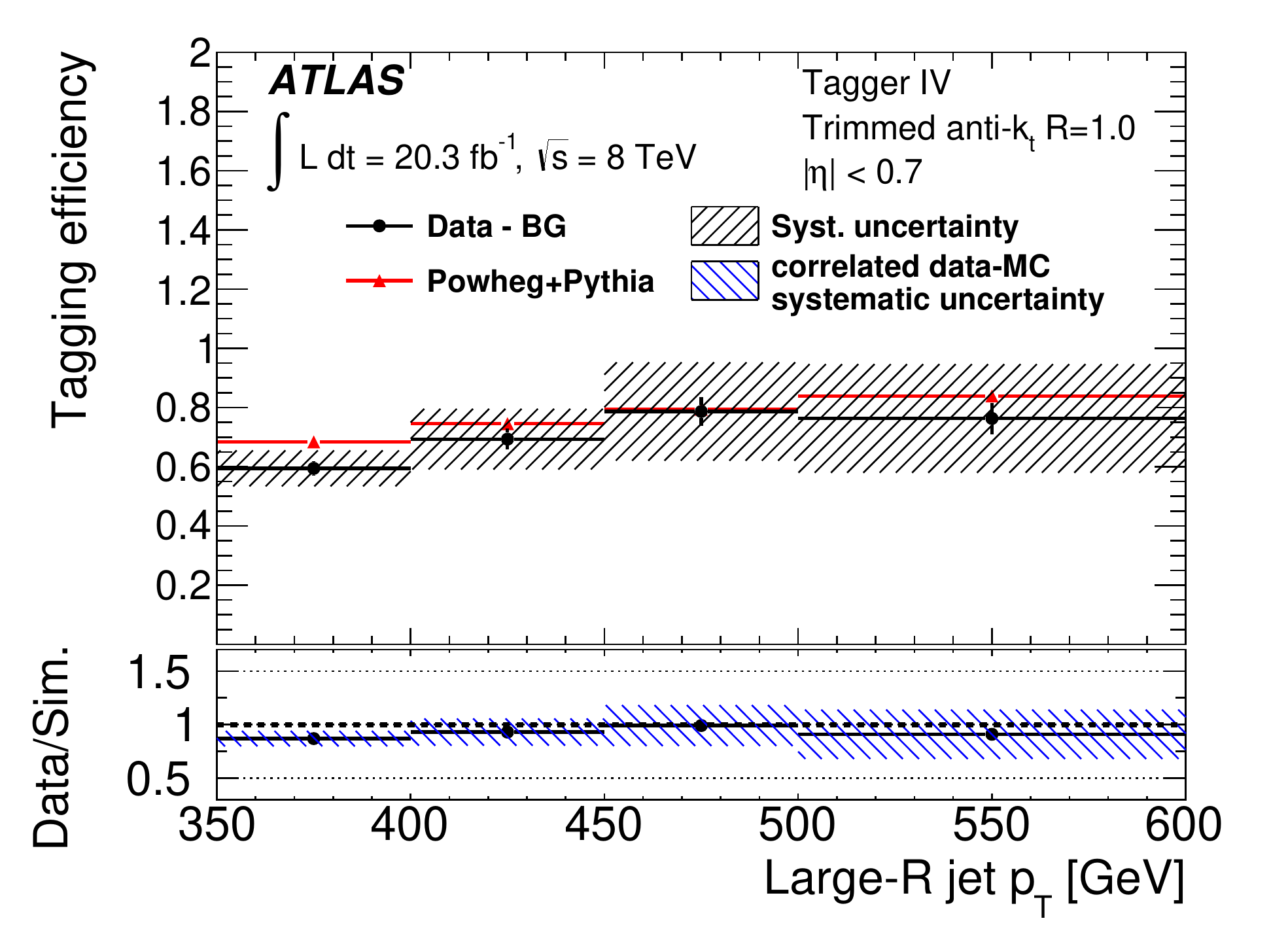}
}
\caption{The efficiency \fdata, as defined in \equref{HTTtaggedfjfraction}, for tagging trimmed \akt $R=1.0$ jets with $|\eta| < 0.7$ with top
taggers based on substructure variables (taggers I--IV) as a function of the \largeR jet \pt. Background (BG)
is statistically subtracted from the data using
simulation. The vertical error bar indicates the statistical uncertainty
of the efficiency measurement and the data uncertainty band shows the systematic
uncertainties. Also shown is the predicted tagging efficiency \fMC,
as defined in \equref{f_MC}, from
\PowhegPythia without systematic uncertainties. The ratio $\fdata/\fMC$ of measured to
predicted efficiency is shown at the bottom of each subfigure and the error bar
gives the statistical uncertainty and the band the systematic uncertainty.
The systematic uncertainty of the ratio is calculated taking into account the
systematic uncertainties in the data and the prediction and their correlation.}
\label{fig:eff_substructure_lowEta}
\end{centering}
\end{figure}

\begin{figure}[!h]
\begin{centering}
\subfigure[]{
\includegraphics[width=0.48\textwidth]{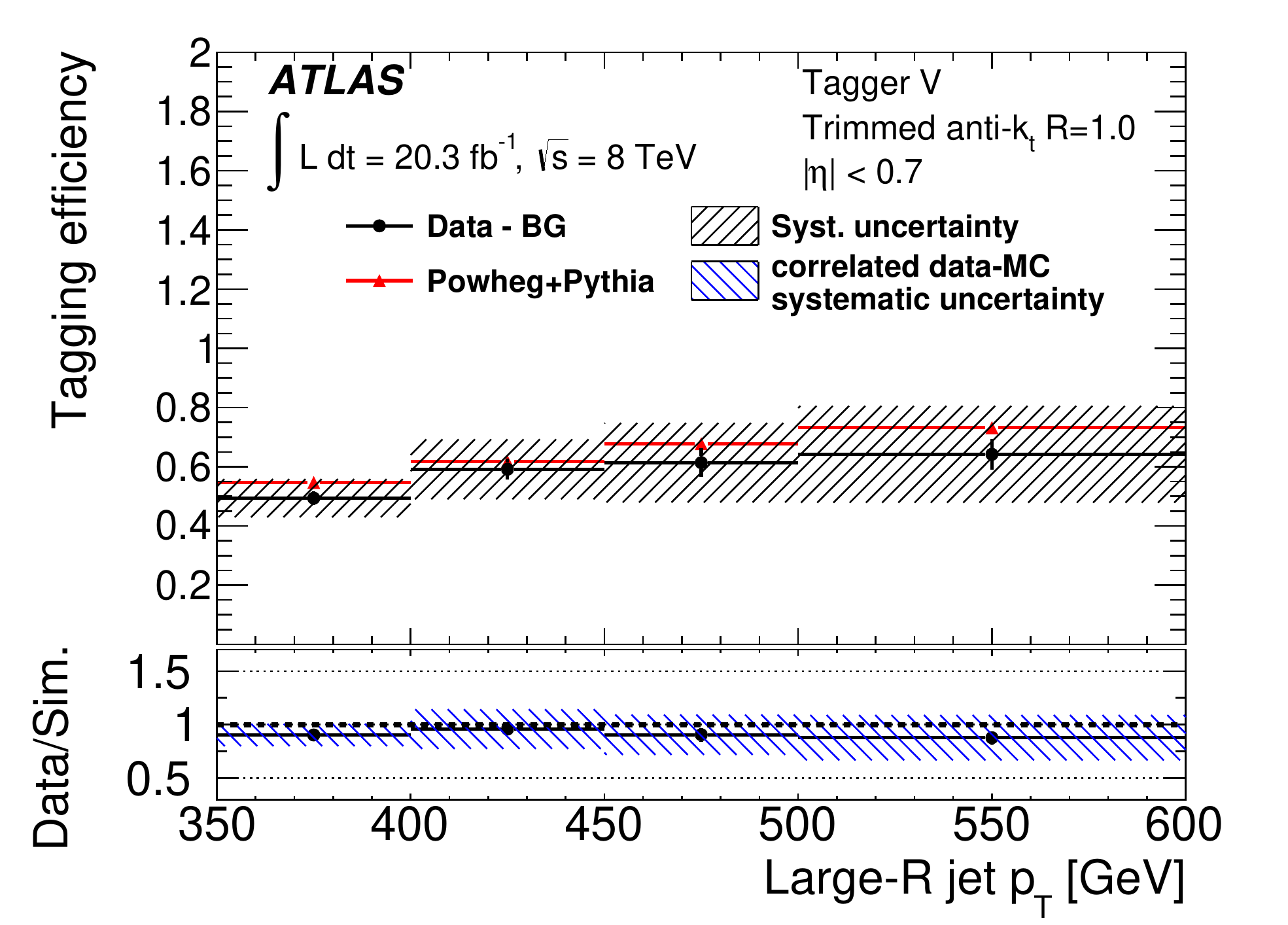}
}
\subfigure[]{
\includegraphics[width=0.48\textwidth]{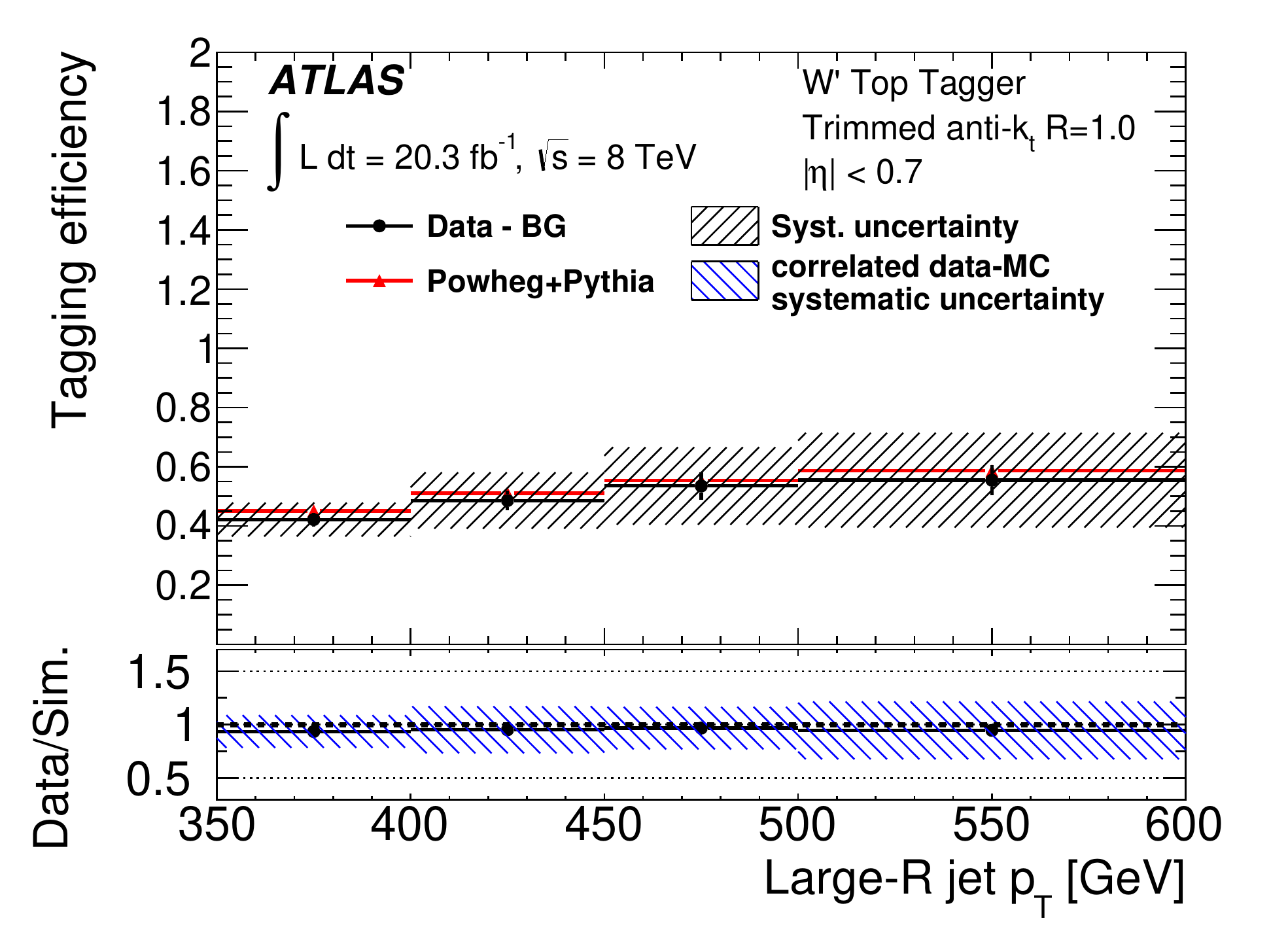}
}
\caption{The efficiency \fdata, as defined in \equref{HTTtaggedfjfraction}, for tagging trimmed \akt $R=1.0$ jets with $|\eta| < 0.7$ with top
taggers based on substructure variables (tagger V and \WPT)
as a function of the \largeR jet \pt.
Background (BG) is statistically subtracted from the data using
simulation. The vertical error bar indicates the statistical uncertainty
of the efficiency measurement and the data uncertainty band shows the systematic
uncertainties.
Also shown is the predicted tagging efficiency \fMC,
as defined in \equref{f_MC}, from
\PowhegPythia without systematic uncertainties. The ratio $\fdata/\fMC$ of measured to
predicted efficiency is shown at the bottom of each subfigure and the error bar
gives the statistical uncertainty and the band the systematic uncertainty.
The systematic uncertainty of the ratio is calculated taking into account the
systematic uncertainties in the data and the prediction and their correlation.}
\label{fig:eff_substructure_lowEtaTwo}
\end{centering}
\end{figure}

\begin{figure}[!h]
\begin{centering}
\subfigure[]{
\includegraphics[width=0.48\textwidth]{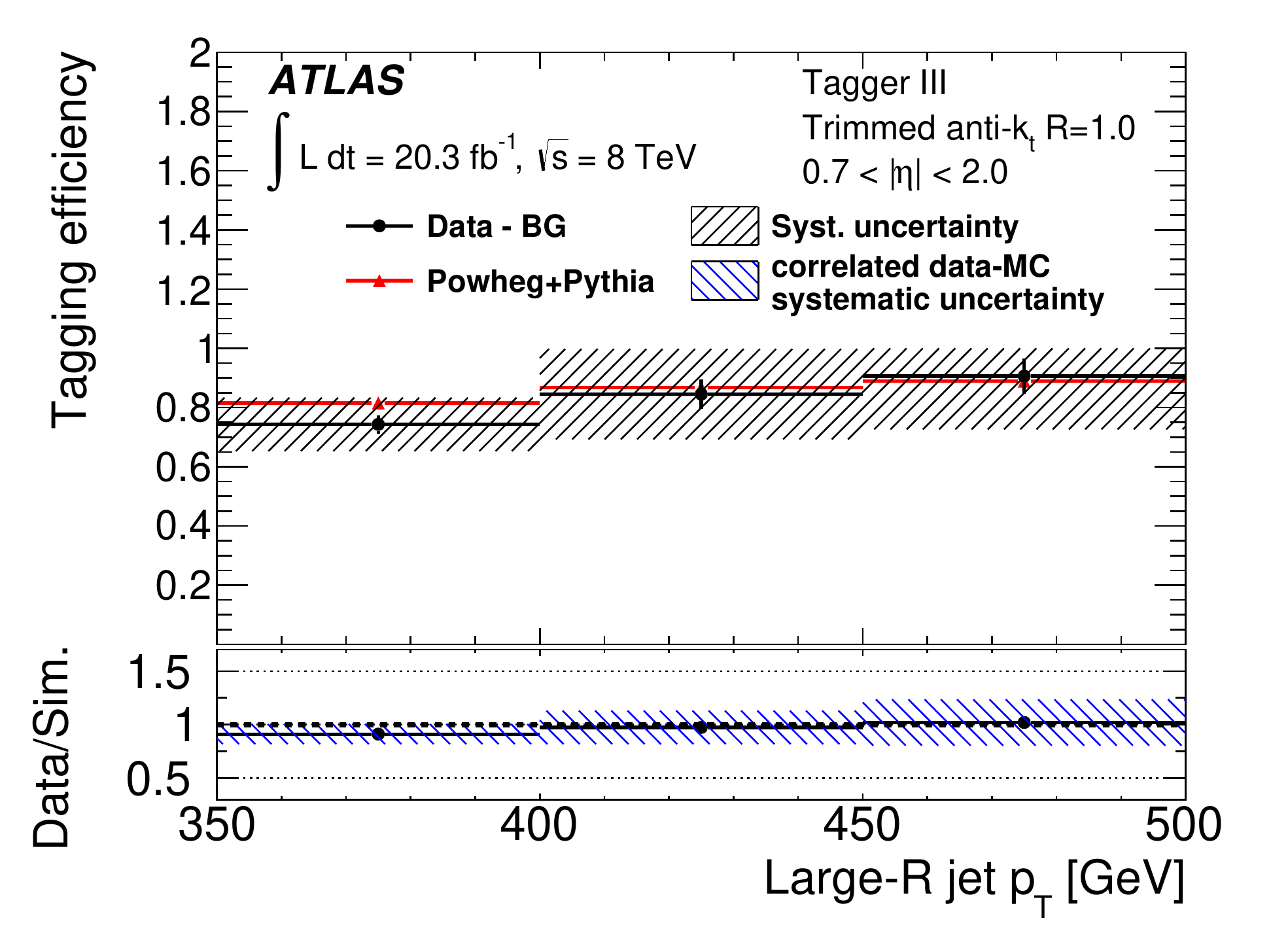}
}
\subfigure[]{
\includegraphics[width=0.48\textwidth]{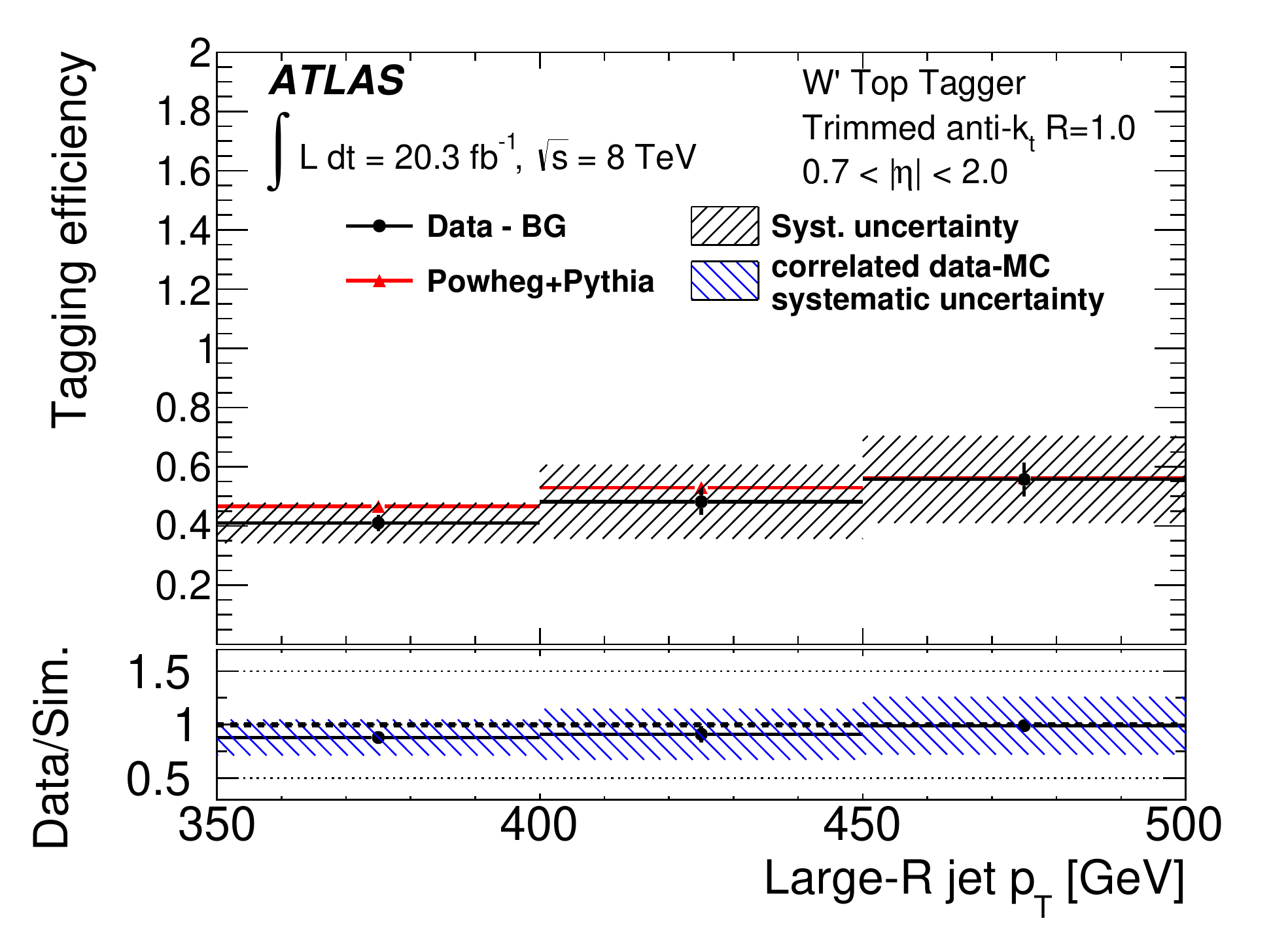}
}
\caption{The efficiency \fdata, as defined in \equref{HTTtaggedfjfraction}, for tagging trimmed \akt $R=1.0$ jets with $0.7 < |\eta| < 2.0$
based on substructure variables (tagger III and \WPT) as a function of the \largeR jet \pt.
Background (BG) is statistically subtracted from the data using
simulation. The vertical error bar indicates the statistical uncertainty
of the efficiency measurement and the data uncertainty band shows the systematic
uncertainties.
Also shown is the predicted tagging efficiency \fMC, as defined in \equref{f_MC}, from
\PowhegPythia without systematic uncertainties. The ratio $\fdata/\fMC$ of measured to
predicted efficiency is shown at the bottom of each subfigure and the error bar
gives the statistical uncertainty and the band the systematic uncertainty.
The systematic uncertainty of the ratio is calculated taking into account the
systematic uncertainties in the data and the prediction and their correlation.}
\label{fig:eff_substructure_highEta}
\end{centering}
\end{figure}

When a \largeR jet is considered matched according to the geometric matching
of the jet axis to the direction of the top quark, this does not necessarily imply that all
decay products of the top quark are contained inside the \largeR jet. Even after subtracting
the not-matched contribution in \equref{HTTtaggedfjfraction}, a significant fraction
of the \largeR jets with lower \pt therefore do not contain all top-quark decay products.
The tagging efficiency is high when all decay products are contained in the
\largeR jet. The efficiency is therefore low for \largeR jets with small \pt and it
rises with \pt because of the tighter collimation.

The efficiency decreases with increasing tagger number from tagger I to tagger V
and the lowest efficiency of the tested taggers based on substructure variables
is found for the \WPT. The efficiencies vary between 40\% and 90\%, depending
on the tagger and the \pt of the \largeR jet.
The efficiencies are similar in the two $\eta$ regions but the measurement is more precise for $|\eta|<0.7$.

The measurement of the efficiency is limited by the systematic uncertainties
resulting from the subtraction of background jets.
The uncertainties in the measured efficiency include uncertainties related to
the choice of generator used for \ttbar production.
In the lowest \largeR jet \pt bin, the relative uncertainties of the
efficiency for $|\eta|<0.7$ are 10\% to 14\%, depending
on the tagger, and for $0.7<|\eta|<2.0$ they vary between 11\% and 17\%.
For $|\eta|<0.7$, the systematic uncertainties in the interval $500$ to $600\GeV$
vary between approximately 17\% and 29\%.
For $0.7<|\eta|<2.0$ the uncertainties from $450$ to $500\GeV$ are
$18$ to $26\%$.
The systematic uncertainty is dominated by the different efficiencies from using
\Powheg or \Mcatnlo for the generation of the \ttbar contribution for
$|\eta|<0.7$.
In the range $0.7<|\eta|<2.0$, the \largeR
JES, the PDF, the parton-shower and the ISR/FSR uncertainties also contribute
significantly to the total systematic uncertainty.

Also shown in the figures is the prediction for $f_\textrm{MC}$ obtained from
the simulated \PowhegPythia \ttbar events using the nominal simulation parameters
and not considering systematic uncertainties. The prediction obtained in this way
is consistent with the measured efficiency within the uncertainties of the measurement.
In the simulation, for which the statistical uncertainty is much smaller than
for the data, the efficiencies continue to rise with \pt, indicating
that a plateau value is not reached in the \pt range studied here.

The ratio $\fdata/\fMC$ is shown in the bottom panels of
\figrange{eff_substructure_lowEta}{eff_substructure_highEta}. The nominal
\PowhegPythia prediction is used for \fMC. For this ratio, the full
systematic uncertainties of \fMC are considered, including the uncertainty
from the choice of \ttbar generator.
The full correlation with the uncertainty
of \fdata is taken into account in the systematic uncertainty of the ratio.
The ratio is consistent with unity within the uncertainty in all measured
\pt and \eta ranges.
For $|\eta|<0.7$, the uncertainty of $\fdata/\fMC$ is $8$--$16\%$ (depending
on the tagger) for \largeR jet \pt from $350$ to $400\GeV$ and
$17$--$28\%$ for $500$--$600\GeV$. For $0.7<|\eta|<2.0$, the uncertainty
is $10$--$19\%$ for $350$--$400\GeV$ and $19$--$28\%$ for $450$--$500\GeV$.

\subsubsection{Efficiency of \sd}

The measurement of the
efficiency for tagging \akt $R=1.0$ jets with SD, using the requirement $\ln(\chi)>2.5$,
is presented in \figref{eff_sd}.
The signal weights are calculated assuming that all top-quark
decay products are included in the \largeR jet. This containment assumption leads to
a rising efficiency with top-quark \pt because of the tighter
collimation at high \pt.
The SD efficiency is approximately 30\% in the region with the lowest \pt of the \largeR jet
($350$--$400\gev$), increases with \pt and
reaches $\approx\!45\%$ for $500$--$600\gev$ in the lower $|\eta|$ range and
for $450$--$500\gev$ in the higher $|\eta|$ range.
Within uncertainties, the measured efficiencies are compatible between
the two \eta regions.

In the lowest measured \pt region, the relative uncertainty is $\approx\!16\%$,
with the largest contributions coming from the difference observed when changing the
\ttbar generator from \Powheg to \Mcatnlo (12\%).
The uncertainties in the subjet energy scale and resolution have a much smaller
impact of 0.6\% and 0.4\%, respectively.
For \pt between $500$ and $600\GeV$ in the lower $|\eta|$ range, the relative
uncertainty is $\approx\!32\%$, with the largest contributions resulting from
the generator choice (27\%).

The efficiency from \PowhegPythia follows the trend of the measured efficiency
and the predicted and measured efficiencies agree within uncertainties, but
the predicted efficiency is systematically higher.
The ratio $\fdata/\fMC$ is approximately $80\%$ throughout
the considered \pt range.
The relative uncertainty of the ratio is
$\approx\!25\%$ for $|\eta|<0.7$.
For $0.7 < |\eta| < 2.0$, the uncertainty varies between $\approx\!25\%$ and $\approx\!35\%$.

\begin{figure}[!h]
\begin{centering}
\subfigure[]{
\includegraphics[width=0.48\textwidth]{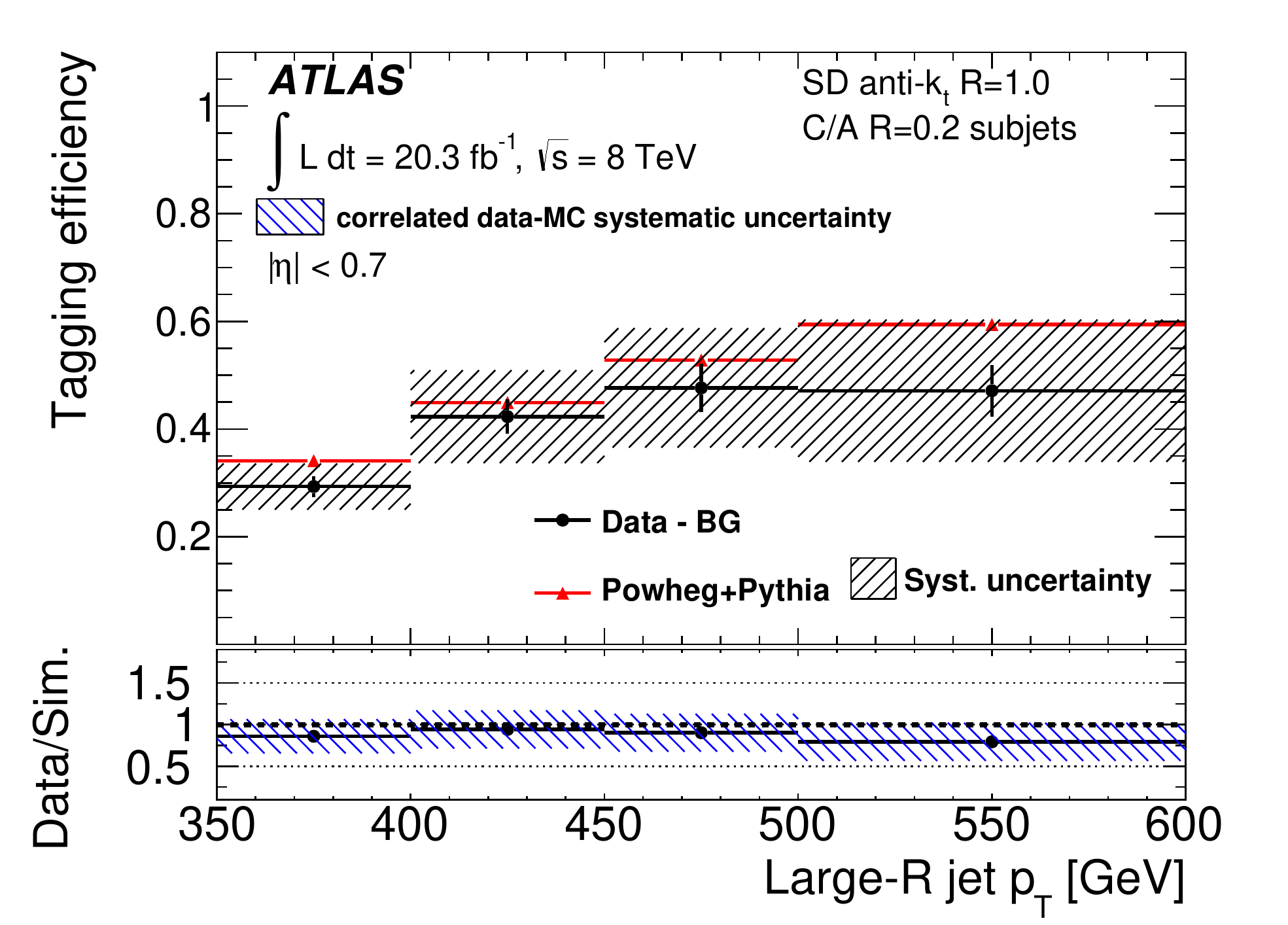}
}
\subfigure[]{
\includegraphics[width=0.48\textwidth]{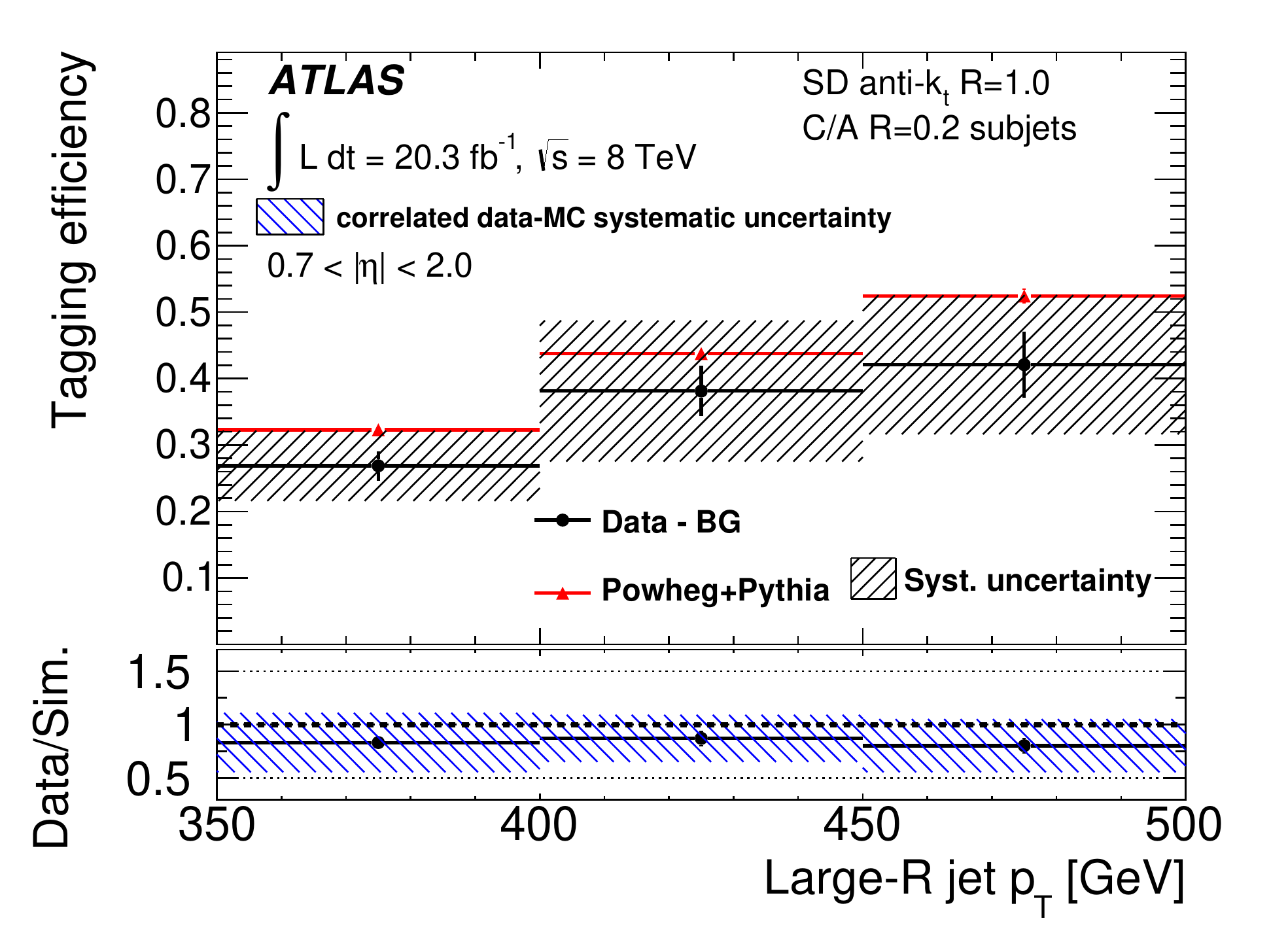}
}
\caption{The efficiency \fdata, as defined in \equref{HTTtaggedfjfraction}, for tagging trimmed \akt $R=1.0$ jets with \sd, using
the requirement $\ln(\chi)>2.5$, as a function
of the \largeR jet \pt. The \largeR jets
are selected in the signal selection and have pseudorapidities (a) $|\eta|<0.7$
and (b) $0.7<|\eta|<2.0$. Background (BG) is statistically subtracted from the data using
simulation. The vertical error bar indicates the statistical uncertainty
of the efficiency measurement and the data uncertainty band shows the systematic
uncertainties.
Also shown is the predicted tagging efficiency \fMC, as defined in \equref{f_MC}, from
\PowhegPythia without systematic uncertainties. The ratio $\fdata/\fMC$ of measured to
predicted efficiency is shown at the bottom of each subfigure and the error bar
gives the statistical uncertainty and the band the systematic uncertainty.
The systematic uncertainty of the ratio is calculated taking into account the
systematic uncertainties in the data and the prediction and their correlation.}
\label{fig:eff_sd}
\end{centering}
\end{figure}

\subsubsection{Efficiency of the \htt}

The efficiency for tagging \CamKt $R=1.5$ jets with the \htt is shown
in \figref{eff_htt} as a function of the \largeR jet \pt. In the lowest \pt interval from
$200$ to $250\GeV$ the efficiency is $\approx\!10\%$. The efficiency increases
with \pt because of the geometric collimation effect and reaches $\approx\!40\%$
for \pt between $350$ and $400\GeV$ and $45$--$50\%$ for $\pt>500\GeV$.
The efficiencies in the two $\eta$ regions are very similar.
The measurement is systematically limited.
In the lowest measured jet \pt interval from $200$ to $250\GeV$, the relative
systematic uncertainty is 8.5\% with similar contributions coming from several sources, the three
largest ones being the difference
between \Powheg and \Mcatnlo as the \ttbar generator (3.9\%), the \largeR jet
energy scale (3.3\%), and the $b$-tagging efficiency (3.3\%). The contributions from
the imperfect knowledge of the subjet energy scale and resolution are 2.5\% and 2.7\%, respectively.
For \largeR jet \pt between $600$ and $700\GeV$, the relative uncertainty
is $54\%$, and the largest contributions are from the generator choice (44\%) and
the \largeR JES (22\%), while the subjet energy scale (2.1\%)
and resolution (0.6\%) have only a small impact.

When clustering objects (particles or clusters of calorimeter cells)
with the \CamKt algorithm using $R=1.5$
and comparing the resulting jet with the jet obtained by clustering the same particles
with the \akt algorithm using $R=1.0$ and then trimming the \akt jet,
the \pt is larger for the \CamKt jet than for the trimmed \akt jet.
In this paper, the \pt
interval $600$--$700\GeV$ for the \CamKt $R=1.5$ jets corresponds approximately to
the interval $500$--$600\GeV$ for the trimmed \akt $R=1.0$ jets. Beyond this
\pt, the statistical and systematic uncertainties become larger than
30\% and 65\%, respectively.

The efficiency predicted by the \PowhegPythia simulation agrees with the
measurement within the uncertainties. The ratio $\fdata/\fMC$ is
consistent with unity, within uncertainties of $\approx\!30\%$ in the lowest
and highest measured \pt intervals and $\approx\!15\%$ between $250$ and $450\GeV$.

\begin{figure}[!h]
\begin{centering}
\subfigure[]{
\includegraphics[width=0.48\textwidth]{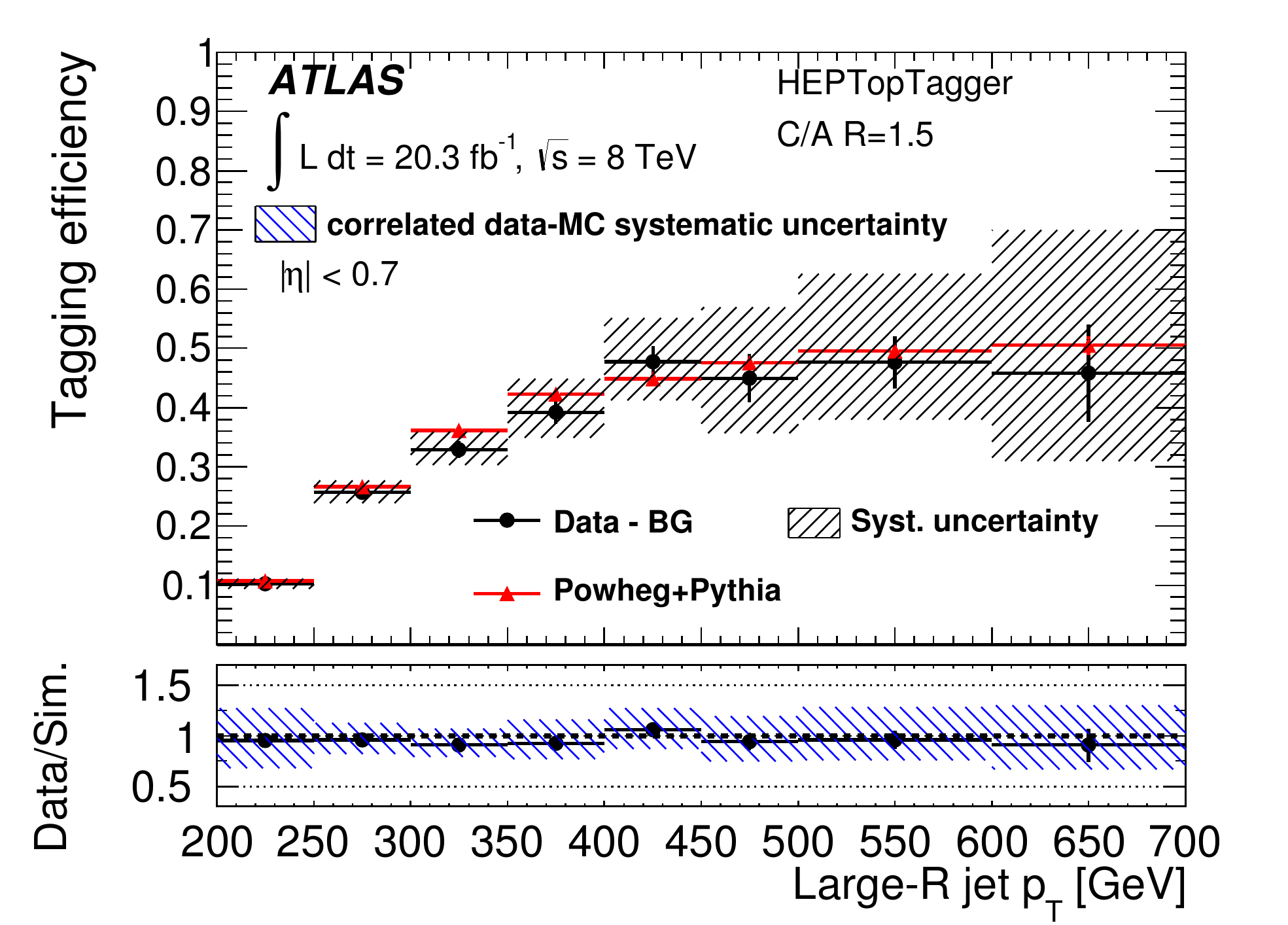}
}
\subfigure[]{
\includegraphics[width=0.48\textwidth]{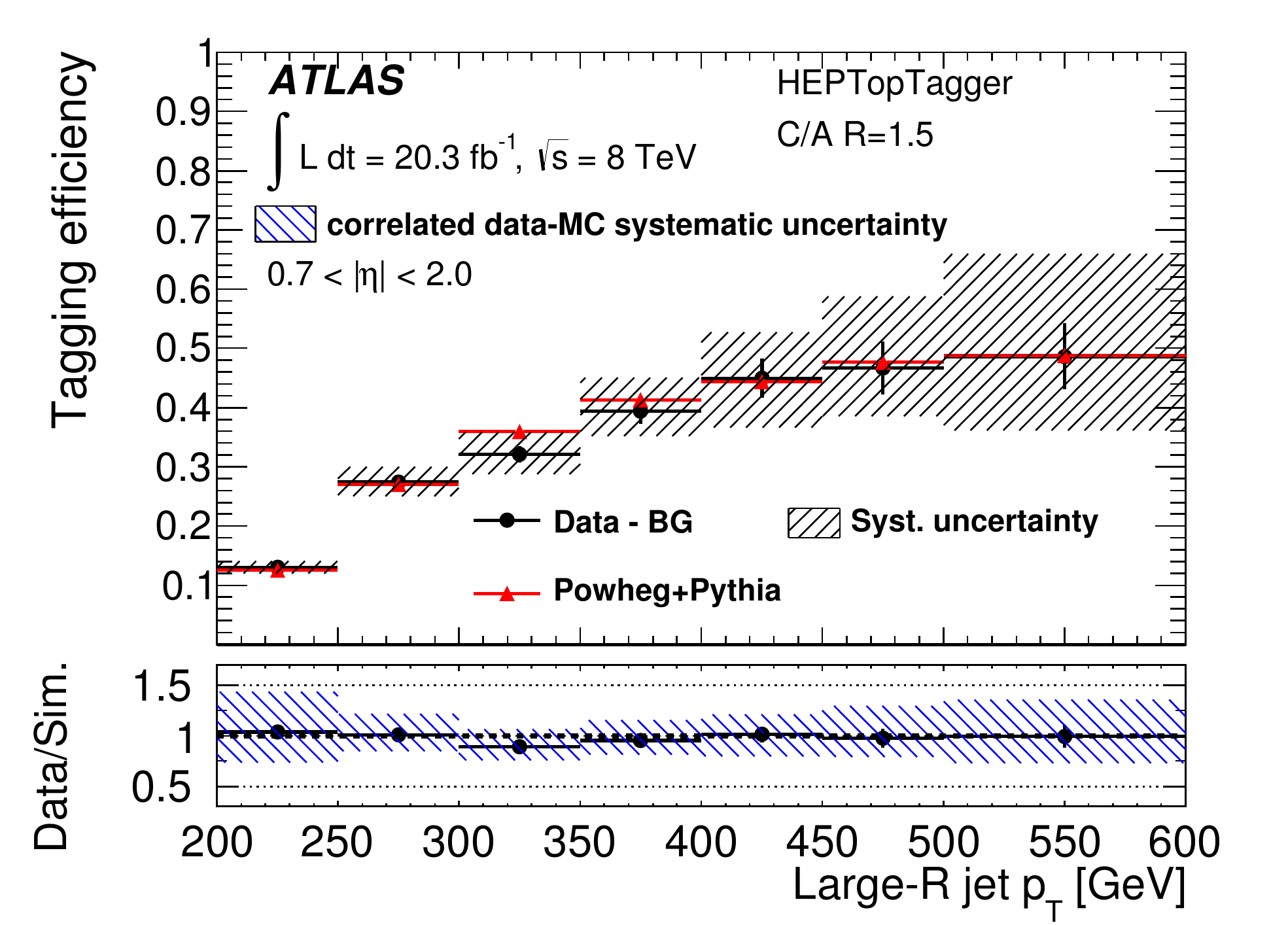}
}
\caption{The efficiency \fdata, as defined in \equref{HTTtaggedfjfraction}, for tagging \CamKt $R=1.5$ jets with the \htt as a function
of the \largeR jet \pt. The \largeR jets
are selected in the signal selection and have pseudorapidities (a) $|\eta|<0.7$
and (b) $0.7<|\eta|<2.0$. Background (BG) is statistically subtracted from the data using
simulation. The vertical error bar indicates the statistical uncertainty
of the efficiency measurement and the data uncertainty band shows the systematic
uncertainties.
Also shown is the predicted tagging efficiency \fMC, as defined in \equref{f_MC}, from
\PowhegPythia without systematic uncertainties. The ratio $\fdata/\fMC$ of measured to
predicted efficiency is shown at the bottom of each subfigure and the error bar
gives the statistical uncertainty and the band the systematic uncertainty.
The systematic uncertainty of the ratio is calculated taking into account the
systematic uncertainties in the data and the prediction and their correlation.}
\label{fig:eff_htt}
\end{centering}
\end{figure}

The total systematic uncertainty of the efficiency measurements when integrating over
the full \pt range and the range $0<|\eta|<2$ is given in
\tabref{effSystematics}. The total uncertainty is $12$--$20\%$ for the
substructure-variable-based taggers, 22\% for SD, and 9.9\% for the \htt.
The largest uncertainty results from the choice of \ttbar generator for the subtraction of the not-matched \ttbar
contribution, which introduces a normalization uncertainty
in the acceptance region of the measurement (high top-quark \pt), because the
\pt-dependence of the cross section is different between \Powheg and \Mcatnlo.
This difference is larger at high \pt, which translates to a larger uncertainty for the substructure-variable-based taggers
and SD, which use trimmed \akt $R=1.0$ jets with $\pt>350\GeV$, whereas the \htt uses
\CamKt $R=1.5$ jets with $\pt>200\GeV$.
For the same reason, the uncertainties in the parton shower and the PDF have a larger
impact for higher \largeR jet \pt.

The \largeR JES uncertainty affects the \htt efficiency less strongly than
the efficiencies of the other taggers (\tabref{effSystematics}). This is due to the requirement placed on the top-quark-candidate transverse momentum ($\pt>200\GeV$).
The \htt algorithm rejects some of the \largeR jet constituents in the
process of finding the hard substructure objects (mass-drop criterion) and
when applying the filtering against underlying-event and \pileup contributions.
The top-quark-candidate \pt is determined by the subjet four-momenta and
is smaller than the \largeR jet \pt, so that the requirement $\pt(\textrm{top-quark candidate})>200\GeV$
is stricter than the requirement $\pt(\textrm{\largeR jet})>200\GeV$.
This is also the reason why the subjet energy-scale uncertainty has a larger
impact on the efficiency of the \htt compared to SD, because for SD no \pt requirement on the
top-quark candidate is included in the signal- and background-hypothesis weights.

\begin{landscape}
\begin{table}
\begin{center}
\begin{tabular}{|c|c|c|c|c|c|c|c|c|}
      \hline
\multirow{3}{*}{Source} & \multicolumn{8}{c|}{Relative uncertainty of top-tagging efficiency (\%)} \\
\cline{2-9}
       & Tagger & Tagger & Tagger & Tagger & Tagger & $\W^\prime$ & Shower & HEPTop- \\
       &  I     &   II   &  III   &  IV    &  V     &  Tagger     & Deconstruction  &  Tagger \\
      \hline
      \LargeR jet energy scale      & 4.4 & 4.5 & 4.8 & 5.3 & 5.8 & 6.0 & 6.7 & 2.9 \\
      \LargeR jet energy resolution &<0.1 & 0.1 & 0.1 & 0.2 & 0.2 & 0.3 & 0.8 & 1.5 \\
      Luminosity                    & 1.0 & 1.0 & 1.1 & 1.2 & 1.3 & 1.4 & 1.5 & 1.3 \\
      $b$-tagging efficiency        & 2.7 & 2.6 & 2.9 & 3.1 & 3.5 & 3.7 & 3.9 & 3.5 \\
      Lepton reconstruction efficiency
                                    & 0.5 & 0.5 & 0.5 & 0.6 & 0.7 & 0.8 & 0.8 & 2.0 \\
      \hline
      \ttbar cross section          & 1.9 & 1.8 & 1.9 & 2.1 & 2.4 & 2.6 & 2.6 & 2.0 \\
      \ttbar ISR/FSR                & 1.4 & 1.3 & 1.4 & 0.5 & 1.6 & 1.6 & 2.2 & 3.2 \\
      \ttbar generator              & 10  & 9.2 & 11  & 12  & 15  & 16  & 18  & 6.7 \\
      \ttbar parton shower          & 4.8 & 4.1 & 4.6 & 4.8 & 4.6 & 5.1 & 5.1 & 1.7 \\
      \ttbar PDF uncertainty        & 4.4 & 3.8 & 4.5 & 4.2 & 5.2 & 6.8 & 8.3 & 2.2 \\
      \ttbar renormalization scale  & 0.8 & 0.8 & 0.8 & 0.9 & 1.0 & 1.1 & 1.0 & 0.6 \\
      \hline
      Trimmed \largeR jet mass scale
                                    & -   & 1.5 & 0.8 & 0.6 & 0.2 & -   & -   & -   \\
      Trimmed \largeR jet mass resolution
                                    & -   & 0.1 & 0.1 & 0.1 &<0.1 & -   & -   & -   \\
      \DOneTwo                      & 1.2 & -   & 0.6 & 0.5 & 0.4 & 0.5 & -   & -   \\
      \DTwoThr                      & -   & -   & -   & 0.7 & 1.1 & -   & -   & -   \\
      \tauTwoOne                    & -   & -   & -   & -   & -   & 0.6 & -   & -   \\
      \tauThrTwo                    & -   & -   & -   & -   & -   & 1.4 & -   & -   \\
      Subjet energy scale           & -   & -   & -   & -   & -   & -   & 0.5 & 1.1 \\
      Subjet energy resolution      & -   & -   & -   & -   & -   & -   & 0.4 & 0.7 \\
      \hline
      Total                         & 13  & 12  & 14  & 15  & 18  & 20  & 22  & 9.9 \\
      \hline
\end{tabular}
\caption{The relative uncertainty of the measured top-tagging efficiency (in percent)
due to different sources of systematic uncertainty and the total systematic
uncertainty obtained by adding the different contributions in quadrature.
}
\label{tab:effSystematics}
\end{center}
\end{table}
\end{landscape}

\subsection{Mistag rate}
\label{sec:misid}

\LargeR jets identified in the background selection are used to measure the
top-tagging misidentification rate (mistag rate).
In each
\largeR jet \pt bin $i$, the mistag rate is defined as
\begin{equation}
\fmisdatai = \left( \frac{N_{\textrm{data}}^\textrm{tag}}
{N_{\textrm{data}}} \right)_i \ ,
\label{eq:fakeRate}
\end{equation}
with $N_{\textrm{data}}^\textrm{(tag)}$ the number of measured (tagged) \largeR jets.
The contamination from \ttbar events is negligible before requiring a tagged top
candidate.
After requiring a \htt-tagged top candidate, the average contamination is
$\approx\!3\%$ ($200<\pt<700\GeV$). It is smaller than 3\%
for $\pt< 350\GeV$.
For larger values of \pt, however, the contamination from \ttbar increases,
as the \largeR jet \pt spectrum falls more steeply for multijet production
than for \ttbar events, leading to a contamination of up to $\approx\!5\%$ for
$350<\pt<600\GeV$ and $\approx\!11\%$ for $600<\pt<700\GeV$.

For SD, the average contamination after requiring a tagged top candidate is $\approx\!8\%$ ($350<\pt<700\GeV$).
Although the \htt gives higher background rejection than SD with
$\ln(\chi)>2.5$, the contamination for SD is larger on average, because the contamination
increases with \largeR jet \pt and the SD is only studied for trimmed \akt $R=1.0$ jets
with $\pt>350\GeV$.
For the substructure-variable taggers, the average contamination is smaller than 1.6\%.
Hence only for the top taggers with high rejection, SD and the \htt, the contribution from \ttbar events is subtracted from the numerator
of \eqref{fakeRate} before calculating the mistag rate.
The systematic uncertainty of the \ttbar contribution is estimated to be $\approx\!50\%$
in each \pt interval. This uncertainty influences the measurement of the mistag rate by a negligible amount compared to the statistical uncertainty
that results from the finite number of tagged \largeR jets in data. Therefore, only the statistical uncertainty is reported.

The measured mistag rate is compared to the mistag rate observed in
multijet events simulated with \Pythia, which is defined as
\begin{equation}
\fmisMCi = \left( \frac{N_\textrm{MC}^\textrm{tag}}{N_\textrm{MC}} \right)_i \ ,
\label{eq:f_fakeMC}
\end{equation}
in which $N_\textrm{MC}^\textrm{(tag)}$ is the number of (tagged) \largeR jets
which pass a looser background selection than required in data.
The electron-trigger requirement, the minimum distance requirement between the electron-trigger
object and the \largeR jet, and the veto on reconstructed electrons are removed.
Including these requirements
for simulation reduces the event yield significantly, which leads to
less predictive power for the mistag rate with the result that
the simulation still describes the measured mistag rates, but
with large statistical uncertainties.

Removing the requirements mentioned above from the background selection
for the simulation
is expected not to bias \fmisMCi.
The low-\pt threshold of the electron trigger avoids biases
towards dijet events with a well defined hard scattering axis, and a possible
trigger bias is reduced by using only \largeR jets away from the trigger object,
i.e. jets with $\Delta R > 1.5$.
The specific requirements applied only for data are therefore designed to allow for
a measurement of the mistag rate in pure multijet events which avoids trigger biases
and can hence be compared to the mistag rate observed in MC simulations.

The electron-trigger requirement is fulfilled preferentially for trigger
objects with high \pt. The \pt of the electron-trigger object and that of the \largeR jet under study for
the mistag-rate determination are correlated through the common hard parton--parton scattering process.
The \largeR jet \pt spectrum is therefore different for events in which the electron-trigger combination is
activated compared to those events in which this trigger combination is inactive. As the trigger requirement
is not applied in simulation, the average \pt of the \largeR jets in simulation is observed to be lower than in data.
The reconstructed MC \pt distribution of the \largeR jets is
therefore reweighted to the \pt distribution observed in data. This reweighting
procedure has only a small impact on the mistag rate, which is
measured in bins of \largeR jet \pt.

\subsubsection{Mistag rate for the substructure-variable taggers}

The mistag rate \fmisdata is shown in \figrange{fr_sub}{fr_sd} for the different
top taggers as a function of the \largeR jet \pt.
Anti-$k_t$ $R=1.0$ jets are used for SD.
The mistag rates rise with the \pt of the \largeR jet, because increased
QCD radiation at higher \pt produces structures inside the jets that resemble
the structures in top jets.
For taggers with high efficiency a larger mistag rate is found than for
those with lower efficiency, because these looser top-tagging criteria are met by a
larger fraction of the background jets.

The mistag rate for trimmed \akt $R=1.0$ jets tagged using substructure-variable
requirements are shown in \figref{fr_sub}.
In the lowest \pt interval from $350$ to $400\GeV$, the mistag rates for
the taggers I--V and the \WPT are approximately 22\%, 20\%, 16\%, 12\%, 6\%,
and 4\%, respectively.
The measured mistag rate increases with \pt and reaches values between
24\% and 36\% for taggers I--IV in the \pt interval $600$--$700\GeV$.
In this highest \pt interval, the mistag rate is $\approx\!16\%$ for tagger V
and $\approx\!6\%$ for the \WPT.
The predicted mistag rate \fmisMC from \Pythia is also shown with an uncertainty
band that includes systematic uncertainties due to the \largeR JES
and resolution uncertainties, and uncertainties of the modelling of the
substructure variables.
Within the uncertainties, the prediction from \Pythia agrees with the
measurement for all taggers. The uncertainties on the ratio $\fdata/\fMC$ are
$5$--$9\%$ for taggers I--IV, and, depending on the \largeR jet \pt, $\approx\!10\%$
for tagger V and $\approx\!20\%$ for the \WPT.
The systematic uncertainties of tagger V and the \WPT are larger than for
taggers I--IV because of the conservative treatment of the correlation between
the variations of the different substructure variables as mentioned in
\secref{systematics}.

\begin{figure}[p]
\begin{centering}
\subfigure[]{
\includegraphics[width=0.48\textwidth]{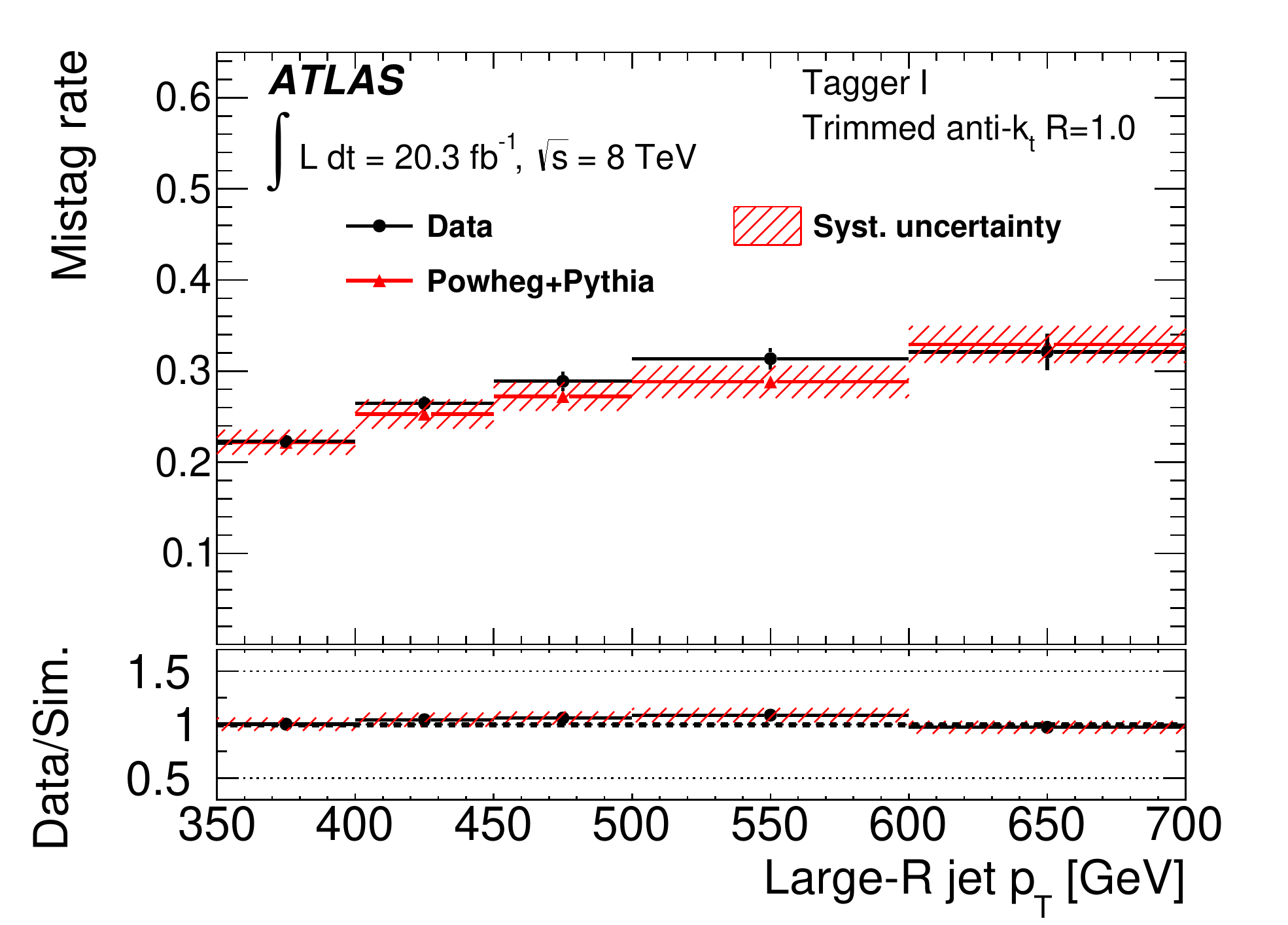}
}
\subfigure[]{
\includegraphics[width=0.48\textwidth]{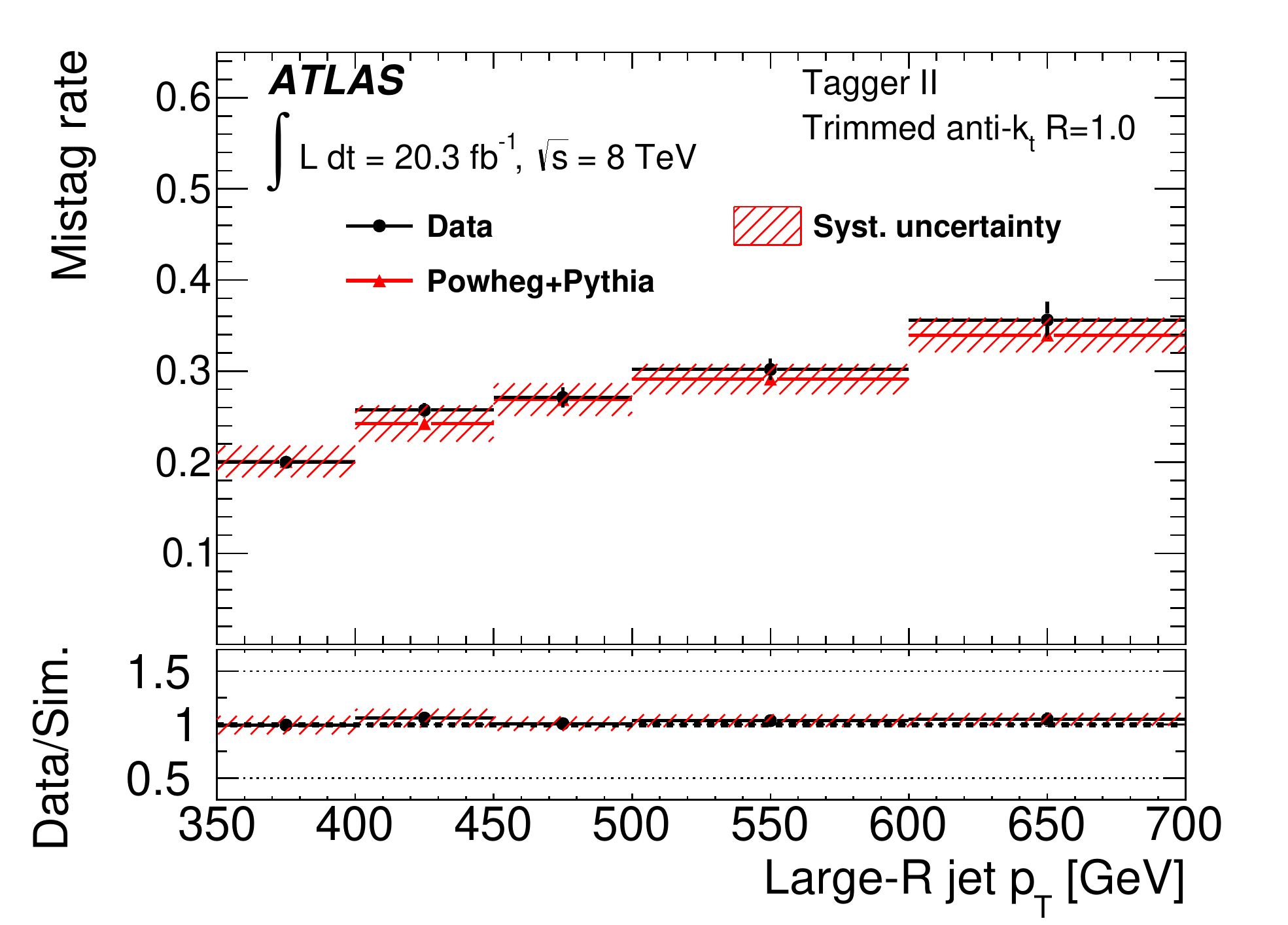}
} \\
\subfigure[]{
\includegraphics[width=0.48\textwidth]{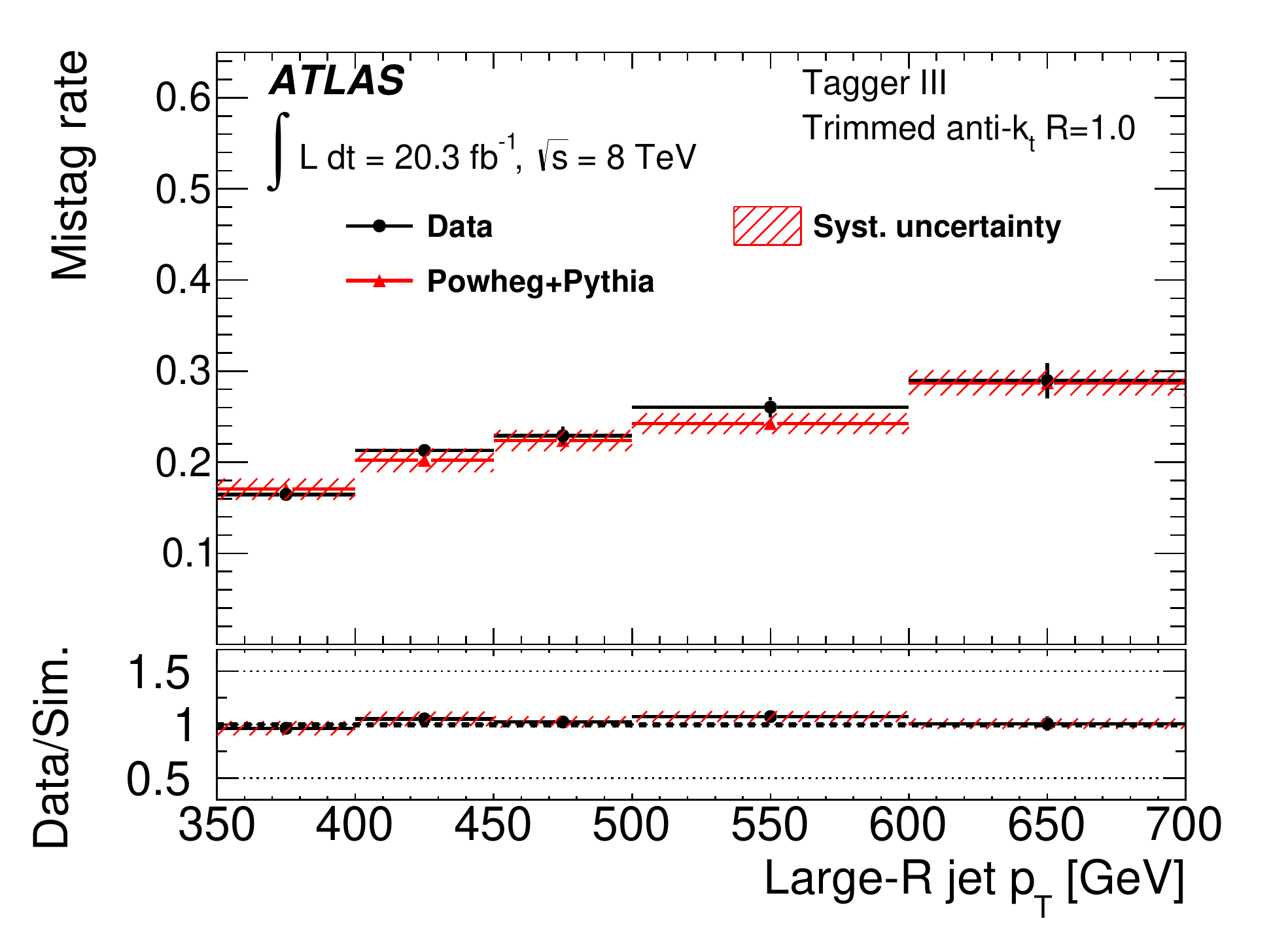}
}
\subfigure[]{
\includegraphics[width=0.48\textwidth]{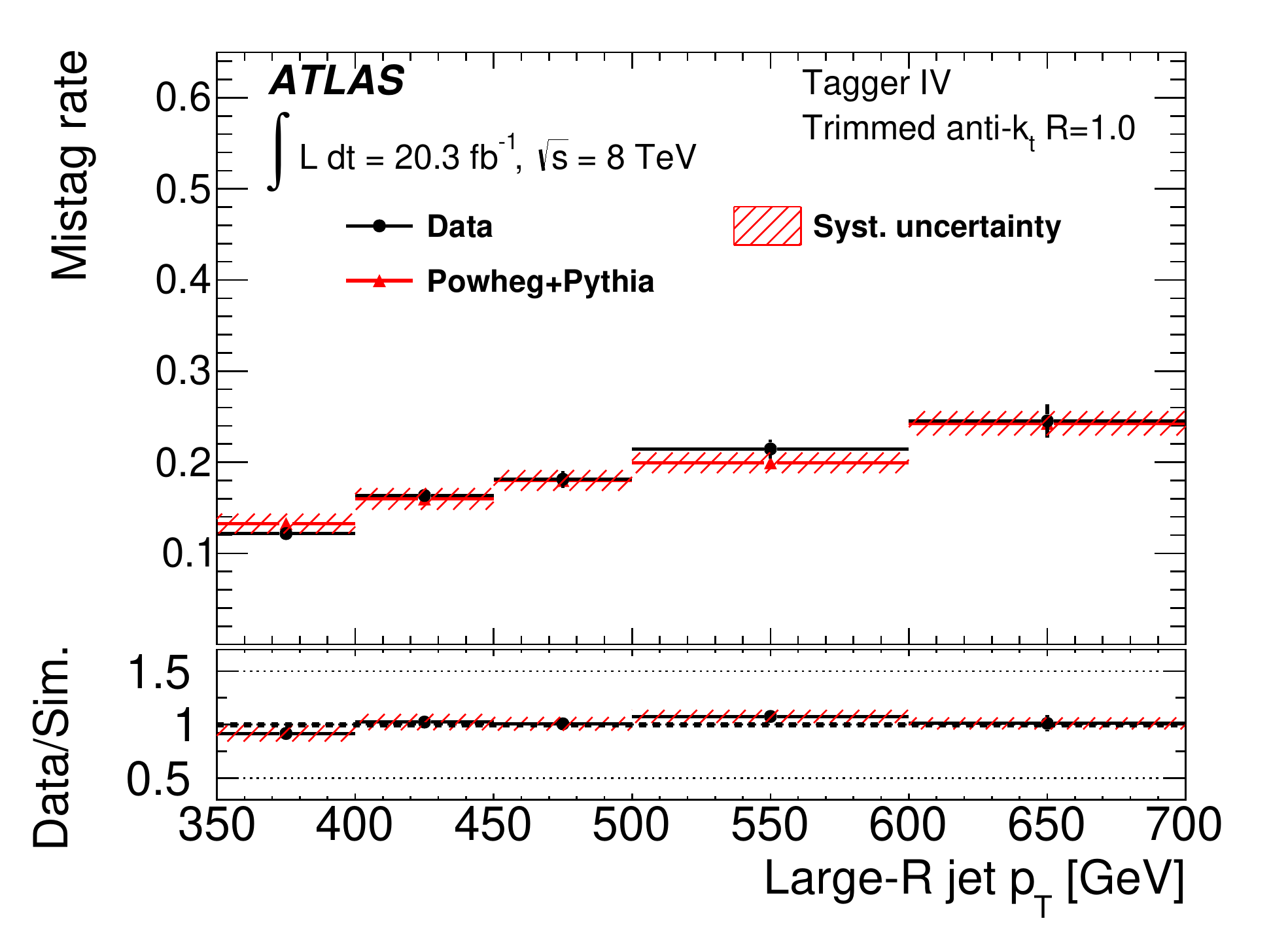}
} \\
\subfigure[]{
\includegraphics[width=0.48\textwidth]{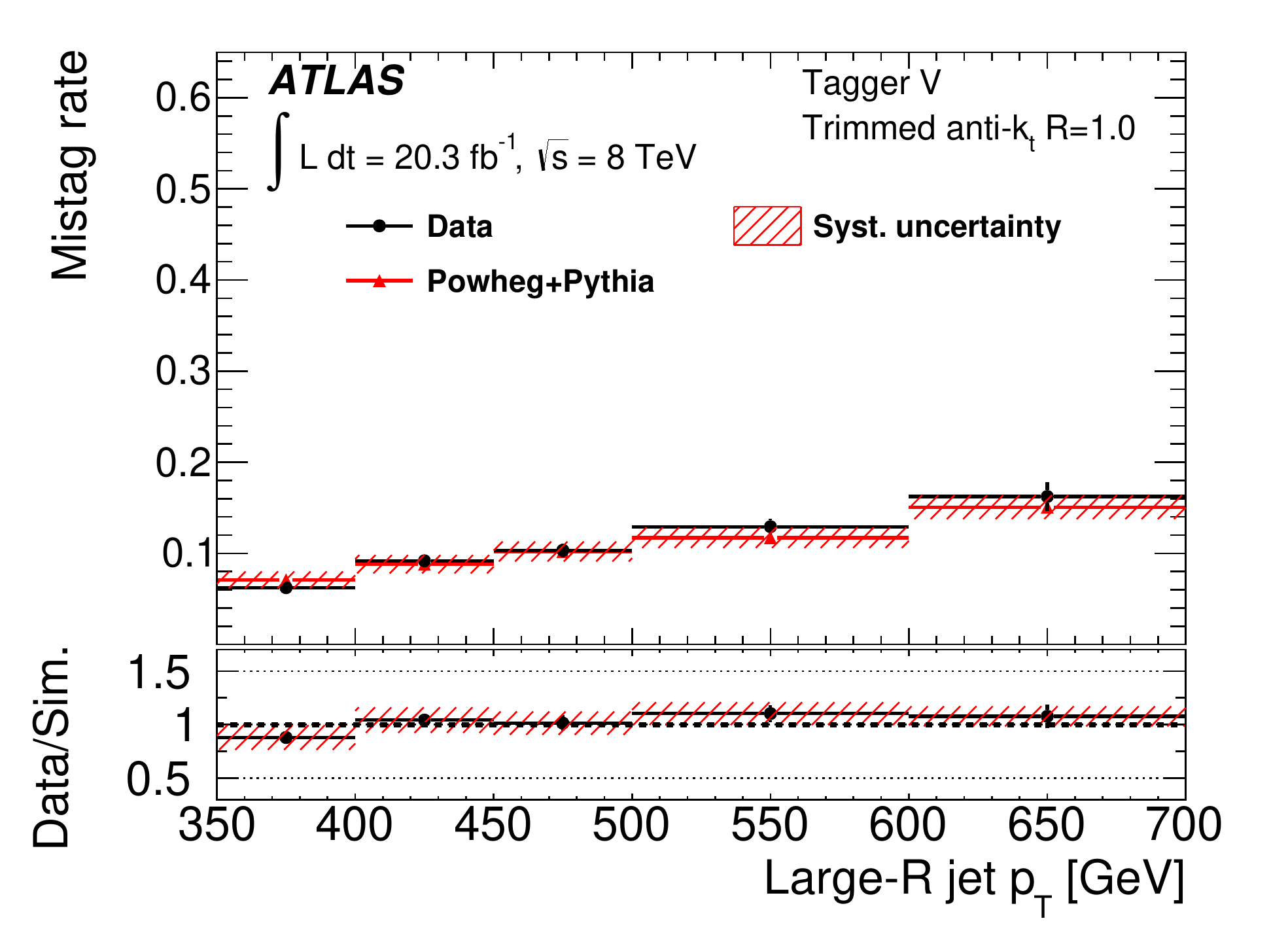}
}
\subfigure[]{
\includegraphics[width=0.48\textwidth]{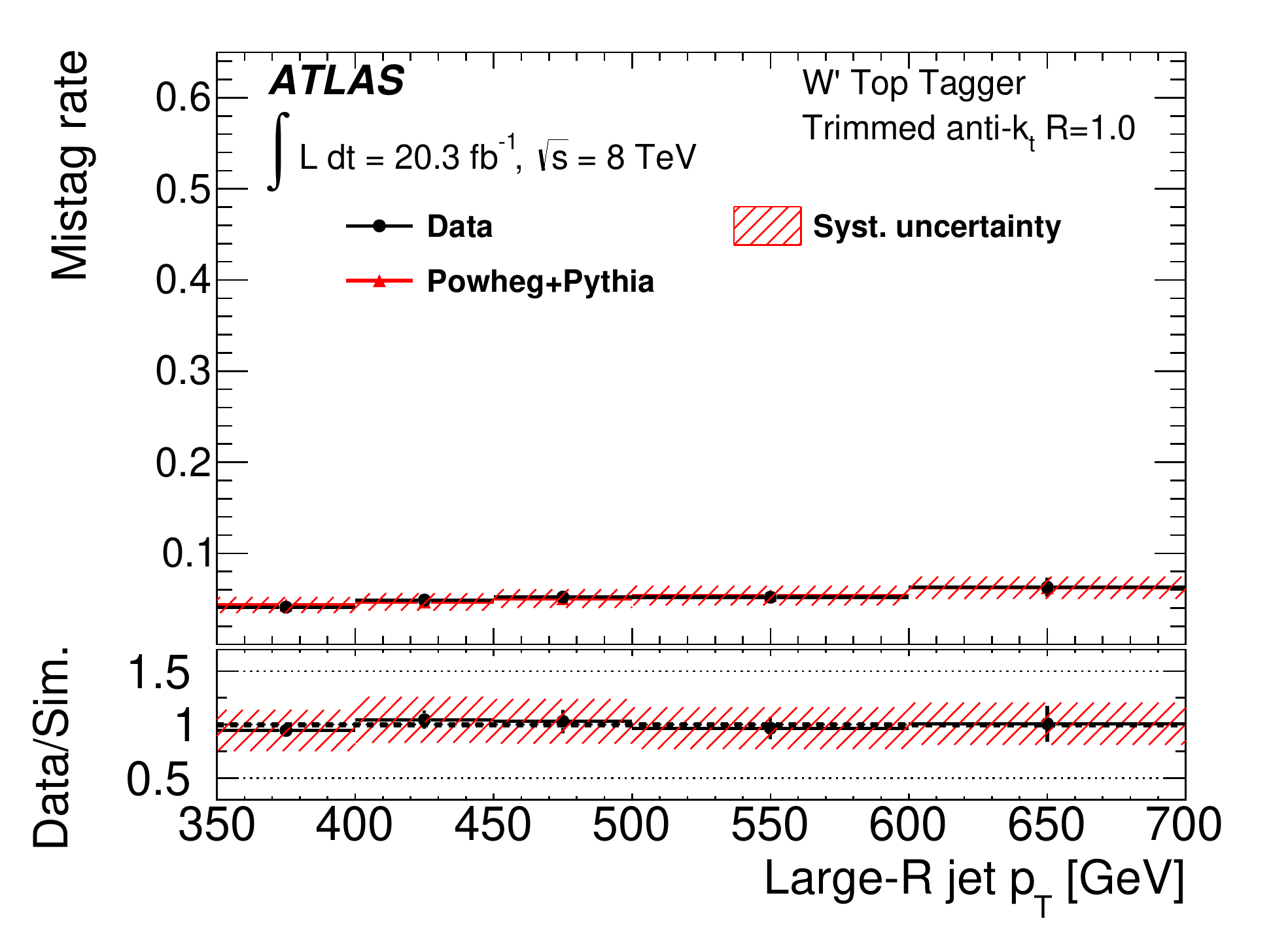}
}
\caption{The mistag rate \fmisdata, as defined in \equref{fakeRate}, for trimmed \akt $R=1.0$ jets
as a function of the \largeR jet \pt using the substructure-variable taggers I--V and the \WPT.
The \largeR jets are selected with the background selection and
have pseudorapidities $|\eta|<2.0$.
The vertical error bar indicates the statistical uncertainty
in the measurement of the mistag rate. Also shown is the
predicted mistag rate \fmisMC, as defined in \equref{f_fakeMC}, from \Pythia with systematic uncertainties included.
The ratio of measured to
predicted mistag rate is shown at the bottom of each subfigure and the error bar
gives the statistical uncertainty of the measurement.}
\label{fig:fr_sub}
\end{centering}
\end{figure}

\subsubsection{Mistag rate for \sd}

For SD, the mistag rate increases from 1\% for \pt between $350$ and $400\GeV$
to $\approx\!4\%$ for $600$--$700\GeV$.
The prediction from \Pythia shows the same trend as in
data and agrees well with the measurement within relative systematic uncertainties between
$\approx\!40\%$ at low \pt and $\approx\!13\%$ at high \pt,
which result from the uncertainties in the energy scales and resolutions of
the subjets and the \largeR jets. Integrated over \pt, the subjet energy-scale
and energy-resolution uncertainties lead to relative uncertainties of 15\% and 13\%,
respectively, while the uncertainty in the \largeR JES contributes 10\%. The
\largeR jet energy-resolution uncertainty has a negligible impact ($<1\%$).

\subsubsection{Mistag rate for the \htt}

For the \htt, the mistag rate increases from 0.5\% for \largeR jet \pt between $200$
and $250\GeV$ to 3\% for $450$--$500\GeV$.
Above $500\GeV$, the statistical uncertainties of the measured rate become
large.
The \Pythia simulation agrees well with the measurement.
The systematic uncertainty of the simulation is given by uncertainties
in the \largeR JES and resolution, and the energy scale and resolution
of the subjets.
The relative systematic uncertainty decreases with \pt:
it is 90\% in the lowest measured \pt bin and 8\% in the highest \pt bin.
This behaviour is driven by the subjet energy-resolution and energy-scale uncertainties, because
at low \largeR jet \pt a larger fraction of the \htt subjets have momenta near
the $20\GeV$ threshold. The mistag-rate uncertainty at low \pt is dominated by the
subjet energy-resolution uncertainty. The impact of the \largeR jet uncertainties is significantly smaller.

\begin{figure}[!h]
\begin{centering}
\subfigure[]{
\includegraphics[width=0.48\textwidth]{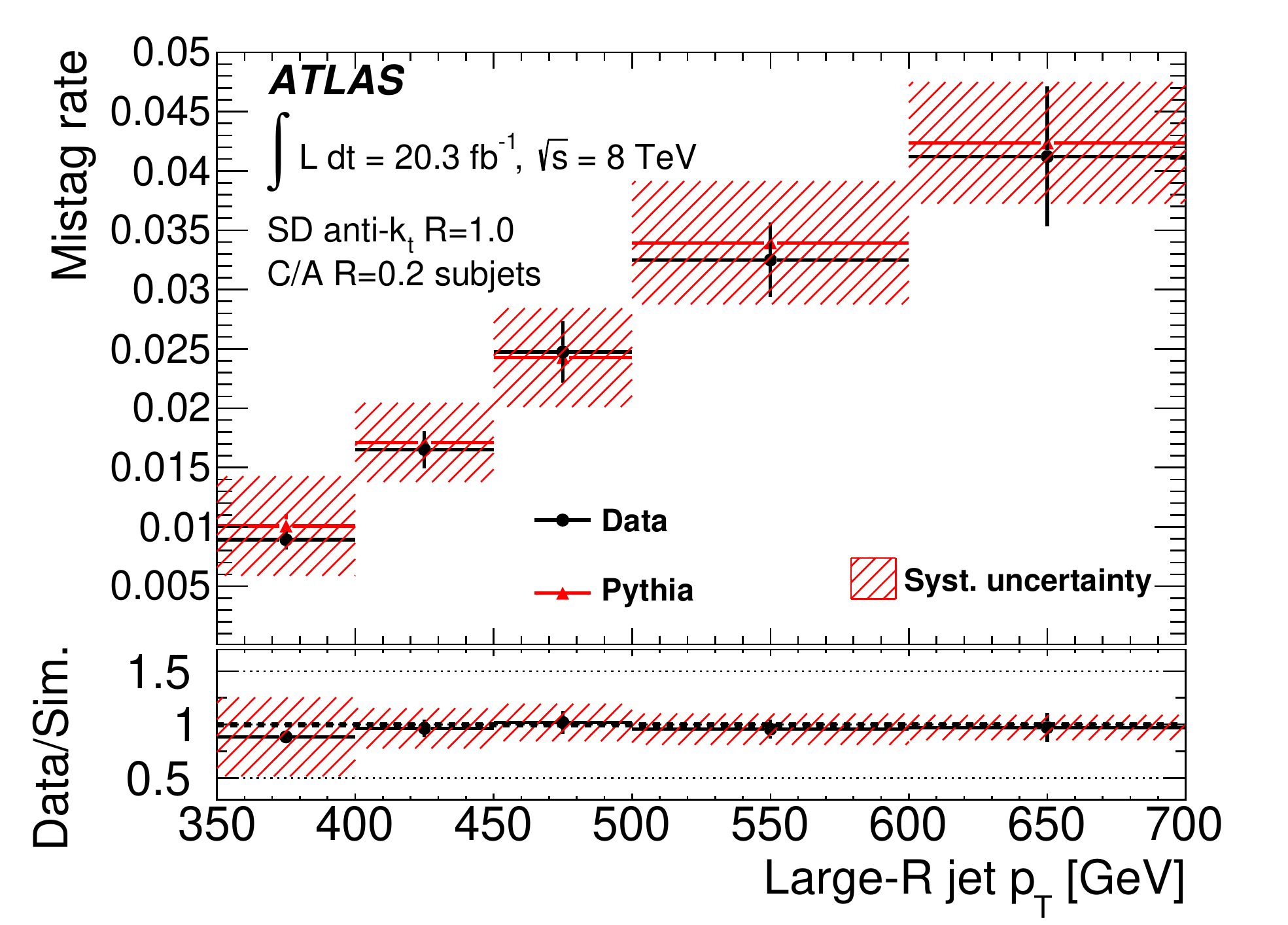}
}
\subfigure[]{
\includegraphics[width=0.48\textwidth]{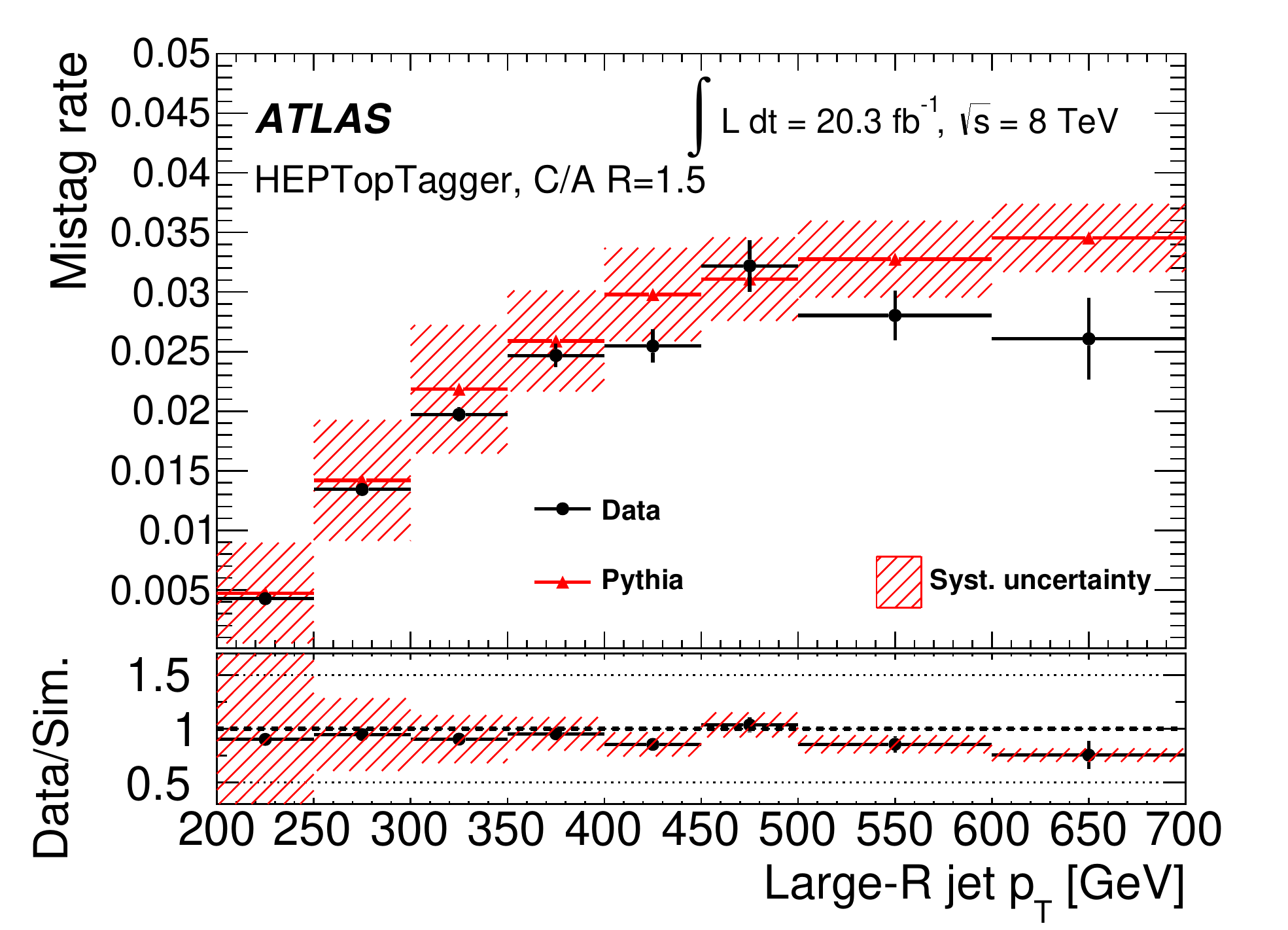}
}
\caption{
The mistag rate \fmisdata, as defined in \equref{fakeRate}, for \largeR jets
with $|\eta|<2.0$ selected with the background selection.
(a) Mistag rate for \akt $R=1.0$ jets
tagged with \sd using the requirement $\ln(\chi)>2.5$ as a function of the trimmed jet \pt.
(b) Mistag rate for \CamKt $R=1.5$ jets tagged with the
\htt as a function of the jet \pt.
The vertical error bar indicates the statistical uncertainty
in the measurement of the mistag rate. Also shown is the
predicted mistag rate \fmisMC, as defined in \equref{f_fakeMC}, from \Pythia with systematic uncertainties included.
The ratio of measured to
predicted mistag rate is shown at the bottom of each subfigure and the error bar
gives the statistical uncertainty of the measurement.}
\label{fig:fr_sd}
\end{centering}
\end{figure}

\clearpage

\FloatBarrier

\section{Summary and conclusions}
\label{sec:conclusion}
Jet substructure techniques are used to identify high-transverse-momentum
top quarks produced in proton--proton collisions at $\sqrt{s}=8\TeV$ at the LHC.
The 2012 ATLAS dataset is used, corresponding to an integrated
luminosity of $20.3\pm0.6~\ifb$.

Jets with a large radius parameter $R$ are reconstructed and their substructure
is analysed using a range of techniques that are sensitive to differences
between hadronic top-quark decay and background processes.
Jets are tagged as top jets
by requirements imposed on the jet mass, splitting scales, and \Nsj, and by
using the more elaborated algorithms of \sd (SD) and the original (not multivariate) \htt.
Six different combinations of requirements on substructure variables are
investigated, five combinations denoted by taggers I--V and the \WPT. For these taggers and for \sd, trimmed
\akt $R=1.0$ jets with $\pt > 350\GeV$ are used.
\ca (\CamKt) $R=0.2$ subjets with $\pt>20\GeV$ are used for SD.
The \htt was designed for, and is used with, ungroomed \CamKt $R=1.5$ jets down
to jet transverse momenta of $200\GeV$. The difference in the jet algorithms,
radii and grooming implies that the same top quark leads to a higher \pt for the
\CamKt $R=1.5$ jet.
A variant of the \htt algorithm is introduced, \httofour, which operates
on the constituents of a set of \akt $R=0.4$ jets instead of one \CamKt $R=1.5$ jet.
This technique is optimized to avoid energy overlap when different types of jets and jet radius parameters
are used to reconstruct the full event final state.
The advantage of this technique compared to a separation requirement applied to
the \CamKt $R=1.5$ jet is studied for simulated events with charged-Higgs-boson decays.

The performance of the
various top-tagging techniques is compared using simulation by matching the
different reconstructed jets to trimmed \akt $R=1.0$ jets formed at the particle level.
The reciprocal of the mistag rate, the background
rejection, is studied as a function of the
efficiency in intervals of the particle-level jet transverse momentum, \pttrue,
ranging from $350$ to $1500\gev$, while
the efficiency and rejection of the \htt is also studied for $200<\pttrue<350\gev$.
For $350<\pttrue<1000\GeV$, SD offers the best rejection up to its maximum
achievable efficiency.
Top-tagging efficiencies above 70\% can be achieved with cuts on substructure variables,
for example, yielding rejections of approximately $3$--$6$ for an efficiency of 80\%.
A rejection of $\approx\!15$--$20$ at an efficiency of $\approx\!50\%$ can be achieved with the
\WPT over the range $450<\pttrue<1000\gev$.
For $1000<\pttrue<1500\gev$, of all the top-tagging methods studied, the
\htt offers the best rejection ($\approx\!30$) at an efficiency of $\approx\!40\%$.

An event sample enriched in top-quark pairs is used to study the
distributions of substructure variables.
Simulations of Standard Model processes describe the relevant distributions well for the
six substructure-variable taggers, SD, \htt and \httofour
within the uncertainties.
The uncertainty in the energy scale of the subjets used by the \htt
is derived by comparing the mass of the top-quark candidate reconstructed
in data and simulation.
The relative subjet \pt uncertainty varies between 1\% and 10\%,
depending on \pt and the functional form chosen to describe the \pt
dependence.

The sample enriched in top-quark pairs is used to measure the efficiency to tag jets containing
a hadronic top-quark decay.
The efficiency is determined for jet \pt between $200$ and $700\gev$ for the
\CamKt $R=1.5$ jets and for $350$--$600\gev$ for the trimmed \akt $R=1.0$ jets.
The reach in \pt is limited by statistical and systematic uncertainties, which
become large at high \pt.
Jets not originating from hadronic top-quark decays are subtracted using simulation and
the subtraction leads to systematic uncertainties in the measured efficiency.
Integrated over the measured \pt range, the relative systematic uncertainty of the efficiency
varies between $\approx\!10$\% and $\approx\!20$\% for the different substructure-variable-based taggers, and
is $\approx\!20$\% for SD and $\approx\!10$\% for the \htt. The dominant source of uncertainty
is the modelling of \ttbar events, and increases with \largeR jet \pt.
The quoted \pt-integrated uncertainties are smaller for the \htt efficiency,
because the measurement extends to smaller \largeR jet \pt.
Simulated events generated with \PowhegPythia,
with the \hdamp parameter set to infinity and the \ttbar and top-quark \pt spectra sequentially reweighted
to describe the \ttbar cross section measured at $7\TeV$, describe the efficiency within the uncertainties of the measurement.

A sample enriched in multijet events is used to measure the
mistag rate of the algorithms.
The misidentification rate increases with the \pt of the \largeR jet
and, in the range of \pt studied,
reaches values of $6$--$36\%$ for the different substructure-variable
taggers, $\approx\!4\%$ for SD, and $\approx\!3\%$ for the
\htt.
The measured mistag rate is well described by simulations using \Pythia
within the modelling uncertainties and the statistical uncertainties of the
measurement.

For top-tagging analyses with a low background level, e.g.\ \ttbar
resonance searches at top quark $\pt>700\GeV$ in the final state with one charged lepton, it is
recommended to use a top tagger with high efficiency, such as the
substructure-variable-based taggers I--IV studied in this paper.
If high rejection is required, e.g.\ for an all-hadronic final state,
then for $\pt>1000\GeV$, one of the following taggers is likely
to give the best sensitivity, depending on the details of the analysis:
the \WPT, the \htt, or SD. For \pt between $450$ and $1000\GeV$, SD
is the tagger of choice if high rejection is required.
Only the performance of the \htt has been studied for \pt down to $200\GeV$.
In final states with high jet multiplicity where the full event needs to be
reconstructed, the \httofour method is a useful approach to avoid
energy sharing between \smallR and \largeR jets.

In analyses, the uncertainty in the top-tagging efficiency for Standard-Model and
beyond-the-Standard-Model predictions comprises detector-related uncertainties
and theoretical modelling uncertainties.
The background in analyses should be determined by employing data-driven
methods, as it was done for the ATLAS Run~1 analyses because the
mistag rate was observed to depend strongly on the choice of
trigger, and small deficiencies in the trigger simulation can have a
large impact on the analysis.

The energy scale of the \htt subjets should be determined using the in situ method
pioneered in this paper. This method takes into account all subjets used by
the \htt, even those with radius parameter $R<0.2$, for which the
MC-based calibrations determined for $R=0.2$ are used.

It is demonstrated in this paper that the substructure of top jets shows
the expected features and that it is well modelled by simulations.
Top tagging has been used in LHC Run~1 analyses
and its importance will increase in Run~2 with more top quarks produced
with high transverse momentum due to the higher centre-of-mass energy.

\section*{Acknowledgements}


We thank CERN for the very successful operation of the LHC, as well as the
support staff from our institutions without whom ATLAS could not be
operated efficiently.

We acknowledge the support of ANPCyT, Argentina; YerPhI, Armenia; ARC, Australia; BMWFW and FWF, Austria; ANAS, Azerbaijan; SSTC, Belarus; CNPq and FAPESP, Brazil; NSERC, NRC and CFI, Canada; CERN; CONICYT, Chile; CAS, MOST and NSFC, China; COLCIENCIAS, Colombia; MSMT CR, MPO CR and VSC CR, Czech Republic; DNRF and DNSRC, Denmark; IN2P3-CNRS, CEA-DSM/IRFU, France; GNSF, Georgia; BMBF, HGF, and MPG, Germany; GSRT, Greece; RGC, Hong Kong SAR, China; ISF, I-CORE and Benoziyo Center, Israel; INFN, Italy; MEXT and JSPS, Japan; CNRST, Morocco; FOM and NWO, Netherlands; RCN, Norway; MNiSW and NCN, Poland; FCT, Portugal; MNE/IFA, Romania; MES of Russia and NRC KI, Russian Federation; JINR; MESTD, Serbia; MSSR, Slovakia; ARRS and MIZ\v{S}, Slovenia; DST/NRF, South Africa; MINECO, Spain; SRC and Wallenberg Foundation, Sweden; SERI, SNSF and Cantons of Bern and Geneva, Switzerland; MOST, Taiwan; TAEK, Turkey; STFC, United Kingdom; DOE and NSF, United States of America. In addition, individual groups and members have received support from BCKDF, the Canada Council, CANARIE, CRC, Compute Canada, FQRNT, and the Ontario Innovation Trust, Canada; EPLANET, ERC, FP7, Horizon 2020 and Marie Sk{\l}odowska-Curie Actions, European Union; Investissements d'Avenir Labex and Idex, ANR, R{\'e}gion Auvergne and Fondation Partager le Savoir, France; DFG and AvH Foundation, Germany; Herakleitos, Thales and Aristeia programmes co-financed by EU-ESF and the Greek NSRF; BSF, GIF and Minerva, Israel; BRF, Norway; Generalitat de Catalunya, Generalitat Valenciana, Spain; the Royal Society and Leverhulme Trust, United Kingdom.

The crucial computing support from all WLCG partners is acknowledged
gratefully, in particular from CERN and the ATLAS Tier-1 facilities at
TRIUMF (Canada), NDGF (Denmark, Norway, Sweden), CC-IN2P3 (France),
KIT/GridKA (Germany), INFN-CNAF (Italy), NL-T1 (Netherlands), PIC (Spain),
ASGC (Taiwan), RAL (UK) and BNL (USA) and in the Tier-2 facilities
worldwide.

\clearpage
\appendix
\part*{Appendix}
\addcontentsline{toc}{part}{Appendix}

\section{Additional distributions for the signal-sample selection}
\label{app:incldistrib}
In this appendix, additional event-level distributions after the signal-sample
selections (\secref{signalsample}) are shown, which complement
\figsref{ctrl_akt_pretag_ept_fj}{ctrl_HTT_pretag_mtw_fj}.

Distributions for the signal selection with
at least one trimmed \akt $R=1.0$ jet with $\pt>350\GeV$
are shown in \figref{app1}.
The lepton transverse momentum (\figref{app1_elpt})
exhibits a falling spectrum for $\pt>50\GeV$. The reduced number of entries
in the bin from $25$ to $45\GeV$ is due to
the fact that the combination of the lepton triggers is not fully efficient below $50\GeV$.
The distribution is well described by simulations of SM processes within
the uncertainties.
The distribution of the distance $\DeltaR$ between the highest-\pt trimmed \akt $R=1.0$ jet and the
highest-\pt $b$-jet within $\Delta R = 1.5$ of the lepton is presented in \figref{app1_dRfjb} and shows that
the \largeR jet and the $b$-jet are well separated.

The dominant systematic uncertainties in \figref{app1}
result from uncertainties in the \largeR jet energy scale, the PDF, and the
\ttbar generator.
The contributions from these sources are approximately equal in size and they
affect mostly the normalization of the distributions.

\begin{figure}[!h]
\begin{centering}
\subfigure[]{
\label{fig:app1_elpt}
\includegraphics[width=0.48\textwidth]{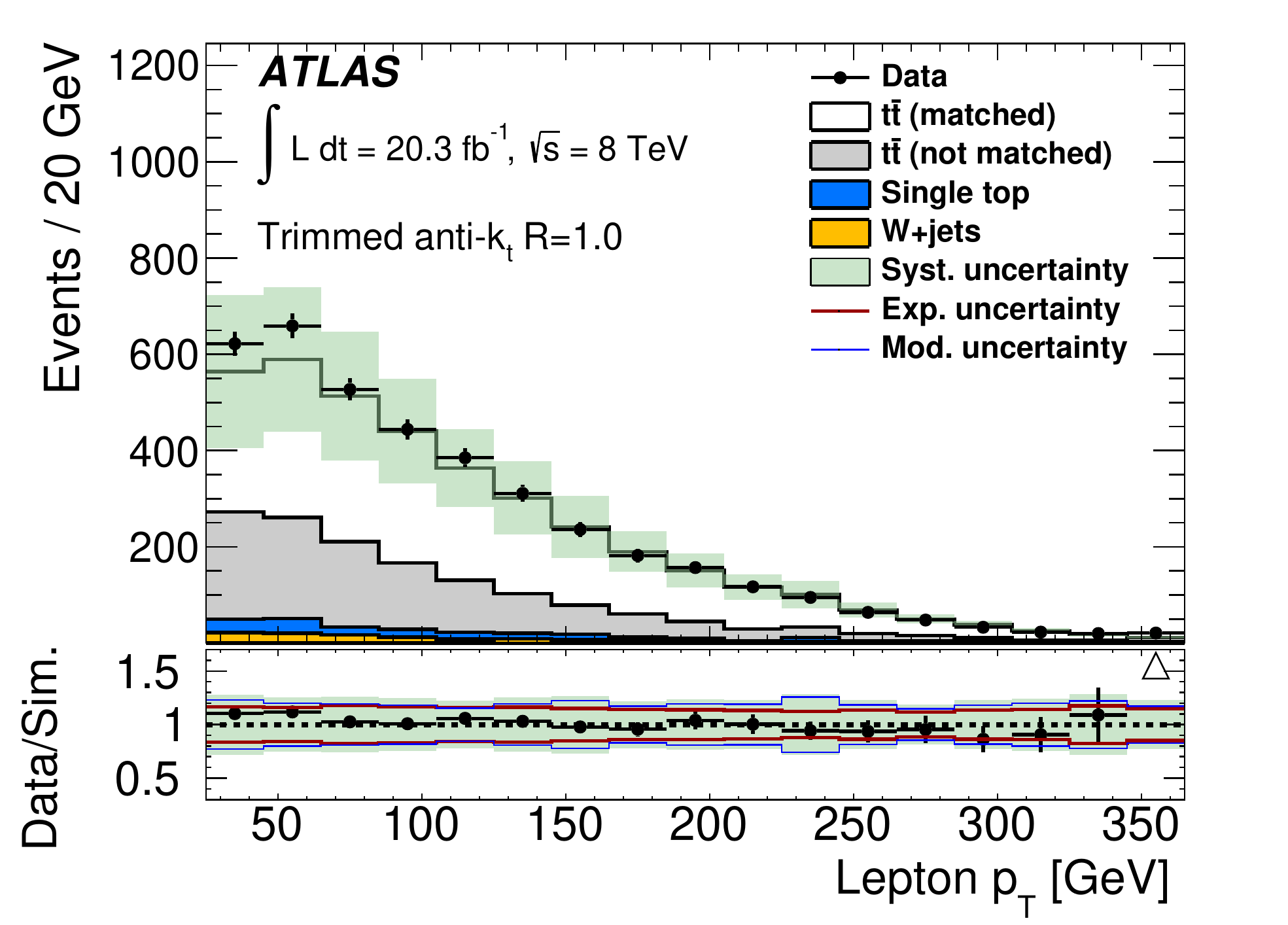}
}
\subfigure[]{
\label{fig:app1_dRfjb}
\includegraphics[width=0.48\textwidth]{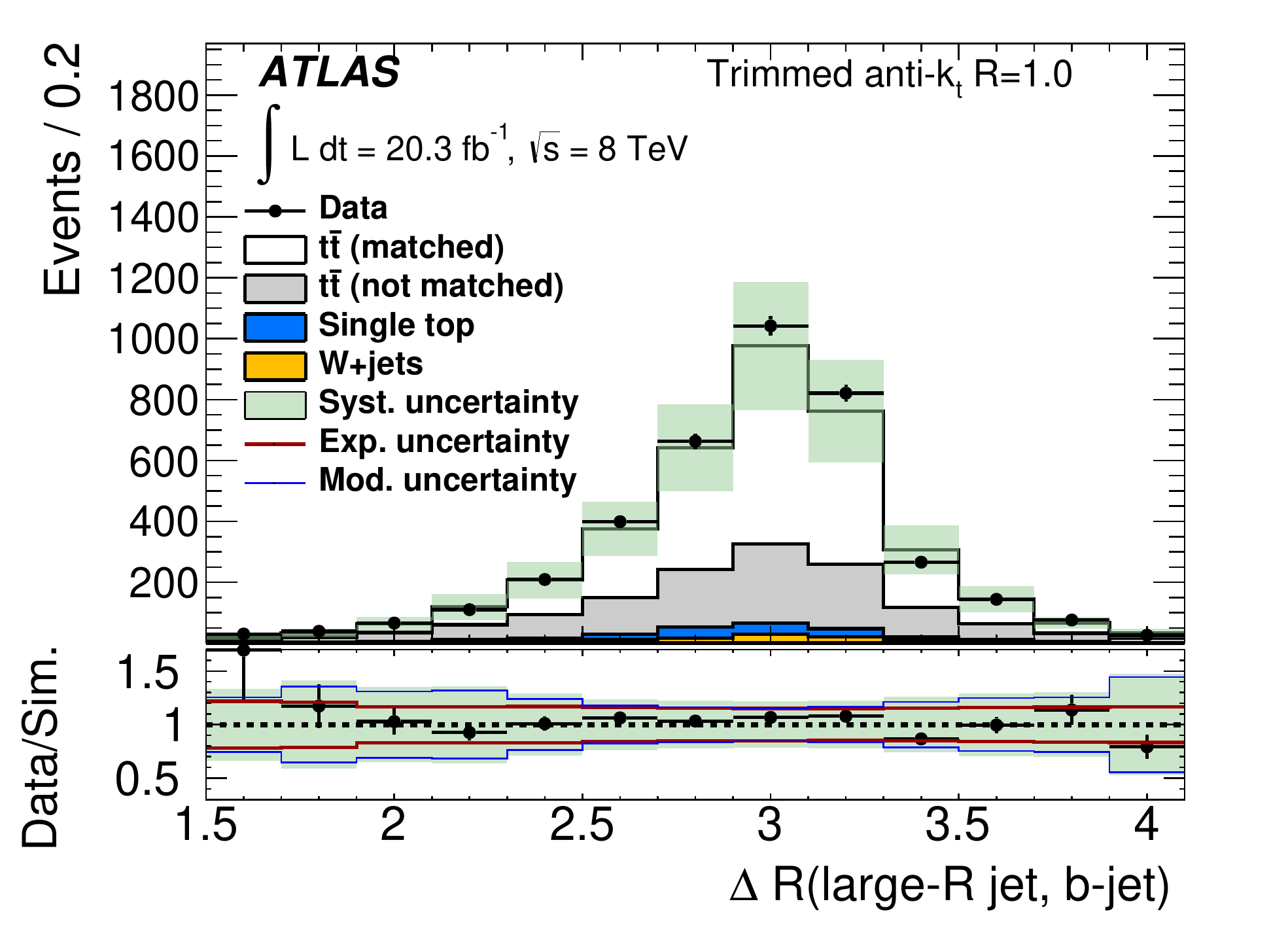}
}
\caption{Detector-level distributions of variables reconstructed in events
passing the signal-sample selection (\ttbar) with
at least one trimmed \akt $R=1.0$ jet with $\pt>350\GeV$.
(a) The transverse momentum of the charged lepton and (b) the
distance in $(\eta, \phi)$ between the highest-\pt $b$-jet within $\Delta R = 1.5$ of the lepton and the highest-\pt trimmed
\akt $R=1.0$ jet.
The vertical error bar indicates the statistical uncertainty
of the measurement. Also shown are distributions for simulated SM contributions
with systematic uncertainties (described in \secref{systematics}) indicated as a band.
The \ttbar prediction is split into a {\em matched} part for which the \largeR
jet axis is within $\Delta R = 0.75$ of the flight direction of a hadronically decaying
top quark and a {\em not matched} part for which this criterion does not hold.
The ratio of measurement to
prediction is shown at the bottom of each subfigure and the error bar and band
give the statistical and systematic uncertainties of the ratio, respectively.
The impacts of experimental and \ttbar modelling uncertainties are
shown separately for the ratio.
}
\label{fig:app1}
\end{centering}
\end{figure}

Distributions for events fulfilling the signal selection with
at least one \CamKt $R=1.5$ jet with $\pt>200\GeV$, as used in the \htt studies,
are shown in \figref{app2}.
The distribution of the transverse mass \mtw is shown in \figref{app2_mtw}.
It exhibits a peak near the $W$-boson mass, which is expected if the reconstructed
charged lepton and the \met correspond to the charged lepton and neutrino from
the \W decay and the momenta of the two particles lie in the transverse plane.
The missing-transverse-momentum distribution (\figref{app2_met}) displays
a peak around $55\GeV$ and a smoothly falling spectrum for larger values.

All distributions are described by the
simulation within the uncertainties.
Important sources of systematic uncertainty for the \mtw and \met distributions are the \largeR JES, the
$b$-tagging efficiency, the prediction of the \ttbar cross section, and \ttbar modelling uncertainties
from the choice of generator, parton shower, and PDF set. None of these uncertainties dominates.

\begin{figure}[!h]
\begin{centering}
\subfigure[]{
\label{fig:app2_mtw}
\includegraphics[width=0.48\textwidth]{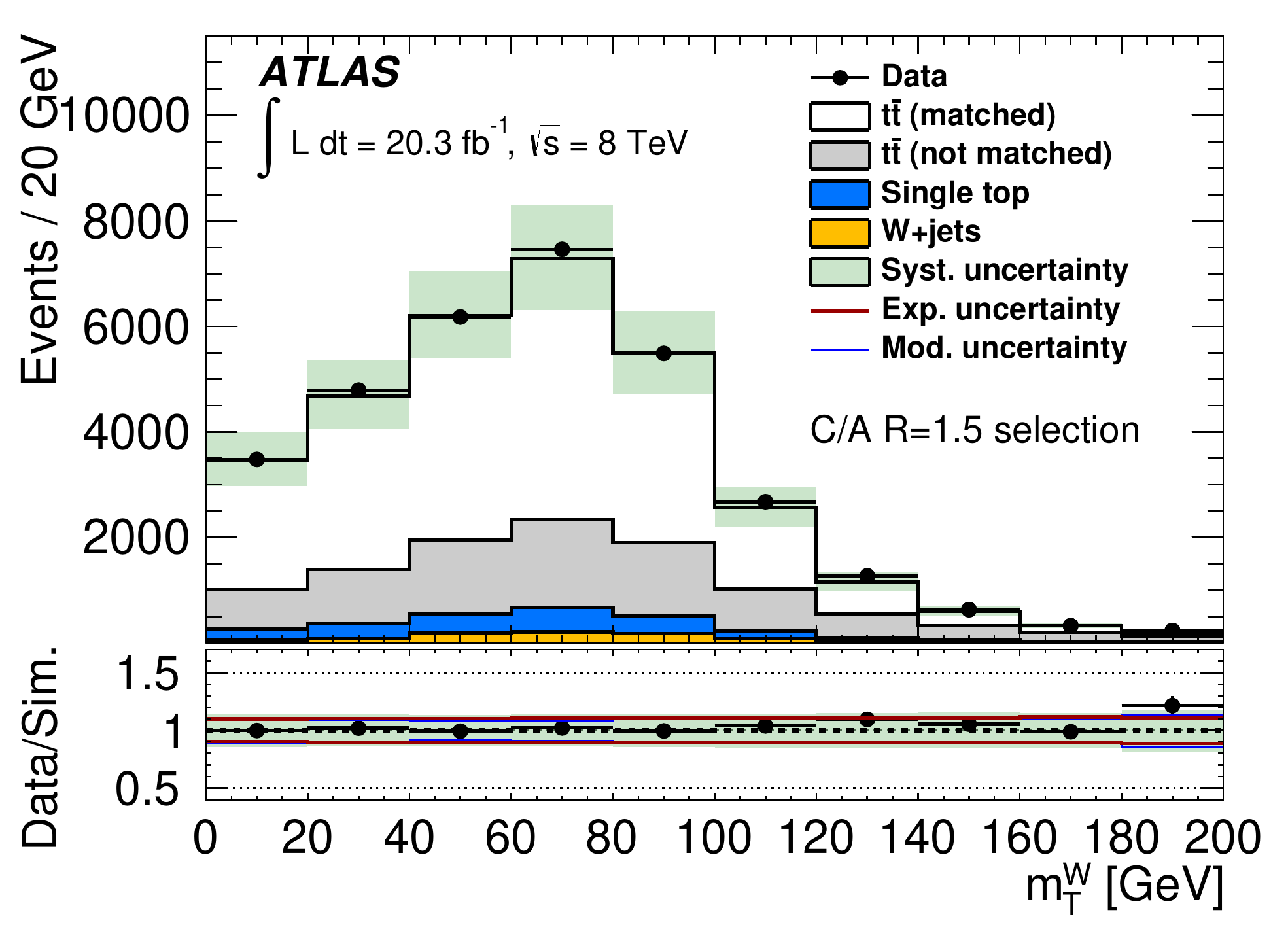}
}
\subfigure[]{
\label{fig:app2_met}
\includegraphics[width=0.48\textwidth]{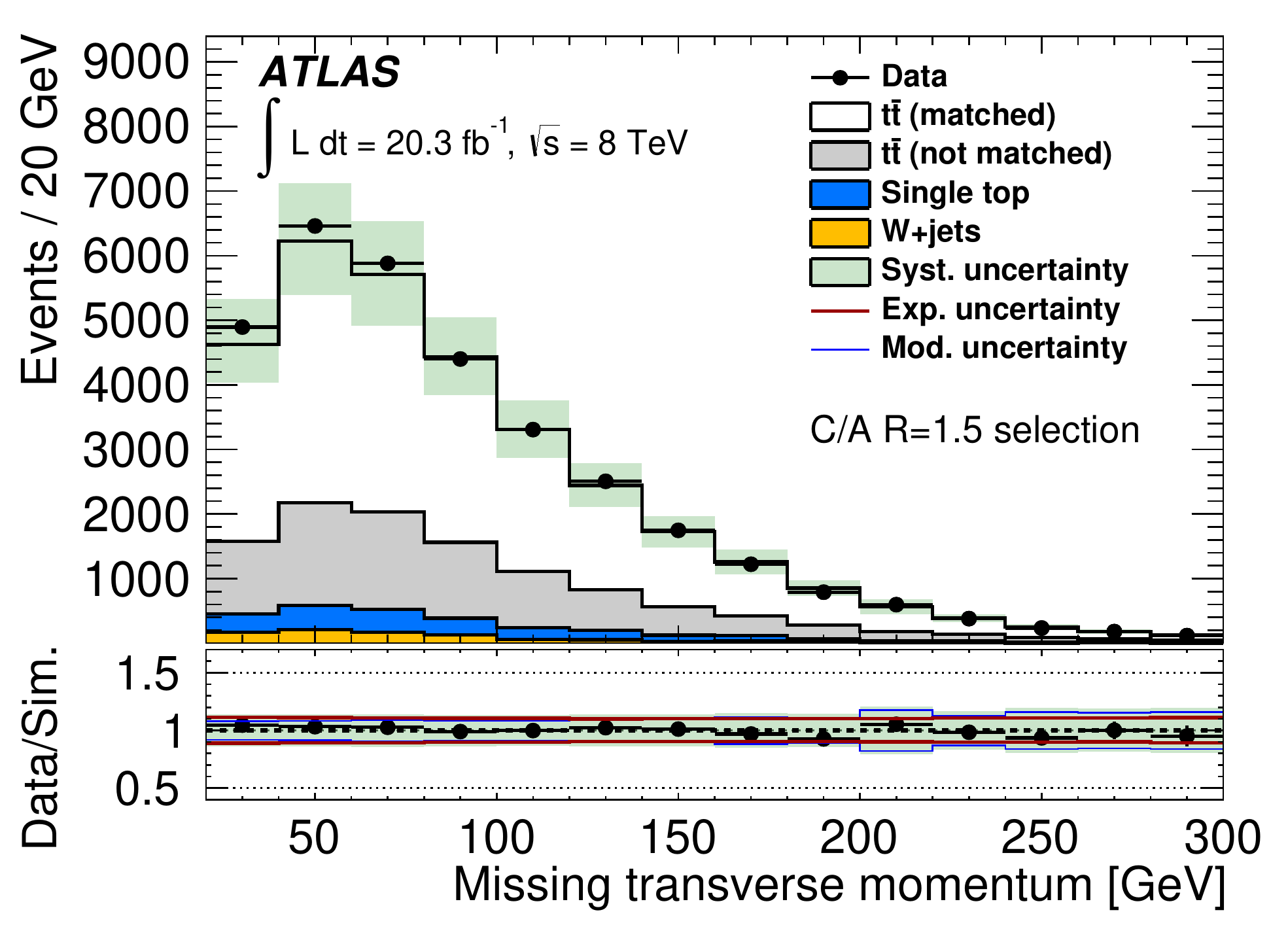}
}
\caption{Detector-level distributions of (a) the transverse mass \mtw and (b) the missing
transverse momentum
\met for events passing the signal selection with
at least one \CamKt $R=1.5$ jet with $\pt>200\GeV$.
The vertical error bar indicates the statistical uncertainty
of the measurement. Also shown are distributions for simulated SM contributions
with systematic uncertainties (described in \secref{systematics}) indicated as a band.
The \ttbar prediction is split into a {\em matched} part for which the \largeR
jet axis is within $\Delta R = 1.0$ of the flight direction of a hadronically decaying
top quark and a {\em not matched} part for which this criterion does not hold.
The ratio of measurement to
prediction is shown at the bottom of each subfigure and the error bar and band
give the statistical and systematic uncertainties of the ratio, respectively.
The impacts of experimental and \ttbar modelling uncertainties are
shown separately for the ratio.
}
\label{fig:app2}
\end{centering}
\end{figure}

\FloatBarrier

\printbibliography

\newpage 
\begin{flushleft}
{\Large The ATLAS Collaboration}

\bigskip

G.~Aad$^\textrm{\scriptsize 86}$,
B.~Abbott$^\textrm{\scriptsize 114}$,
J.~Abdallah$^\textrm{\scriptsize 152}$,
O.~Abdinov$^\textrm{\scriptsize 11}$,
R.~Aben$^\textrm{\scriptsize 108}$,
M.~Abolins$^\textrm{\scriptsize 91}$,
O.S.~AbouZeid$^\textrm{\scriptsize 159}$,
H.~Abramowicz$^\textrm{\scriptsize 154}$,
H.~Abreu$^\textrm{\scriptsize 153}$,
R.~Abreu$^\textrm{\scriptsize 117}$
% ...
\bigskip
\\
% ...
$^{11}$ Institute of Physics, Azerbaijan Academy of Sciences, Baku, Azerbaijan\\
% ...
$^{86}$ CPPM, Aix-Marseille Universit{\'e} and CNRS/IN2P3, Marseille, France\\
% ...
$^{91}$ Department of Physics and Astronomy, Michigan State University, East Lansing MI, United States of America\\
% ...
$^{108}$ Nikhef National Institute for Subatomic Physics and University of Amsterdam, Amsterdam, Netherlands\\
% ...
$^{114}$ Homer L. Dodge Department of Physics and Astronomy, University of Oklahoma, Norman OK, United States of America\\
% ...
$^{117}$ Center for High Energy Physics, University of Oregon, Eugene OR, United States of America\\
% ...
$^{152}$ Institute of Physics, Academia Sinica, Taipei, Taiwan\\
$^{153}$ Department of Physics, Technion: Israel Institute of Technology, Haifa, Israel\\
$^{154}$ Raymond and Beverly Sackler School of Physics and Astronomy, Tel Aviv University, Tel Aviv, Israel\\
% ...
$^{159}$ Department of Physics, University of Toronto, Toronto ON, Canada\\
% ...
\end{flushleft}


\end{document}